\documentclass[aip]{revtex4-2}

\usepackage{graphicx}
\usepackage[dvipsnames]{xcolor}
\usepackage{subcaption} 
\usepackage{bm}
\usepackage{natbib}
\usepackage{appendix}
\usepackage{amsbsy}
\usepackage{psfrag}
\usepackage{amsmath,amssymb}
\usepackage{tikz}
\usepackage{cases}
\usepackage{soul}
\usepackage{comment}
\usepackage{enumitem}
\usepackage{verbatim}
\usepackage{fdsymbol}
\usepackage{multirow}
\usepackage{booktabs}
\usepackage[colorlinks=true,
            linkcolor=Maroon,
            urlcolor=blue,
            citecolor=blue,
            hypertexnames=false]{hyperref}
\usepackage{diagbox}
\usepackage{caption}

\begin{document}

\title{Effects of the Coriolis force on the coherent structures in conventionally neutral atmospheric boundary layers}

\author{Changlong Wang}
\affiliation{Center for Particle-Laden Turbulence, Key Laboratory of Mechanics on Disaster and Environment in Western China, Ministry of Education, and College of Civil Engineering and Mechanics, Lanzhou University, Lanzhou 730000, Gansu, China}

\author{Luoqin Liu}
\affiliation{Department of Modern Mechanics, University of Science and Technology of China, Hefei 230027, Anhui, China}

\author{Xiang I. A. Yang}
\affiliation{Department of Mechanical Engineering, Pennsylvania State University, State College, PA 16802, USA}

\author{Ruifeng Hu}
\email{hurf@lzu.edu.cn}
\affiliation{Center for Particle-Laden Turbulence, Key Laboratory of Mechanics on Disaster and Environment in Western China, Ministry of Education, and College of Civil Engineering and Mechanics, Lanzhou University, Lanzhou 730000, Gansu, China}

\date{\today}

\begin{abstract}
It is well known that the Coriolis force due to Earth's rotation can induce wind veer in the mean flow velocity of an atmospheric boundary layer (ABL), but much less is known about its effects on turbulent coherent structures. 
In this work, large-eddy simulation (LES) is employed to investigate the effects of the Coriolis force on the characteristics of turbulent coherent structures in the conventionally neutral atmospheric boundary layers (CNBL). Variation of the Coriolis force is realized by changing latitude or geostrophic wind speed.
We found that the Coriolis force causes distinct deflection of coherent velocity structures to the geostrophic wind direction, which is not aligned with the direction of either the mean wind or the mean shear. 
When a relative structure deflection angle is plotted against a relative local wind veer angle, a universal relationship between them can be established regardless of latitude or geostrophic wind speed. Moreover, we also studied the effect of the Coriolis force on the inclination of large-scale turbulent structures. It is found that the inclination angle in the surface layer varies under different conditions, which is probably due to the deflection of turbulent structures caused by the Coriolis force.
\end{abstract}

\maketitle

\newpage
\section{\label{sec:level1}Introduction}

The atmospheric boundary layer (ABL) encompasses the lowest portion of a planet's atmosphere, playing a crucial role in wind power harvesting and the transport of heat and particulate matter \citep{moninAtmosphericBoundaryLayer1970,wyngaardAtmosphericTurbulence1992,garratt1994atmospheric,mahrtStablyStratifiedAtmospheric2014,zhangLargescaleCoherentStructures2018a,veersGrandChallengesScience2019,liuLargescaleStructuresWallbounded2021,xiao2023long}. 
The flow dynamics of the ABL are prominently affected by thermal stratification and the Coriolis force due to the planet's rotation \citep{,wyngaard2010turbulence,stull2012introduction}. In the context of thermal stratification, the ABL can be classified as neutrally, convectively, or stably stratified \citep{moninAtmosphericBoundaryLayer1970,wyngaard2010turbulence,stull2012introduction}. The celebrated Monin-Obukhov similarity theory (MOST) provides a unified theoretical framework for the mean flow quantities in the lower part of the ABL (atmospheric surface layer, ASL) \citep{moninBasicLawsTurbulent,foken50YearsMonin2006,liOKEYPSEquation602021,stiperskiGeneralizingMoninObukhovSimilarity2023}.

When the heat flux on the ground surface is approximately negligible and the flow is capped by an inversion of potential temperature against a stable background stratification in the free atmosphere, the ABL is considered conventionally neutral \citep{csanadyEquilibriumTheoryPlanetary1974,zilitinkevich2002integral}.
The conventional neutral boundary layer (CNBL) is commonly observed over the sea, large lakes, or land during the transition period after sunset or on cloudy days with strong winds \citep{allaerts2017boundary}.
Owing to its physical and practical importance, the CNBL has been continually studied in the literature. One central effort is developing analytical expressions for the mean wind profile and global quantities, such as the geostrophic drag law. This issue has a long history \citep{blackadar1962vertical,taylorPlanetaryBoundaryLayer1969,csanadyEquilibriumTheoryPlanetary1974,nieuwstadtSolutionStationaryBaroclinic1983,zilitinkevichVelocityProfilesResistance1989,byunDeterminationSimilarityFunctions1991,zilitinkevichSimilaritytheoryModelWind1998,hessEvaluatingModelsNeutral2002,hessEvaluatingModelsNeutral2002a,zilitinkevich2002integral,esauParameterizationSurfaceDrag2004,zilitinkevich2005resistance,gryning2007extension,kelly2010long} and was also remarkably advanced in recent studies \citep{kelly2019universal,liuUniversalWindProfile2021,liuGeostrophicDragLaw2021,liu2022vertical,liuGeostrophicDragLaw2024,narasimhan2024analytical}.

The Earth's rotation causes veering of the mean wind velocity in the wall-normal direction, as the well-known Ekman spiral \citep{ekman1905influence,ellison1955ekman}, which has been thoroughly investigated in the aforementioned studies. 
{Moreover, the wake of a wind turbine or wind farm in ABL can also be profoundly affected by the Coriolis force \citep{luLargeeddySimulationVery2011,abkarInfluenceCoriolisForce2016,narasimhan2022effects}.} 
However, the effects of the Coriolis force due to Earth's rotation on the coherent structures of the CNBL are much less known and not fully determined.
{The coherent structure of a turbulent flow is usually believed to be a three-dimensional region of the flow over which at least one fundamental flow quantity exhibits significant correlation with itself or another quantity over a range of space and/or time that is significantly larger than the smallest scales of the flow \citep{robinson1991coherent,Jimenez2018coherent,massaroInterpretingCoherenceWall2026}. {There are certainly different views on the definition of coherent turbulent structures. For instance, Hussain \citep{hussainCoherentStructuresReality1983,hussainCoherentStructuresTurbulence1986} suggested a coherent structure as a connected turbulent fluid mass with instantaneously phase-correlated vorticity over its spatial extent.
In this study, we primarily focus on coherent structures defined by flow velocities, such as streaks, because they directly contribute to turbulent kinetic energy.}
Observations of velocity structures in the CNBL have been obtained using large-eddy simulations (LES). 
{Deardorff \citep{deardorffThreedimensionalNumericalInvestigation1970,deardorff1972numerical} performed pioneering LES studies of the ABL and found that the orientation of elongated longitudinal velocity structures is closer to the mean shear vector than to the mean wind direction.} 
In contrast, Moeng and Sullivan \citep{moengComparisonShearBuoyancyDriven1994} proposed that the structure orientation is aligned with the mean flow velocity direction. 
Furthermore, Zikanov \emph{et al.} \citep{zikanov2003large} found that the deflection direction of turbulent structures is close to that of the mean shear in their LES study of turbulent Ekman flow, in agreement with Deardorff.
{Shingai and Kawamura \citep{shingaiStudyTurbulenceStructure2004a} quantified the deflection of turbulent structures in Ekman flows and suggested that it is along the decreasing direction of the mean Exner function in the near-wall region.}
The deflection of turbulent structures in the CNBL or the Ekman boundary layer has also been observed in other studies \citep{masonLargeEddySimulationsNeutralstaticstability1987,linCoherentStructuresDynamics1996a,sullivanSubgridscaleModelLargeeddy1994,khannaThreeDimensionalBuoyancyShearInduced1998,deusebioNumericalStudyUnstratified2014,bergLargeEddySimulationConditionally2020}.
As mentioned above, although numerous studies have reported deflected coherent turbulent structures in the CNBL or the Ekman boundary layer, there is a significant lack of its quantification and the dependence on the Coriolis force (varying latitude or geostrophic wind speed), as well as the precise relationship with the directions of the mean wind and mean shear. 

Another distinct feature of the large-scale turbulent structures in high-Reynolds-number wall-bounded turbulent flows is the shallow inclination to the horizontal plane \citep{brownLargeStructureTurbulent1977a,boppeLargeScaleMotionsMarine1999,carper2004role,marusic2007reynolds,chauhan2013structure,liu2017variation,saleskyBuoyancyEffectsLargescale2018,saleskyRevisitingInclinationLargescale2020,liScaledependentInclinationAngle2022,gibbsInclinationAnglesTurbulent2023,huangTheoreticalModelStructure2023}. 
The inclination angle in neutral flows is well within 10$^\circ$ to 20$^\circ$ and believed to be a signature of hairpin vortex packets \citep{christensenStatisticalEvidenceHairpin2001}. However, the situation may be much more complex in ABL flows due to thermal stratification and the Coriolis force. For the latter, Marusic and Heuer \citep{marusic2007reynolds} discovered a Reynolds-number-invariant property of the inclination angle in a neutral ASL flow compared to laboratory measurements.
However, Liu \emph{et al.} \citep{liu2017variation} reported a dependence of the inclination angle on the wind friction velocity based on neutral ASL measurements. {This discrepancy may be attributed to the effect of the Coriolis force, which can be characterized by the Rossby number, while it has not been fully revealed yet.}

According to the aforementioned knowledge gap and controversy in the literature, the present study aims to examine the effects of the Coriolis force due to Earth's rotation on the characteristics of coherent turbulent structures in the CNBL. 
High-fidelity LES is employed as it has become a state-of-the-art tool in numerical simulations of ABL \citep{stollLargeEddySimulationAtmospheric2020b}. The variation of the Coriolis force is realized by changing latitude or geostrophic wind speed. We focus on the basic statistics, instantaneous flow, the scales, deflections, and inclinations of the coherent turbulent structures in the CNBL and compare them to the truly neutral boundary layer (TNBL), in which the Coriolis force is neglected \citep{narasimhan2022effects}. It is noted that sometimes TNBL refers to the unstratified turbulent Ekman boundary layer, in which the Coriolis force is not ignored \citep{zilitinkevich2002integral}.

The paper is organized as follows.
In \S \ref{level:2}, the computational models for the LES of CNBL, as well as the simulation settings and validations, are described. 
In \S \ref{sec3}, we present the simulation results and analyze the effects of the Coriolis force by varying latitude and geostrophic wind speed. 
The conclusions are drawn in \S \ref{level:5}.

\section{\label{level:2}COMPUTATIONAL MODELS}

\subsection{Governing equations and numerical method}
We adopt the open-source LES code LESGO
\href{https://github.com/lesgo-jhu/lesgo}{(https://github.com/lesgo-jhu/lesgo)}
to solve the filtered Navier-Stokes equations (with the Boussinesq approximation for buoyancy effects) and the scalar potential temperature transport equation:
\begin{equation}\label{eq 1}
    \frac{\partial \widetilde{u}_{i}}{\partial x_{i}}=0,
\end{equation}
\begin{equation}\label{eq 2}
    \frac{\partial \tilde{T}}{\partial t} + \widetilde{u}_j \frac{\partial \tilde{T}}{\partial x_j} = -\frac{\partial \Pi_j}{\partial x_j},
\end{equation}
\begin{equation}\label{eq 3}
        \frac{\partial \widetilde{u}_{i}}{\partial t}+\widetilde{u}_{j}\left(\frac{\partial \widetilde{u}_{i}}{\partial x_{j}}-\frac{\partial \widetilde{u}_{j}}{\partial x_{i}}\right)=-\frac{1}{\rho_{0}} \frac{\partial p_{\infty}}{\partial x_{i}}-\frac{\partial \widetilde{p}_*}{\partial x_{i}}-\frac{\partial \tau_{i j}}{\partial x_{j}}
        -f_{c} \widetilde{u} \delta_{i 2}+f_{c} \widetilde{v} \delta_{i 1}+\frac{g}{{T}_{0}}\left(\widetilde{T}-{T}_{0}\right) \delta_{i 3},
\end{equation}
where the tilde $(\widetilde{\cdot})$ represents the spatial filtering operation such that $\widetilde{u}_{i} = (\widetilde{u} ,\widetilde{v},\widetilde{w}$) are the filtered flow velocity components in the longitudinal ($x$), lateral ($y$) and vertical ($z$) directions, respectively, and $\widetilde{T}$ is the filtered potential temperature.
{The $p_\infty$ stands for the mean pressure at infinity. In addition, we impose the geostrophic wind at the top of the simulation domain to drive the flow, which finally balances the mean pressure gradient.}
The term ${\tau}_{ij} = {\sigma}_{ij} - (1/3){\sigma}_{kk}{\delta}_{ij}$ is the deviatoric part of the stress tensor ${\sigma_{ij}}=\widetilde{u_{i} u_{j}} -\widetilde{u_i}\widetilde{u_j} $ at the subgrid scale (SGS).
In Eq.~(\ref{eq 2}), the term $\Pi_j=\widetilde{u_j T}-\widetilde{u_j} \widetilde{T}$ is the SGS heat flux.
{The molecular viscosity and diffusivity terms are neglected due to very high Reynolds number of the ABL flows.}
The quantity $\widetilde{p}_*=\widetilde{p}/{\rho_0}+(1/3) {\sigma_{kk}}+(1/2)\widetilde{u}_k\widetilde{u}_k$ is the modified pressure, where the actual filtered pressure $\widetilde{p}$ is divided by the ambient fluid density ${\rho_0}$, plus the trace of the SGS stress tensor and the kinematic pressure generated by expressing the nonlinear convection term in rotational form. The $\delta_{ij}$ in Eq.~(\ref{eq 3}) is the Kronecker delta function that determines the directions of buoyancy and the Coriolis force. 
The Coriolis parameter is $f_c=2\Omega \sin\psi$, where $\Omega=7.292\times10^{-5}$ s$^{-1}$ is the rotational angular velocity of the Earth, and $\psi$ is latitude. It can be seen from Eq.~(\ref{eq 3}) that the traditional or the ``$f$-plane'' approximation is invoked as the horizontal components of the Earth's rotation are neglected, which is believed to be a good approximation at mid-to-high latitudes.
In the buoyancy term, $g=9.81$ m/s$^2$ is the gravitational acceleration, and $T_0$ is the reference potential temperature, taken as 288 K in the CNBL case. 

{When the horizontal mean pressure gradient and the Coriolis force in the CNBL are in balance, they obey the following geostrophic balance relationship:}
\begin{equation}\label{eq:geostr_bal}
    \begin{split}
      \frac{1}{\rho_0} \frac{\partial p_{\infty}}{\partial x} = f_c V_g, \quad 
    \frac{1}{\rho_0} \frac{\partial p_{\infty}}{\partial y} = -f_c U_g,
    \end{split}
\end{equation}
where the geostrophic velocity component is specified as $U_g=G\cos\theta$, $V_g=G\sin\theta$, where $G=\sqrt{{U_g}^2+{V_g}^2}$ is the magnitude of the geostrophic wind, and $\theta $ is the wind veer angle of the composite mean wind speed vector relative to the $x$ direction. 
{For the TNBL, we preserve almost the same settings as CNBL while just ignoring the Coriolis force terms in momentum Eq.~(\ref{eq 3}).}
 It should be noted again that the present definition of the TNBL is different from some studies in which the Coriolis force is retained \citep{zilitinkevich2002integral}.
{In this study, the $x$ and $y$ directions are denoted as the longitudinal and lateral directions, respectively. The streamwise direction is defined as the mean wind velocity direction, while the spanwise direction is perpendicular to the streamwise and forms a right-hand coordinate system. Thus, the streamwise and spanwise directions vary with height due to the Coriolis force in CNBL. In contrast, they remain unchanged with height and align with the $x$ and $y$ directions in TNBL, respectively.}

LESGO is a parallel high-fidelity LES code. The filtered Navier-Stokes equations in the high-Reynolds-number limit are numerically solved on a Cartesian grid. 
The pseudo-spectral scheme is adopted in the horizontal directions.
The second-order central difference scheme is used in the vertical direction.
The second-order Adams-Bashforth scheme is employed for time advancement.
The SGS stress and heat flux are modeled with the eddy-viscosity and eddy-diffusivity assumptions.
The Lagrangian scale-dependent dynamic model (LASD) \cite{Bou-Zeid2005} is adopted to model the SGS stress, which is quite accurate among several SGS models \citep{wang2020comparative}. 
The SGS heat flux is determined by assuming that the eddy diffusivity is proportional to the eddy viscosity \citep{narasimhan2022effects,narasimhan2024analytical}.
An equilibrium wall model is employed at the ground surface \citep{Moeng1984large,porte2000scale,Bou-Zeid2005,fengLargeEddySimulation2020,yang2022logarithmic}.

\subsection{Simulation setup and validation}

{In the present study, the computational domain size is 24 km $\times$ 3 km $\times$ 2 km = $24\delta$ $\times$ $3\delta$ $\times$ $2\delta$ in the $x$, $y$ and $z$ directions with $\delta=1$ km, respectively.}
{The length and width of the domain are more than 24 and 3 times the boundary-layer height, respectively, which is adequate to capture large-scale horizontal flow structures in all cases \citep{esauStructuringTurbulenceIts2013,Fang2015b,wang2020comparative}.}
In the $x$ and $y$ directions, periodic boundary conditions are used. The positive $z$ direction is opposite to the gravity vector, the positive $x$ direction is aligned with the direction of the geostrophic wind at the top of the domain, and the definition of the $y$ direction forms an orthogonal right-hand coordinate system. 

In the simulation, a capping inversion layer is set at 900-1000 m with a thickness of 100 m. In the capping inversion layer, the potential temperature rapidly increases with height, which helps to form stable stratification and inhibit vertical turbulent mixing. The reference potential temperature at the bottom of the inversion layer is set to 288 K, and that at the top of the inversion layer is 290.5 K.
Furthermore, a sponge (Rayleigh damping) layer is placed 500 m above the top of the free atmosphere. Within the designated sponge region, damping terms are added to the governing equations, relaxing flow variables toward a target state with increasing damping strength across the layer. It absorbs outgoing perturbations or waves to prevent unphysical reflections into the computational domain. This ensures that disturbances leaving the main domain are dissipated, preserving the integrity of turbulent structures in the boundary layer \citep{albertson1999surface,sescu2014control,stevens2014large}. The present generation method for the CNBL is the same as Narasimhan \emph{et al.} \citep{narasimhan2022effects}.

\begin{table}[!htb]
    \centering
    \caption{Summary of the simulation cases and parameters.}
    \label{table1}
    \begin{tabular}{cccccccc}
        \hline
        Case & Domain (km)& Grid & $Zi$& $Ro$ & $G$ (m/s) & $U_\tau$ (m/s) &  {$h$} (m)\\
        \hline
        V1 & 6$\times$3$\times$2 & 150$\times$76$\times$128 & 56.6 & 4.10 & 8 & 0.343 &  835.9 \\
        V2 & 6$\times$3$\times$2 & 300$\times$150$\times$256 & 56.6 & 4.83 & 8 & 0.360 &  746.1 \\
        V3 & 6$\times$3$\times$2 & 600$\times$300$\times$512 & 56.6 & 4.66 & 8 & 0.374  & 802.7 \\
        V4 & 3.75$\times$1.5$\times$2 & 360$\times$144$\times$432 & 56.6 & 4.20 & 15 & 0.332 & 791.0 \\
        N25 & 24$\times$3$\times$2 & 1200$\times$150$\times$256 & 94.7 & 6.89 & 8 & 0.343 &  808.6 \\
        N30 & 24$\times$3$\times$2 & 1200$\times$150$\times$256 & 80.0 & 6.08 & 8 & 0.345 &  783.3 \\
        N45 & 24$\times$3$\times$2 & 1200$\times$150$\times$256 & 56.6 & 4.37 & 8 & 0.347 &  769.5 \\
        N70 & 24$\times$3$\times$2 & 1200$\times$150$\times$256 & 42.6 & 3.75 & 8 & 0.367 &  714.8 \\
        N90 & 24$\times$3$\times$2 & 1200$\times$150$\times$256 & 40.0 & 3.60 & 8 & 0.368 &  699.2 \\
        S45 & 24$\times$3$\times$2 & 1200$\times$150$\times$256 & 56.6 & 4.43 & 8 & 0.345 &  769.7 \\
        G12 & 24$\times$3$\times$2 & 1200$\times$150$\times$256 & 56.6 & 5.49 & 12 & 0.485 &  855.5 \\
        G16 & 24$\times$3$\times$2 & 1200$\times$150$\times$256 & 56.6 & 6.97 & 16 & 0.699 &  972.7 \\
        {TNBL} & {24$\times$3$\times$2} & {1200$\times$150$\times$256} & {$\frac{0}{0}$ } & {$\infty$} & {8} & {0.369} &  {799.2} \\
        {HN45} & {24$\times$3$\times$2} & {1200$\times$150$\times$256} & {56.6} & 7.52 & {8} & {0.333} &  {429.7} \\
        \hline
    \end{tabular}
\end{table}

The key dimensionless parameters of the CNBL problem include the Rossby number $Ro$ and the Zilitinkevich number $Zi$ \citep{zilitinkevich2002integral,liuUniversalWindProfile2021,liu2022vertical}, namely $Ro=U_\tau/(f_c h)$ and $Zi$ = $N/|f_c|$, where $U_\tau$ is the friction velocity, {$z_0$ is the surface roughness height, which is set to $z_0=0.1$ m following \citet{narasimhan2022effects} and corresponds to the roughly-open type according to the Davenport roughness classification \citep{davenport1960rationale,wieringa1992updating},} $N=\sqrt{\Gamma g/\theta_0}$ is the free-atmosphere Brunt–V\"ais\"al\"a frequency, and $\Gamma=0.001$ K/m is the temperature increase lapse rate above the capping inversion layer. 

Table~\ref{table1} summarises the current simulation cases and their corresponding parameters. Cases V1 to V4 are validation cases. In particular, V1 to V3 are performed to examine grid convergence. Additionally, V4 is used to validate flow statistics against other LES studies. Cases N25 to S45 are designed to investigate the effects of the Coriolis force by varying latitude; `N' indicates a northern latitude, `S' indicates a southern latitude, and the following number represents the corresponding latitude. For example, `N45' denotes a northern latitude of 45$^\circ$, while `S45' denotes a southern latitude of 45$^\circ$. Cases `G12' and `G16' examine the effects of geostrophic wind speed, whereas `TNBL' refers to the truly neutral boundary layer without Coriolis force. In all simulation cases, the geostrophic wind angle to the $x$ direction is 0$^\circ$, which means $V_g=0$ m/s.
The boundary layer height {$h$} is determined as the height at which $5\%$ of the maximum value of turbulent shear stress is reached for the first time \cite{blackadar1962vertical,tennekes1972first}.
{To study the effect of the boundary-layer height, we also specifically include an HN45 case at a northern latitude of 45$^\circ$ for comparison purposes, whose inversion layer is within 400-500 m, \emph{i.e.} half of that in N45.}

\begin{figure}
\centering   
\centerline{\includegraphics[width=0.5\linewidth]{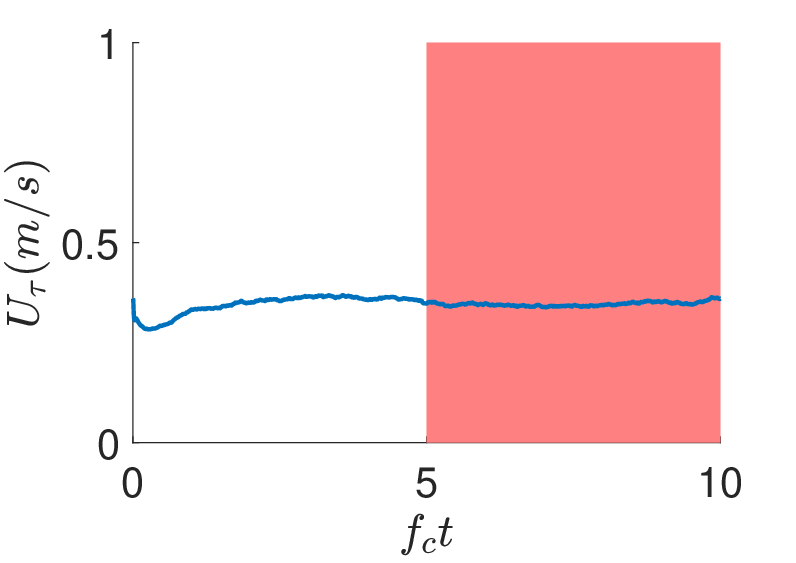}}
 \caption{Temporal evolution of the friction velocity $U_\tau$ with dimensionless time $f_c t$, and the shaded area represents the time-averaging window.}
 \label{fig1}
\end{figure}

\begin{figure}
    \centering
    \subfloat[\label{fig2a}]{
        \begin{tikzpicture}
        \node[anchor=north west, inner sep=0] (image) at (0,0) {    \includegraphics[width=0.33\textwidth]{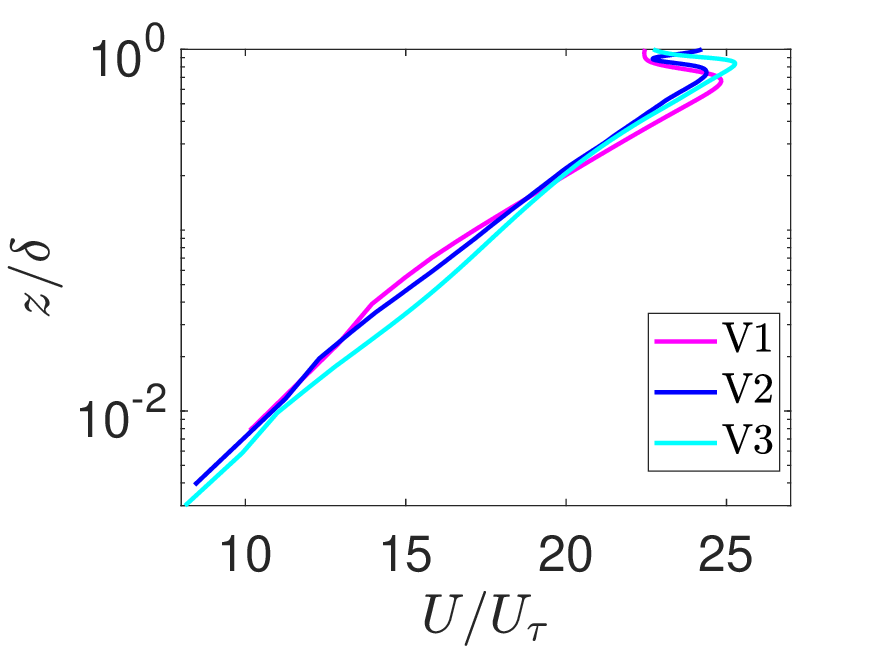}};
        \node[anchor=north west,
        xshift=-2mm,yshift=-2mm] at (image.north west) {{\rmfamily\fontsize{12}{13}\fontseries{l}\selectfont(a)}};
        \end{tikzpicture}}
     \subfloat[\label{fig2b}]{
        \begin{tikzpicture}
        \node[anchor=north west, inner sep=0] (image) at (0,0) {    \includegraphics[width=0.33\textwidth]{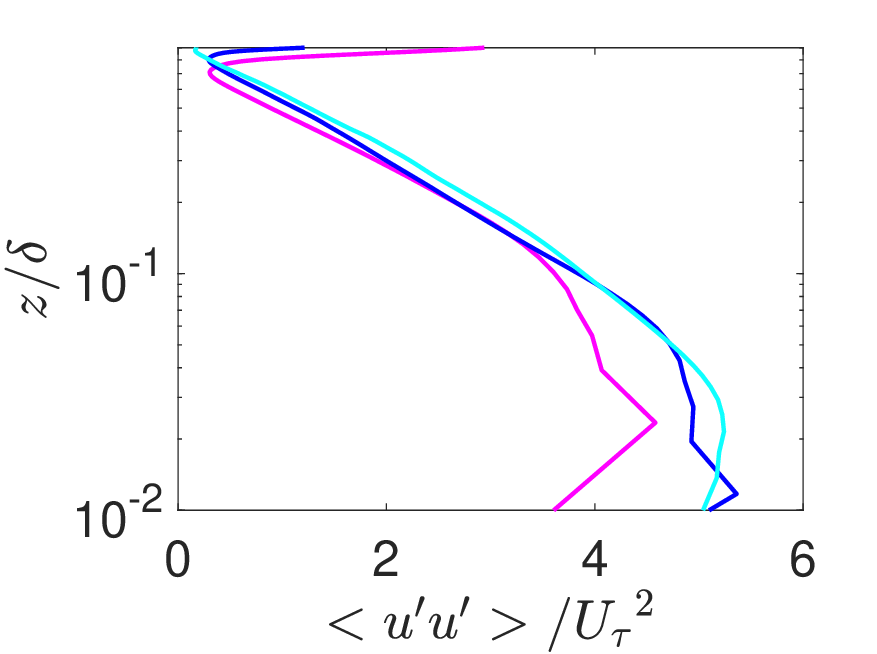}};
        \node[anchor=north west,
        xshift=-2mm,yshift=-2mm] at (image.north west) {{\rmfamily\fontsize{12}{13}\fontseries{l}\selectfont(b)}};
        \end{tikzpicture}}
     \subfloat[\label{fig2c}]{
        \begin{tikzpicture}
        \node[anchor=north west, inner sep=0] (image) at (0,0) {    \includegraphics[width=0.33\textwidth]{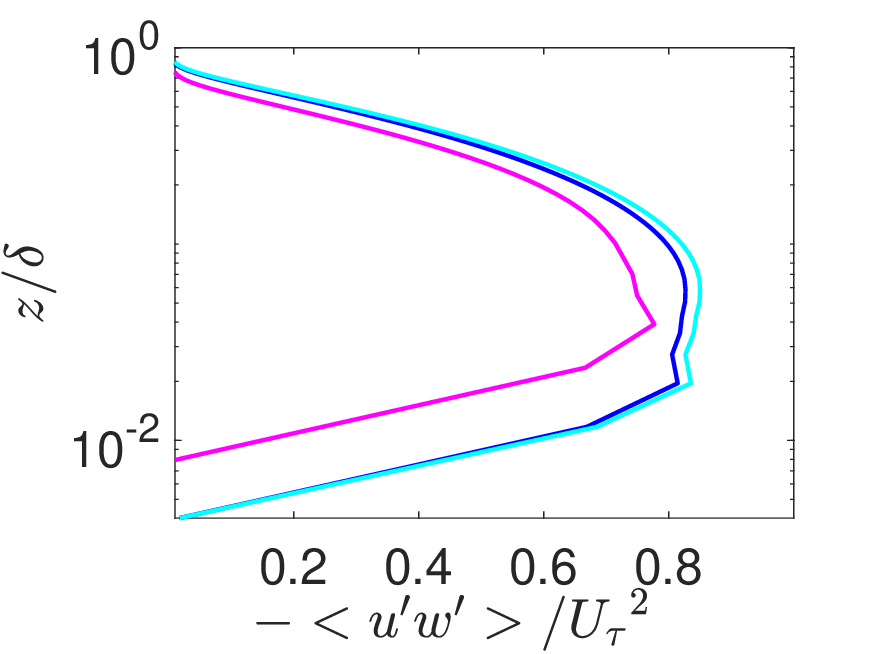}};
        \node[anchor=north west,
        xshift=-2mm,yshift=-2mm] at (image.north west) {{\rmfamily\fontsize{12}{13}\fontseries{l}\selectfont(c)}};
        \end{tikzpicture}}
    \caption{{Comparison of the CNBL statistics with different grids. (a) Mean longitudinal velocity, (b) Reynolds normal stress, and (c) Reynolds shear stress.}}
    \label{fig2}
\end{figure}
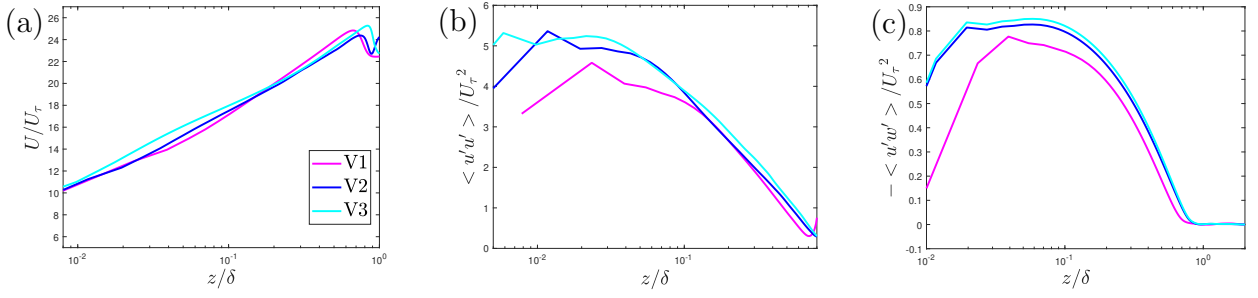

For collecting turbulence statistics, Stull \citep{stull2012introduction} proposed a time average period of $2\pi/f_c$ to ensure that statistics are not affected by inertial oscillations. 
We basically follow this criterion in the present study.
Fig.~\ref{fig1} shows the temporal evolution of the wind friction velocity of the CNBL at a northern latitude $\psi=25^\circ$. It is seen that a quasi-equilibrium state has been established once $f_c t \geq 5$, after which we calculated the statistics in a time average window of $ \Delta t = 5/f_c$. 

To check the effect of grid resolution, we use three different grids for CNBL simulation at northern 45$^\circ$, \emph{i.e.} 150 $\times$ 76 $\times$ 128 (V1), 300 $\times$ 150 $\times$ 256 (V2) and 600 $\times$ 300 $\times$ 512 (V3), in a relatively small computational domain of 6 km $\times$ 3 km $\times$ 2 km. Fig.~\ref{fig2} shows the profiles of the mean longitudinal ($x$) velocity, Reynolds normal stress, and Reynolds shear stress obtained at different grid resolutions. 
The results of V2 and V3 are shown to be quite close. Therefore, considering both accuracy and computational efficiency, we choose to use the medium grid resolution of V2, with a horizontal resolution of 20 m and a vertical resolution of 7.8 m. It is noted that this resolution is close to that used by Liu \emph{et al.} \citep{liuUniversalWindProfile2021,liu2022vertical}.

\begin{figure}
    \centering
    \subfloat[\label{fig3a}]{
        \begin{tikzpicture}
        \node[anchor=north west, inner sep=0] (image) at (0,0) {    \includegraphics[width=0.48\textwidth]{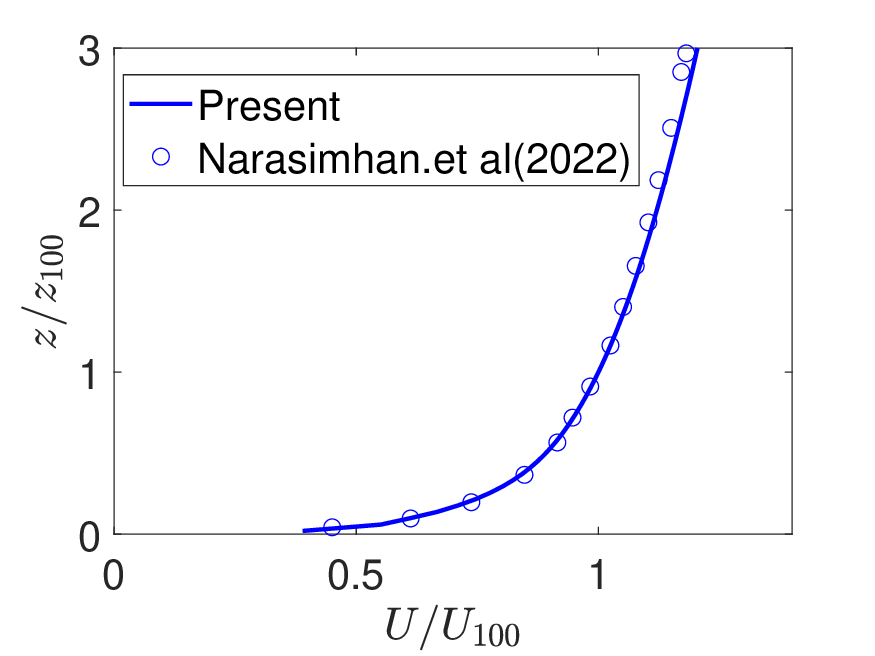}};
        \node[anchor=north west,
        xshift=-2mm,yshift=-2mm] at (image.north west) {{\rmfamily\fontsize{12}{13}\fontseries{l}\selectfont(a)}};
        \end{tikzpicture}}
    \subfloat[\label{fig3b}]{
        \begin{tikzpicture}
        \node[anchor=north west, inner sep=0] (image) at (0,0) {    \includegraphics[width=0.48\textwidth]{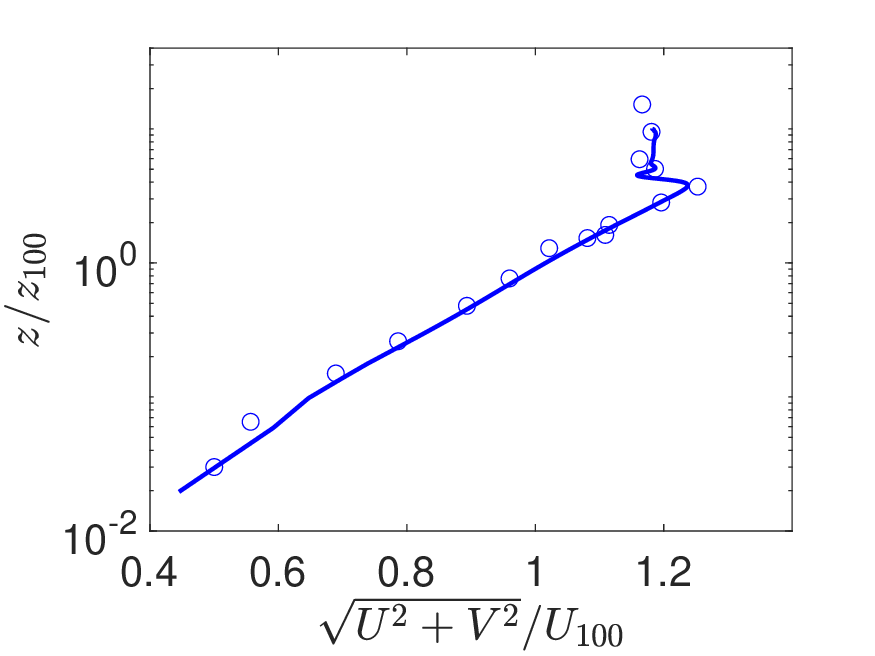}};
        \node[anchor=north west,
        xshift=-2mm,yshift=-2mm] at (image.north west) {{\rmfamily\fontsize{12}{13}\fontseries{l}\selectfont(b)}};
        \end{tikzpicture}}
    \caption{Comparison of the present simulation and Narasimhan \emph{et al.} \citep{narasimhan2022effects}. (a) Mean longitudinal velocity; (b) mean total velocity. {$z_{100}$ is the height of 100 m, and $U_{100}$ is the mean longitudinal wind velocity at $z=100\,\text{m}$.}}
    \label{fig3}
\end{figure}

For the validation of the solver, {a} LES simulation case (V4) of a CNBL that is the same as that in Narasimhan \emph{et al.} \citep{narasimhan2022effects} is performed, in which the domain extends 3.7 km $\times$ 1.5 km $\times$ 2.0 km and the grid is 360 $\times$ 144 $\times$ 432.
The results are shown in Fig.~\ref{fig3}. It can be seen that the results of the present simulation are in excellent agreement with those of Narasimhan \emph{et al.} \citep{narasimhan2022effects}.

\section{Results}
\label{sec3}


To investigate the effects of the Coriolis force due to the variation of latitude and geostrophic wind, we compare the flow structures, energy spectra, correlation structures, and structure deflection and inclination angles of the ABL flows in cases N25 to S45 in Table~\ref{table1}. The basic statistics can be found in the Appendix.

\subsection{Coherent structures}

\begin{figure}
\centering
   \subfloat[\label{fig7a}]{
        \begin{tikzpicture}
        \node[anchor=north west, inner sep=0] (image) at (0,0) {
    \includegraphics[width=0.48\textwidth]{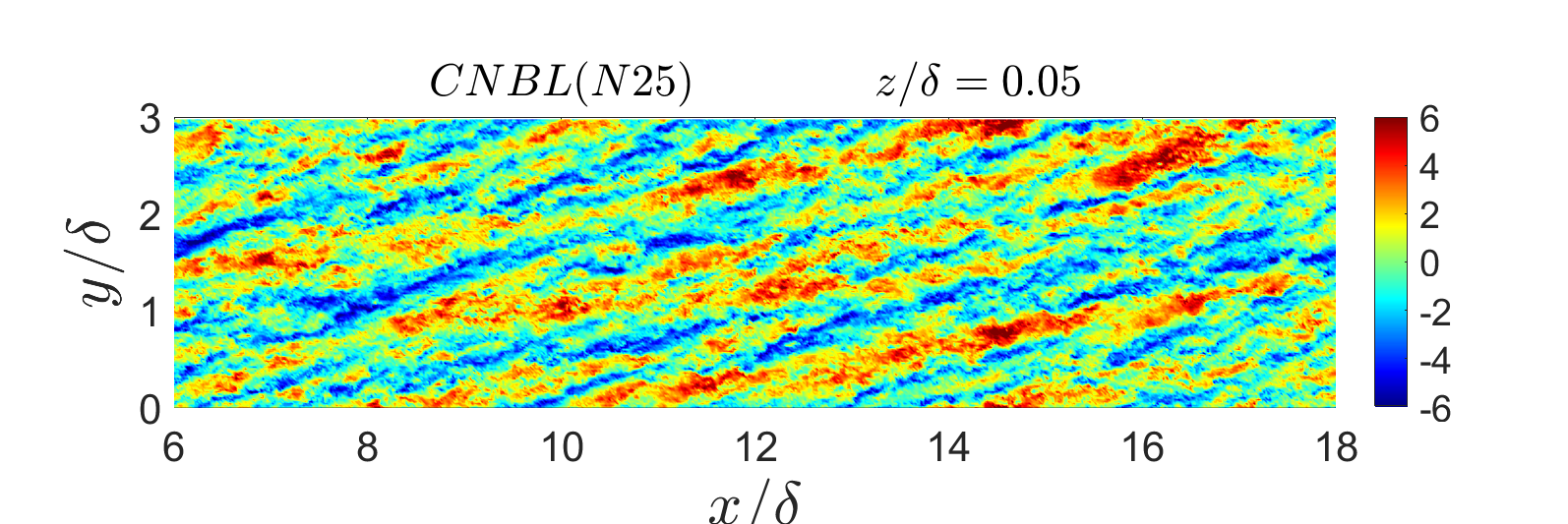}
            };
    \node[anchor=north west,
        xshift=-2mm,yshift=-2mm] at (image.north west) {{\rmfamily\fontsize{12}{13}\fontseries{l}\selectfont(a)}};
        \end{tikzpicture}}
    \subfloat[\label{fig7b}]{
        \begin{tikzpicture}
        \node[anchor=north west, inner sep=0] (image) at (0,0) {
    \includegraphics[width=0.48\textwidth]{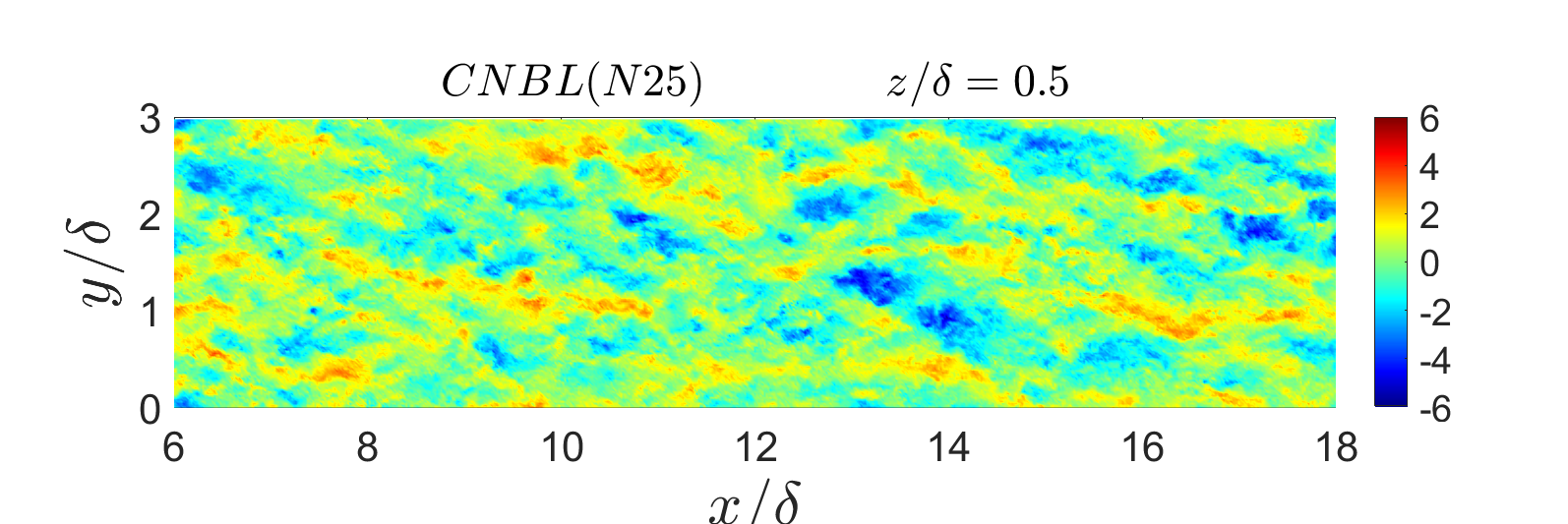}
            };
    \node[anchor=north west,
        xshift=-2mm,yshift=-2mm] at (image.north west) {{\rmfamily\fontsize{12}{13}\fontseries{l}\selectfont(b)}};
        \end{tikzpicture}}

\vspace{-1.35cm} 
\centering
   \subfloat[\label{fig7c}]{
        \begin{tikzpicture}
        \node[anchor=north west, inner sep=0] (image) at (0,0) {
    \includegraphics[width=0.48\textwidth]{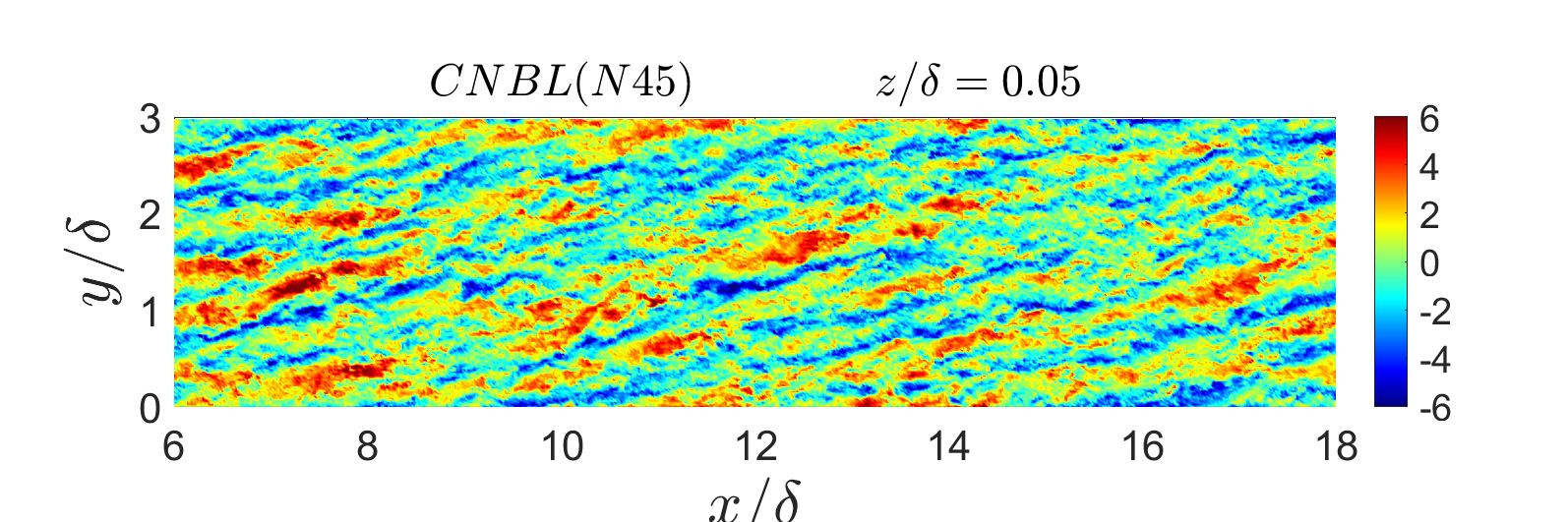}
            };
    \node[anchor=north west,
        xshift=-2mm,yshift=-2mm] at (image.north west) {{\rmfamily\fontsize{12}{13}\fontseries{l}\selectfont(c)}};
        \end{tikzpicture}}
    \subfloat[\label{fig7d}]{
        \begin{tikzpicture}
        \node[anchor=north west, inner sep=0] (image) at (0,0) {
    \includegraphics[width=0.48\textwidth]{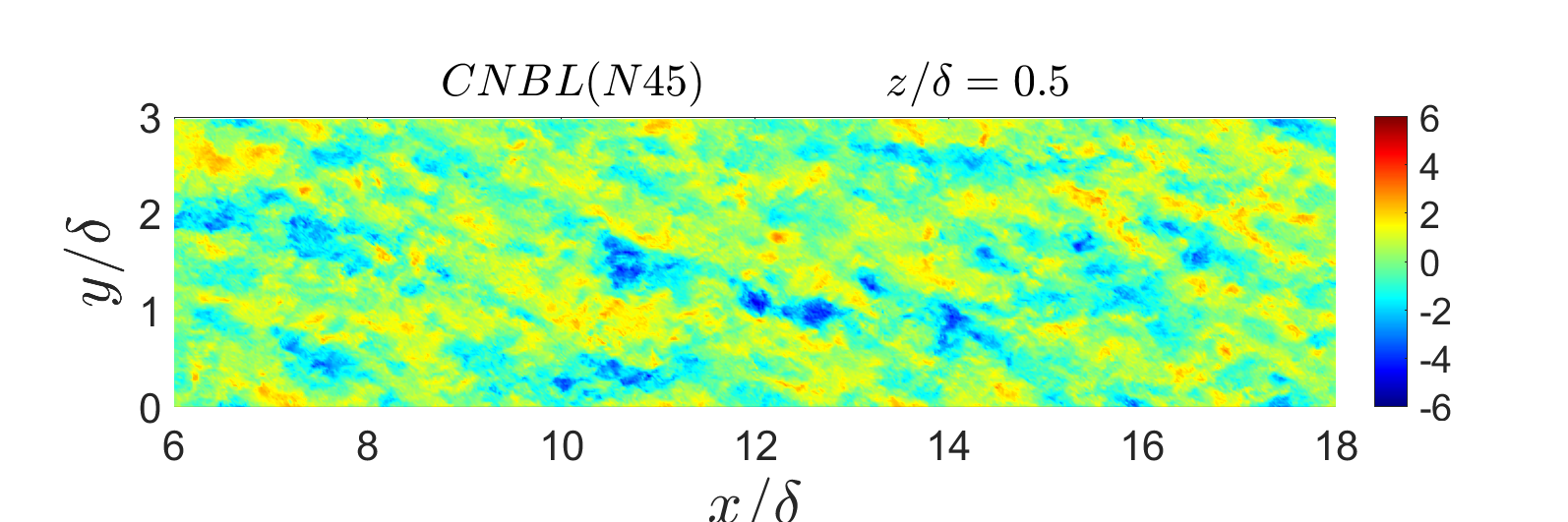}
            };
    \node[anchor=north west,
        xshift=-2mm,yshift=-2mm] at (image.north west) {{\rmfamily\fontsize{12}{13}\fontseries{l}\selectfont(d)}};
        \end{tikzpicture}}
        
   \vspace{-1.35cm}     
   \centering
   \subfloat[\label{fig7e}]{
        \begin{tikzpicture}
        \node[anchor=north west, inner sep=0] (image) at (0,0) {
    \includegraphics[width=0.48\textwidth]{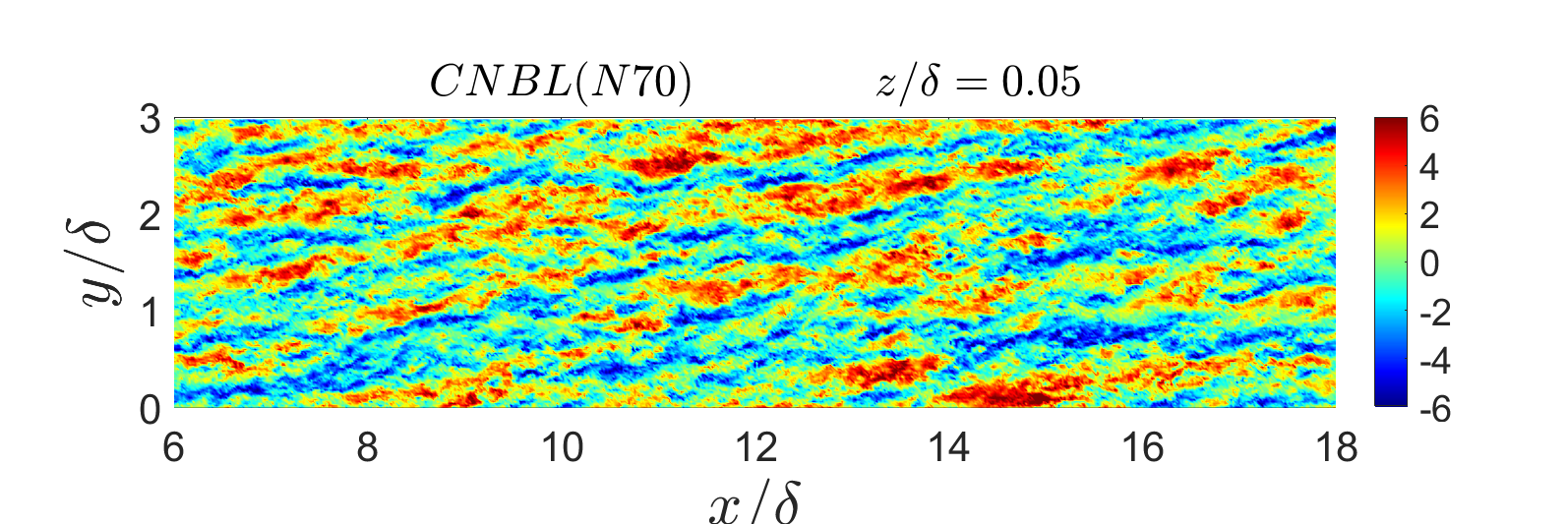}
            };
    \node[anchor=north west,
        xshift=-2mm,yshift=-2mm] at (image.north west) {{\rmfamily\fontsize{12}{13}\fontseries{l}\selectfont(e)}};
        \end{tikzpicture}}
    \subfloat[\label{fig7f}]{
        \begin{tikzpicture}
        \node[anchor=north west, inner sep=0] (image) at (0,0) {
    \includegraphics[width=0.48\textwidth]{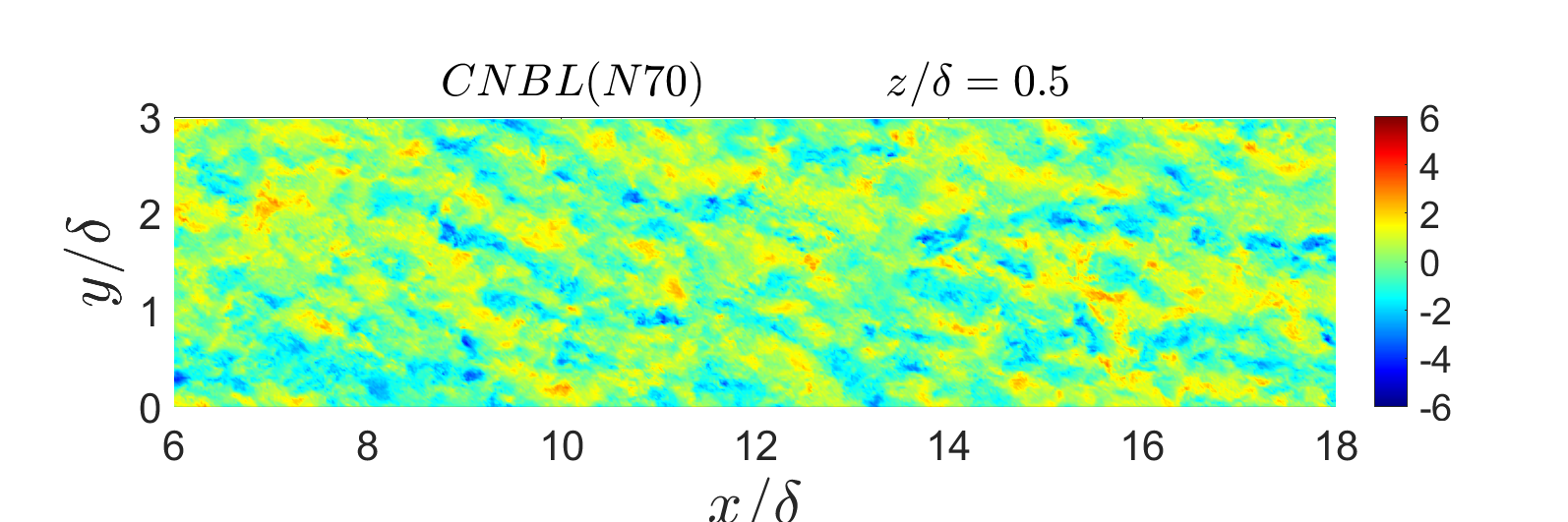}
            };
    \node[anchor=north west,
        xshift=-2mm,yshift=-2mm] at (image.north west) {{\rmfamily\fontsize{12}{13}\fontseries{l}\selectfont(f)}};
        \end{tikzpicture}}
        
 \vspace{-1.35cm}  
    \centering
   \subfloat[\label{fig7g}]{
        \begin{tikzpicture}
        \node[anchor=north west, inner sep=0] (image) at (0,0) {
    \includegraphics[width=0.48\textwidth]{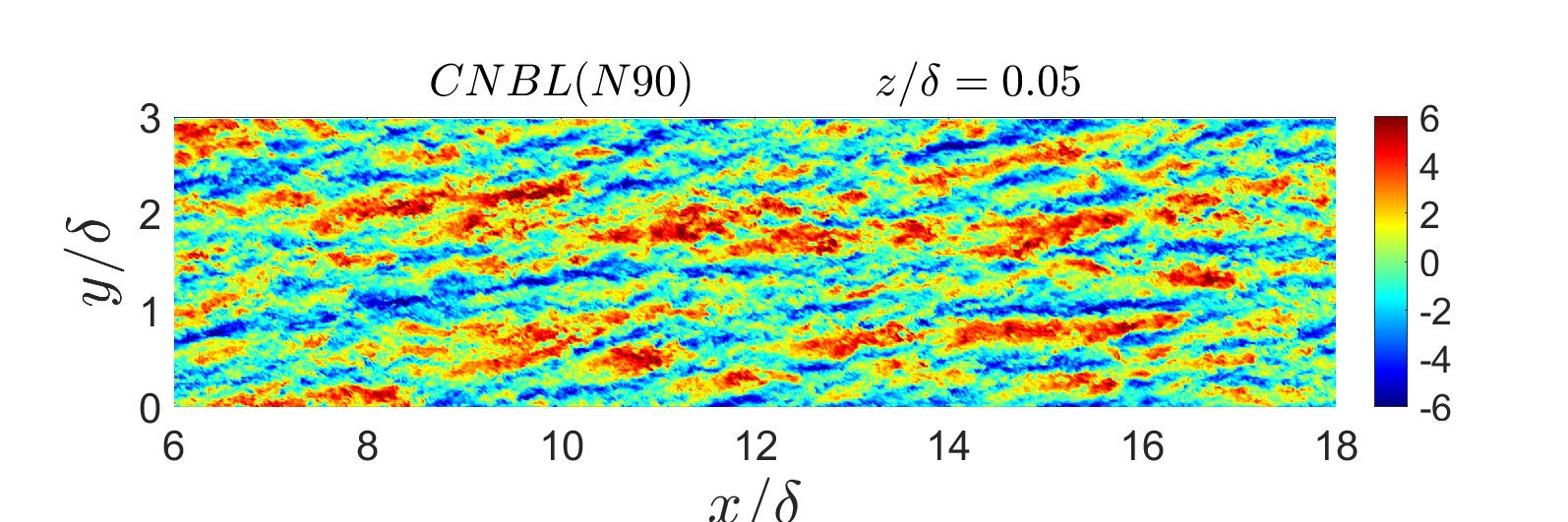}
            };
    \node[anchor=north west,
        xshift=-2mm,yshift=-2mm] at (image.north west) {{\rmfamily\fontsize{12}{13}\fontseries{l}\selectfont(g)}};
        \end{tikzpicture}}
    \subfloat[\label{fig7h}]{
        \begin{tikzpicture}
        \node[anchor=north west, inner sep=0] (image) at (0,0) {
    \includegraphics[width=0.48\textwidth]{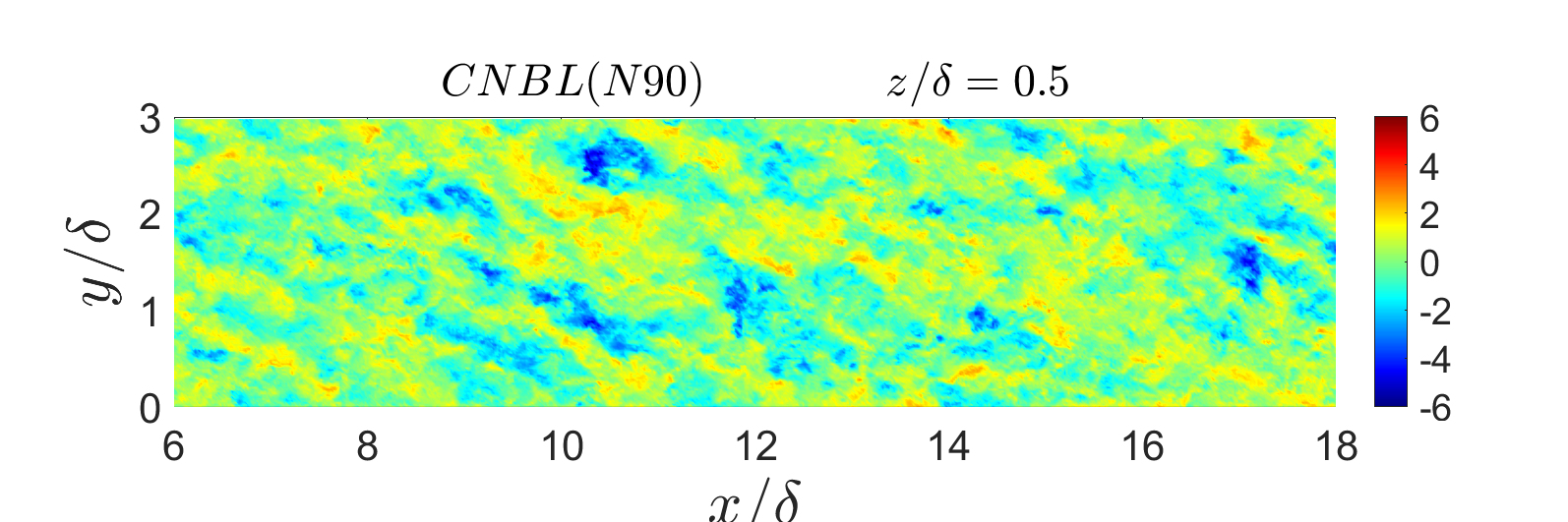}
            };
    \node[anchor=north west,
        xshift=-2mm,yshift=-2mm] at (image.north west) {{\rmfamily\fontsize{12}{13}\fontseries{l}\selectfont(h)}};
        \end{tikzpicture}}

 \vspace{-1.35cm}  
    \centering
   \subfloat[\label{fig7i}]{
        \begin{tikzpicture}
        \node[anchor=north west, inner sep=0] (image) at (0,0) {
    \includegraphics[width=0.48\textwidth]{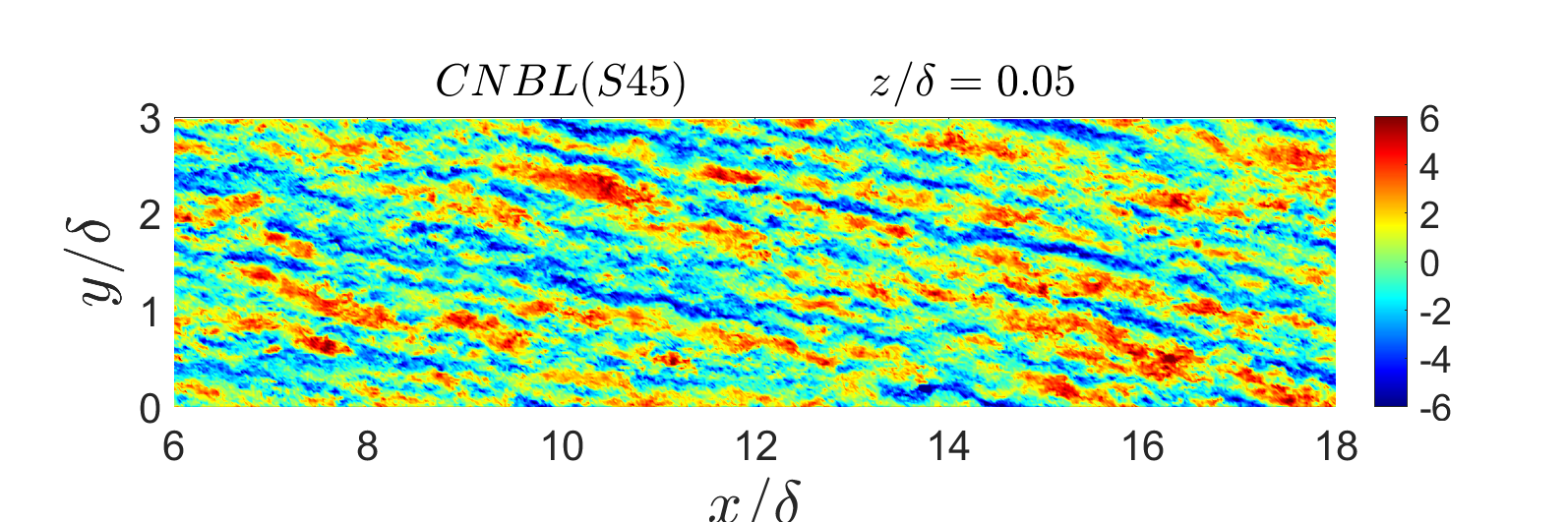}
            };
    \node[anchor=north west,
        xshift=-2mm,yshift=-2mm] at (image.north west) {{\rmfamily\fontsize{12}{13}\fontseries{l}\selectfont(i)}};
        \end{tikzpicture}}
    \subfloat[\label{fig7j}]{
        \begin{tikzpicture}
        \node[anchor=north west, inner sep=0] (image) at (0,0) {
    \includegraphics[width=0.48\textwidth]{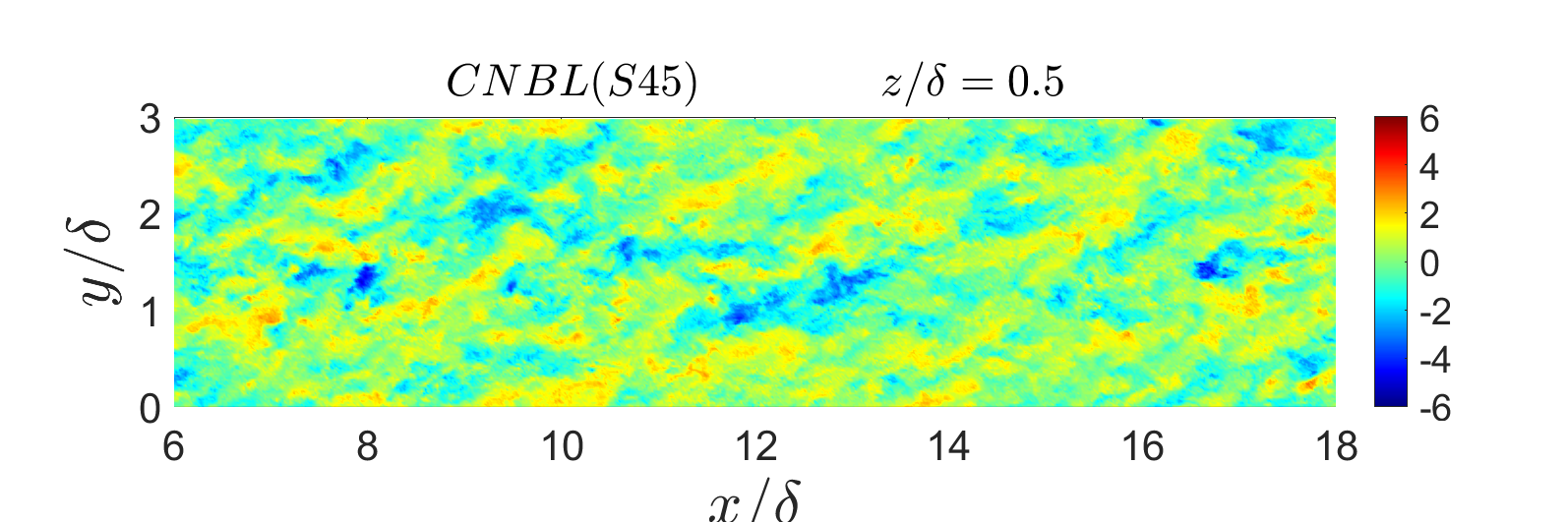}
            };
    \node[anchor=north west,
        xshift=-2mm,yshift=-2mm] at (image.north west) {{\rmfamily\fontsize{12}{13}\fontseries{l}\selectfont(j)}};
        \end{tikzpicture}}

 \vspace{-1.35cm}  
    \centering
   \subfloat[\label{fig7k}]{
        \begin{tikzpicture}
        \node[anchor=north west, inner sep=0] (image) at (0,0) {
    \includegraphics[width=0.48\textwidth]{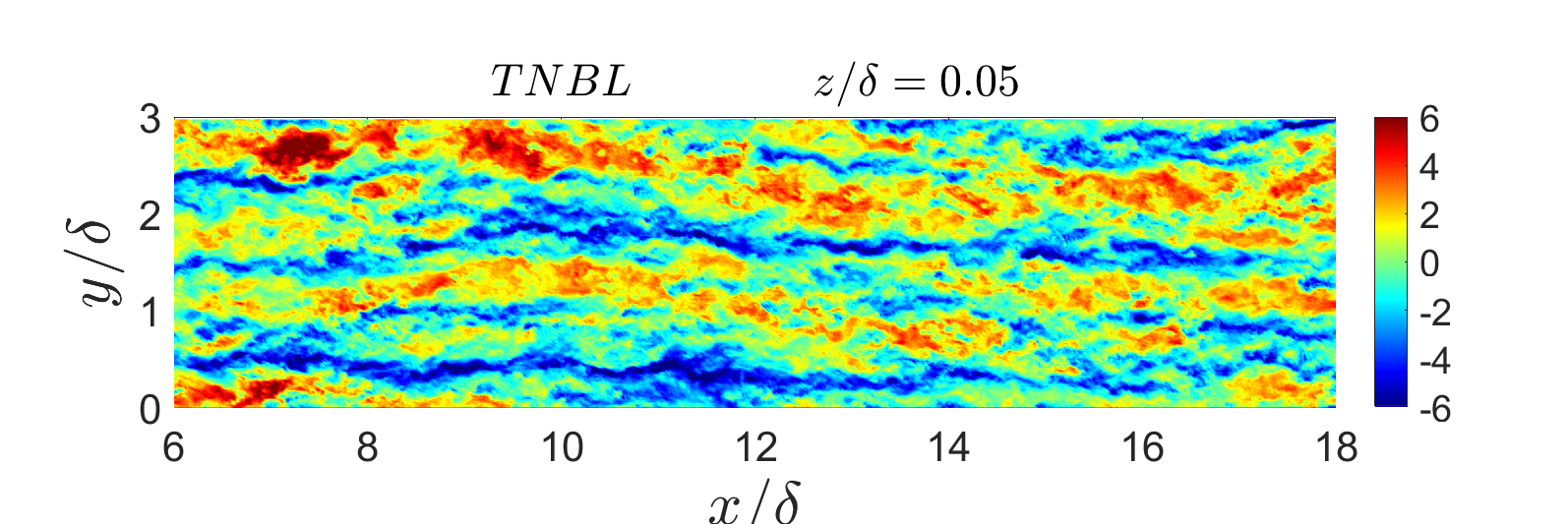}
            };
    \node[anchor=north west,
        xshift=-2mm,yshift=-2mm] at (image.north west) {{\rmfamily\fontsize{12}{13}\fontseries{l}\selectfont(k)}};
        \end{tikzpicture}}
    \subfloat[\label{fig7l}]{
        \begin{tikzpicture}
        \node[anchor=north west, inner sep=0] (image) at (0,0) {
    \includegraphics[width=0.48\textwidth]{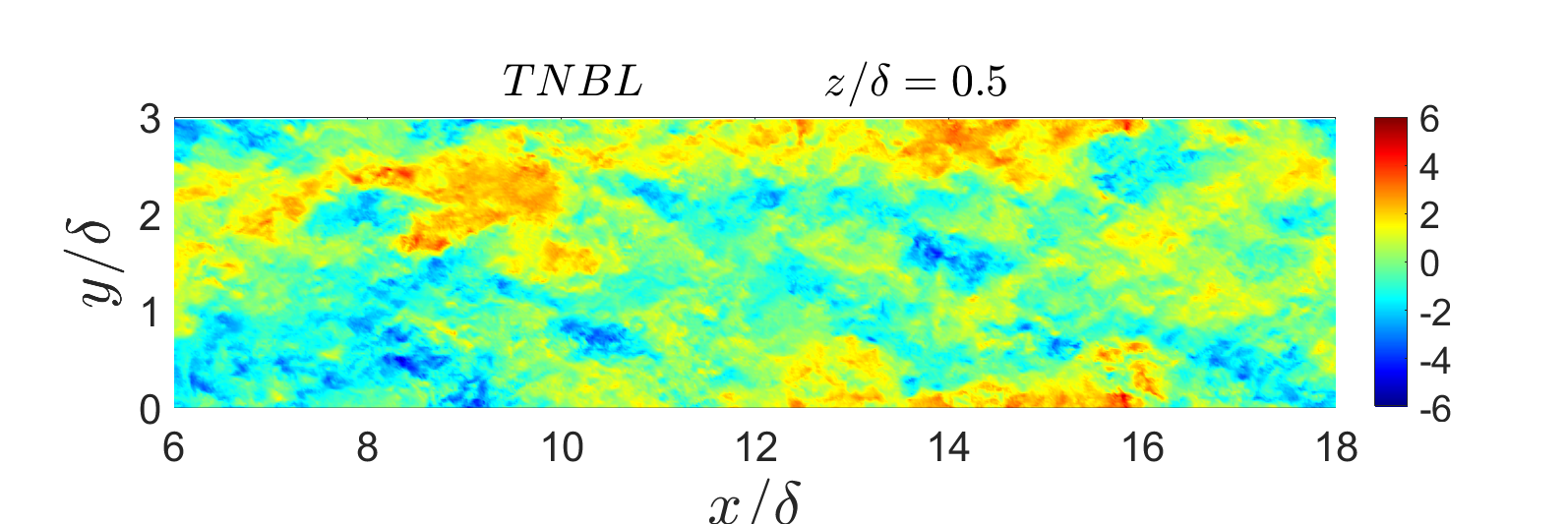}
            };
    \node[anchor=north west,
        xshift=-2mm,yshift=-2mm] at (image.north west) {{\rmfamily\fontsize{12}{13}\fontseries{l}\selectfont(l)}};
        \end{tikzpicture}}
    \caption{Spatial distributions of longitudinal flow velocity fluctuations at $z/\delta=0.05$ (left column) and 0.5 (right column) in the CNBLs at different latitudes and the TNBL.}
    \label{fig7}
\end{figure}

\begin{figure}
\centering
   \subfloat[\label{fig8a}]{
        \begin{tikzpicture}
        \node[anchor=north west, inner sep=0] (image) at (0,0) {
    \includegraphics[width=0.48\textwidth]{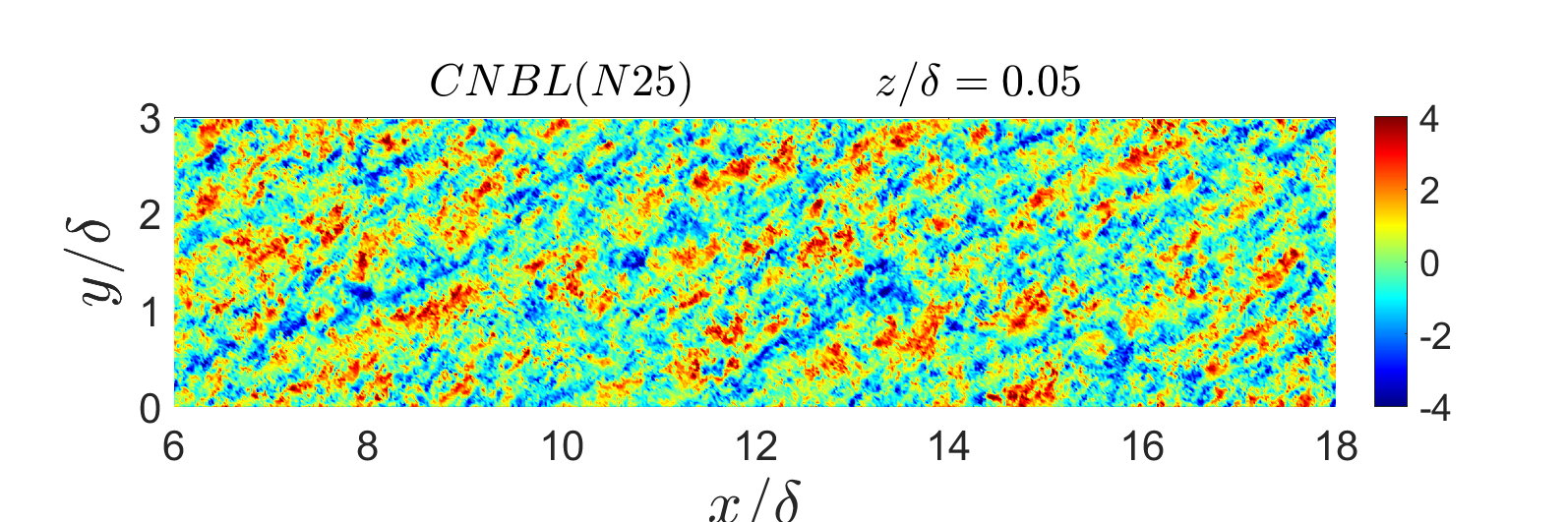}
            };
    \node[anchor=north west,
        xshift=-2mm,yshift=-2mm] at (image.north west) {{\rmfamily\fontsize{12}{13}\fontseries{l}\selectfont(a)}};
        \end{tikzpicture}}
    \subfloat[\label{fig8b}]{
        \begin{tikzpicture}
        \node[anchor=north west, inner sep=0] (image) at (0,0) {
    \includegraphics[width=0.48\textwidth]{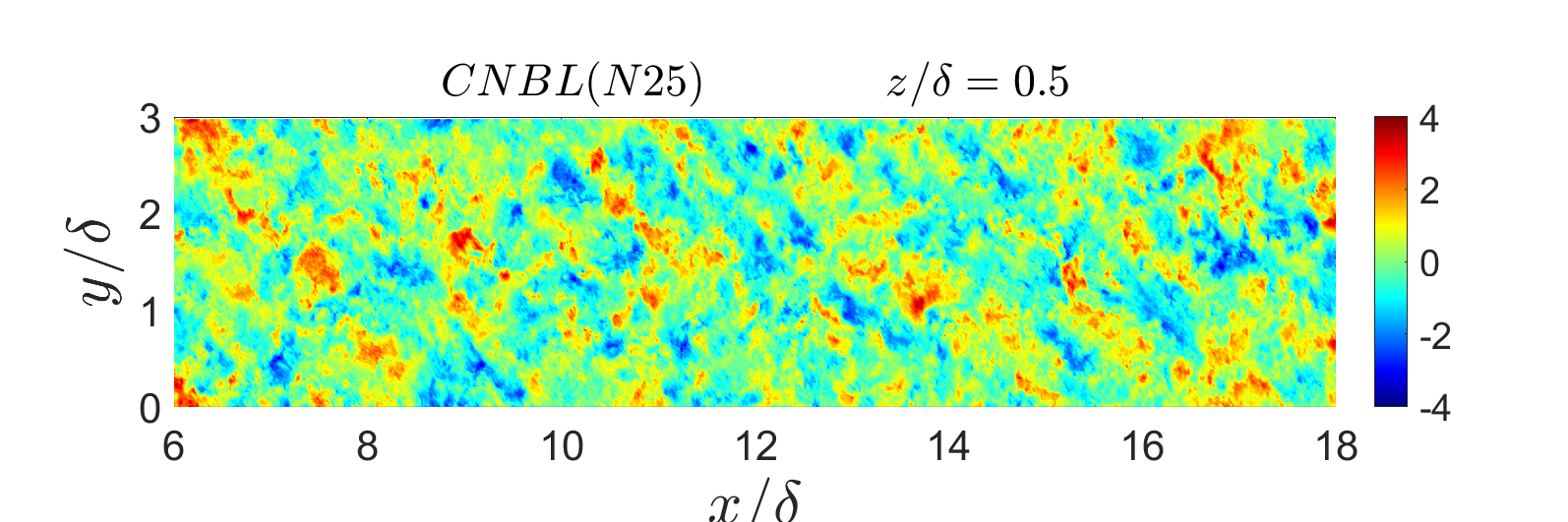}
            };
    \node[anchor=north west,
        xshift=-2mm,yshift=-2mm] at (image.north west) {{\rmfamily\fontsize{12}{13}\fontseries{l}\selectfont(b)}};
        \end{tikzpicture}}

\vspace{-1.35cm} 
\centering
   \subfloat[\label{fig8c}]{
        \begin{tikzpicture}
        \node[anchor=north west, inner sep=0] (image) at (0,0) {
    \includegraphics[width=0.48\textwidth]{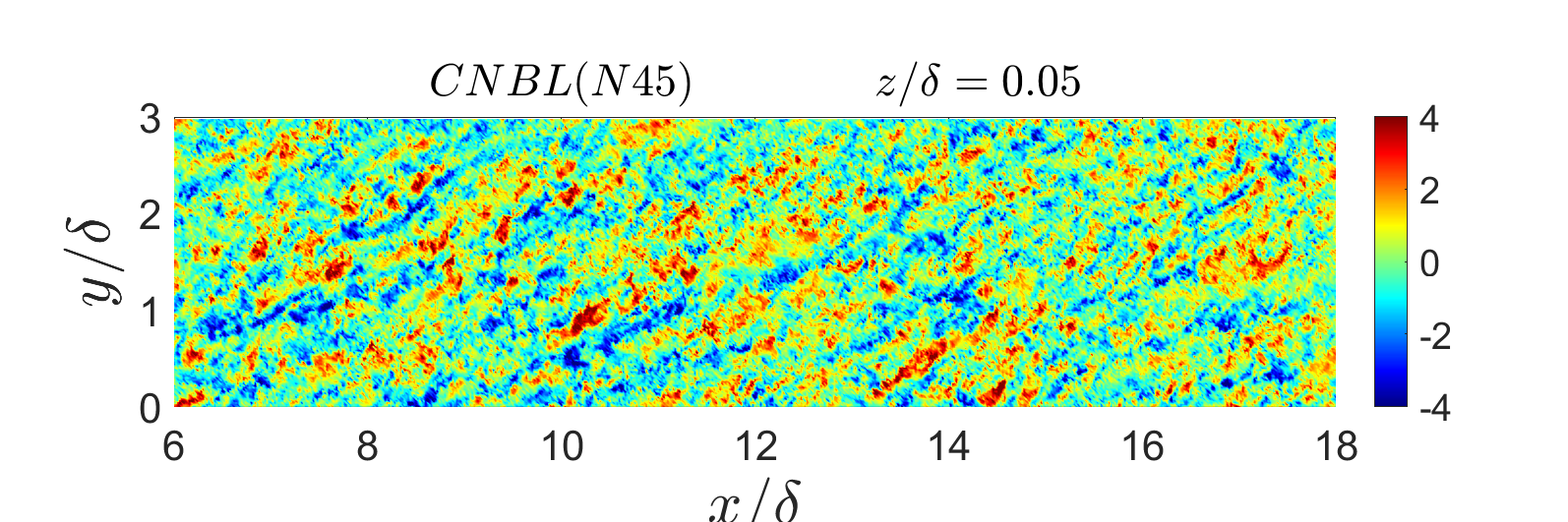}
            };
    \node[anchor=north west,
        xshift=-2mm,yshift=-2mm] at (image.north west) {{\rmfamily\fontsize{12}{13}\fontseries{l}\selectfont(c)}};
        \end{tikzpicture}}
    \subfloat[\label{fig8d}]{
        \begin{tikzpicture}
        \node[anchor=north west, inner sep=0] (image) at (0,0) {
    \includegraphics[width=0.48\textwidth]{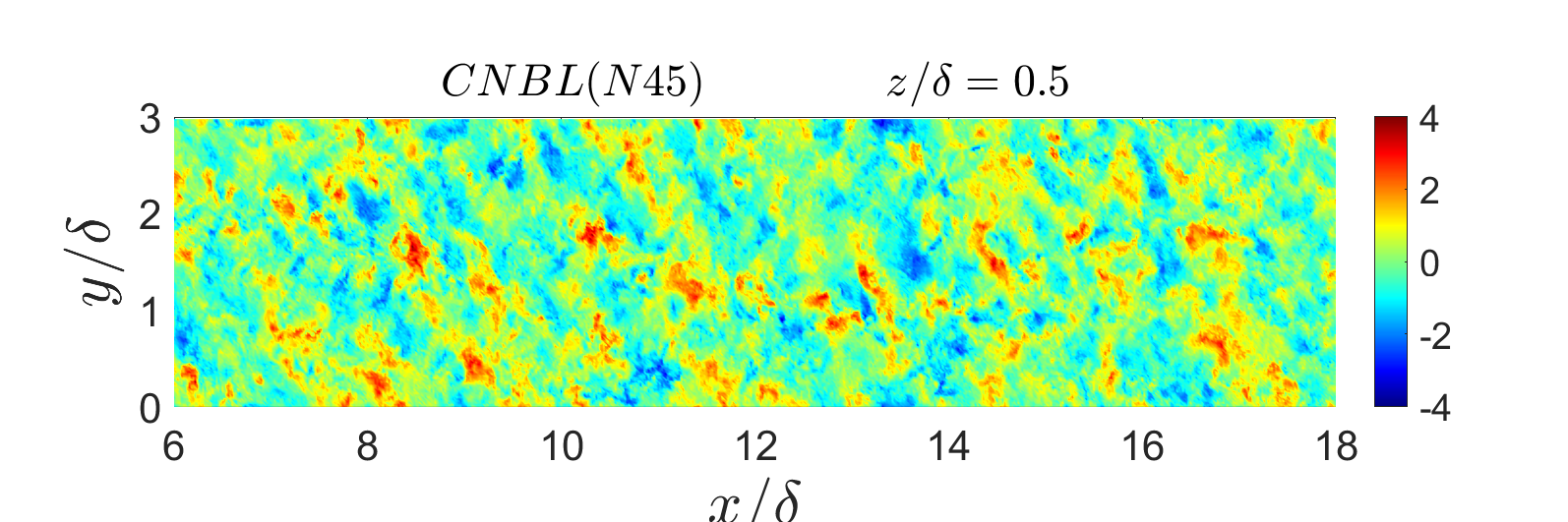}
            };
    \node[anchor=north west,
        xshift=-2mm,yshift=-2mm] at (image.north west) {{\rmfamily\fontsize{12}{13}\fontseries{l}\selectfont(d)}};
        \end{tikzpicture}}
        
   \vspace{-1.35cm}     
   \centering
   \subfloat[\label{fig8e}]{
        \begin{tikzpicture}
        \node[anchor=north west, inner sep=0] (image) at (0,0) {
    \includegraphics[width=0.48\textwidth]{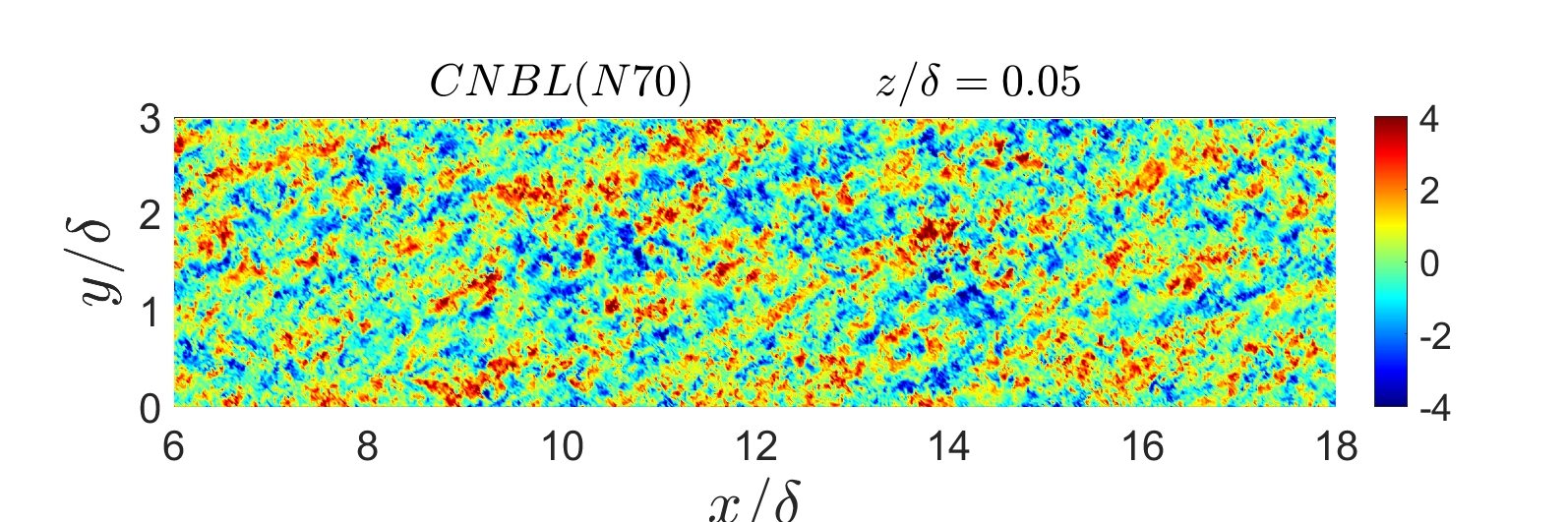}
            };
    \node[anchor=north west,
        xshift=-2mm,yshift=-2mm] at (image.north west) {{\rmfamily\fontsize{12}{13}\fontseries{l}\selectfont(e)}};
        \end{tikzpicture}}
    \subfloat[\label{fig8f}]{
        \begin{tikzpicture}
        \node[anchor=north west, inner sep=0] (image) at (0,0) {
    \includegraphics[width=0.48\textwidth]{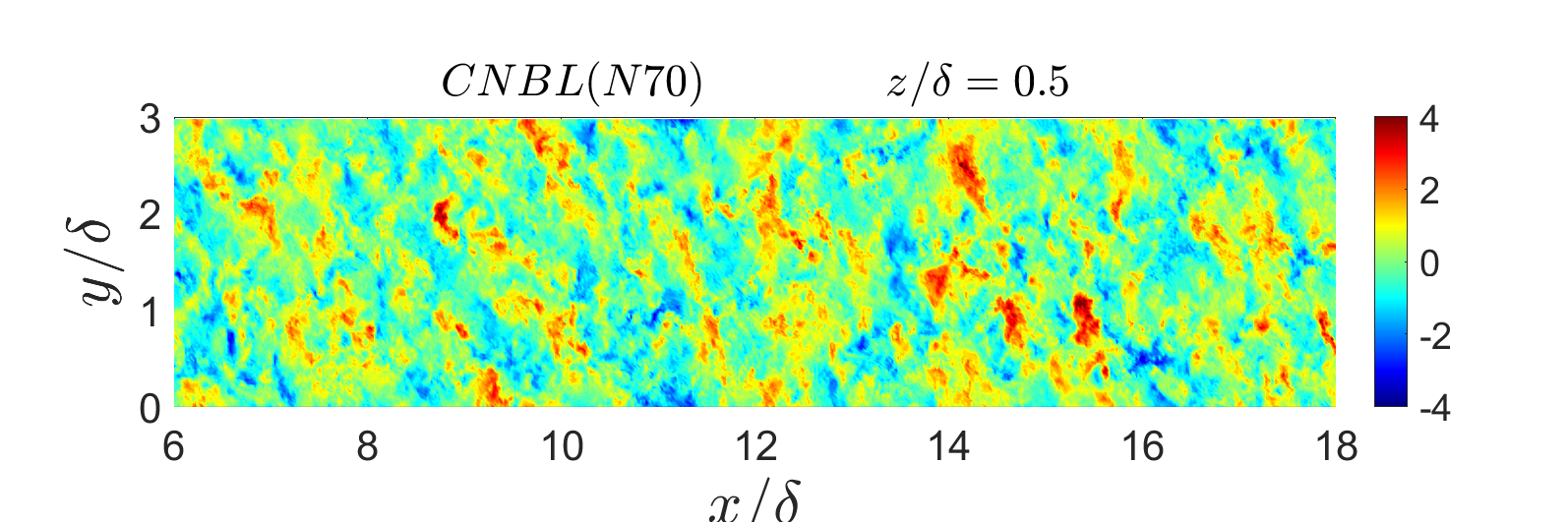}
            };
    \node[anchor=north west,
        xshift=-2mm,yshift=-2mm] at (image.north west) {{\rmfamily\fontsize{12}{13}\fontseries{l}\selectfont(f)}};
        \end{tikzpicture}}
        
 \vspace{-1.35cm}  
    \centering
   \subfloat[\label{fig8g}]{
        \begin{tikzpicture}
        \node[anchor=north west, inner sep=0] (image) at (0,0) {
    \includegraphics[width=0.48\textwidth]{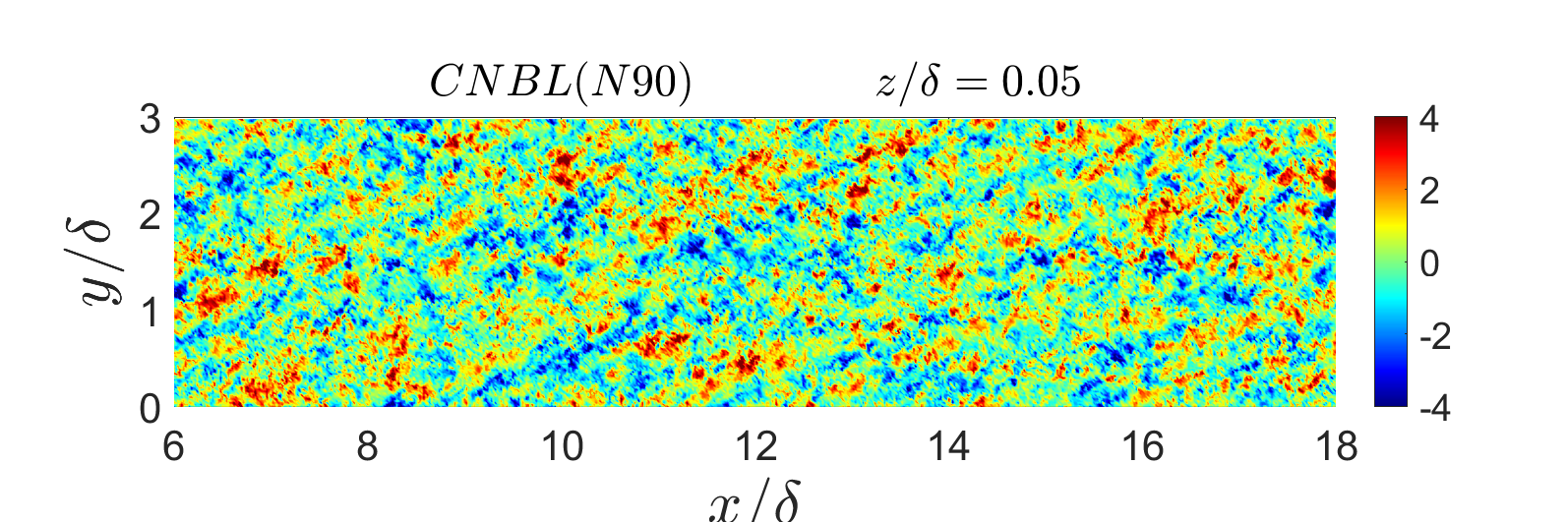}
            };
    \node[anchor=north west,
        xshift=-2mm,yshift=-2mm] at (image.north west) {{\rmfamily\fontsize{12}{13}\fontseries{l}\selectfont(g)}};
        \end{tikzpicture}}
    \subfloat[\label{fig8h}]{
        \begin{tikzpicture}
        \node[anchor=north west, inner sep=0] (image) at (0,0) {
    \includegraphics[width=0.48\textwidth]{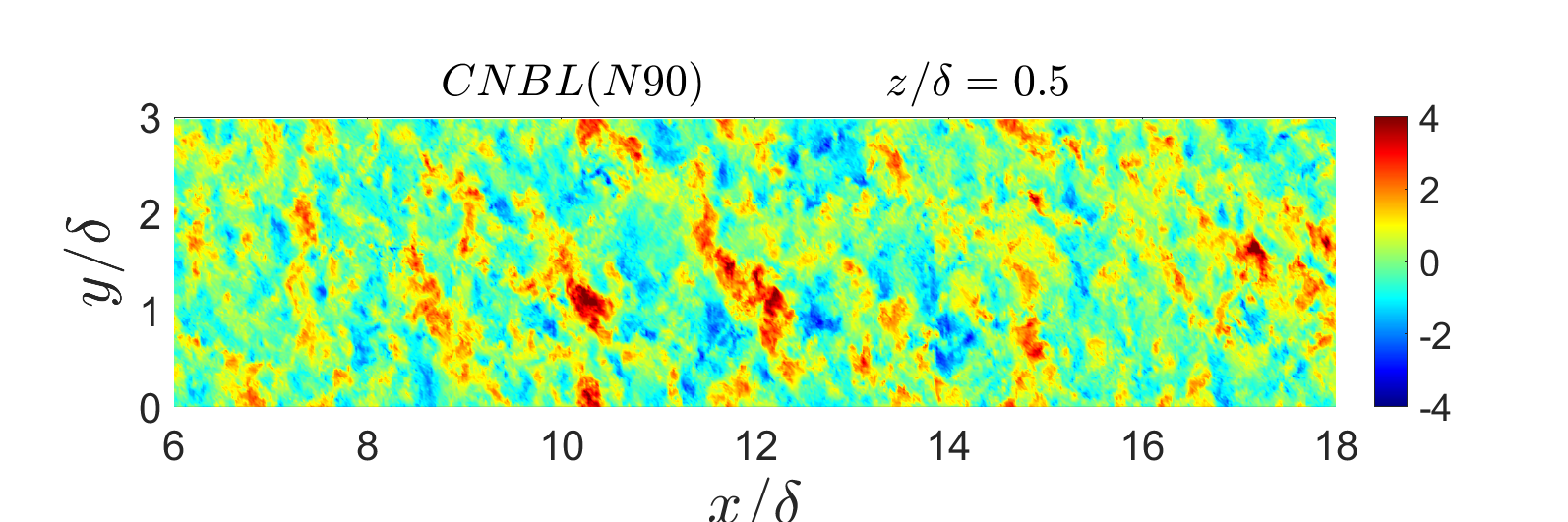}
            };
    \node[anchor=north west,
        xshift=-2mm,yshift=-2mm] at (image.north west) {{\rmfamily\fontsize{12}{13}\fontseries{l}\selectfont(h)}};
        \end{tikzpicture}}

 \vspace{-1.35cm}  
    \centering
   \subfloat[\label{fig8i}]{
        \begin{tikzpicture}
        \node[anchor=north west, inner sep=0] (image) at (0,0) {
    \includegraphics[width=0.48\textwidth]{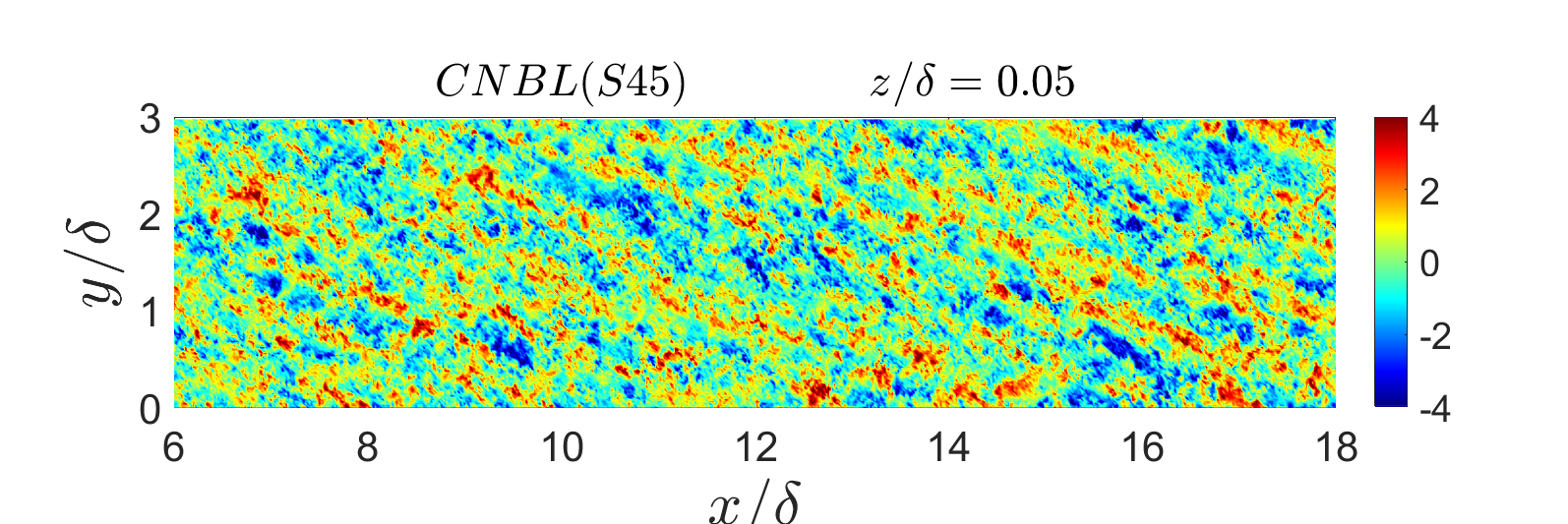}
            };
    \node[anchor=north west,
        xshift=-2mm,yshift=-2mm] at (image.north west) {{\rmfamily\fontsize{12}{13}\fontseries{l}\selectfont(i)}};
        \end{tikzpicture}}
    \subfloat[\label{fig8j}]{
        \begin{tikzpicture}
        \node[anchor=north west, inner sep=0] (image) at (0,0) {
    \includegraphics[width=0.48\textwidth]{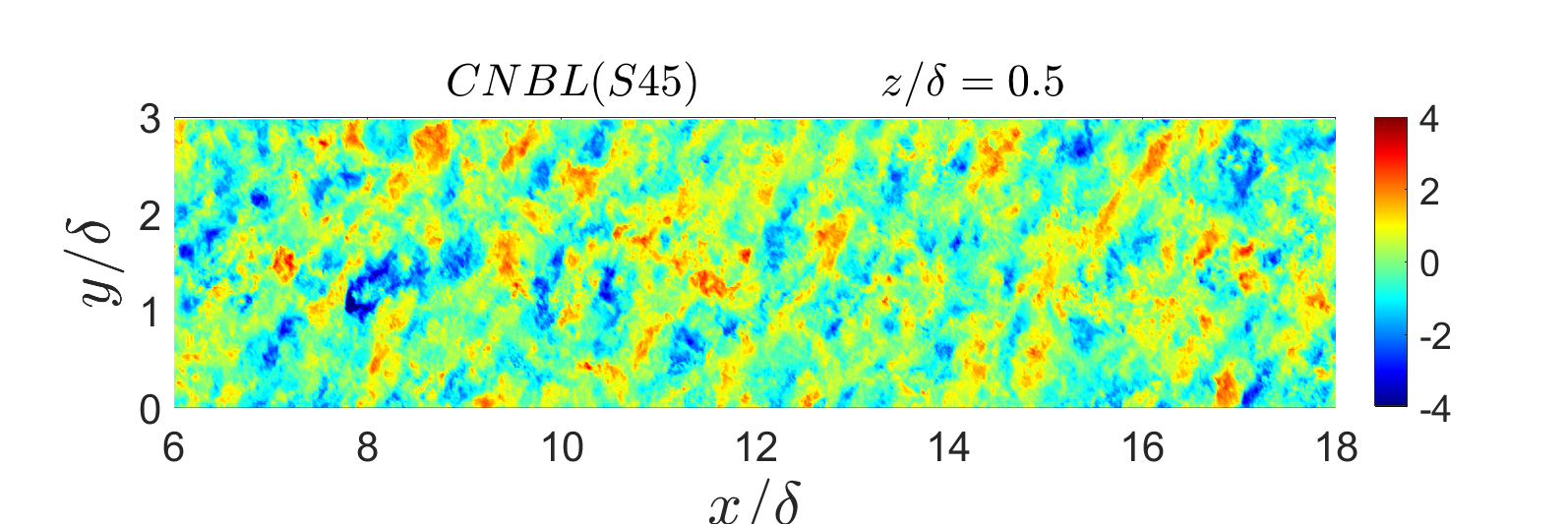}
            };
    \node[anchor=north west,
        xshift=-2mm,yshift=-2mm] at (image.north west) {{\rmfamily\fontsize{12}{13}\fontseries{l}\selectfont(j)}};
        \end{tikzpicture}}

 \vspace{-1.35cm}  
    \centering
   \subfloat[\label{fig8k}]{
        \begin{tikzpicture}
        \node[anchor=north west, inner sep=0] (image) at (0,0) {
    \includegraphics[width=0.48\textwidth]{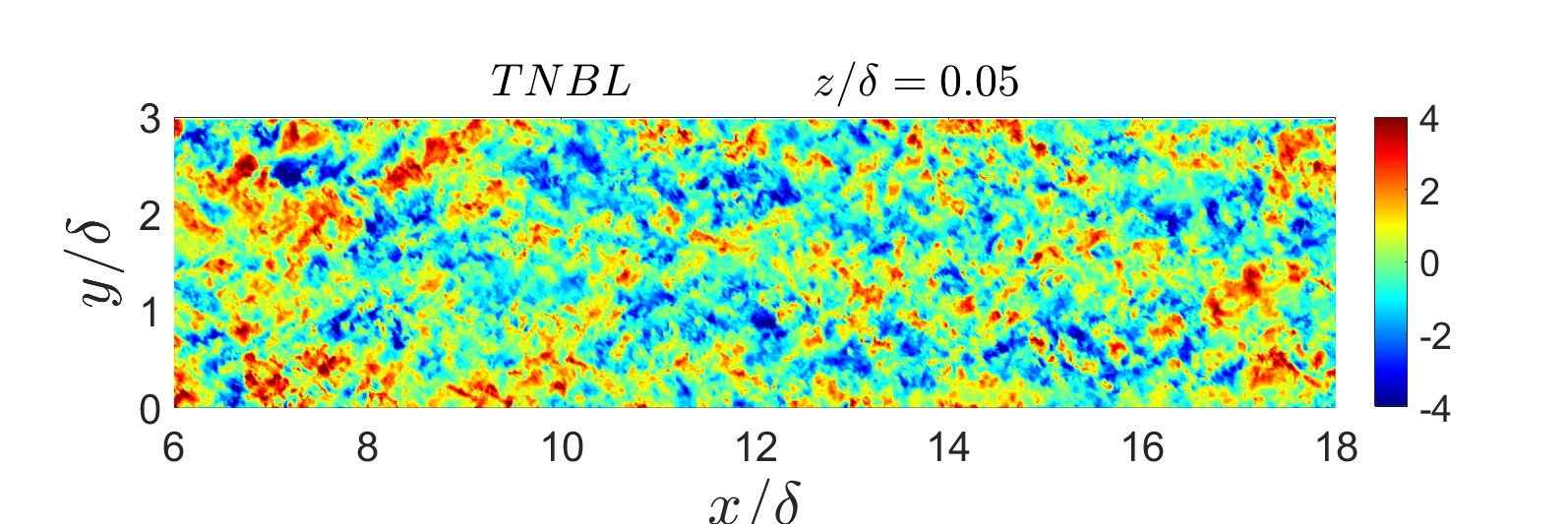}
            };
    \node[anchor=north west,
        xshift=-2mm,yshift=-2mm] at (image.north west) {{\rmfamily\fontsize{12}{13}\fontseries{l}\selectfont(k)}};
        \end{tikzpicture}}
    \subfloat[\label{fig8l}]{
        \begin{tikzpicture}
        \node[anchor=north west, inner sep=0] (image) at (0,0) {
    \includegraphics[width=0.48\textwidth]{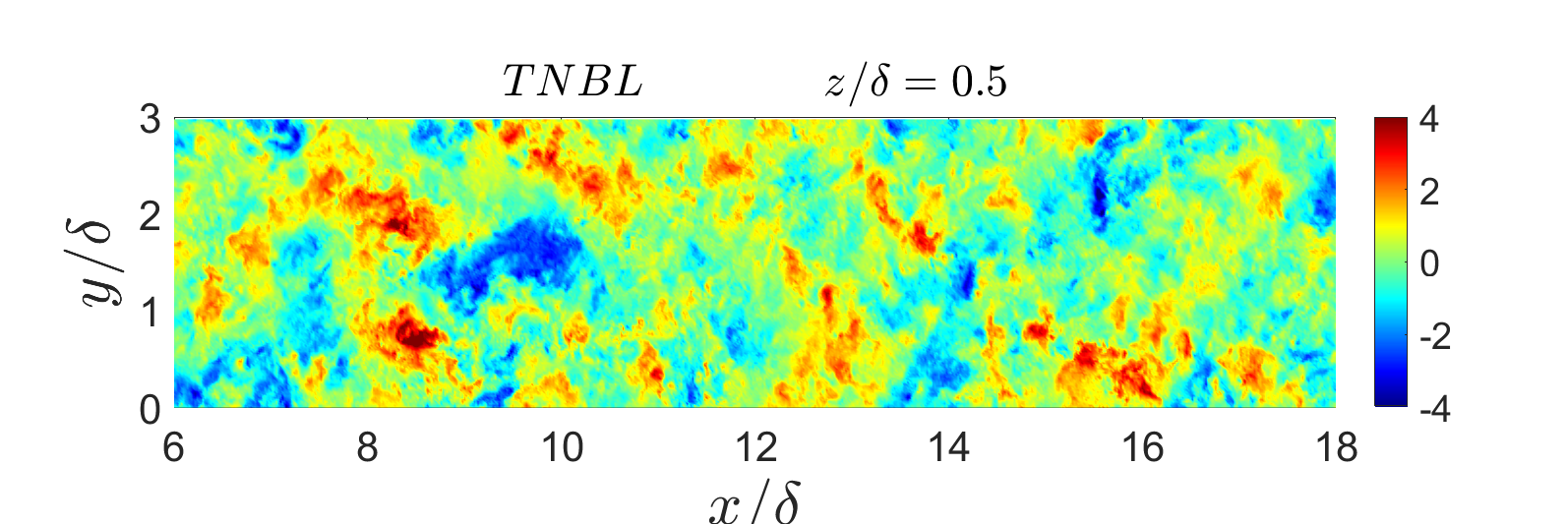}
            };
    \node[anchor=north west,
        xshift=-2mm,yshift=-2mm] at (image.north west) {{\rmfamily\fontsize{12}{13}\fontseries{l}\selectfont(l)}};
        \end{tikzpicture}}
    \caption{Spatial distributions of lateral flow velocity fluctuations at $z/\delta=0.05$ (left column) and 0.5 (right column) in the CNBLs at different latitudes and the TNBL.}
    \label{fig8}
\end{figure}

Fig.~\ref{fig7} shows the spatial distributions of longitudinal flow velocity fluctuations in the CNBLs at different latitudes and the TNBL. The flow velocity is normalized by the friction velocity $U_\tau$.
In general, one can observe deflected and elongated streaks of positive and negative longitudinal velocity fluctuations in the boundary layers, similar to the findings in previous studies \citep{deardorff1972numerical,moengComparisonShearBuoyancyDriven1994,masonLargeEddySimulationsNeutralstaticstability1987,colemanNumericalStudyTurbulent1990a,zikanov2003large,shingaiStudyTurbulenceStructure2004a}. The velocity streaks seem narrower near the ground at $z/\delta=0.05$, while they are much wider away from the ground surface at $z/\delta=0.5$. 
Quantitative analysis of the length scales of the turbulent structures will be given through the premultiplied energy spectra afterwards.
More importantly, it can be seen that there is an obvious deflection trend of the velocity structures in the CNBLs due to the Coriolis force. 
In the Northern Hemisphere, the streaks are anticlockwise deflected at $z/\delta=0.05$ to the longitudinal direction (the geostrophic wind direction) while being clockwise rotated at $z/\delta=0.5$. The situation is the opposite in the Southern Hemisphere. There is no deflection of velocity streaks in the TNBL due to the absence of the Coriolis force. 
Moreover, with the increase of latitude, the degree of the deflection decreases slightly at $z/\delta=0.05$ while increasing at $z/\delta=0.5$.
{It should be noted that the mean wind veer is different at different latitudes. Thus, one needs to study the structure deflection with respect to the mean wind veer, which will be presented afterwards.}

We display the spatial distributions of lateral velocity fluctuations in the CNBLs at different latitudes and the TNBL in Fig.~\ref{fig8}. It can be seen that there are also streaks or patches of the lateral velocity fluctuations in the flow. However, the lateral velocity structures are significantly shorter than the longitudinal velocity structures. 
Similarly, we also find that there is an anticlockwise deflection of the lateral velocity structures at $z/\delta=0.05$, while there is a clockwise deflection at $z/\delta=0.5$ in the Northern Hemisphere, which is opposite in the Southern Hemisphere. 
The deflection of the lateral velocity structures seems greater than that of the longitudinal velocity at the same height. 
Meanwhile, the lateral velocity structures are still aligned with the longitudinal direction in the TNBL.

\begin{figure}
\centering
   \subfloat[\label{fig9a}]{
        \begin{tikzpicture}
        \node[anchor=north west, inner sep=0] (image) at (0,0) {
    \includegraphics[width=0.48\textwidth]{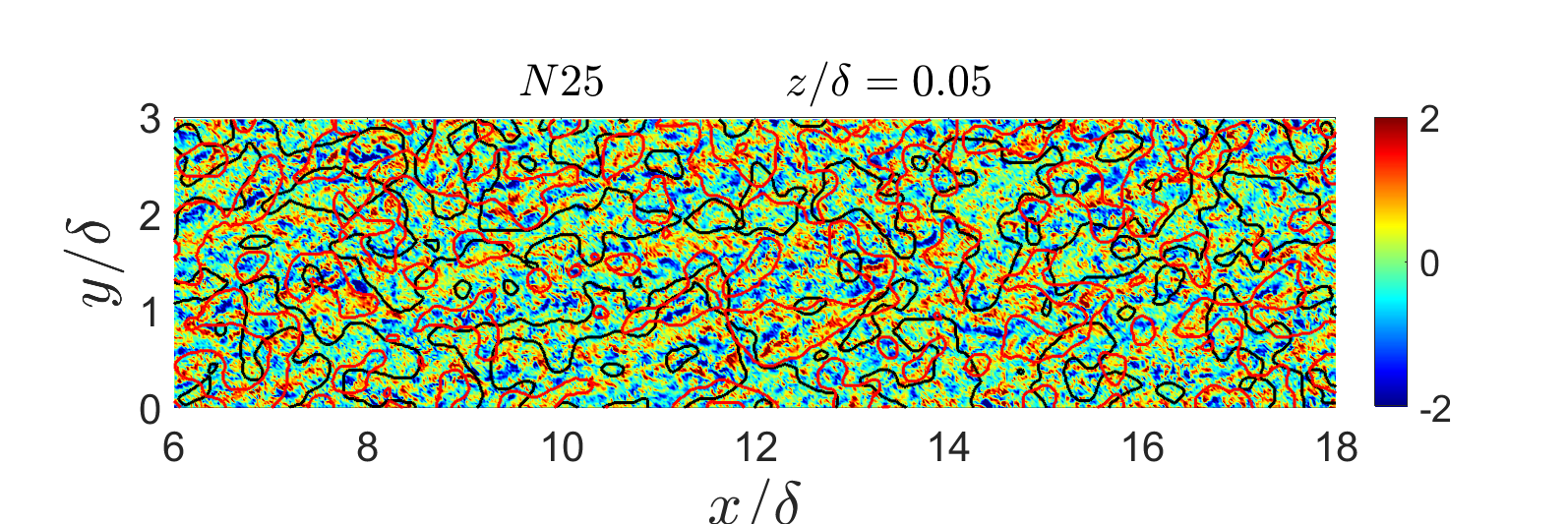}
            };
    \node[anchor=north west,
        xshift=-2mm,yshift=-2mm] at (image.north west) {{\rmfamily\fontsize{12}{13}\fontseries{l}\selectfont(a)}};
        \end{tikzpicture}}
    \subfloat[\label{fig9b}]{
        \begin{tikzpicture}
        \node[anchor=north west, inner sep=0] (image) at (0,0) {
    \includegraphics[width=0.48\textwidth]{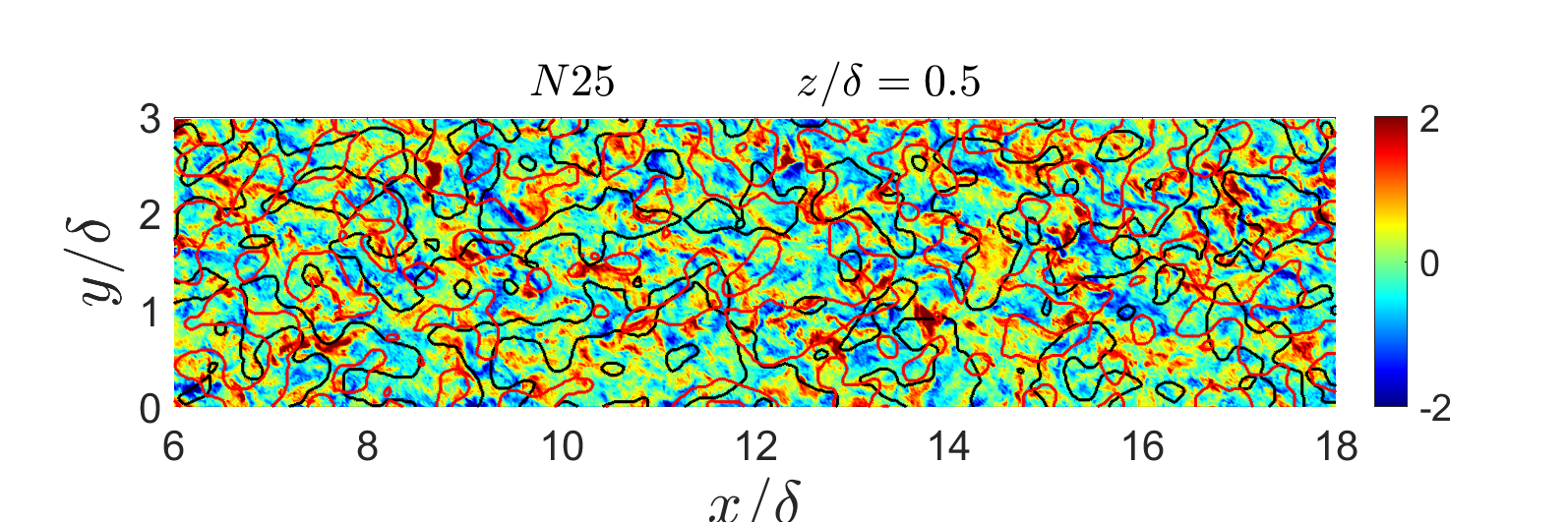}
            };
    \node[anchor=north west,
        xshift=-2mm,yshift=-2mm] at (image.north west) {{\rmfamily\fontsize{12}{13}\fontseries{l}\selectfont(b)}};
        \end{tikzpicture}}

\vspace{-1.35cm} 
\centering
   \subfloat[\label{fig9c}]{
        \begin{tikzpicture}
        \node[anchor=north west, inner sep=0] (image) at (0,0) {
    \includegraphics[width=0.48\textwidth]{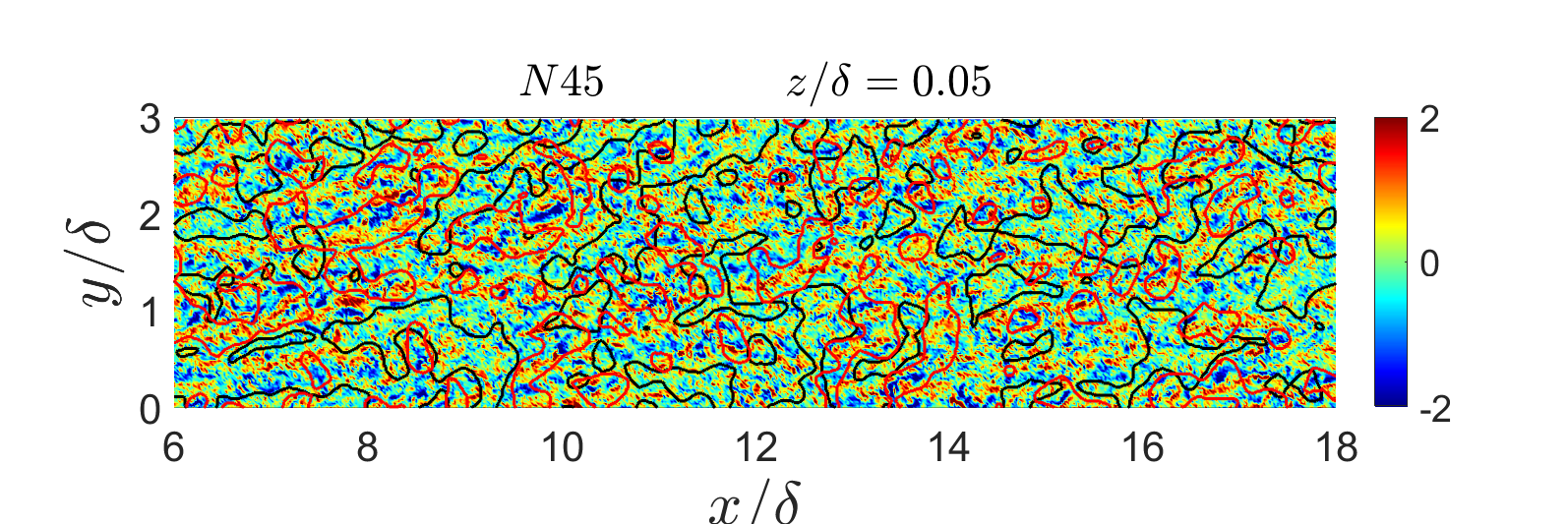}
            };
    \node[anchor=north west,
        xshift=-2mm,yshift=-2mm] at (image.north west) {{\rmfamily\fontsize{12}{13}\fontseries{l}\selectfont(c)}};
        \end{tikzpicture}}
    \subfloat[\label{fig9d}]{
        \begin{tikzpicture}
        \node[anchor=north west, inner sep=0] (image) at (0,0) {
    \includegraphics[width=0.48\textwidth]{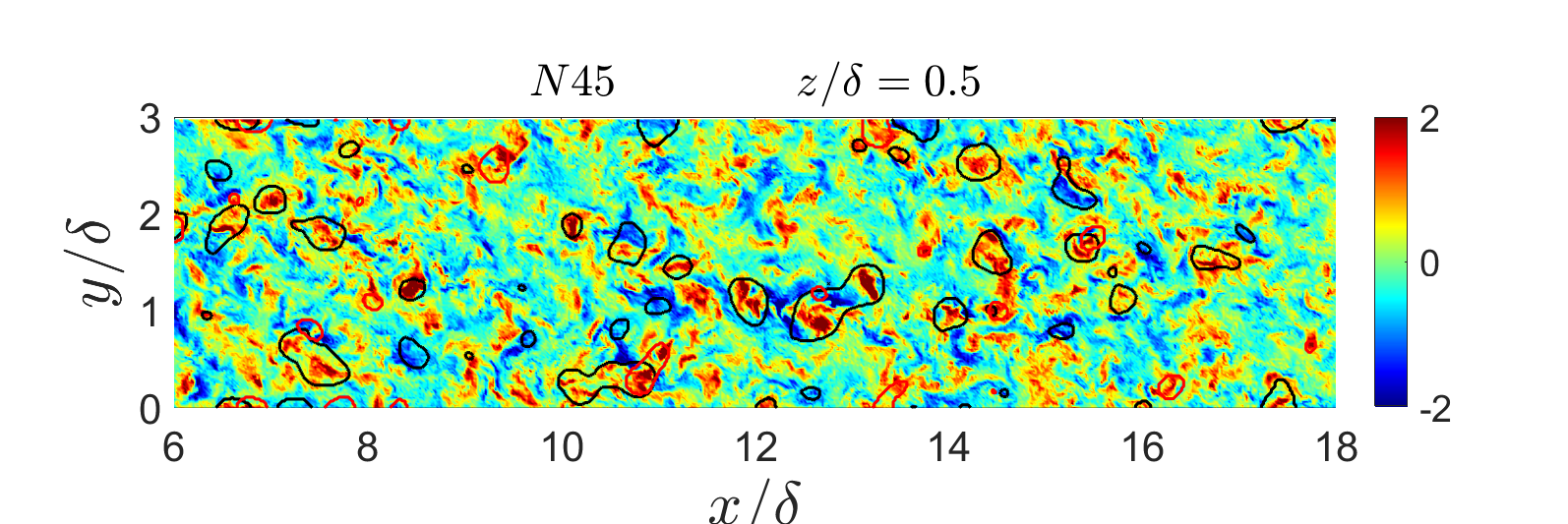}
            };
    \node[anchor=north west,
        xshift=-2mm,yshift=-2mm] at (image.north west) {{\rmfamily\fontsize{12}{13}\fontseries{l}\selectfont(d)}};
        \end{tikzpicture}}
        
   \vspace{-1.35cm}     
   \centering
   \subfloat[\label{fig9e}]{
        \begin{tikzpicture}
        \node[anchor=north west, inner sep=0] (image) at (0,0) {
    \includegraphics[width=0.48\textwidth]{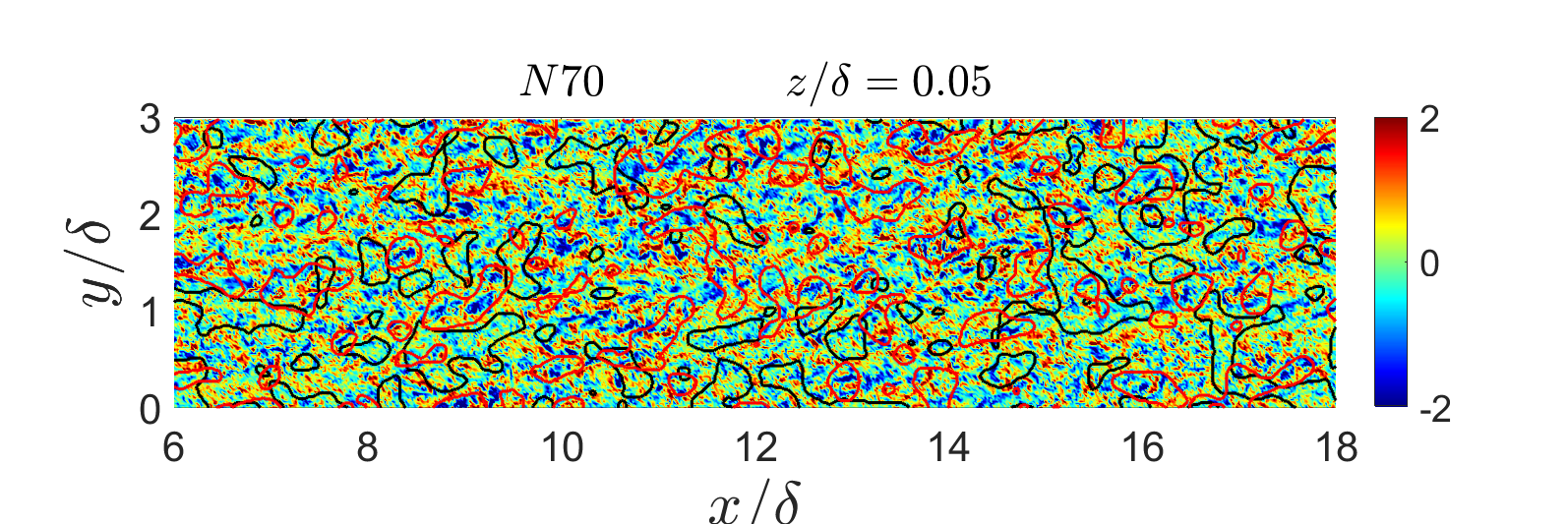}
            };
    \node[anchor=north west,
        xshift=-2mm,yshift=-2mm] at (image.north west) {{\rmfamily\fontsize{12}{13}\fontseries{l}\selectfont(e)}};
        \end{tikzpicture}}
    \subfloat[\label{fig9f}]{
        \begin{tikzpicture}
        \node[anchor=north west, inner sep=0] (image) at (0,0) {
    \includegraphics[width=0.48\textwidth]{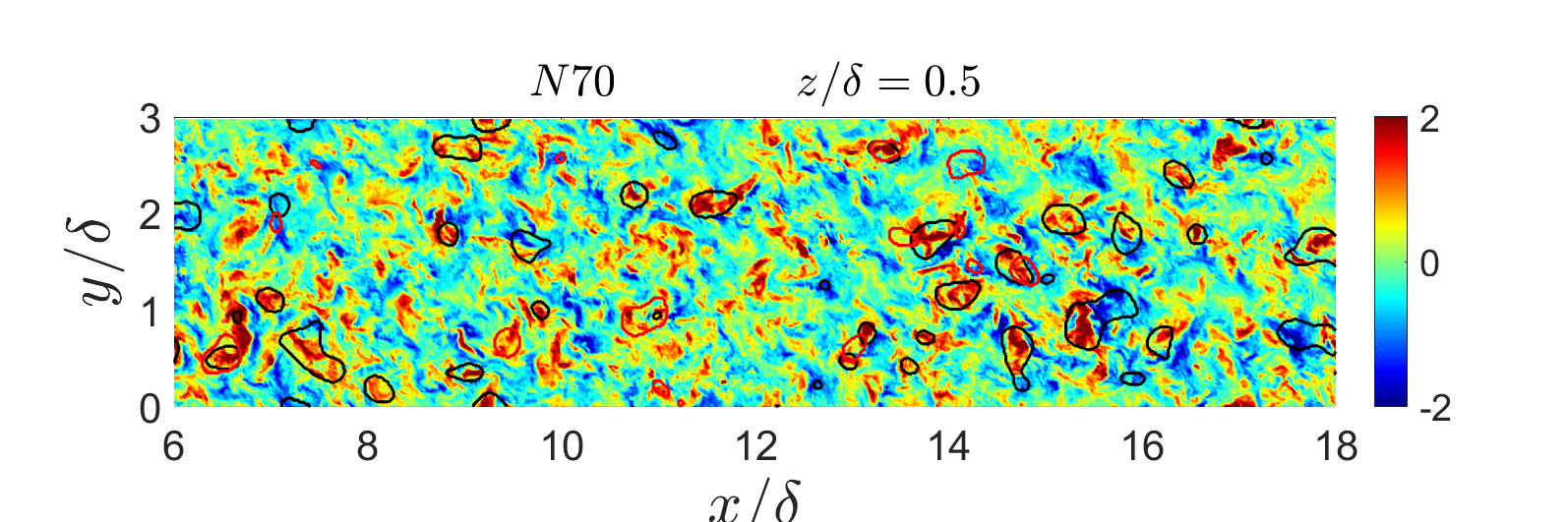}
            };
    \node[anchor=north west,
        xshift=-2mm,yshift=-2mm] at (image.north west) {{\rmfamily\fontsize{12}{13}\fontseries{l}\selectfont(f)}};
        \end{tikzpicture}}
        
 \vspace{-1.35cm}  
    \centering
   \subfloat[\label{fig9g}]{
        \begin{tikzpicture}
        \node[anchor=north west, inner sep=0] (image) at (0,0) {
    \includegraphics[width=0.48\textwidth]{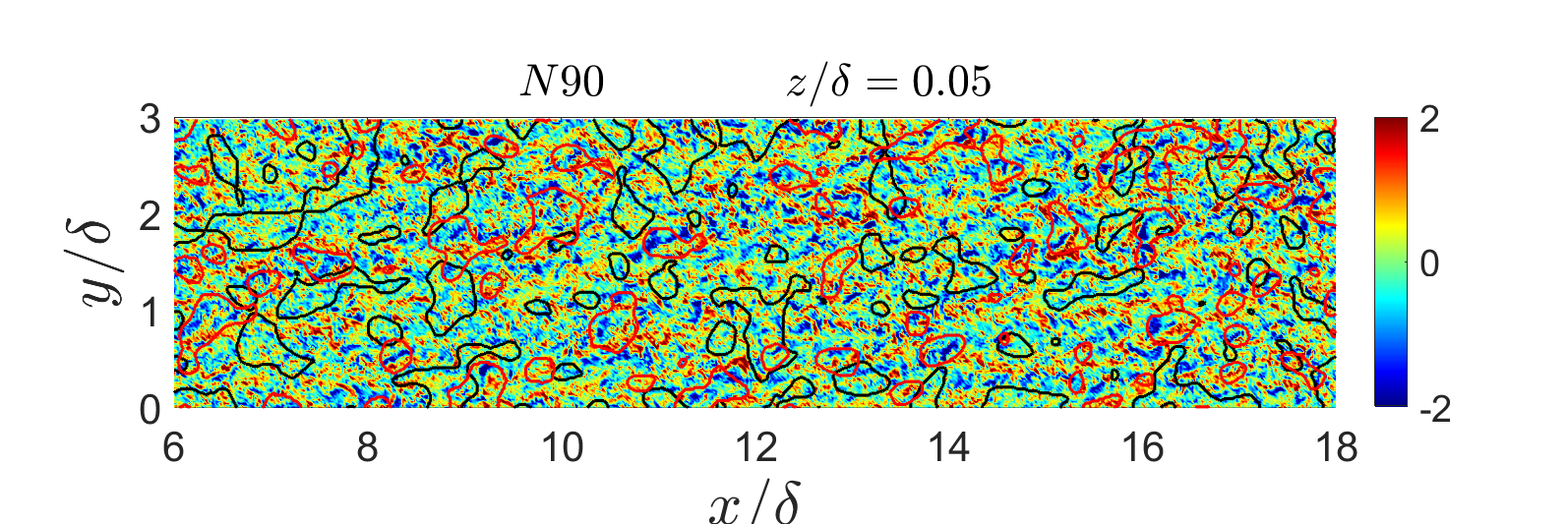}
            };
    \node[anchor=north west,
        xshift=-2mm,yshift=-2mm] at (image.north west) {{\rmfamily\fontsize{12}{13}\fontseries{l}\selectfont(g)}};
        \end{tikzpicture}}
    \subfloat[\label{fig9h}]{
        \begin{tikzpicture}
        \node[anchor=north west, inner sep=0] (image) at (0,0) {
    \includegraphics[width=0.48\textwidth]{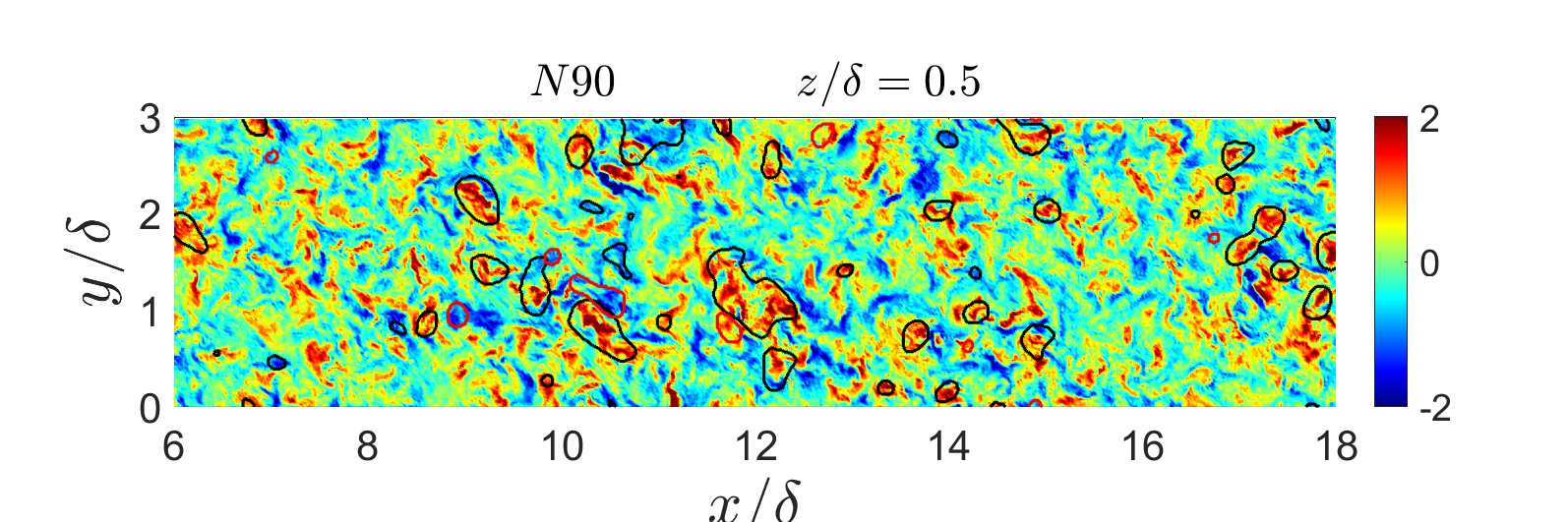}
            };
    \node[anchor=north west,
        xshift=-2mm,yshift=-2mm] at (image.north west) {{\rmfamily\fontsize{12}{13}\fontseries{l}\selectfont(h)}};
        \end{tikzpicture}}

 \vspace{-1.35cm}  
    \centering
   \subfloat[\label{fig9i}]{
        \begin{tikzpicture}
        \node[anchor=north west, inner sep=0] (image) at (0,0) {
    \includegraphics[width=0.48\textwidth]{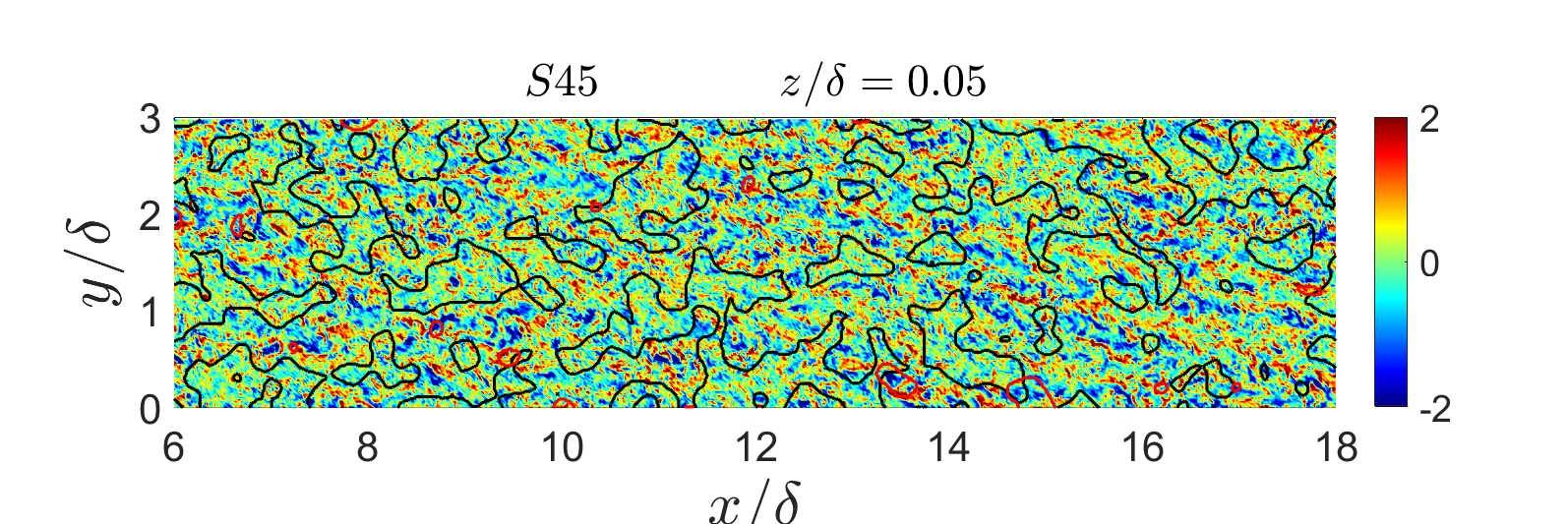}
            };
    \node[anchor=north west,
        xshift=-2mm,yshift=-2mm] at (image.north west) {{\rmfamily\fontsize{12}{13}\fontseries{l}\selectfont(i)}};
        \end{tikzpicture}}
    \subfloat[\label{fig9j}]{
        \begin{tikzpicture}
        \node[anchor=north west, inner sep=0] (image) at (0,0) {
    \includegraphics[width=0.48\textwidth]{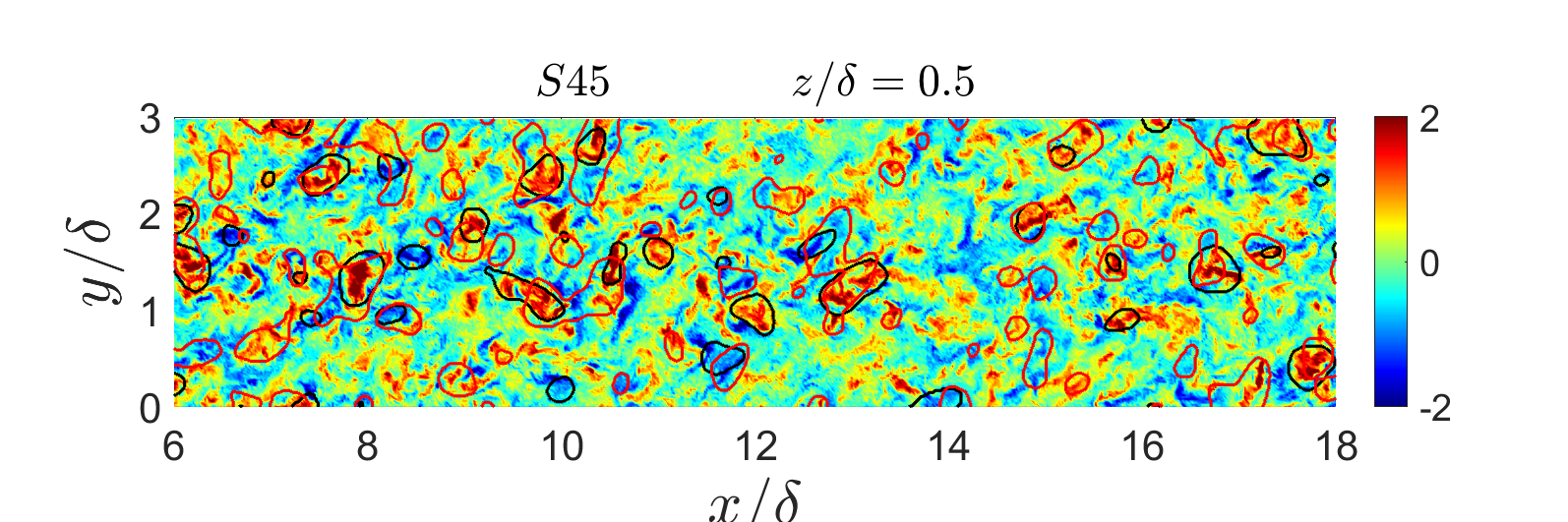}
            };
    \node[anchor=north west,
        xshift=-2mm,yshift=-2mm] at (image.north west) {{\rmfamily\fontsize{12}{13}\fontseries{l}\selectfont(j)}};
        \end{tikzpicture}}

 \vspace{-1.35cm}  
    \centering
   \subfloat[\label{fig9k}]{
        \begin{tikzpicture}
        \node[anchor=north west, inner sep=0] (image) at (0,0) {
    \includegraphics[width=0.48\textwidth]{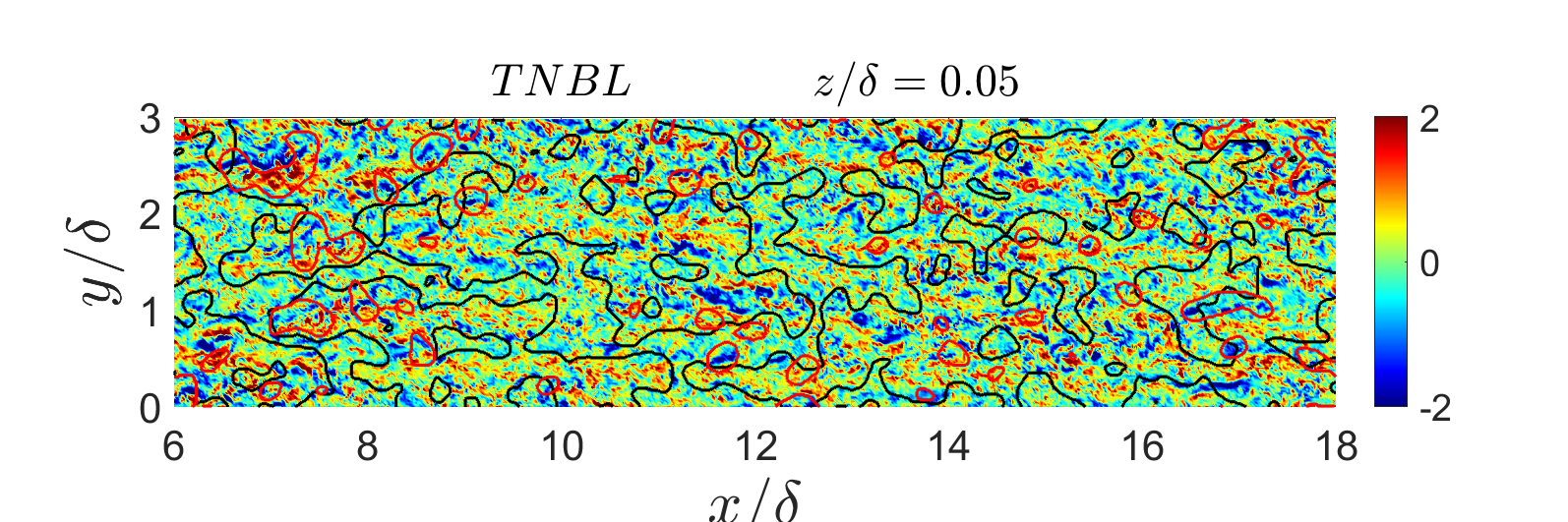}
            };
    \node[anchor=north west,
        xshift=-2mm,yshift=-2mm] at (image.north west) {{\rmfamily\fontsize{12}{13}\fontseries{l}\selectfont(k)}};
        \end{tikzpicture}}
    \subfloat[\label{fig9l}]{
        \begin{tikzpicture}
        \node[anchor=north west, inner sep=0] (image) at (0,0) {
    \includegraphics[width=0.48\textwidth]{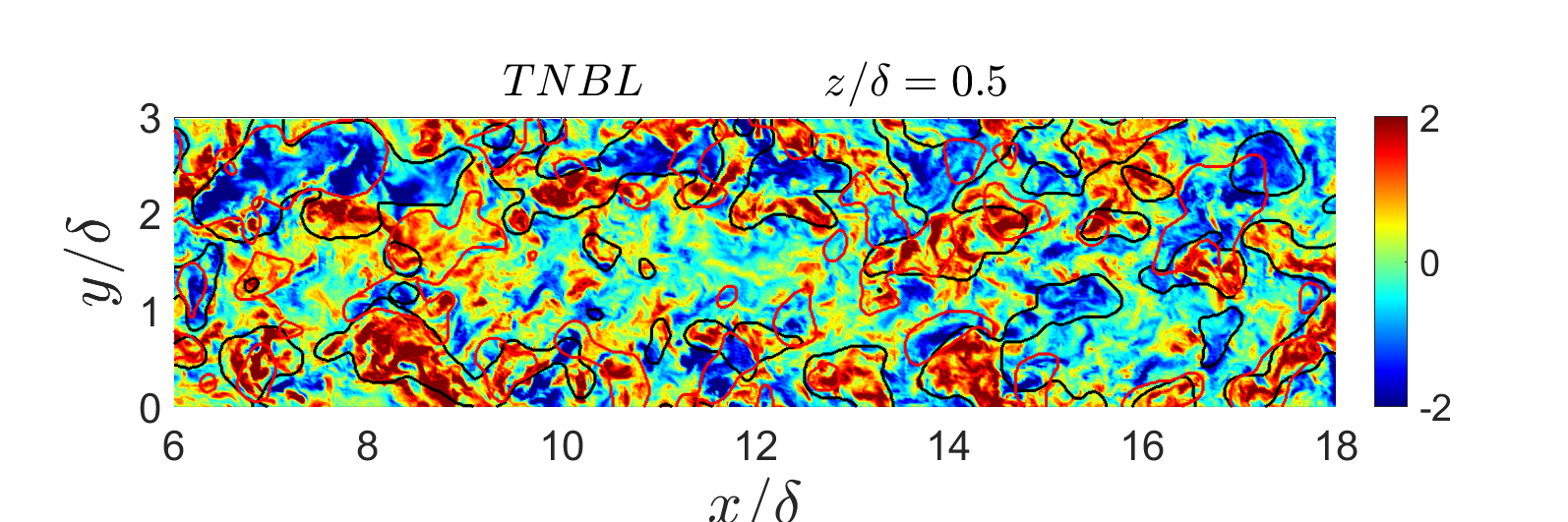}
            };
    \node[anchor=north west,
        xshift=-2mm,yshift=-2mm] at (image.north west) {{\rmfamily\fontsize{12}{13}\fontseries{l}\selectfont(l)}};
        \end{tikzpicture}}
    \caption{Spatial distributions of vertical flow velocity fluctuations at $z/\delta=0.05$ (left column) and 0.5 (right column) in the CNBLs at different latitudes and the TNBL. {The black contour lines represent $-u'w'=0.6$, and the red ones represent $-v'w'=0.3$.}}
    \label{fig9}
\end{figure}

Fig.~\ref{fig9} shows the spatial distributions of the vertical flow velocity fluctuations in the CNBLs at different latitudes and the TNBL. Compared to the longitudinal and lateral velocities, the length scales of the vertical velocity structures are much smaller. Despite these differences, one can still roughly observe the deflection of vertical velocity structures at $z/\delta=0.05$ in the CNBLs. {The characteristics of the deflection resemble those of the longitudinal and lateral velocities, \emph{i.e.} an anticlockwise deflection at $z/\delta=0.05$ in the Northern Hemisphere, which is opposite in the Southern Hemisphere.} 
Moreover, we also show the contour lines of $-u'w'=0.6$ (black) and $-v'w'=0.3$ (red) as they represent Reynolds-shear-stress events manifested by downdraft/updraft motions (or the so-called sweeping/ejection motions) \citep{alcayagaLargeScaleCoherentTurbulence2022}. It is found that these events always coincide with strong $w'$ events, signifying their evident role in the vertical momentum transport. {It should be noted that the Reynolds-shear-stress events themselves cannot distinguish updrafts/downdrafts unless the sign of $w'$ is known ($w'>0$ is an updraft and $w'<0$ is a downdraft).}
As latitude increases, the stronger Coriolis force reduces the occurrence and size of the $u'w'$ and $v'w'$ events. 

\begin{figure}
\centering
    \subfloat[\label{u0.05g12}]{
        \begin{tikzpicture}
        \node[anchor=north west, inner sep=0] (image) at (0,0) {
    \includegraphics[width=0.48\textwidth]{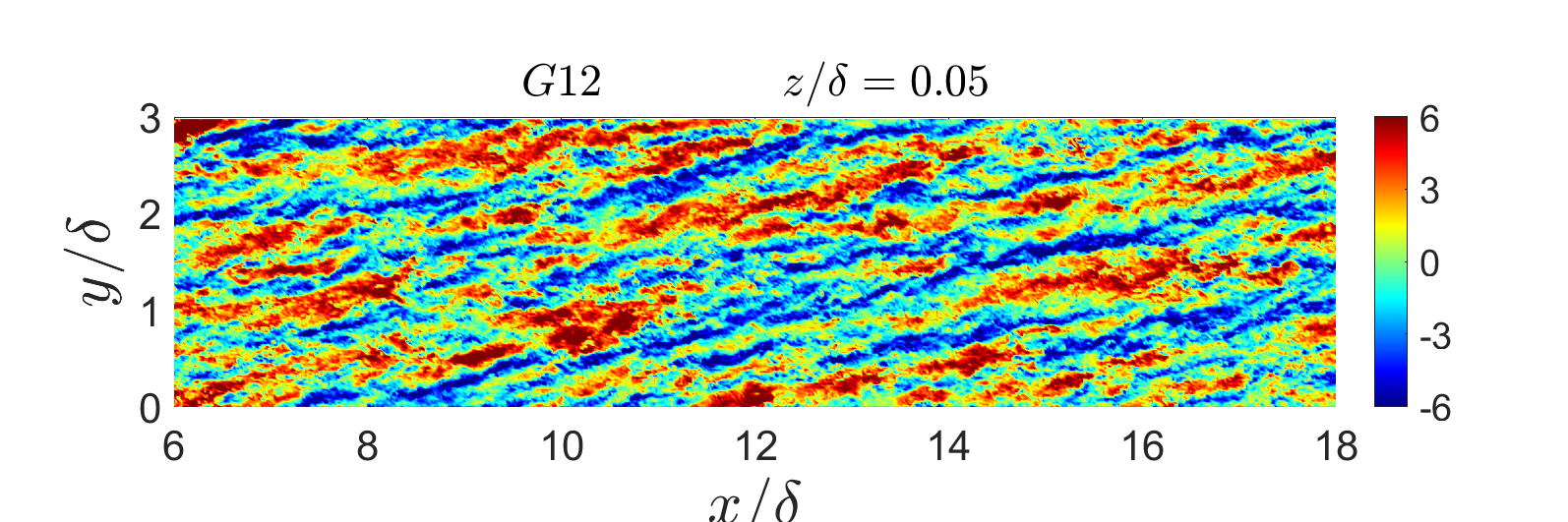}
            };
    \node[anchor=north west,
        xshift=-2mm,yshift=-2mm] at (image.north west) {{\rmfamily\fontsize{12}{13}\fontseries{l}\selectfont(a)}};
        \end{tikzpicture}}
    \subfloat[\label{u0.05g16}]{
        \begin{tikzpicture}
        \node[anchor=north west, inner sep=0] (image) at (0,0) {
    \includegraphics[width=0.48\textwidth]{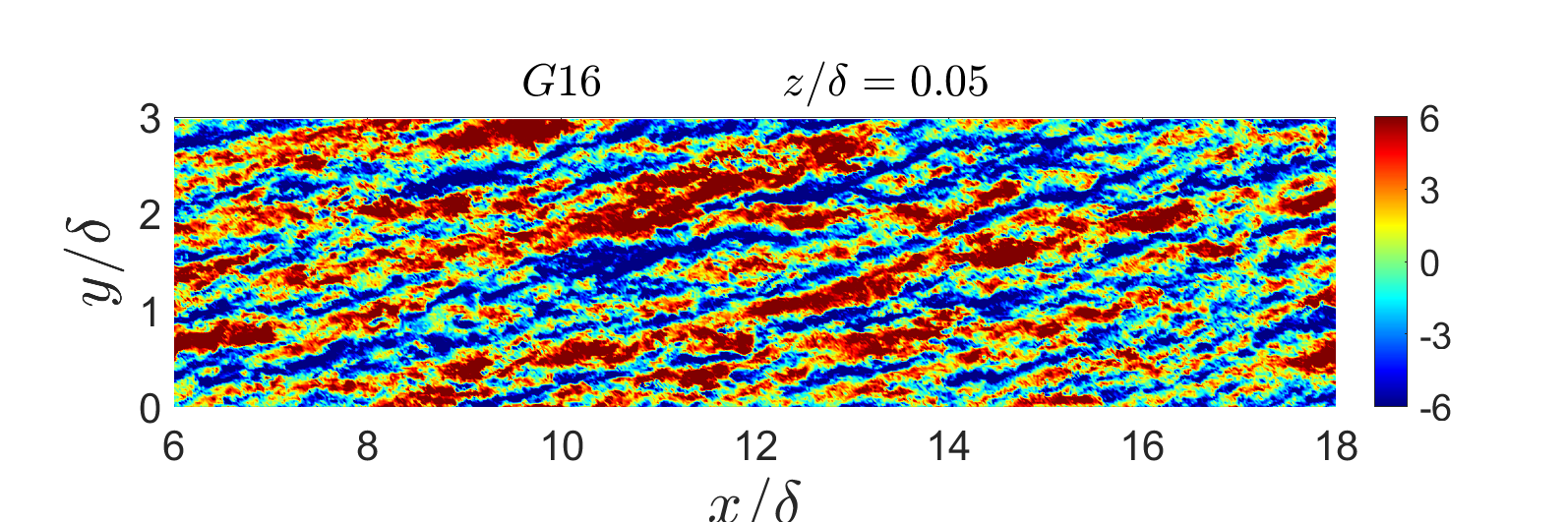}
            };
    \node[anchor=north west,
        xshift=-2mm,yshift=-2mm] at (image.north west) {{\rmfamily\fontsize{12}{13}\fontseries{l}\selectfont(b)}};
        \end{tikzpicture}}
        
  \vspace{-1.35cm}      
 \centering
    \subfloat[\label{u0.25g12}]{
        \begin{tikzpicture}
        \node[anchor=north west, inner sep=0] (image) at (0,0) {
    \includegraphics[width=0.48\textwidth]{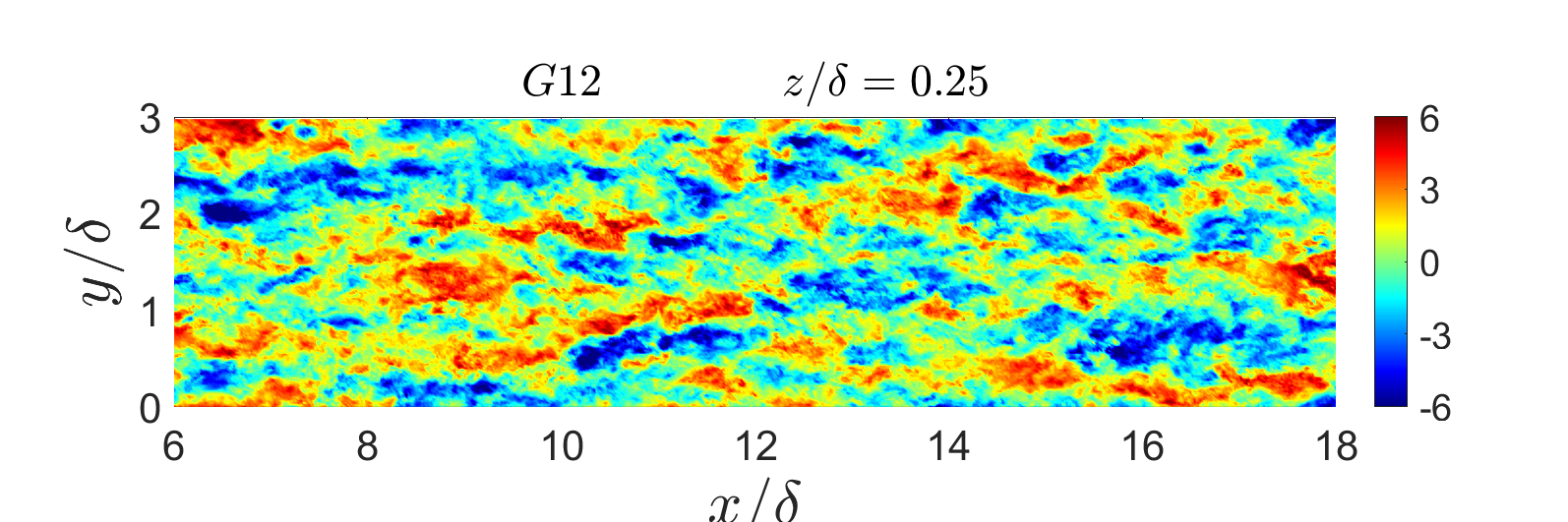}
            };
    \node[anchor=north west,
        xshift=-2mm,yshift=-2mm] at (image.north west) {{\rmfamily\fontsize{12}{13}\fontseries{l}\selectfont(c)}};
        \end{tikzpicture}}
    \subfloat[\label{u0.25g16}]{
        \begin{tikzpicture}
        \node[anchor=north west, inner sep=0] (image) at (0,0) {
    \includegraphics[width=0.48\textwidth]{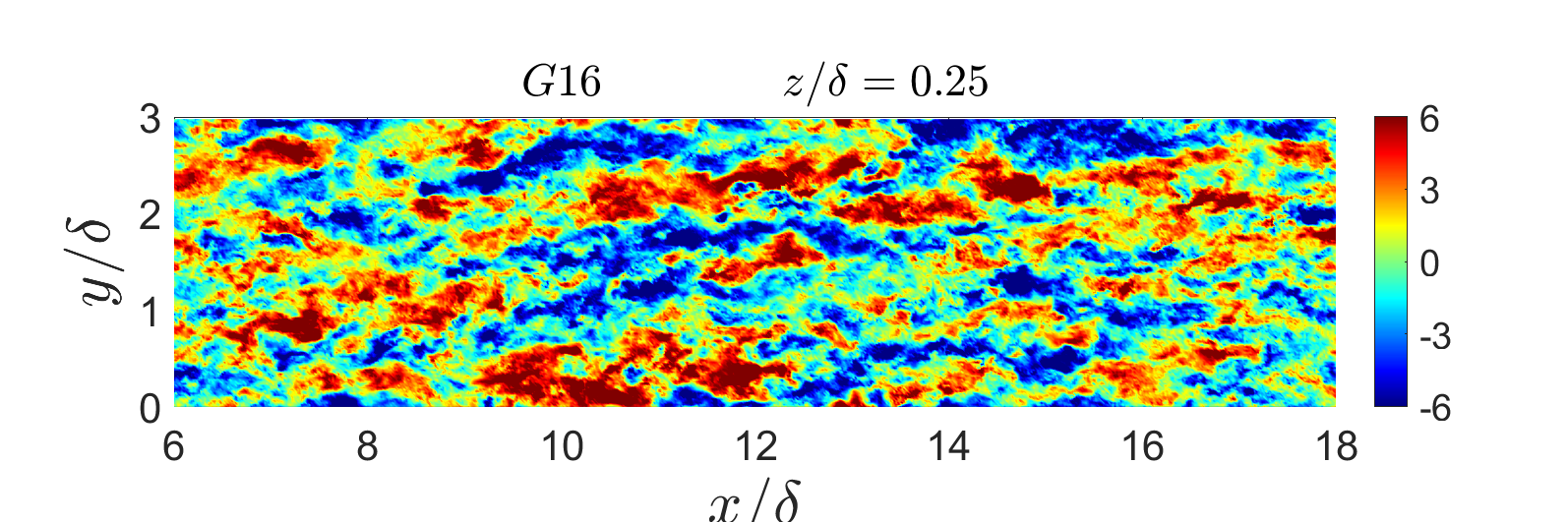}
            };
    \node[anchor=north west,
        xshift=-2mm,yshift=-2mm] at (image.north west) {{\rmfamily\fontsize{12}{13}\fontseries{l}\selectfont(d)}};
        \end{tikzpicture}}
        
    \vspace{-1.35cm}         
    \centering
    \subfloat[\label{u0.5g12}]{
        \begin{tikzpicture}
        \node[anchor=north west, inner sep=0] (image) at (0,0) {
    \includegraphics[width=0.48\textwidth]{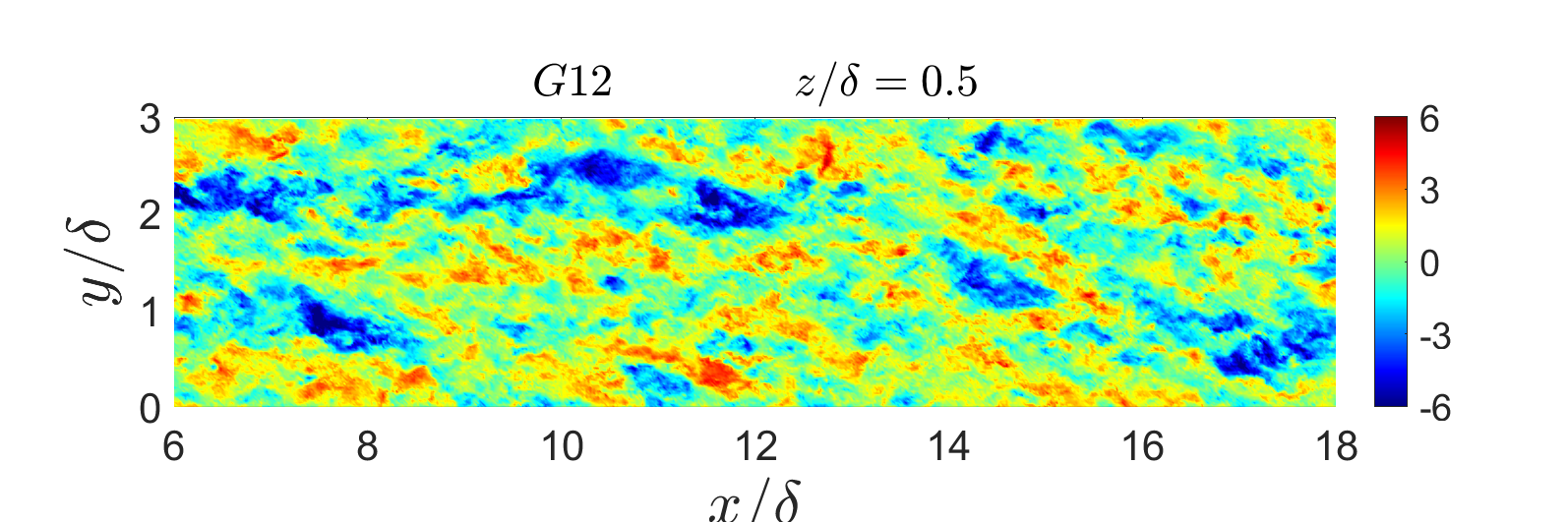}
            };
    \node[anchor=north west,
        xshift=-2mm,yshift=-2mm] at (image.north west) {{\rmfamily\fontsize{12}{13}\fontseries{l}\selectfont(e)}};
        \end{tikzpicture}}
    \subfloat[\label{u0.5g16}]{
        \begin{tikzpicture}
        \node[anchor=north west, inner sep=0] (image) at (0,0) {
    \includegraphics[width=0.48\textwidth]{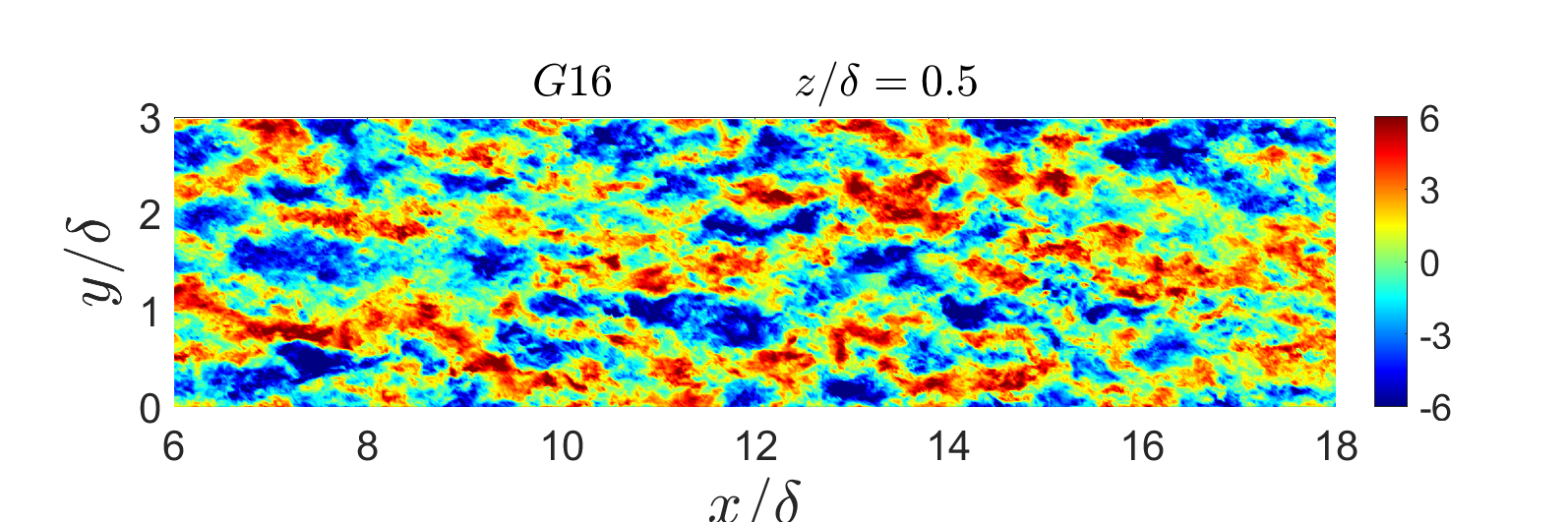}
            };
    \node[anchor=north west,
        xshift=-2mm,yshift=-2mm] at (image.north west) {{\rmfamily\fontsize{12}{13}\fontseries{l}\selectfont(f)}};
        \end{tikzpicture}}
    \caption{Spatial distributions of longitudinal flow velocity fluctuations in the CNBLs with different geostrophic wind speeds ($U_g = 12$ m/s and 16 m/s).}
    \label{fig:G streamwise fluctuation}
\end{figure}

\begin{figure}[!htb]
\centering
    \subfloat[\label{v0.05g12}]{
        \begin{tikzpicture}
        \node[anchor=north west, inner sep=0] (image) at (0,0) {
    \includegraphics[width=0.48\textwidth]{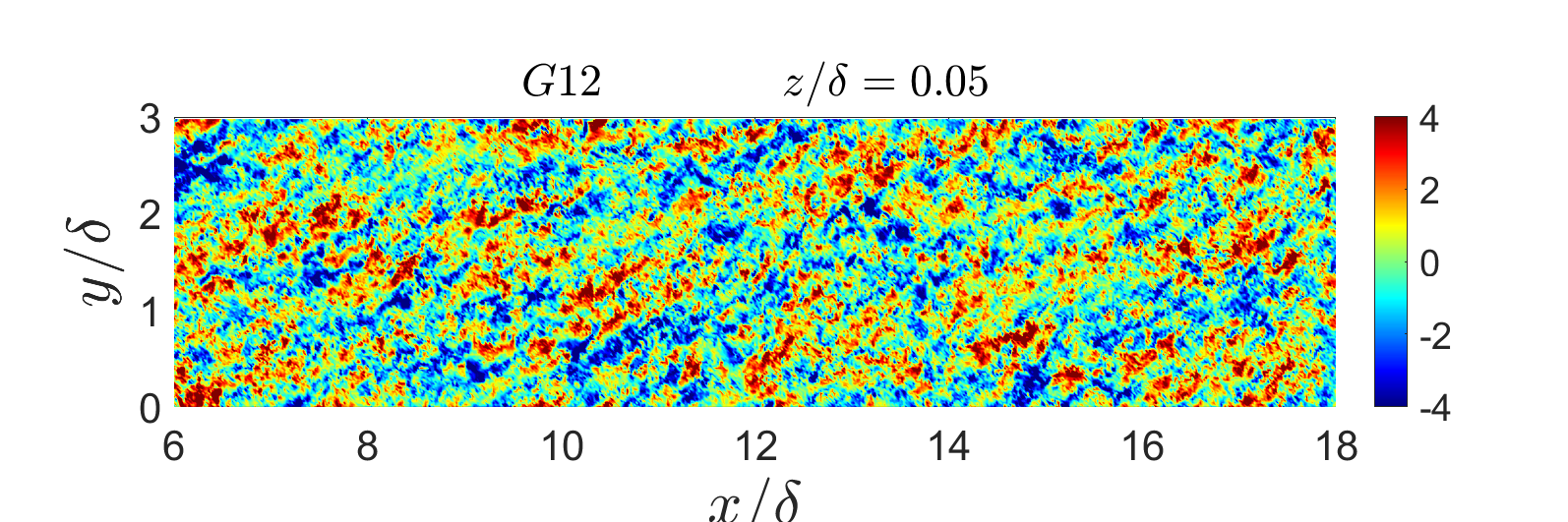}
            };
    \node[anchor=north west,
        xshift=-2mm,yshift=-2mm] at (image.north west) {{\rmfamily\fontsize{12}{13}\fontseries{l}\selectfont(a)}};
        \end{tikzpicture}}
    \subfloat[\label{v0.05g16}]{
        \begin{tikzpicture}
        \node[anchor=north west, inner sep=0] (image) at (0,0) {
    \includegraphics[width=0.48\textwidth]{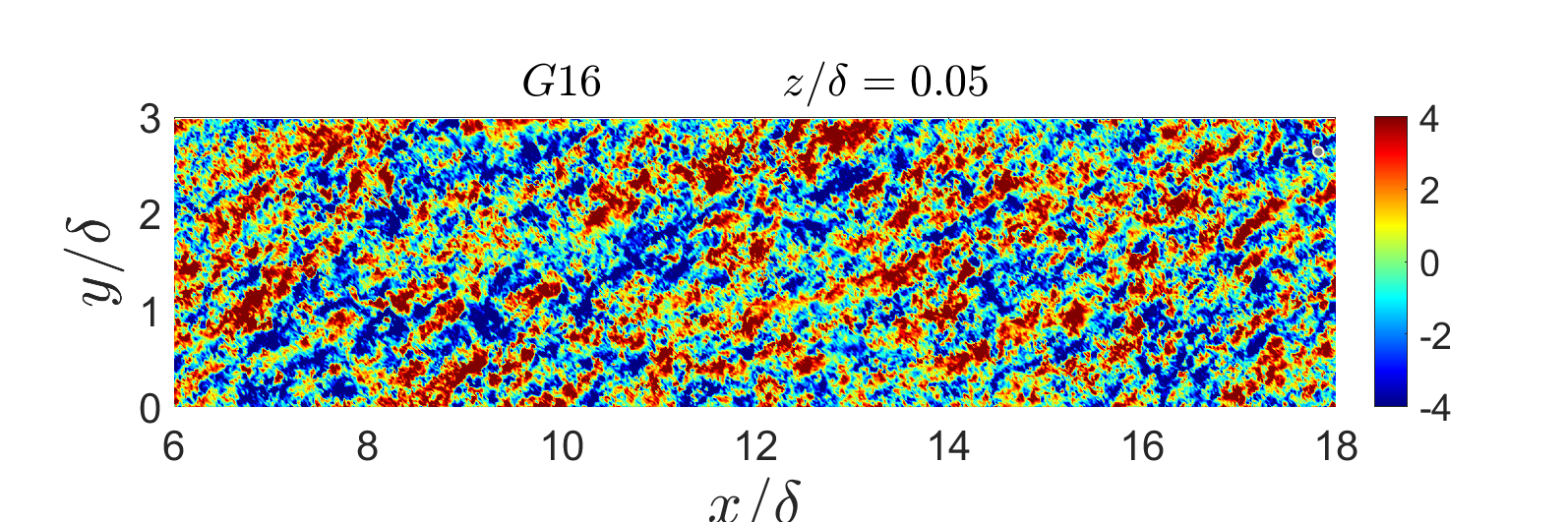}
            };
    \node[anchor=north west,
        xshift=-2mm,yshift=-2mm] at (image.north west) {{\rmfamily\fontsize{12}{13}\fontseries{l}\selectfont(b)}};
        \end{tikzpicture}}
        
  \vspace{-1.35cm}      
 \centering
    \subfloat[\label{v0.25g12}]{
        \begin{tikzpicture}
        \node[anchor=north west, inner sep=0] (image) at (0,0) {
    \includegraphics[width=0.48\textwidth]{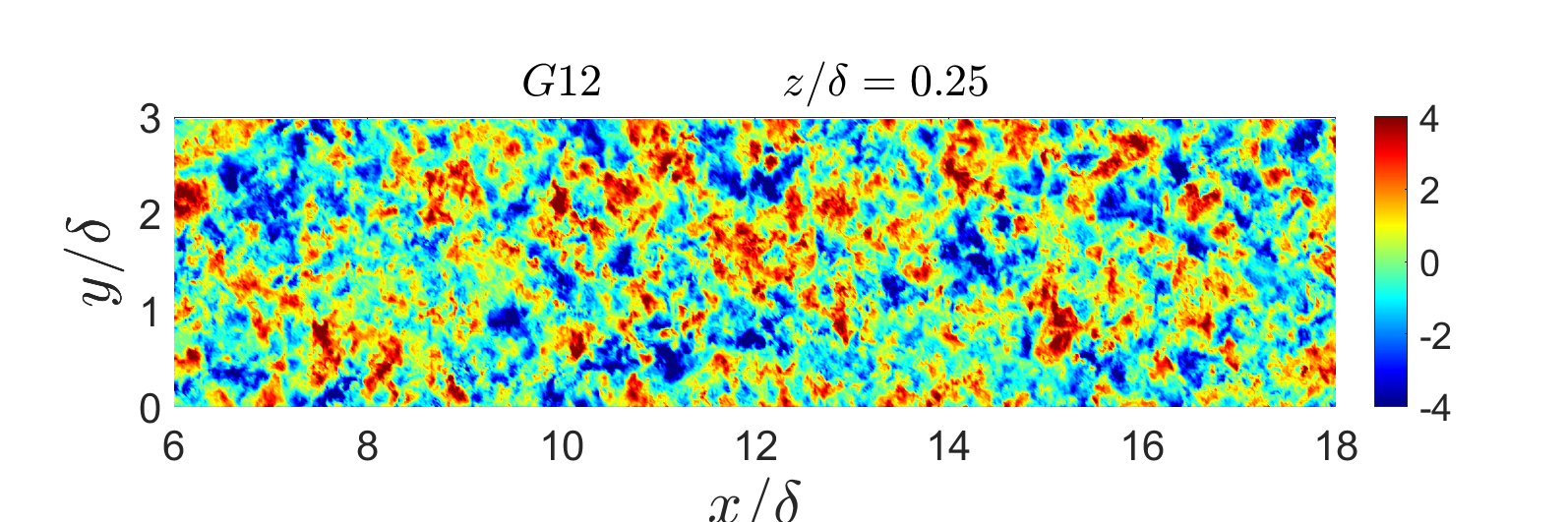}
            };
    \node[anchor=north west,
        xshift=-2mm,yshift=-2mm] at (image.north west) {{\rmfamily\fontsize{12}{13}\fontseries{l}\selectfont(c)}};
        \end{tikzpicture}}
    \subfloat[\label{v0.25g16}]{
        \begin{tikzpicture}
        \node[anchor=north west, inner sep=0] (image) at (0,0) {
    \includegraphics[width=0.48\textwidth]{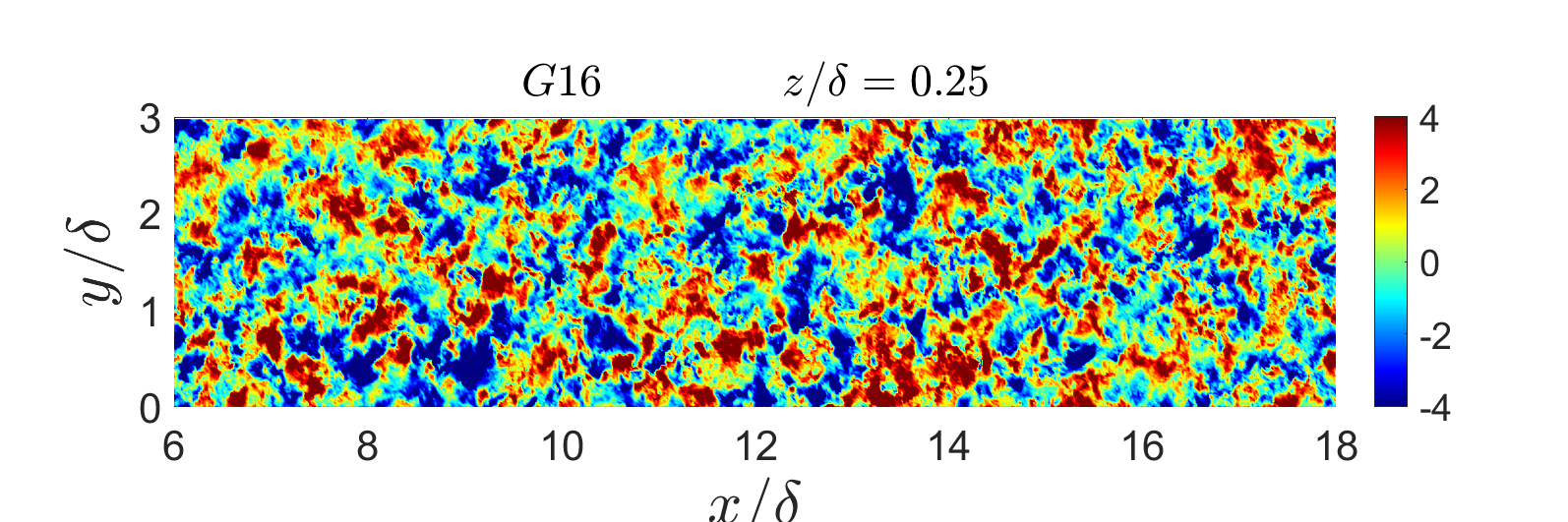}
            };
    \node[anchor=north west,
        xshift=-2mm,yshift=-2mm] at (image.north west) {{\rmfamily\fontsize{12}{13}\fontseries{l}\selectfont(d)}};
        \end{tikzpicture}}
        
    \vspace{-1.35cm}         
    \centering
    \subfloat[\label{v0.5g12}]{
        \begin{tikzpicture}
        \node[anchor=north west, inner sep=0] (image) at (0,0) {
    \includegraphics[width=0.48\textwidth]{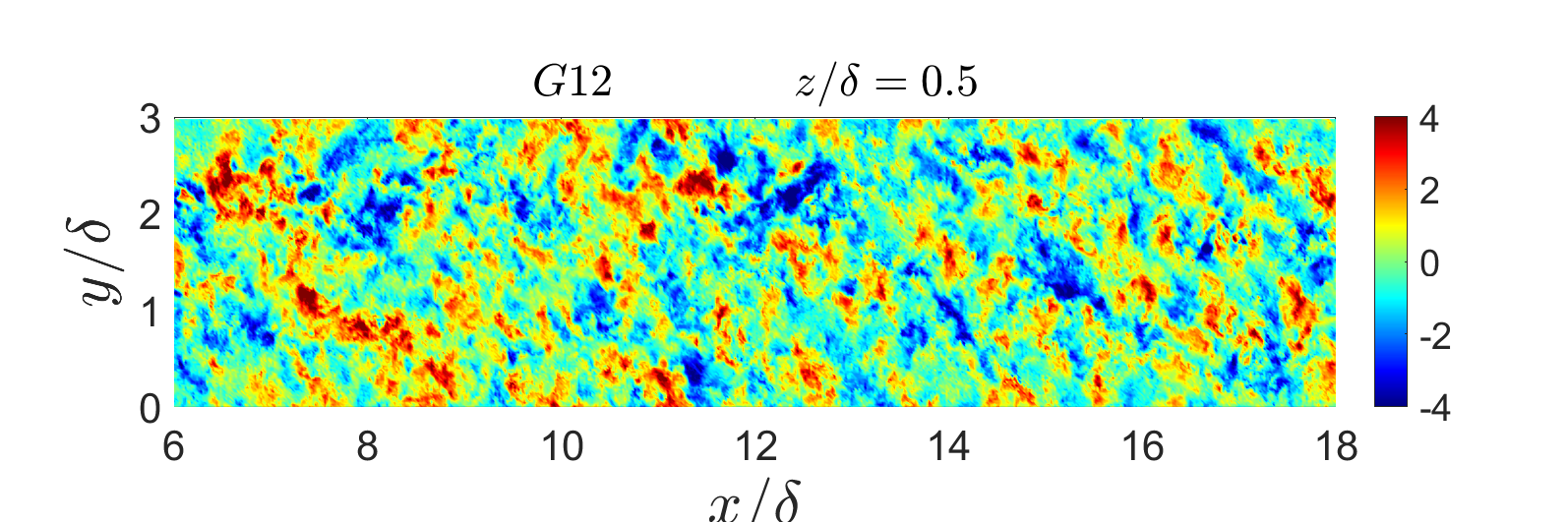}
            };
    \node[anchor=north west,
        xshift=-2mm,yshift=-2mm] at (image.north west) {{\rmfamily\fontsize{12}{13}\fontseries{l}\selectfont(e)}};
        \end{tikzpicture}}
    \subfloat[\label{v0.5g16}]{
        \begin{tikzpicture}
        \node[anchor=north west, inner sep=0] (image) at (0,0) {
    \includegraphics[width=0.48\textwidth]{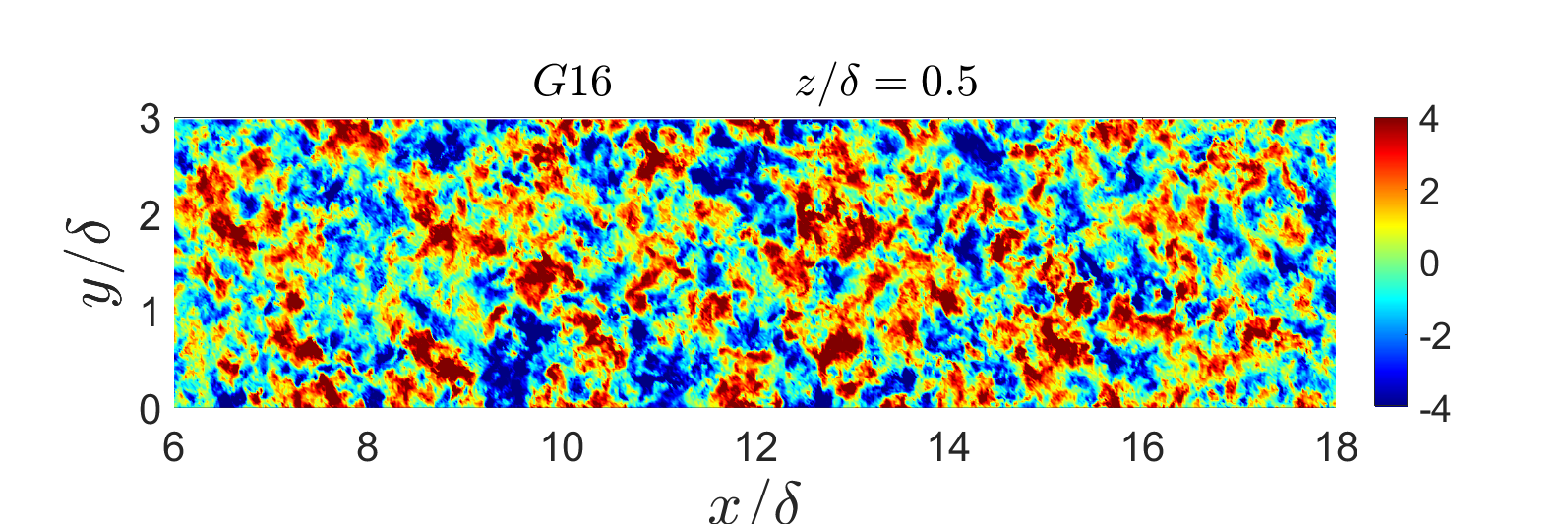}
            };
    \node[anchor=north west,
        xshift=-2mm,yshift=-2mm] at (image.north west) {{\rmfamily\fontsize{12}{13}\fontseries{l}\selectfont(f)}};
        \end{tikzpicture}}
    \caption{Spatial distributions of lateral flow velocity fluctuations in the CNBLs with different geostrophic wind speeds ($U_g = 12$ m/s and 16 m/s).}
    \label{fig:G spanwise fluctuation}
\end{figure}

\begin{figure}[!htb]
\centering
    \subfloat[\label{w0.05g12}]{
        \begin{tikzpicture}
        \node[anchor=north west, inner sep=0] (image) at (0,0) {
    \includegraphics[width=0.48\textwidth]{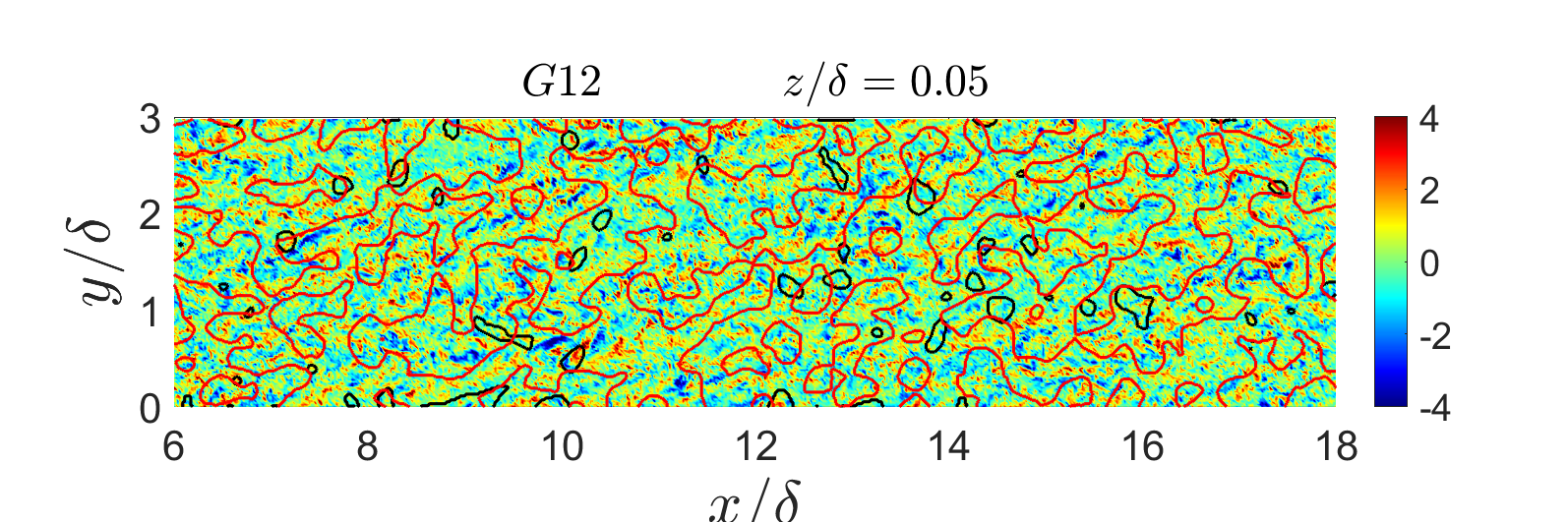}
            };
    \node[anchor=north west,
        xshift=-2mm,yshift=-2mm] at (image.north west) {{\rmfamily\fontsize{12}{13}\fontseries{l}\selectfont(a)}};
        \end{tikzpicture}}
    \subfloat[\label{w0.05g16}]{
        \begin{tikzpicture}
        \node[anchor=north west, inner sep=0] (image) at (0,0) {
    \includegraphics[width=0.48\textwidth]{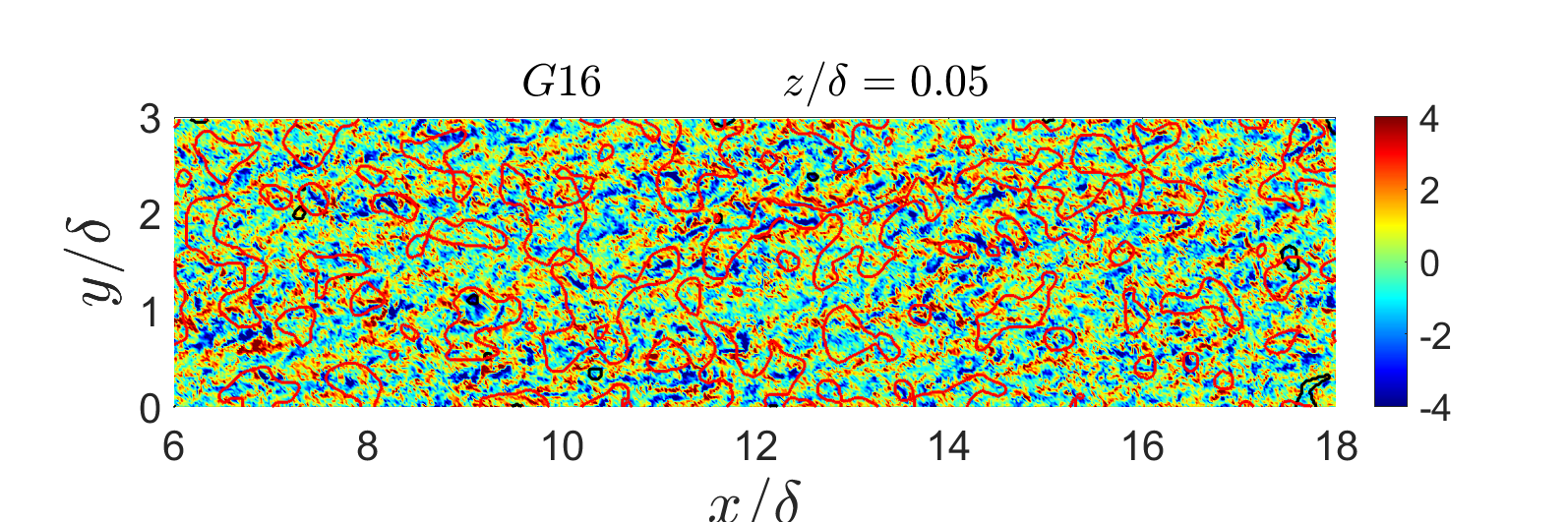}
            };
    \node[anchor=north west,
        xshift=-2mm,yshift=-2mm] at (image.north west) {{\rmfamily\fontsize{12}{13}\fontseries{l}\selectfont(b)}};
        \end{tikzpicture}}
        
  \vspace{-1.35cm}      
 \centering
    \subfloat[\label{w0.25g12}]{
        \begin{tikzpicture}
        \node[anchor=north west, inner sep=0] (image) at (0,0) {
    \includegraphics[width=0.48\textwidth]{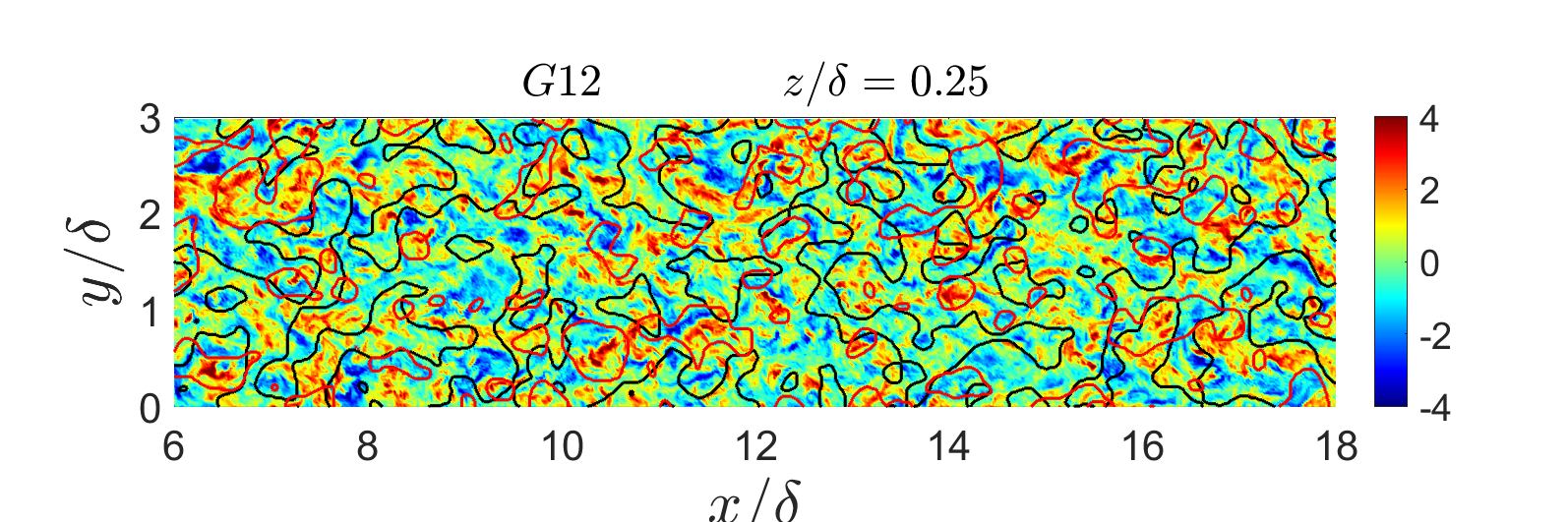}
            };
    \node[anchor=north west,
        xshift=-2mm,yshift=-2mm] at (image.north west) {{\rmfamily\fontsize{12}{13}\fontseries{l}\selectfont(c)}};
        \end{tikzpicture}}
    \subfloat[\label{w0.25g16}]{
        \begin{tikzpicture}
        \node[anchor=north west, inner sep=0] (image) at (0,0) {
    \includegraphics[width=0.48\textwidth]{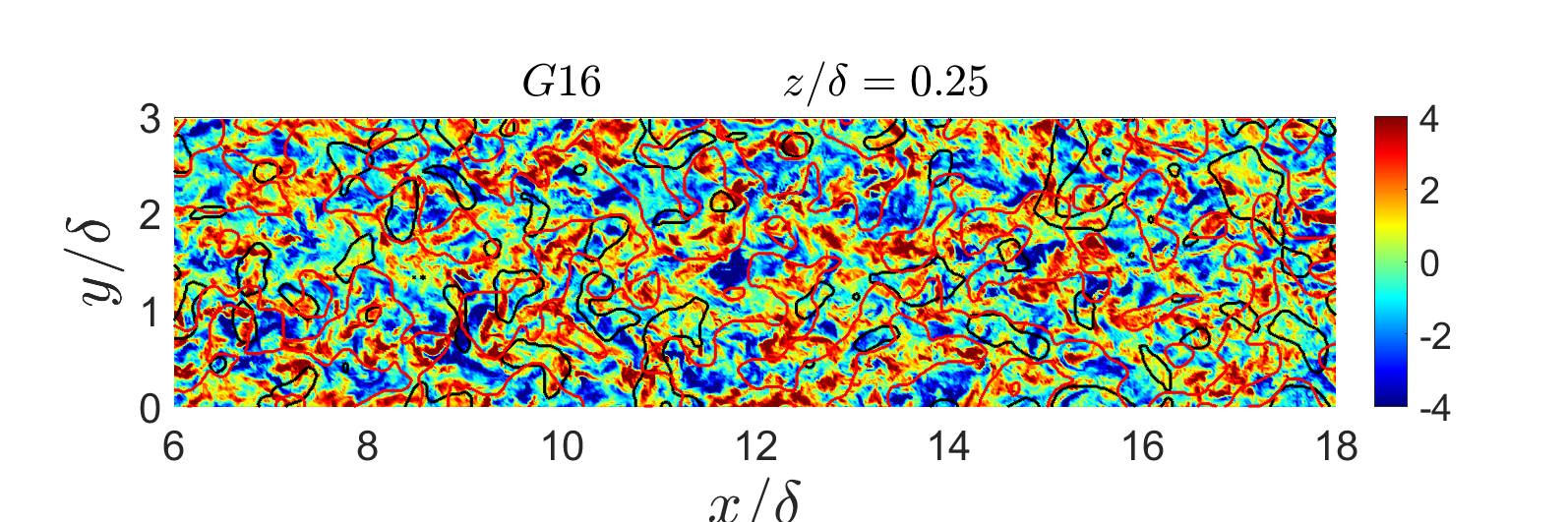}
            };
    \node[anchor=north west,
        xshift=-2mm,yshift=-2mm] at (image.north west) {{\rmfamily\fontsize{12}{13}\fontseries{l}\selectfont(d)}};
        \end{tikzpicture}}
        
    \vspace{-1.35cm}         
    \centering
    \subfloat[\label{w0.5g12}]{
        \begin{tikzpicture}
        \node[anchor=north west, inner sep=0] (image) at (0,0) {
    \includegraphics[width=0.48\textwidth]{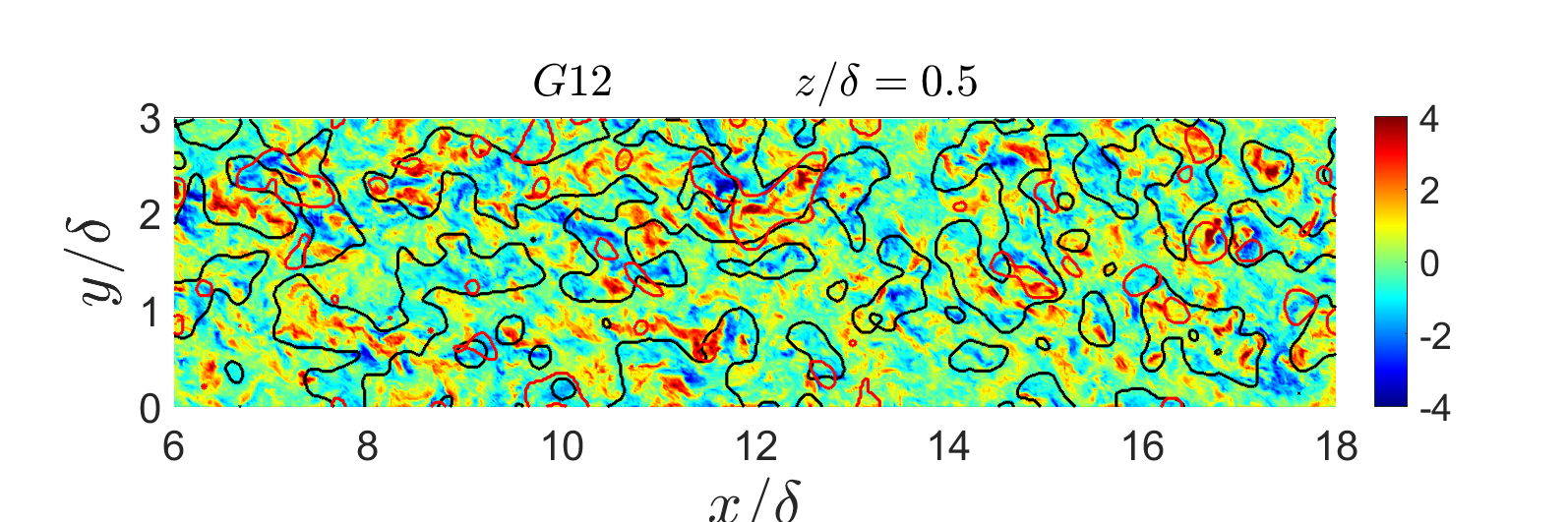}
            };
    \node[anchor=north west,
        xshift=-2mm,yshift=-2mm] at (image.north west) {{\rmfamily\fontsize{12}{13}\fontseries{l}\selectfont(e)}};
        \end{tikzpicture}}
    \subfloat[\label{w0.5g16}]{
        \begin{tikzpicture}
        \node[anchor=north west, inner sep=0] (image) at (0,0) {
    \includegraphics[width=0.48\textwidth]{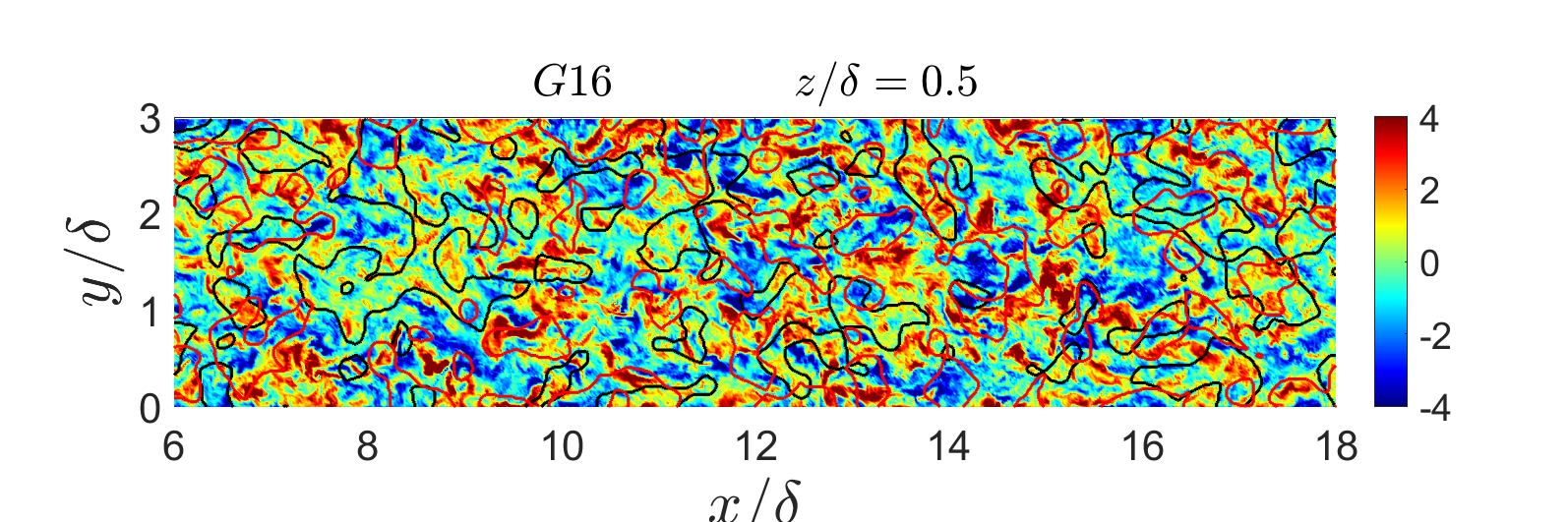}
            };
    \node[anchor=north west,
        xshift=-2mm,yshift=-2mm] at (image.north west) {{\rmfamily\fontsize{12}{13}\fontseries{l}\selectfont(f)}};
        \end{tikzpicture}}
    \caption{Spatial distributions of vertical flow velocity fluctuations in the CNBLs with different geostrophic wind speeds ($U_g = 12$ m/s and 16 m/s). {The black contour lines represent $-u'w'=0.6$, and the red ones represent $-v'w'=0.3$.}}
    \label{fig:G vertical fluctuation}
\end{figure}

The spatial distributions of the longitudinal flow velocity fluctuations with two geostrophic wind speeds ($U_g = 12$ m/s and 16 m/s) are shown in Fig.~\ref{fig:G streamwise fluctuation}. 
It can be seen that as the geostrophic wind speed increases, the flow velocity streaks or patches in the boundary layer are also evidently strengthened, \emph{i.e.}, the longitudinal flow velocity fluctuation intensity increases. 
The deflection of the velocity structures can also be observed.
However, the length scales and deflection angles cannot be clearly identified from the contours, and will be quantified afterwards through premultiplied spectra and correlation maps. 
Figs.~\ref{fig:G spanwise fluctuation} and \ref{fig:G vertical fluctuation} show the spatial distributions of lateral and vertical flow velocity fluctuations, respectively. Similar to the longitudinal velocity, the magnitudes of the lateral and vertical flow velocity structures are also intensified with the increase in geostrophic wind speed.

\begin{figure}
    \centering
     \subfloat[\label{fig12a}]{
        \begin{tikzpicture}
        \node[anchor=north west, inner sep=0] (image) at (0,0) {
    \includegraphics[width=0.45\textwidth]{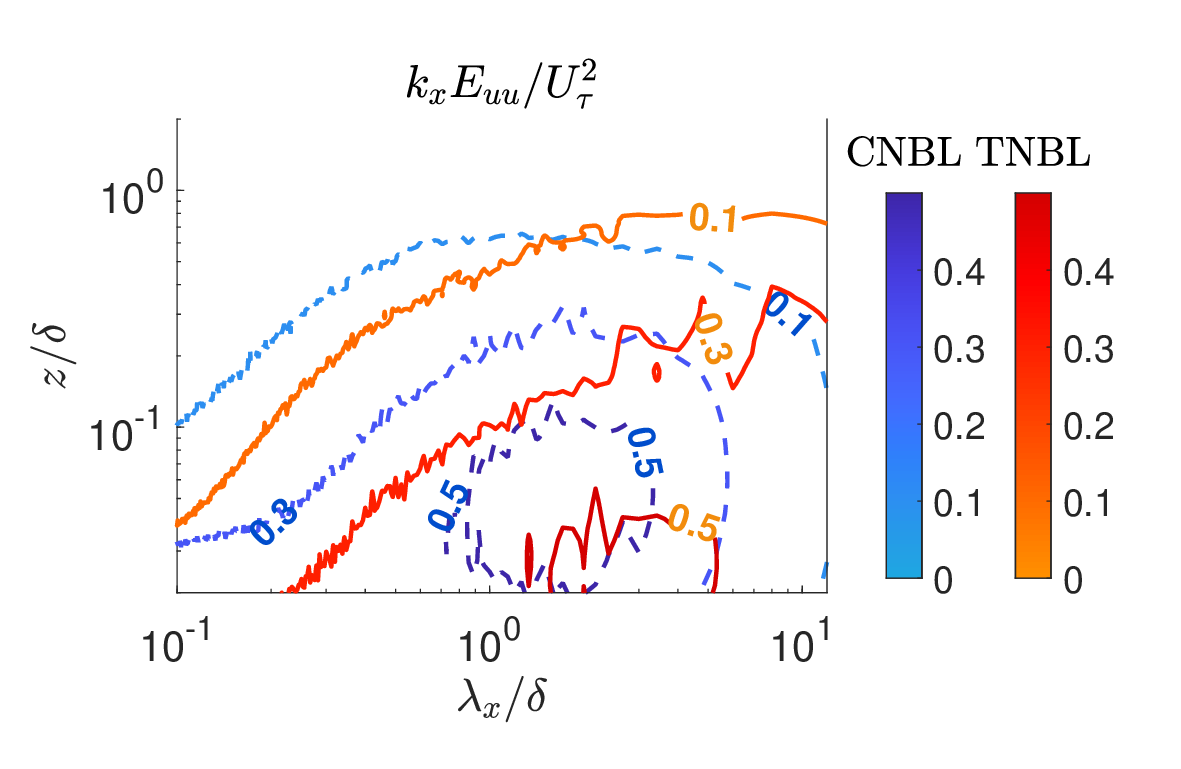}
            };
    \node[anchor=north west,
        xshift=-2mm,yshift=-2mm] at (image.north west) {{\rmfamily\fontsize{12}{13}\fontseries{l}\selectfont(a)}};
        \end{tikzpicture}}
    \subfloat[\label{fig12b}]{
        \begin{tikzpicture}
        \node[anchor=north west, inner sep=0] (image) at (0,0) {
    \includegraphics[width=0.45\textwidth]{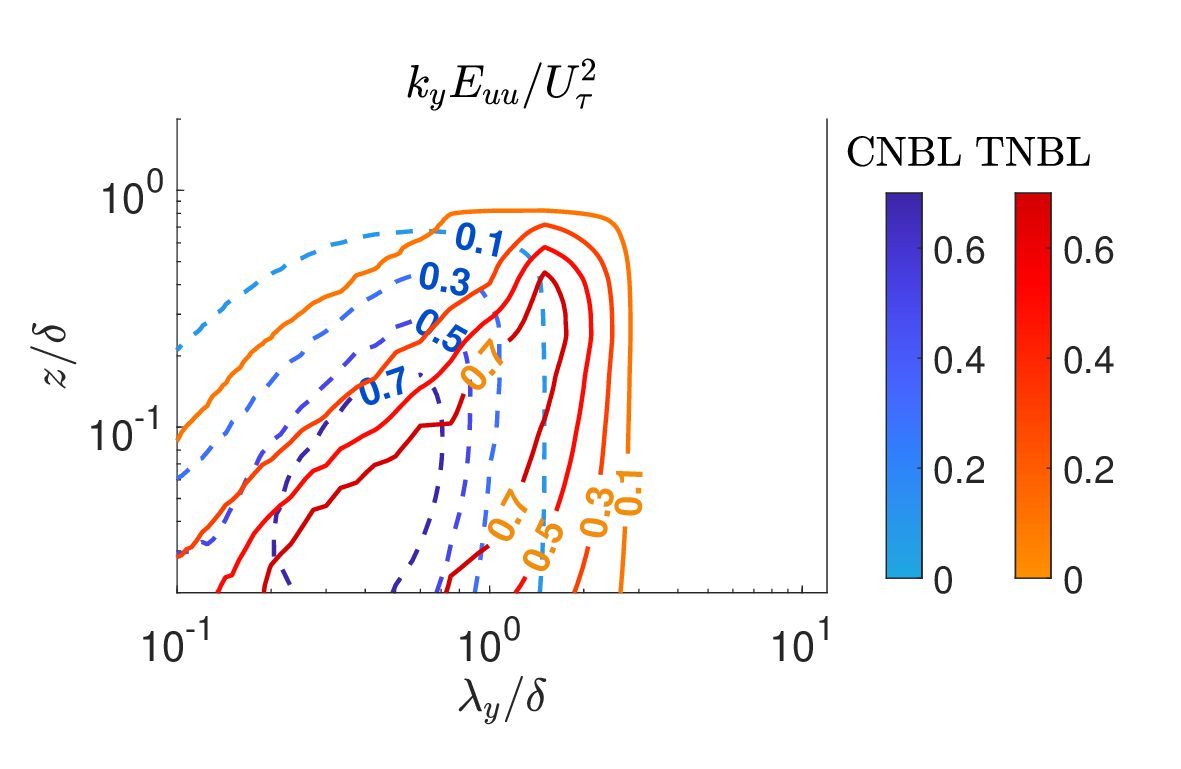}
            };
    \node[anchor=north west,
        xshift=-2mm,yshift=-2mm] at (image.north west) {{\rmfamily\fontsize{12}{13}\fontseries{l}\selectfont(b)}};
        \end{tikzpicture}}
        \vfill
        \vspace{-1.5cm}  
    \subfloat[\label{fig12c}]{
        \begin{tikzpicture}
        \node[anchor=north west, inner sep=0] (image) at (0,0) {
    \includegraphics[width=0.45\textwidth]{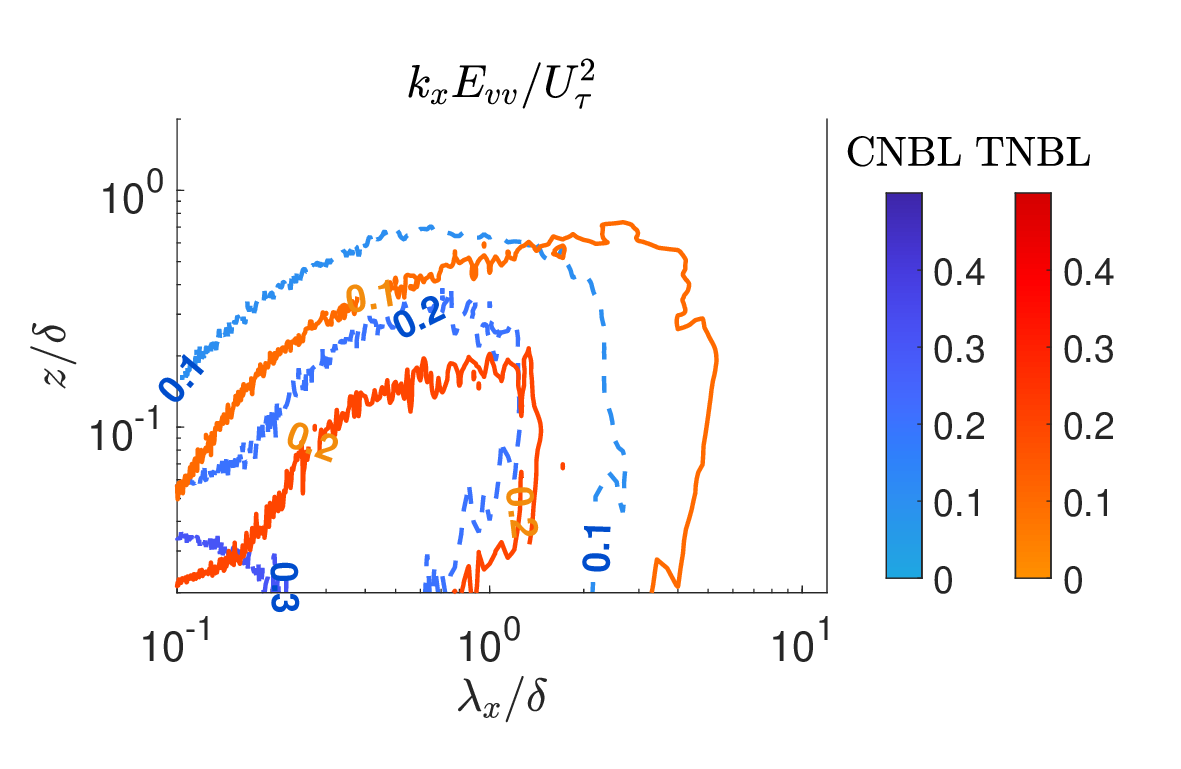}
            };
    \node[anchor=north west,
        xshift=-2mm,yshift=-2mm] at (image.north west) {{\rmfamily\fontsize{12}{13}\fontseries{l}\selectfont(c)}};
        \end{tikzpicture}}
    \subfloat[\label{fig12d}]{
        \begin{tikzpicture}
        \node[anchor=north west, inner sep=0] (image) at (0,0) {
    \includegraphics[width=0.45\textwidth]{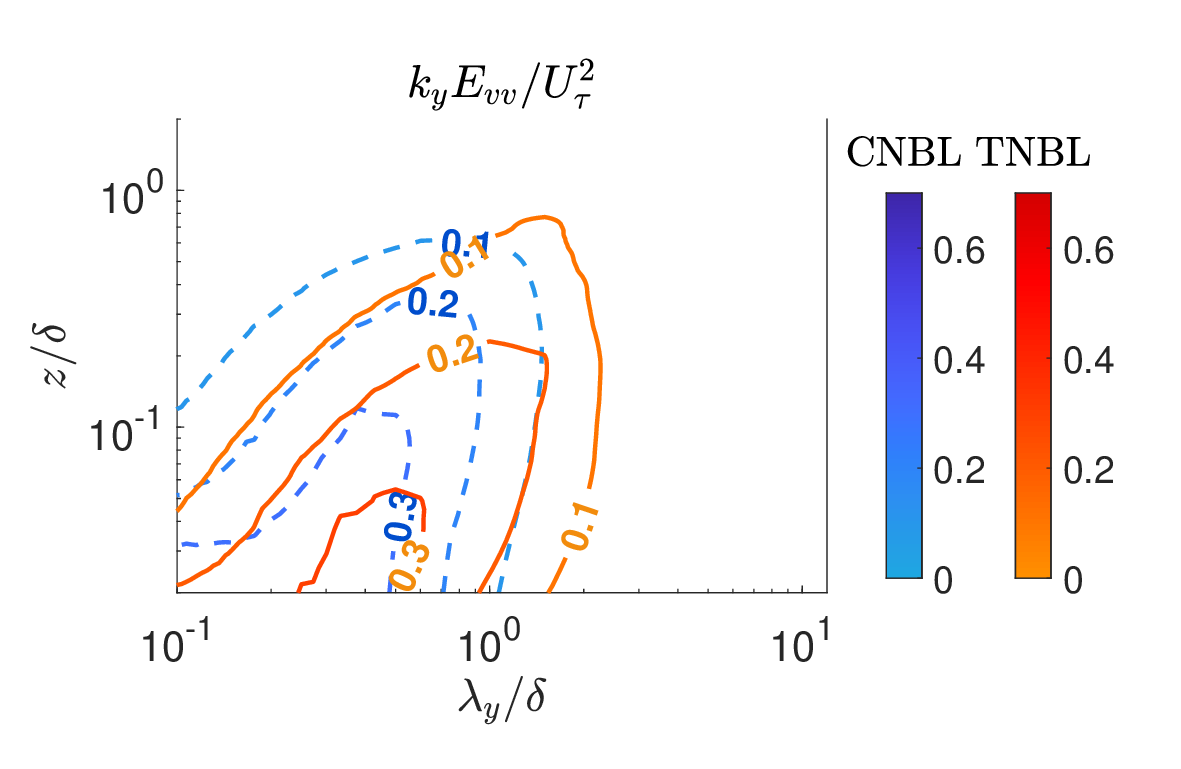}
            };
    \node[anchor=north west,
        xshift=-2mm,yshift=-2mm] at (image.north west) {{\rmfamily\fontsize{12}{13}\fontseries{l}\selectfont(d)}};
        \end{tikzpicture}}
    \vfill
    \vspace{-1.5cm}  
    \subfloat[\label{fig12e}]{
        \begin{tikzpicture}
        \node[anchor=north west, inner sep=0] (image) at (0,0) {
    \includegraphics[width=0.45\textwidth]{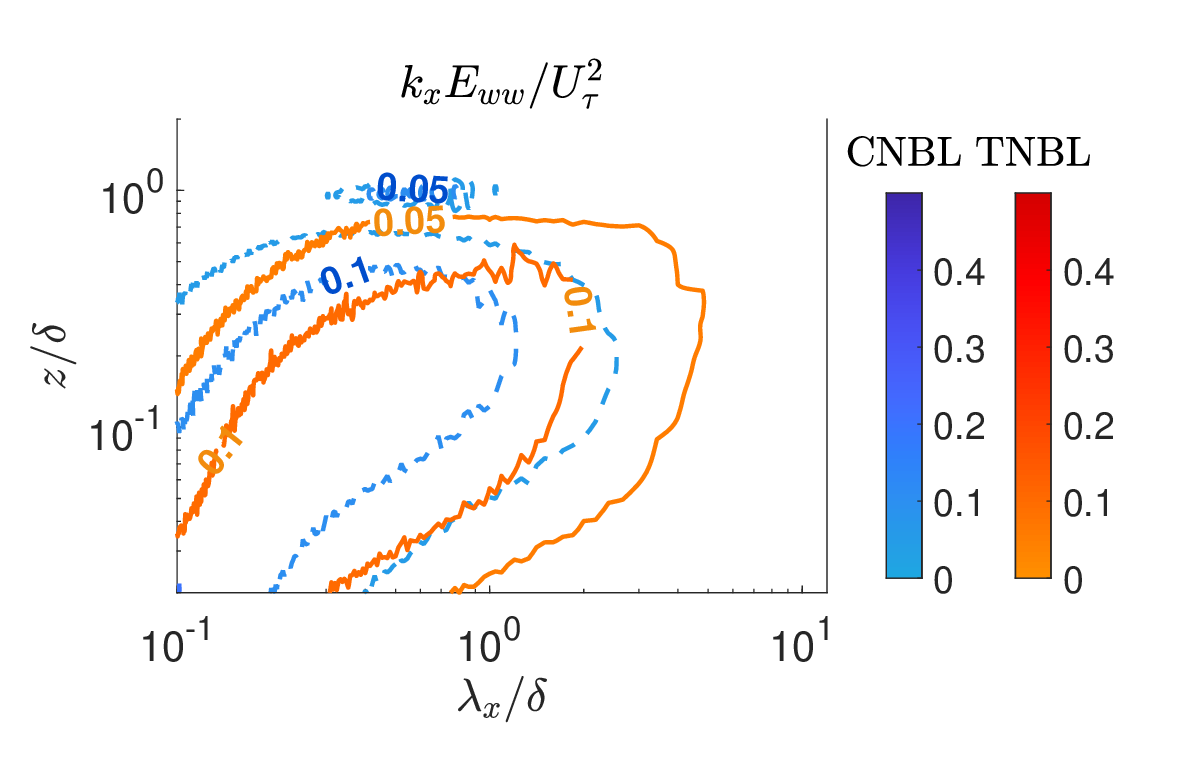}
            };
    \node[anchor=north west,
        xshift=-2mm,yshift=-2mm] at (image.north west) {{\rmfamily\fontsize{12}{13}\fontseries{l}\selectfont(e)}};
        \end{tikzpicture}}
    \subfloat[\label{fig12f}]{
        \begin{tikzpicture}
        \node[anchor=north west, inner sep=0] (image) at (0,0) {
    \includegraphics[width=0.45\textwidth]{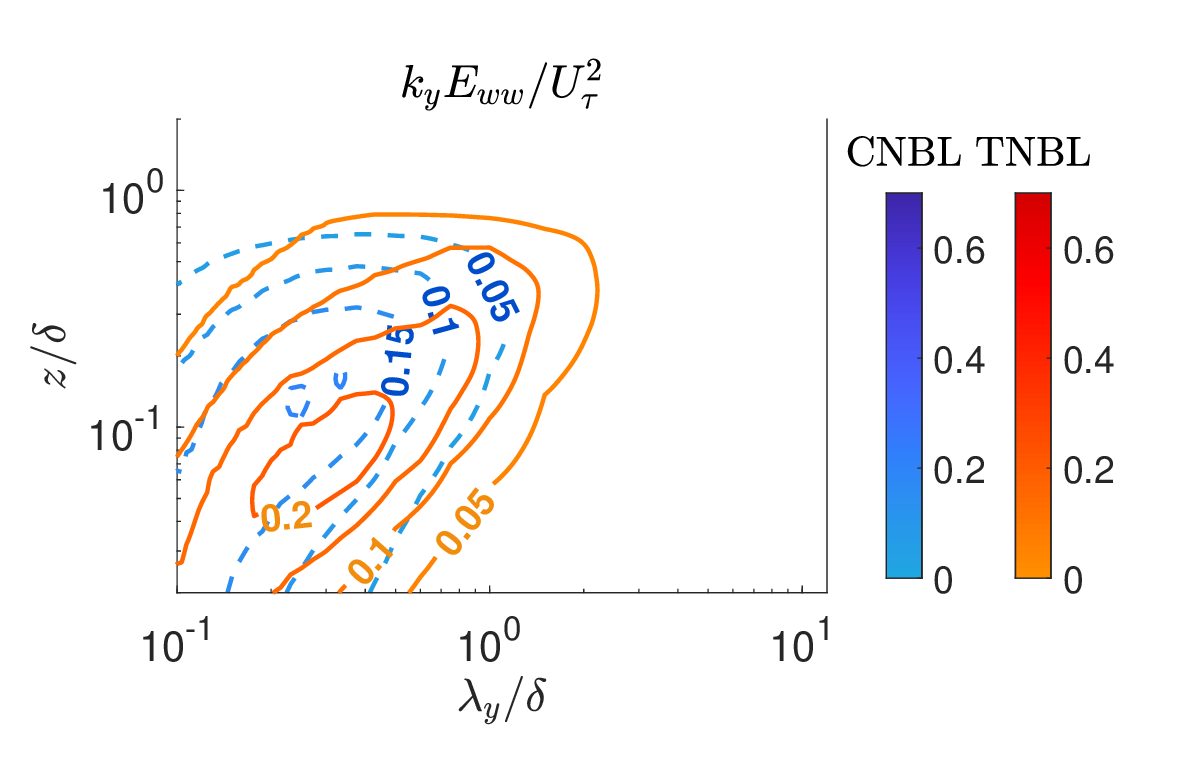}
            };
    \node[anchor=north west,
        xshift=-2mm,yshift=-2mm] at (image.north west) {{\rmfamily\fontsize{12}{13}\fontseries{l}\selectfont(f)}};
        \end{tikzpicture}}
        \vspace{-0.7cm}  
    \caption{{Contours of the longitudinal (a,c,e) and lateral (b,d,f) premultiplied energy spectra of the longitudinal (a,b), lateral (c,d) and vertical (e,f) velocities in the CNBL (N45) and the TNBL.}}
    \label{fig12}
\end{figure}

\begin{figure}
    \centering
     \subfloat[\label{Eplane(2)}]{
        \begin{tikzpicture}
        \node[anchor=north west, inner sep=0] (image) at (0,0) {
    \includegraphics[width=0.45\textwidth]{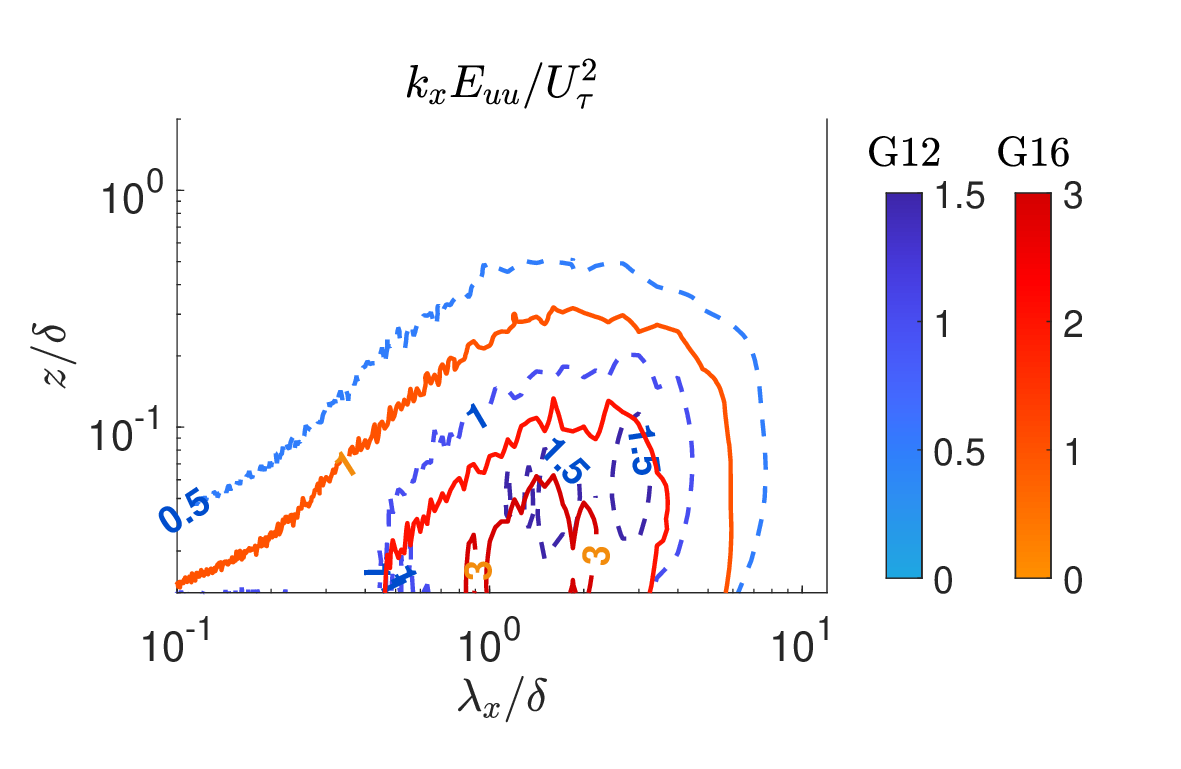}
            };
    \node[anchor=north west,
        xshift=-2mm,yshift=-2mm] at (image.north west) {{\rmfamily\fontsize{12}{13}\fontseries{l}\selectfont(a)}};
        \end{tikzpicture}}
    \subfloat[\label{Eyplane(2)}]{
        \begin{tikzpicture}
        \node[anchor=north west, inner sep=0] (image) at (0,0) {
    \includegraphics[width=0.45\textwidth]{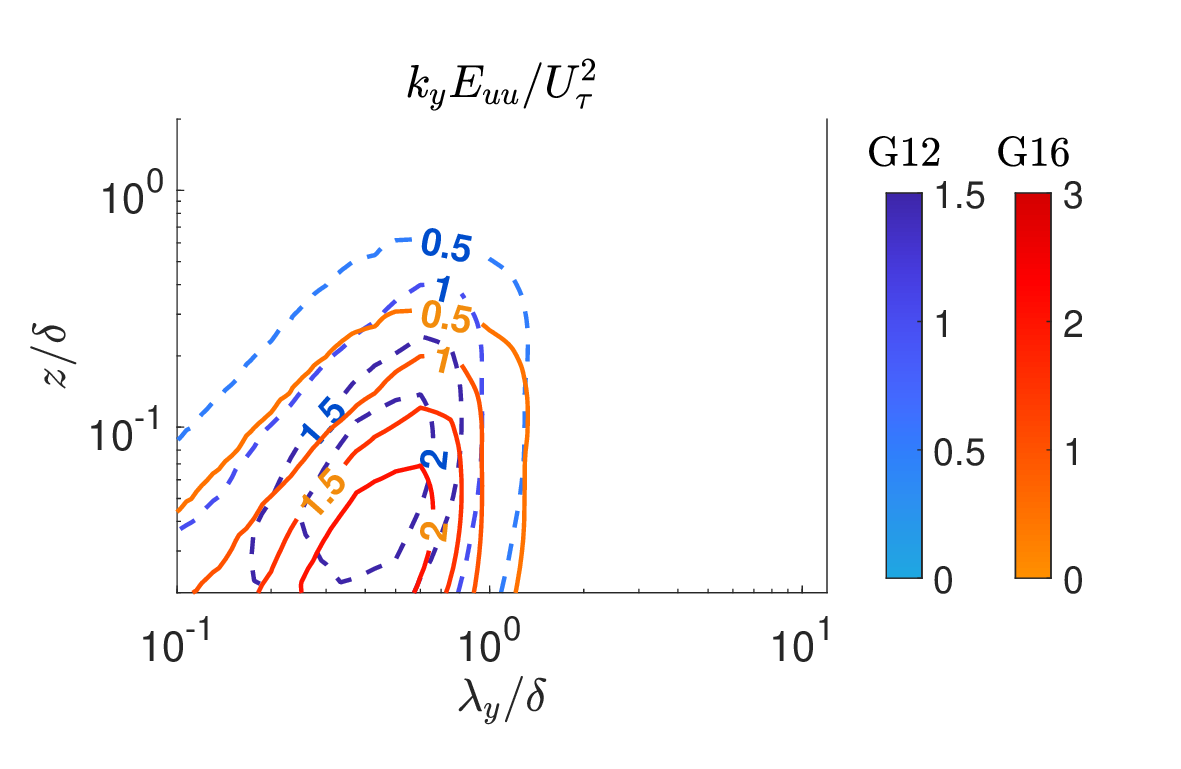}
            };
    \node[anchor=north west,
        xshift=-2mm,yshift=-2mm] at (image.north west) {{\rmfamily\fontsize{12}{13}\fontseries{l}\selectfont(b)}};
        \end{tikzpicture}}
        \vfill
        \vspace{-1.5cm}  
    \subfloat[\label{Evplane(2)}]{
        \begin{tikzpicture}
        \node[anchor=north west, inner sep=0] (image) at (0,0) {
    \includegraphics[width=0.45\textwidth]{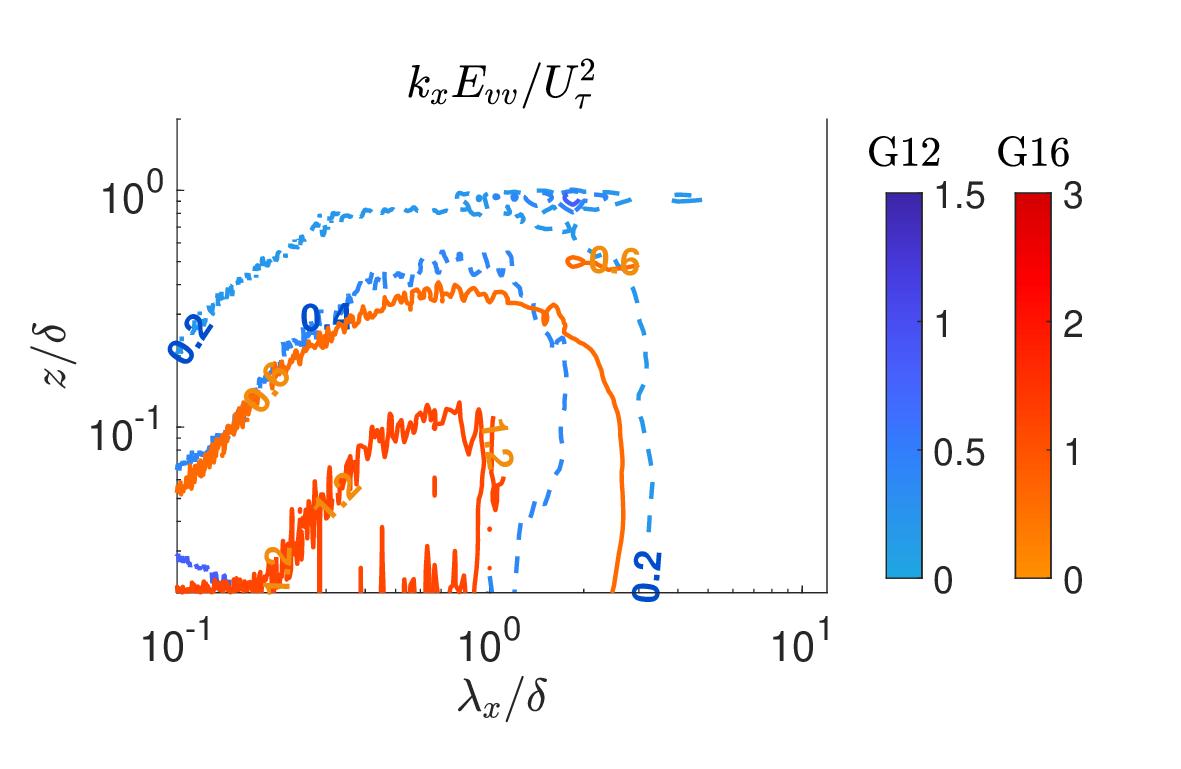}
            };
    \node[anchor=north west,
        xshift=-2mm,yshift=-2mm] at (image.north west) {{\rmfamily\fontsize{12}{13}\fontseries{l}\selectfont(c)}};
        \end{tikzpicture}}
    \subfloat[\label{Eyvplane(2)}]{
        \begin{tikzpicture}
        \node[anchor=north west, inner sep=0] (image) at (0,0) {
    \includegraphics[width=0.45\textwidth]{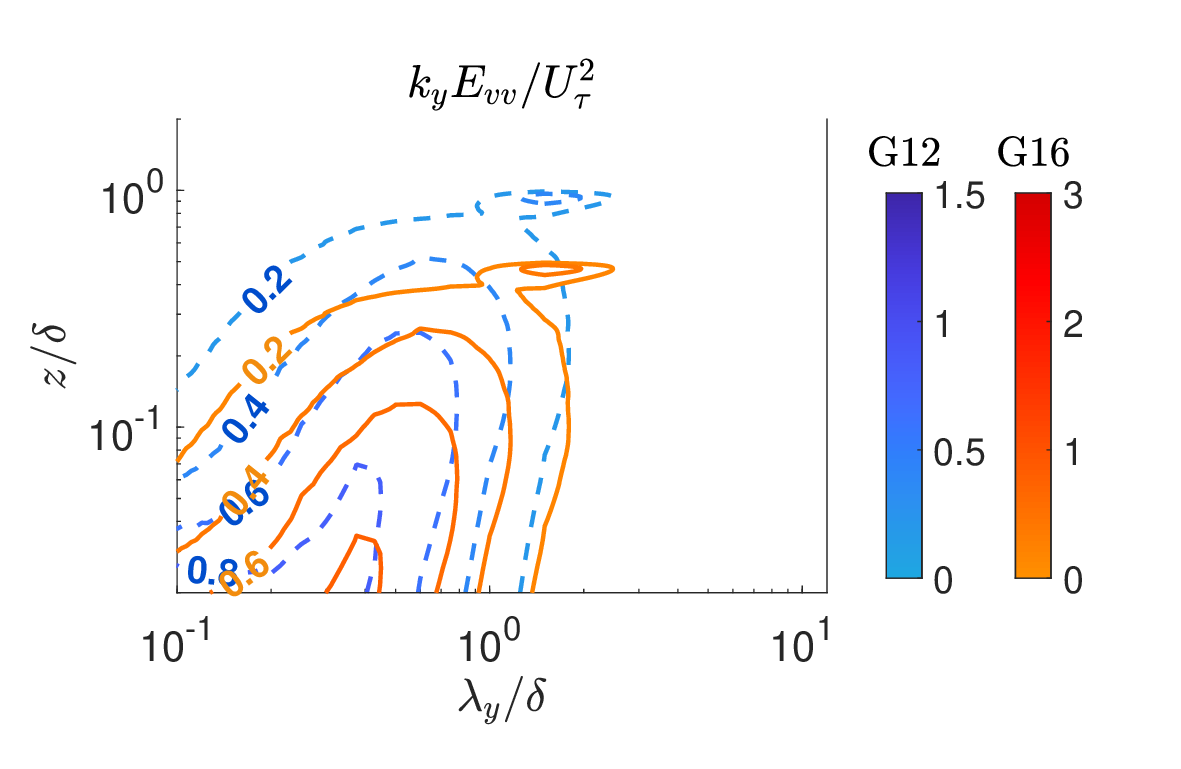}
            };
    \node[anchor=north west,
        xshift=-2mm,yshift=-2mm] at (image.north west) {{\rmfamily\fontsize{12}{13}\fontseries{l}\selectfont(d)}};
        \end{tikzpicture}}
    \vfill
    \vspace{-1.5cm}  
    \subfloat[\label{Ewplane(2)}]{
        \begin{tikzpicture}
        \node[anchor=north west, inner sep=0] (image) at (0,0) {
    \includegraphics[width=0.45\textwidth]{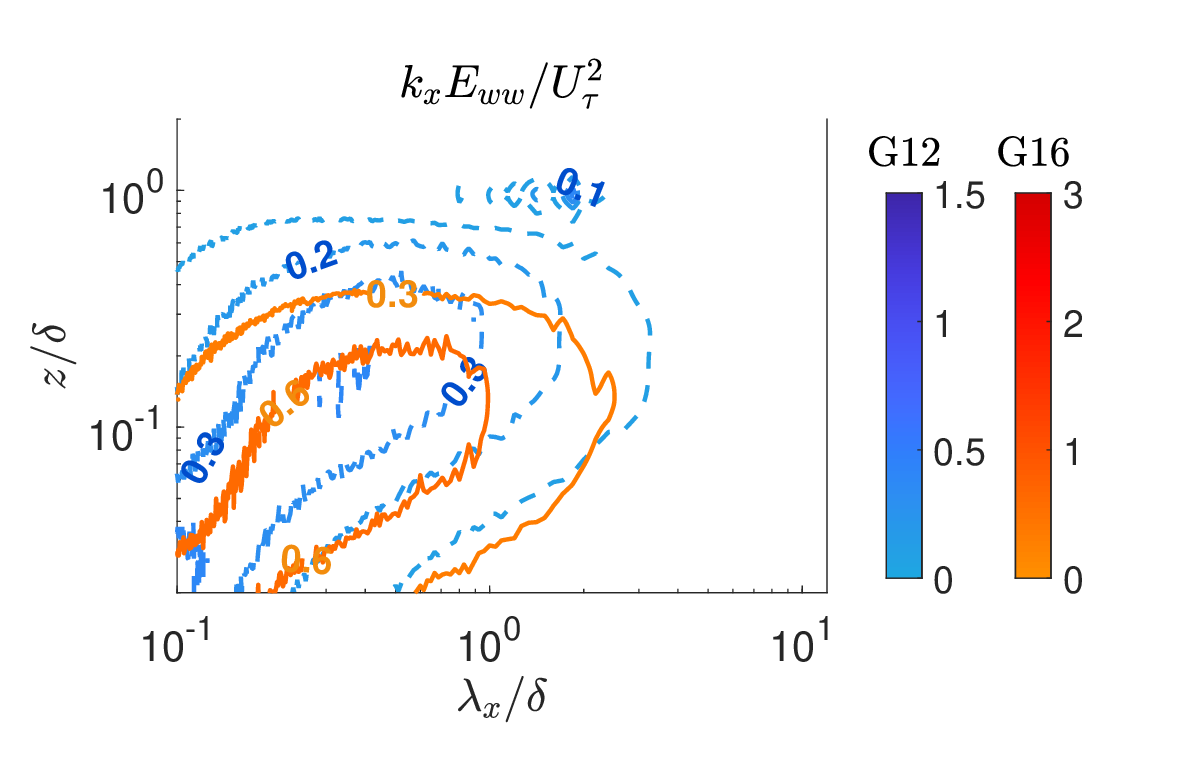}
            };
    \node[anchor=north west,
        xshift=-2mm,yshift=-2mm] at (image.north west) {{\rmfamily\fontsize{12}{13}\fontseries{l}\selectfont(e)}};
        \end{tikzpicture}}
    \subfloat[\label{Eywplane(2)}]{
        \begin{tikzpicture}
        \node[anchor=north west, inner sep=0] (image) at (0,0) {
    \includegraphics[width=0.45\textwidth]{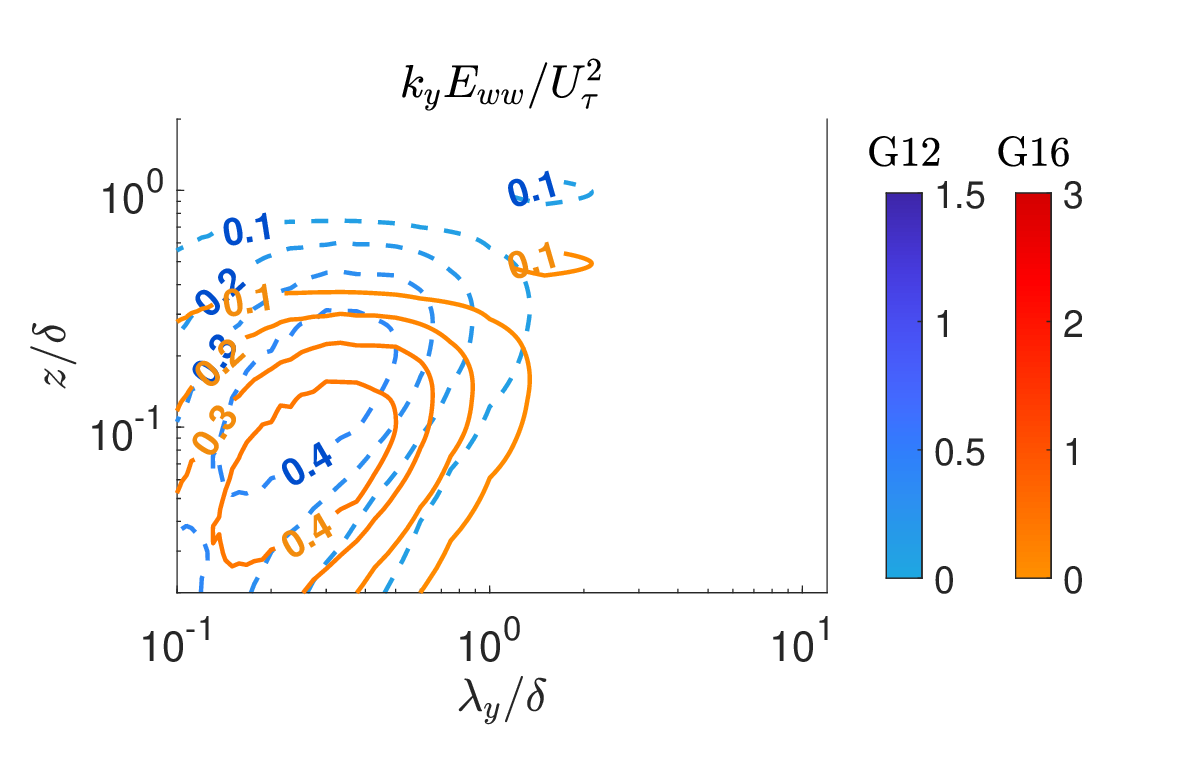}
            };
    \node[anchor=north west,
        xshift=-2mm,yshift=-2mm] at (image.north west) {{\rmfamily\fontsize{12}{13}\fontseries{l}\selectfont(f)}};
        \end{tikzpicture}}
        \vspace{-0.7cm}  
    \caption{{Contours of the longitudinal (a,c,e) and lateral (b,d,f) premultiplied energy spectra of the longitudinal (a,b), lateral (c,d) and vertical (e,f) velocities in the G12 and G16 cases.}}
    \label{fig:EGplane}
\end{figure}

Next, the premultiplied energy spectrum is leveraged to quantitatively demonstrate the spectral energy distributions and the prominent length scales of the turbulent structures.
Fig.~\ref{fig12} shows the contours of the premultiplied energy spectra in the N45 CNBL and the TNBL as functions of the wavelength and wall-normal height, from which one can clearly identify the location and length scale of the spectral peak.
For the TNBL, the spectral contour maps of the longitudinal velocity have a peak at $\lambda_x/\delta \approx 3\sim 4$ and $\lambda_y/\delta \approx 1$, which are typical length scales of very-large-scale motions or superstructures in wall-bounded turbulent flows \citep{Kim1999,hutchins2007evidence,wangVeryLargeScale2016,liuSpatialLengthScales2017} and can be well predicted by the large-eddy simulation \citep{Fang2015b,wang2020comparative}.
From the premultiplied spectra of the longitudinal velocity in the N45 CNBL, one can find that the spectral peak is at $\lambda_x/\delta \approx 1 \sim 2$ and $\lambda_y/\delta \approx 0.6 \sim 0.8$, which are smaller than those in the TNBL.
{The reduced longitudinal length scales by the Coriolis force can be attributed to the deflection of the coherent structures. In this sense, the lateral length scales of the coherent structures in the CNBLs should be wider if their sizes are retained, which is not true according to Fig.~\ref{fig12}. This implies that there must be some other acting mechanisms, \emph{e.g.} the Coriolis force may attenuate specific scales of the coherent structures. 
Another observation is that the spectral energies of the CNBLs shift to smaller scales and greater heights in both the longitudinal and lateral directions. The underlying physical mechanism is not intuitive and requires further analysis using spectral turbulent kinetic energy budgets \citep{leeSpectralAnalysisBudget2019}. }

Fig.~\ref{fig:EGplane} shows the contours of the longitudinal and lateral premultiplied energy spectra of flow velocities as functions of the length scales and wall-normal height with two geostrophic wind speeds ($U_g$= 12 m/s and 16 m/s). It can be seen that the wall-normal height of the spectral peak is lower with a larger geostrophic wind speed. In contrast, the length scales of the velocity structures do not seem to depend strongly on the geostrophic wind speed.
{It is also observed that there may exist a second peak in the spectra at $z \sim \delta$, which is located within the capping inversion layer. This spectral peak may signify a very long structure, while its analysis is out of the scope of the present study.}

{Moreover, Figs.~\ref{fig12} and ~\ref{fig:EGplane} show that the peaks of both the longitudinal and lateral premultiplied spectra are well captured in the present simulations, and the spectral densities approximately decay to zero at the largest scales, indicating that the length and width of the domain are adequate to resolve the large-scale motions in the CNBLs and TNBL \citep{wang2020comparative,Fang2015b}.}

\subsection{Structure deflections}
\label{Two-dimensional correlation structures}

\begin{figure}[!htb]
    \centering
    \subfloat[\label{selfc0.05}]{
        \begin{tikzpicture}
        \node[anchor=north west, inner sep=0] (image) at (0,0) {
    \includegraphics[width=0.3\textwidth]{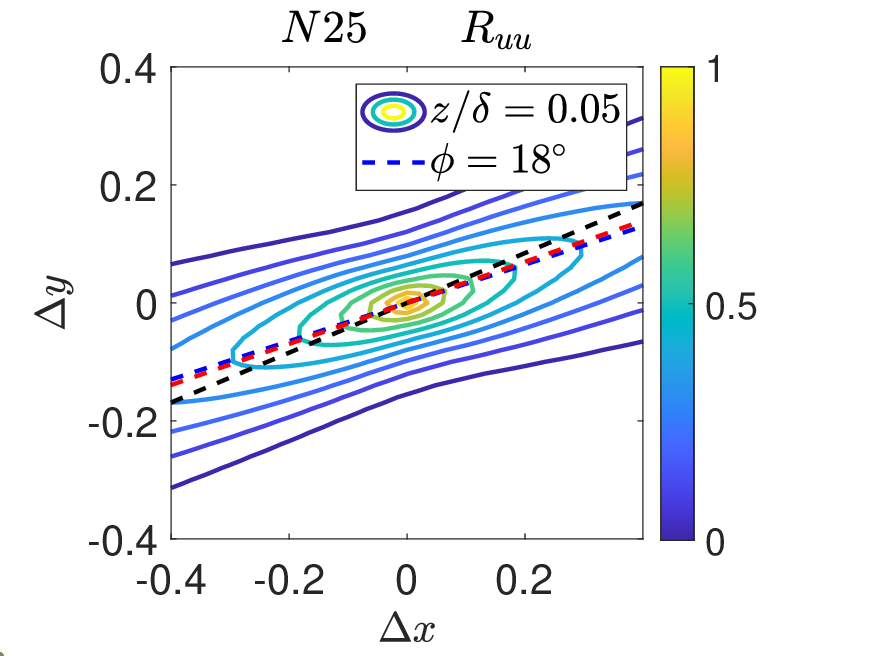}
            };
    \node[anchor=north west,
        xshift=-2mm,yshift=-2mm] at (image.north west) {{\rmfamily\fontsize{12}{13}\fontseries{l}\selectfont(a)}};
        \end{tikzpicture}}
    \subfloat[\label{selfc0.25}]{
        \begin{tikzpicture}
        \node[anchor=north west, inner sep=0] (image) at (0,0) {
    \includegraphics[width=0.3\textwidth]{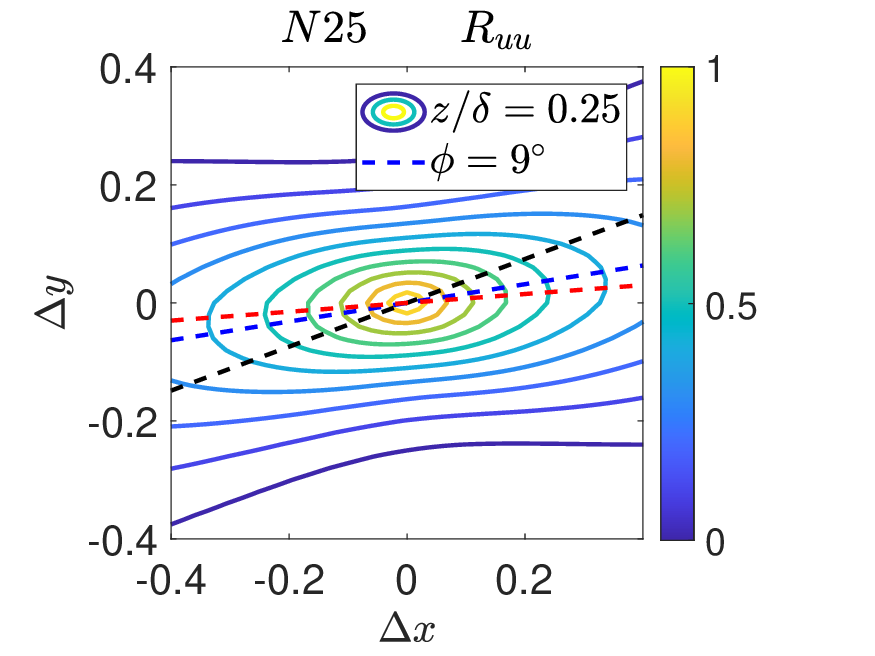}
            };
    \node[anchor=north west,
        xshift=-2mm,yshift=-2mm] at (image.north west) {{\rmfamily\fontsize{12}{13}\fontseries{l}\selectfont(b)}};
        \end{tikzpicture}}
    \subfloat[\label{selfc0.5}]{
        \begin{tikzpicture}
        \node[anchor=north west, inner sep=0] (image) at (0,0) {
    \includegraphics[width=0.3\textwidth]{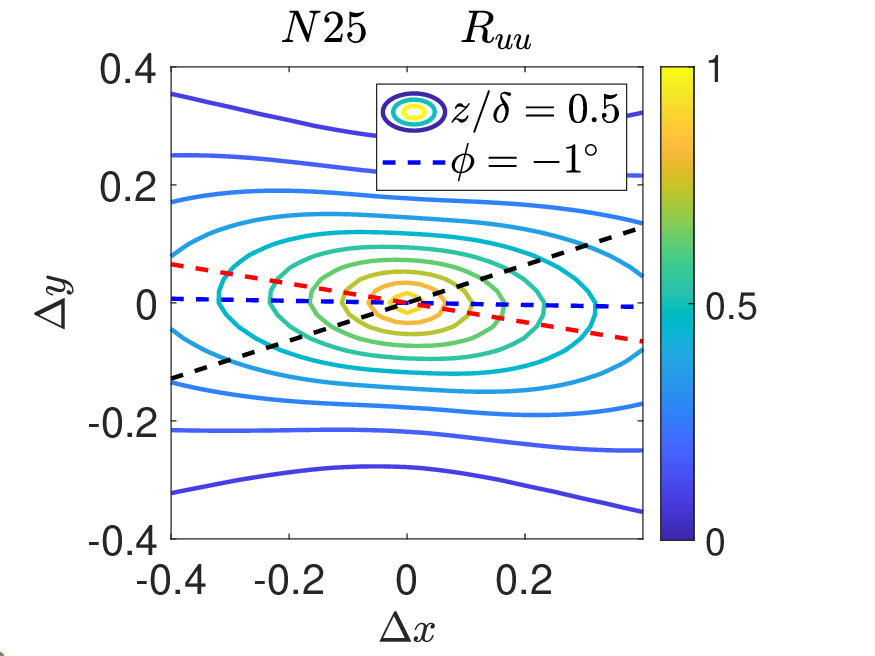}
            };
    \node[anchor=north west,
        xshift=-2mm,yshift=-2mm] at (image.north west) {{\rmfamily\fontsize{12}{13}\fontseries{l}\selectfont(c)}};
        \end{tikzpicture}}
\vspace{-1.45cm}  

 \centering
    \subfloat[\label{selfc0.05n45}]{
        \begin{tikzpicture}
        \node[anchor=north west, inner sep=0] (image) at (0,0) {
    \includegraphics[width=0.3\textwidth]{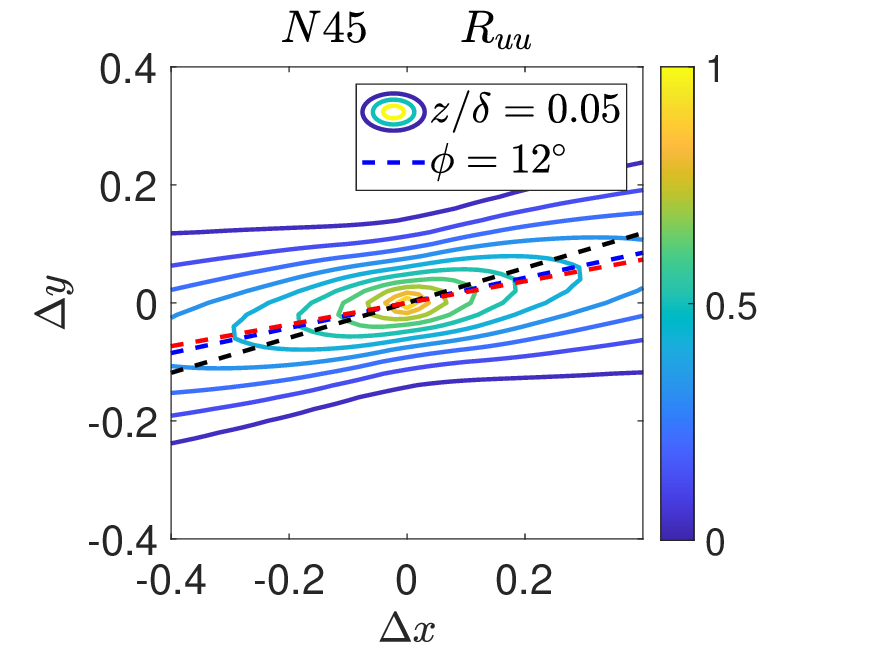}
            };
    \node[anchor=north west,
        xshift=-2mm,yshift=-2mm] at (image.north west) {{\rmfamily\fontsize{12}{13}\fontseries{l}\selectfont(d)}};
        \end{tikzpicture}}
   \subfloat[\label{selfc0.25n45}]{
        \begin{tikzpicture}
        \node[anchor=north west, inner sep=0] (image) at (0,0) {
    \includegraphics[width=0.3\textwidth]{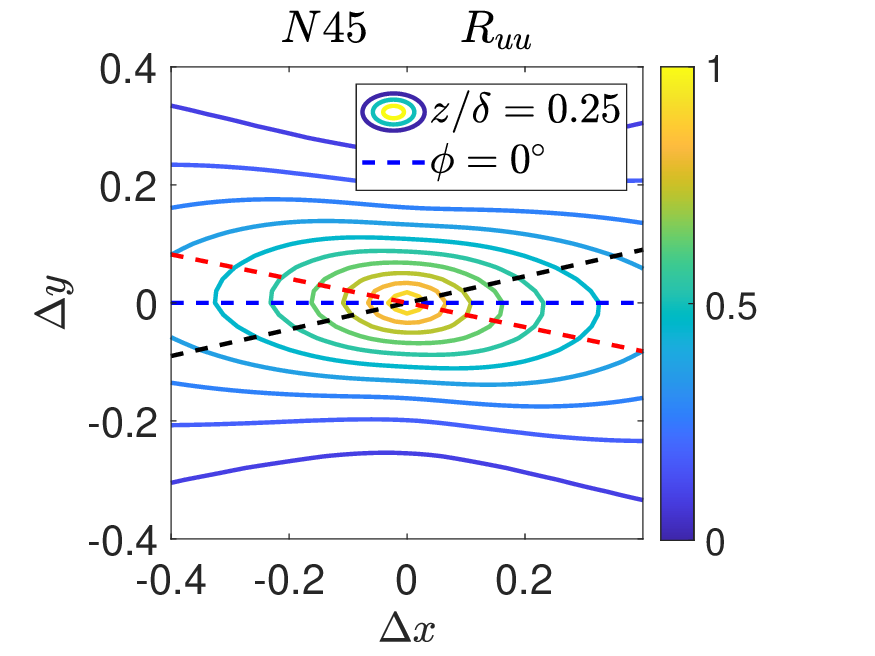}
            };
    \node[anchor=north west,
        xshift=-2mm,yshift=-2mm] at (image.north west) {{\rmfamily\fontsize{12}{13}\fontseries{l}\selectfont(e)}};
        \end{tikzpicture}}
    \subfloat[\label{selfc0.5n45}]{
        \begin{tikzpicture}
        \node[anchor=north west, inner sep=0] (image) at (0,0) {
    \includegraphics[width=0.3\textwidth]{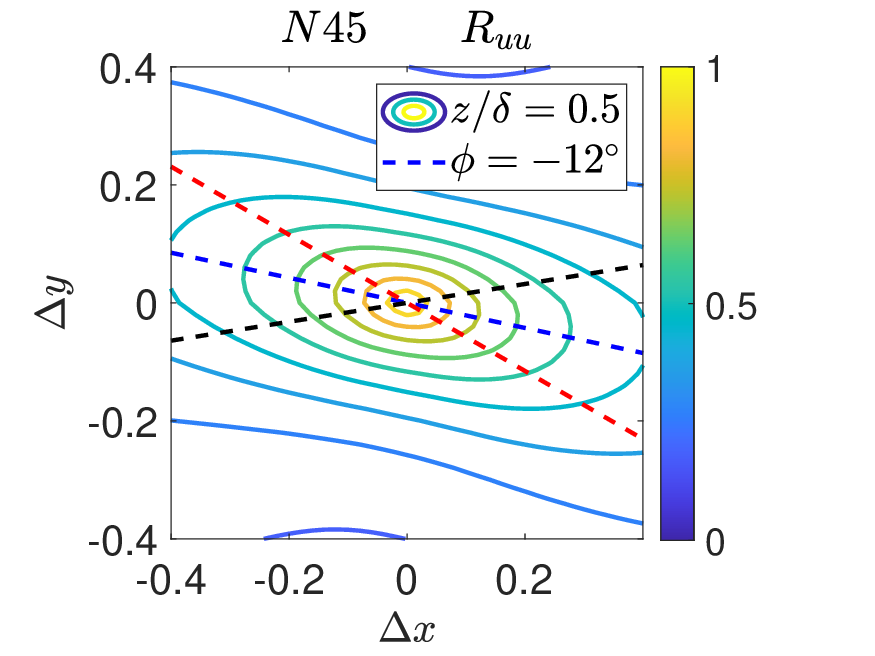}
            };
    \node[anchor=north west,
        xshift=-2mm,yshift=-2mm] at (image.north west) {{\rmfamily\fontsize{12}{13}\fontseries{l}\selectfont(f)}};
        \end{tikzpicture}}
  \vspace{-1.45cm}

  \centering
   \subfloat[\label{selfc0.05n70}]{
        \begin{tikzpicture}
        \node[anchor=north west, inner sep=0] (image) at (0,0) {
    \includegraphics[width=0.3\textwidth]{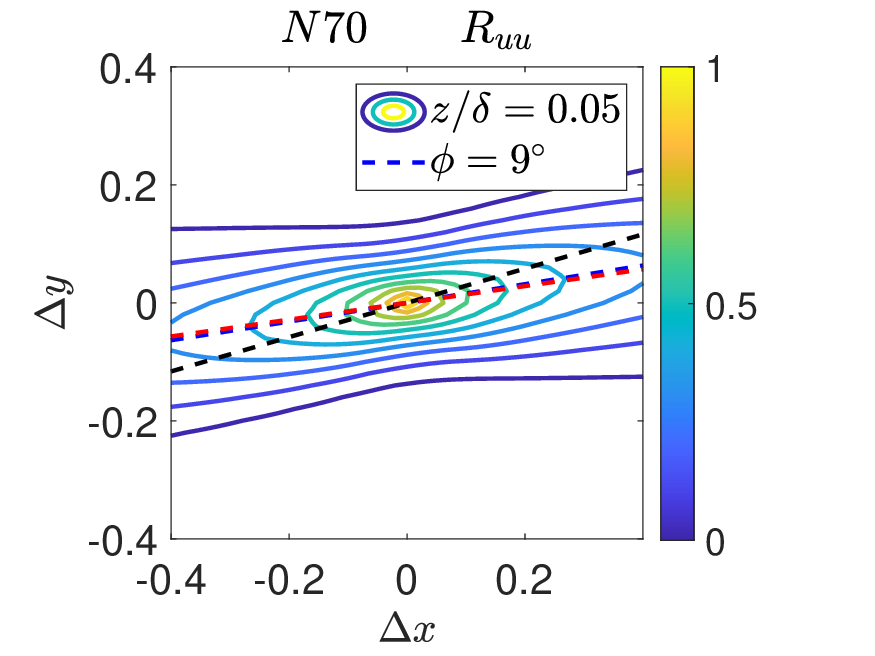}
            };
    \node[anchor=north west,
        xshift=-2mm,yshift=-2mm] at (image.north west) {{\rmfamily\fontsize{12}{13}\fontseries{l}\selectfont(g)}};
        \end{tikzpicture}}
     \subfloat[\label{selfc0.25n70}]{
        \begin{tikzpicture}
        \node[anchor=north west, inner sep=0] (image) at (0,0) {
    \includegraphics[width=0.3\textwidth]{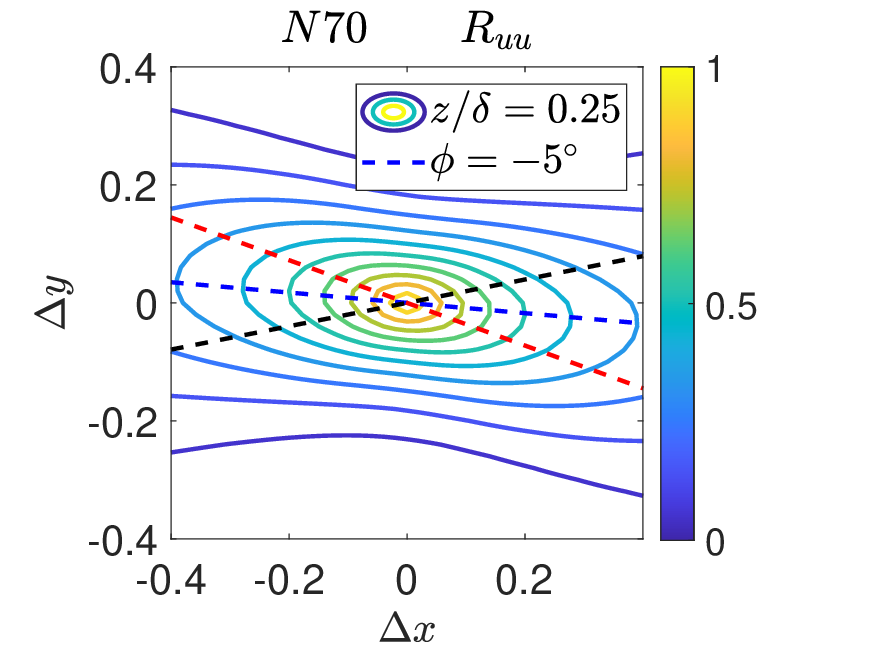}
            };
    \node[anchor=north west,
        xshift=-2mm,yshift=-2mm] at (image.north west) {{\rmfamily\fontsize{12}{13}\fontseries{l}\selectfont(h)}};
        \end{tikzpicture}}
     \subfloat[\label{selfc0.5n70}]{
        \begin{tikzpicture}
        \node[anchor=north west, inner sep=0] (image) at (0,0) {
    \includegraphics[width=0.3\textwidth]{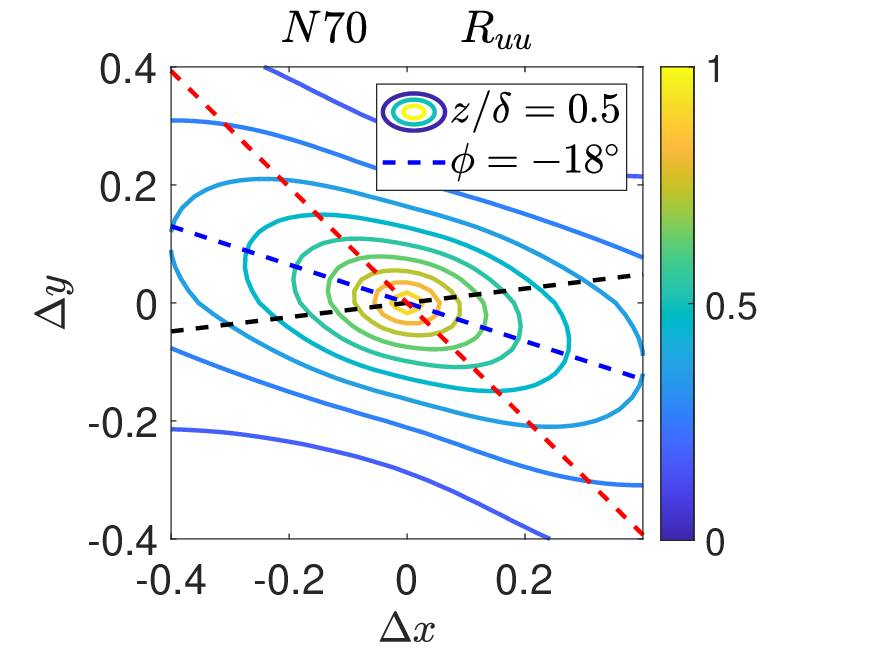}
            };
    \node[anchor=north west,
        xshift=-2mm,yshift=-2mm] at (image.north west) {{\rmfamily\fontsize{12}{13}\fontseries{l}\selectfont(i)}};
        \end{tikzpicture}}
 \vspace{-1.45cm}
   
         \centering
     \subfloat[\label{selfc0.05n90}]{
        \begin{tikzpicture}
        \node[anchor=north west, inner sep=0] (image) at (0,0) {
    \includegraphics[width=0.3\textwidth]{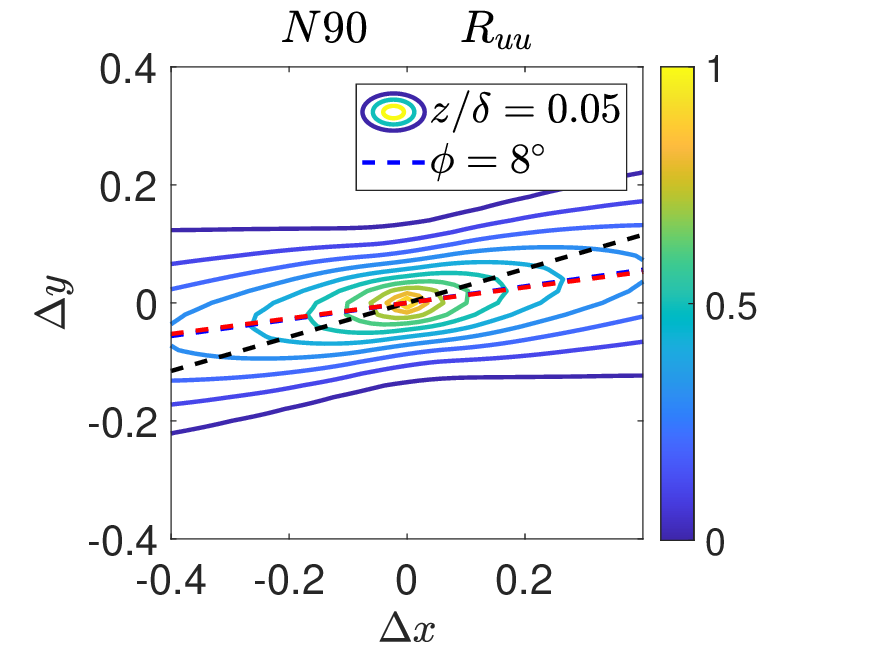}
            };
    \node[anchor=north west,
        xshift=-2mm,yshift=-2mm] at (image.north west) {{\rmfamily\fontsize{12}{13}\fontseries{l}\selectfont(j)}};
        \end{tikzpicture}}
    \subfloat[\label{selfc0.25n90}]{
        \begin{tikzpicture}
        \node[anchor=north west, inner sep=0] (image) at (0,0) {
    \includegraphics[width=0.3\textwidth]{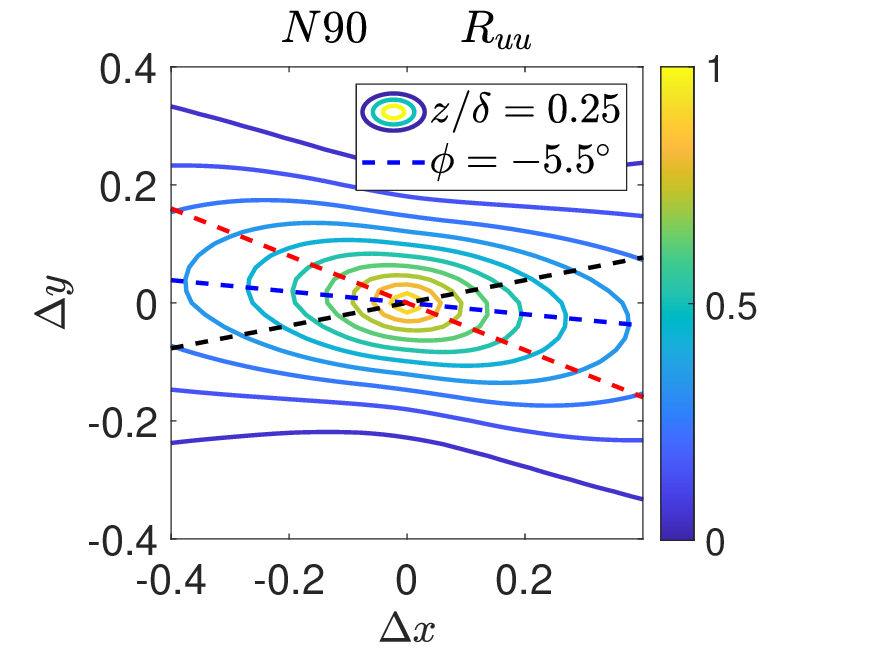}
            };
    \node[anchor=north west,
        xshift=-2mm,yshift=-2mm] at (image.north west) {{\rmfamily\fontsize{12}{13}\fontseries{l}\selectfont(k)}};
        \end{tikzpicture}}
     \subfloat[\label{selfc0.5n90}]{
        \begin{tikzpicture}
        \node[anchor=north west, inner sep=0] (image) at (0,0) {
    \includegraphics[width=0.3\textwidth]{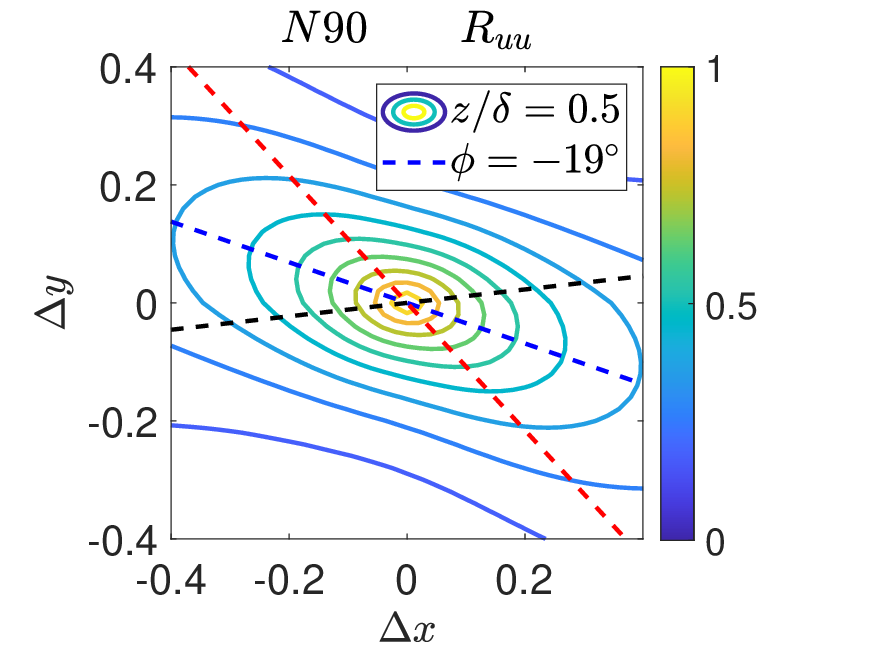}
            };
    \node[anchor=north west,
        xshift=-2mm,yshift=-2mm] at (image.north west) {{\rmfamily\fontsize{12}{13}\fontseries{l}\selectfont(l)}};
        \end{tikzpicture}}
  \vspace{-1.45cm}  
  
    \subfloat[\label{selfc0.05s45}]{
        \begin{tikzpicture}
        \node[anchor=north west, inner sep=0] (image) at (0,0) {
    \includegraphics[width=0.3\textwidth]{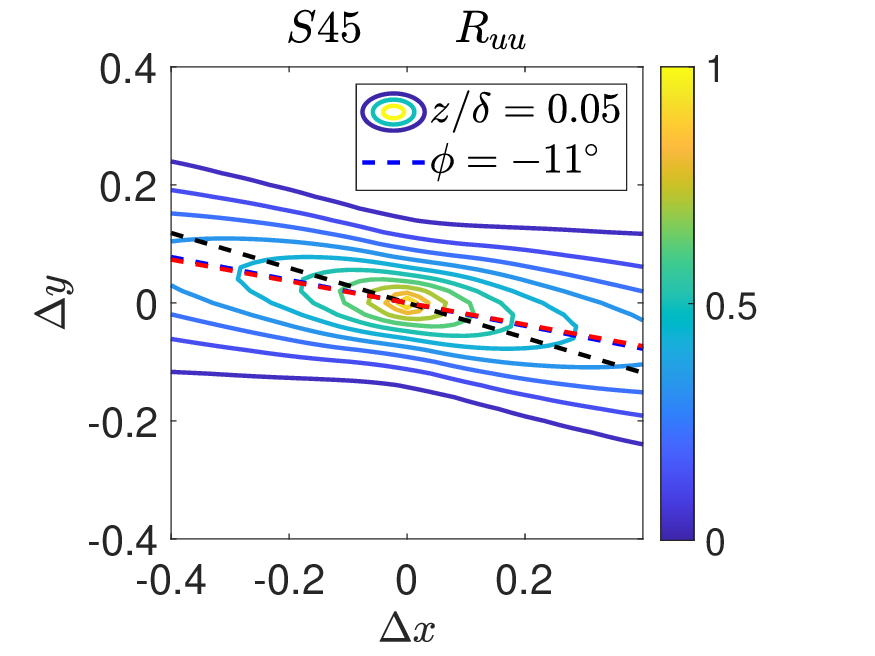}
            };
    \node[anchor=north west,
        xshift=-2mm,yshift=-2mm] at (image.north west) {{\rmfamily\fontsize{12}{13}\fontseries{l}\selectfont(m)}};
        \end{tikzpicture}}
    \subfloat[\label{selfc0.25s45}]{
        \begin{tikzpicture}
        \node[anchor=north west, inner sep=0] (image) at (0,0) {
    \includegraphics[width=0.3\textwidth]{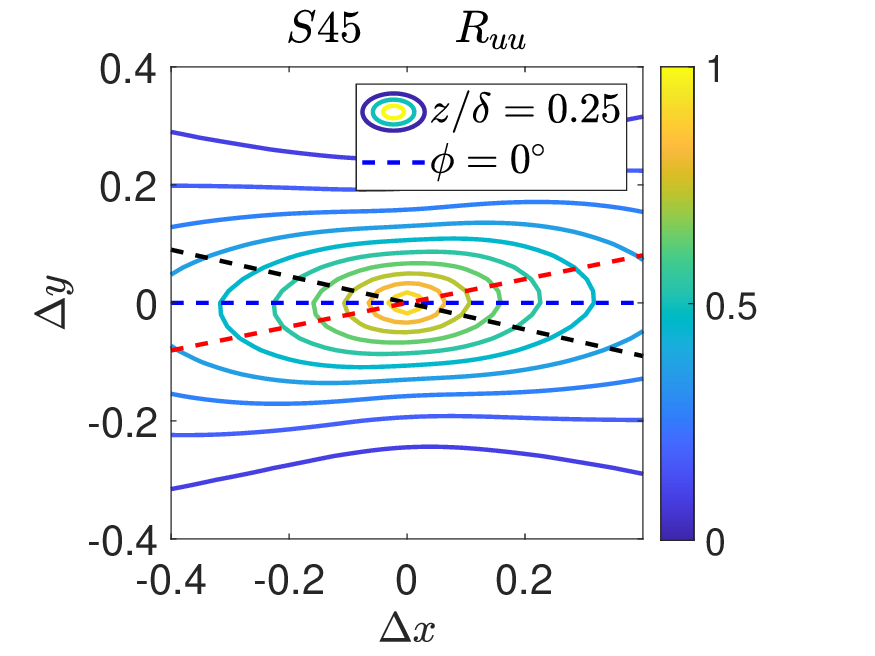}
            };
    \node[anchor=north west,
        xshift=-2mm,yshift=-2mm] at (image.north west) {{\rmfamily\fontsize{12}{13}\fontseries{l}\selectfont(n)}};
        \end{tikzpicture}}
    \subfloat[\label{selfc0.5s45}]{
        \begin{tikzpicture}
        \node[anchor=north west, inner sep=0] (image) at (0,0) {
    \includegraphics[width=0.3\textwidth]{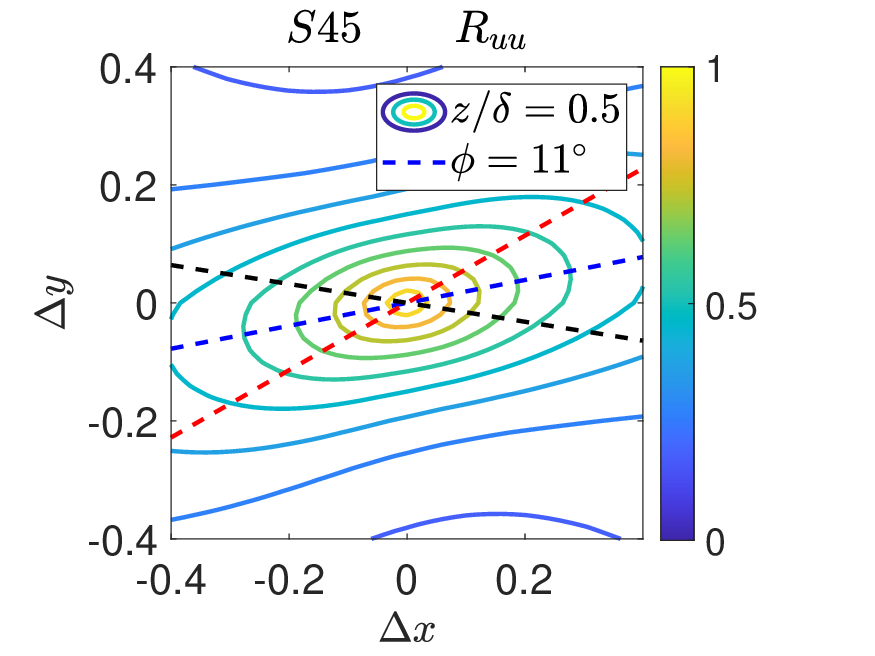}
            };
    \node[anchor=north west,
        xshift=-2mm,yshift=-2mm] at (image.north west) {{\rmfamily\fontsize{12}{13}\fontseries{l}\selectfont(o)}};
        \end{tikzpicture}}
    \caption{Two-point correlation contour maps of the streamwise velocity fluctuations in the $x-y$ planes at various heights in the CNBLs at different latitudes and the TNBL. The blue dashed line indicates the deflection direction. The black dashed line represents the mean wind direction, and the red dashed line is the mean shear direction {at the plane height}.}
    \label{fig：streamwise coherence}
\end{figure}

\begin{figure}
    \centering
    \subfloat[\label{selfc0.05v}]{
        \begin{tikzpicture}
        \node[anchor=north west, inner sep=0] (image) at (0,0) {
    \includegraphics[width=0.3\textwidth]{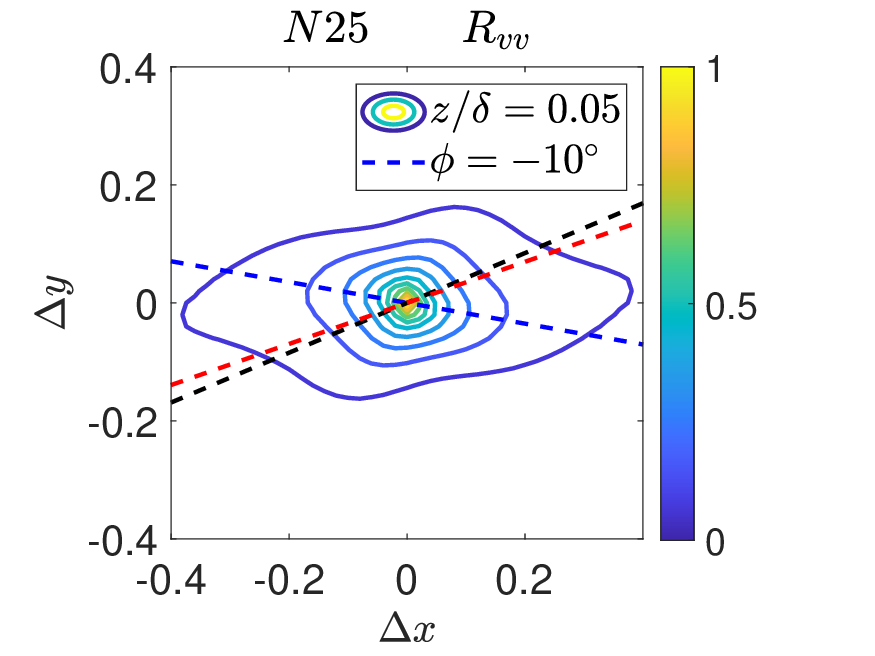}
            };
    \node[anchor=north west,
        xshift=-2mm,yshift=-2mm] at (image.north west) {{\rmfamily\fontsize{12}{13}\fontseries{l}\selectfont(a)}};
        \end{tikzpicture}}
    \subfloat[\label{selfc0.15v}]{
        \begin{tikzpicture}
        \node[anchor=north west, inner sep=0] (image) at (0,0) {
    \includegraphics[width=0.3\textwidth]{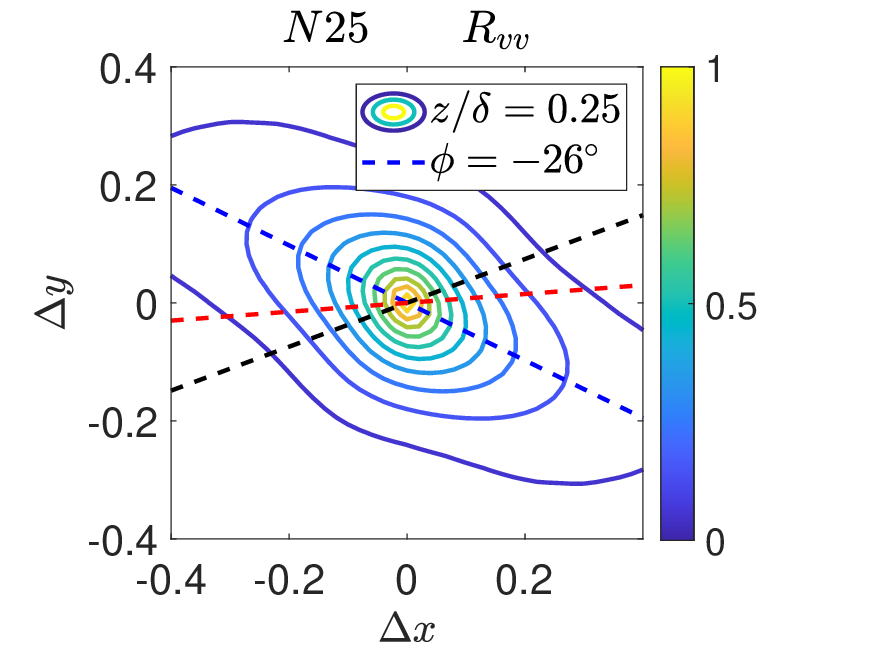}
            };
    \node[anchor=north west,
        xshift=-2mm,yshift=-2mm] at (image.north west) {{\rmfamily\fontsize{12}{13}\fontseries{l}\selectfont(b)}};
        \end{tikzpicture}}
    \subfloat[\label{selfc0.25v}]{
        \begin{tikzpicture}
        \node[anchor=north west, inner sep=0] (image) at (0,0) {
    \includegraphics[width=0.3\textwidth]{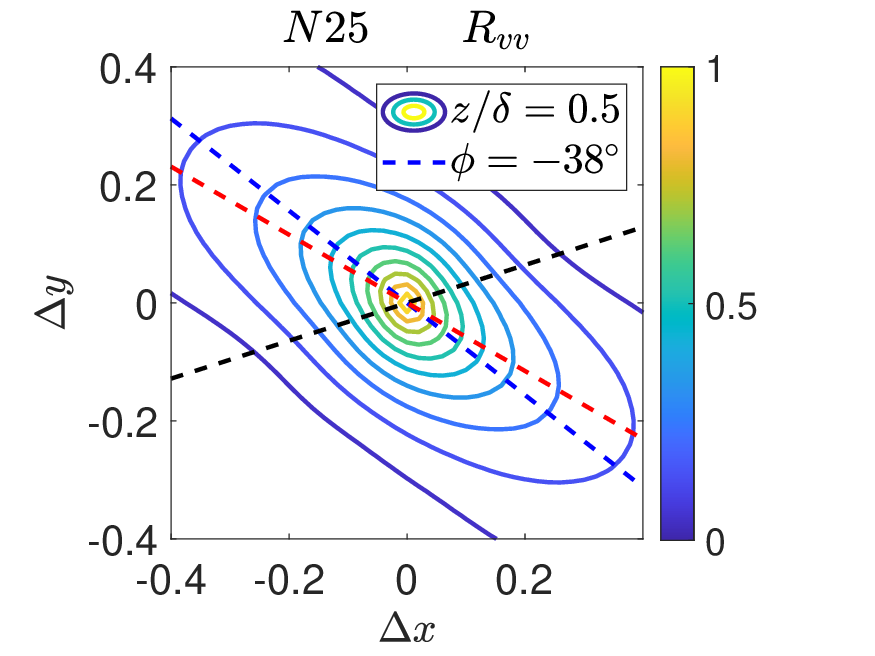}
            };
    \node[anchor=north west,
        xshift=-2mm,yshift=-2mm] at (image.north west) {{\rmfamily\fontsize{12}{13}\fontseries{l}\selectfont(c)}};
        \end{tikzpicture}}
\vspace{-1.45cm}  

 \centering
    \subfloat[\label{selfc0.05vn45}]{
        \begin{tikzpicture}
        \node[anchor=north west, inner sep=0] (image) at (0,0) {
    \includegraphics[width=0.3\textwidth]{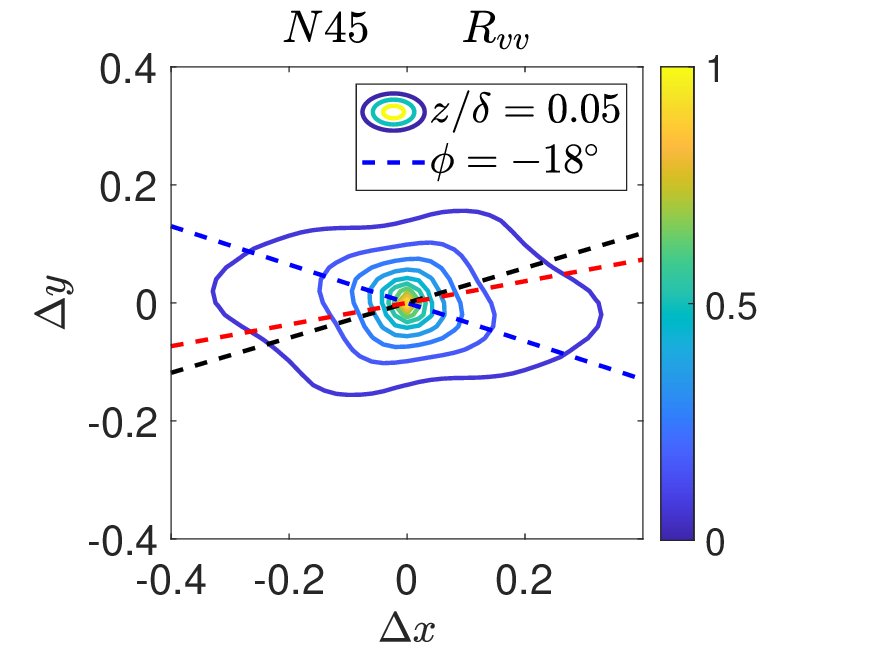}
            };
    \node[anchor=north west,
        xshift=-2mm,yshift=-2mm] at (image.north west) {{\rmfamily\fontsize{12}{13}\fontseries{l}\selectfont(d)}};
        \end{tikzpicture}}
   \subfloat[\label{selfc0.25vn45}]{
        \begin{tikzpicture}
        \node[anchor=north west, inner sep=0] (image) at (0,0) {
    \includegraphics[width=0.3\textwidth]{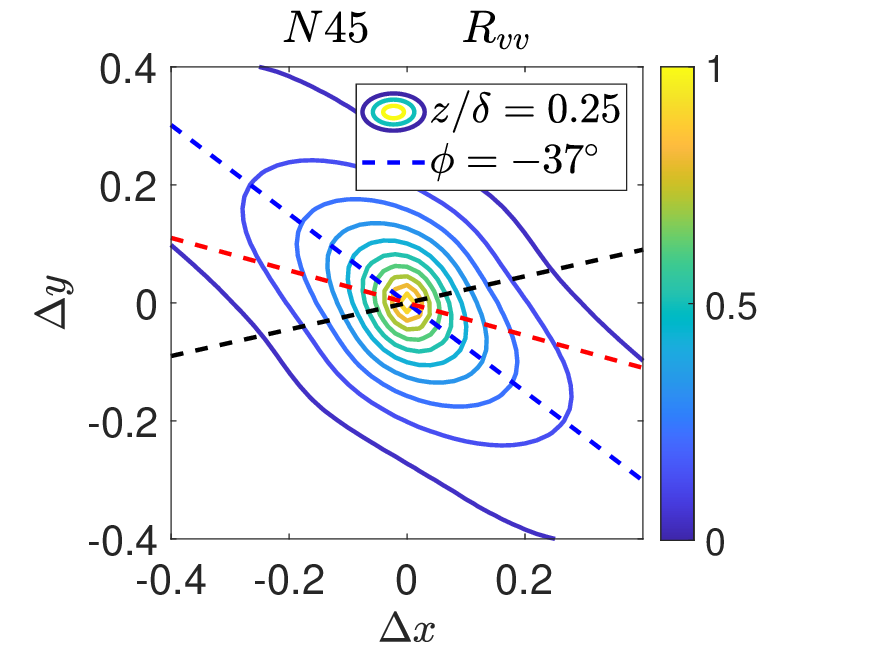}
            };
    \node[anchor=north west,
        xshift=-2mm,yshift=-2mm] at (image.north west) {{\rmfamily\fontsize{12}{13}\fontseries{l}\selectfont(e)}};
        \end{tikzpicture}}
    \subfloat[\label{selfc0.5vn45}]{
        \begin{tikzpicture}
        \node[anchor=north west, inner sep=0] (image) at (0,0) {
    \includegraphics[width=0.3\textwidth]{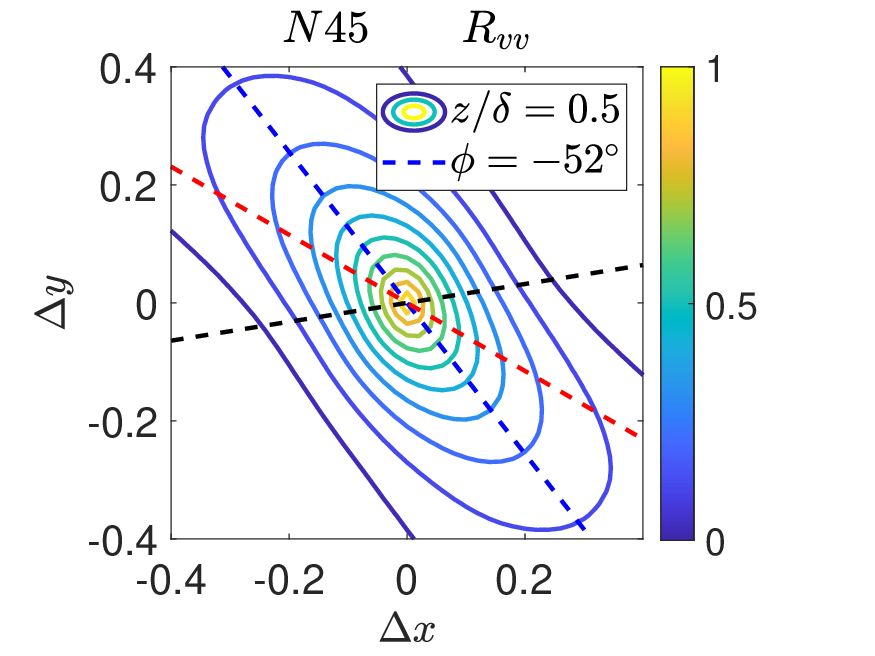}
            };
    \node[anchor=north west,
        xshift=-2mm,yshift=-2mm] at (image.north west) {{\rmfamily\fontsize{12}{13}\fontseries{l}\selectfont(f)}};
        \end{tikzpicture}}
  \vspace{-1.45cm}

  \centering
   \subfloat[\label{selfc0.05vn70}]{
        \begin{tikzpicture}
        \node[anchor=north west, inner sep=0] (image) at (0,0) {
    \includegraphics[width=0.3\textwidth]{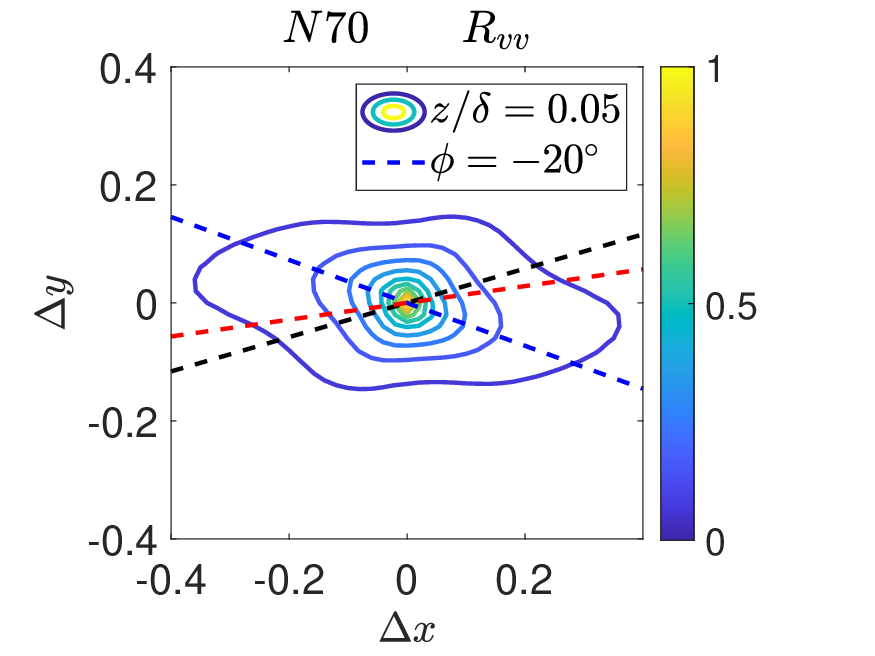}
            };
    \node[anchor=north west,
        xshift=-2mm,yshift=-2mm] at (image.north west) {{\rmfamily\fontsize{12}{13}\fontseries{l}\selectfont(g)}};
        \end{tikzpicture}}
     \subfloat[\label{selfc0.25vn70}]{
        \begin{tikzpicture}
        \node[anchor=north west, inner sep=0] (image) at (0,0) {
    \includegraphics[width=0.3\textwidth]{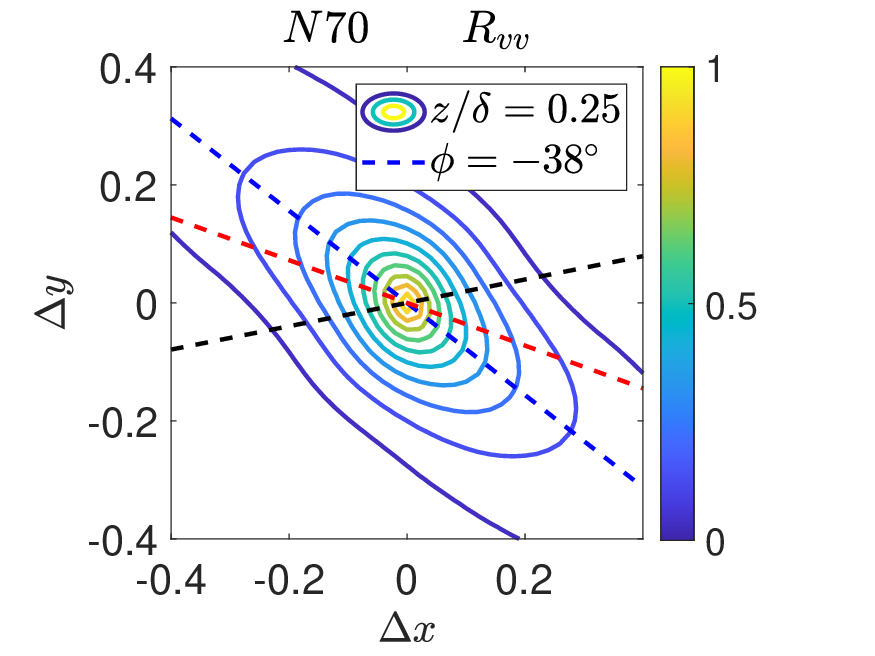}
            };
    \node[anchor=north west,
        xshift=-2mm,yshift=-2mm] at (image.north west) {{\rmfamily\fontsize{12}{13}\fontseries{l}\selectfont(h)}};
        \end{tikzpicture}}
     \subfloat[\label{selfc0.5vn70}]{
        \begin{tikzpicture}
        \node[anchor=north west, inner sep=0] (image) at (0,0) {
    \includegraphics[width=0.3\textwidth]{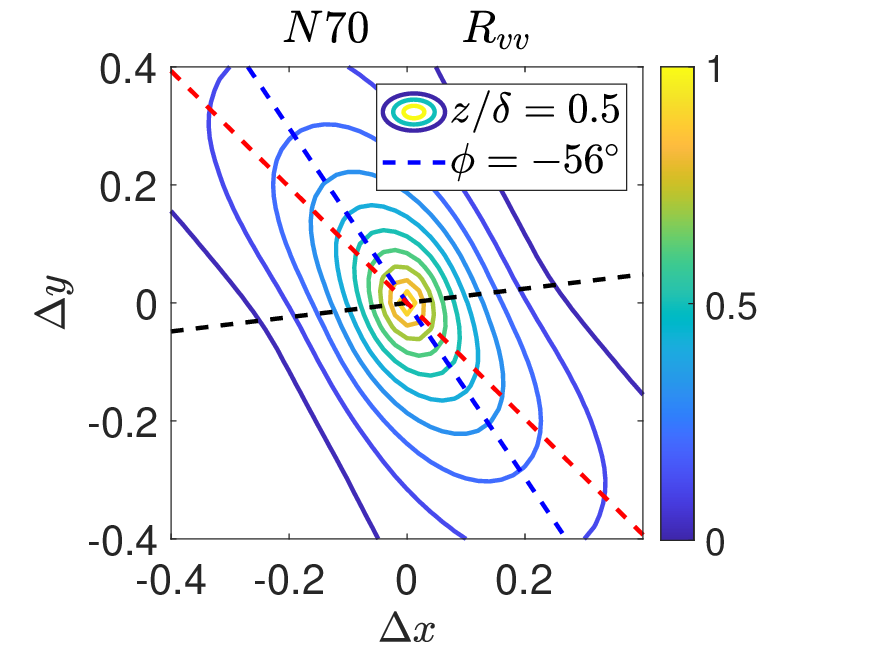}
            };
    \node[anchor=north west,
        xshift=-2mm,yshift=-2mm] at (image.north west) {{\rmfamily\fontsize{12}{13}\fontseries{l}\selectfont(i)}};
        \end{tikzpicture}}
 \vspace{-1.45cm}
   
         \centering
     \subfloat[\label{selfc0.05vn90}]{
        \begin{tikzpicture}
        \node[anchor=north west, inner sep=0] (image) at (0,0) {
    \includegraphics[width=0.3\textwidth]{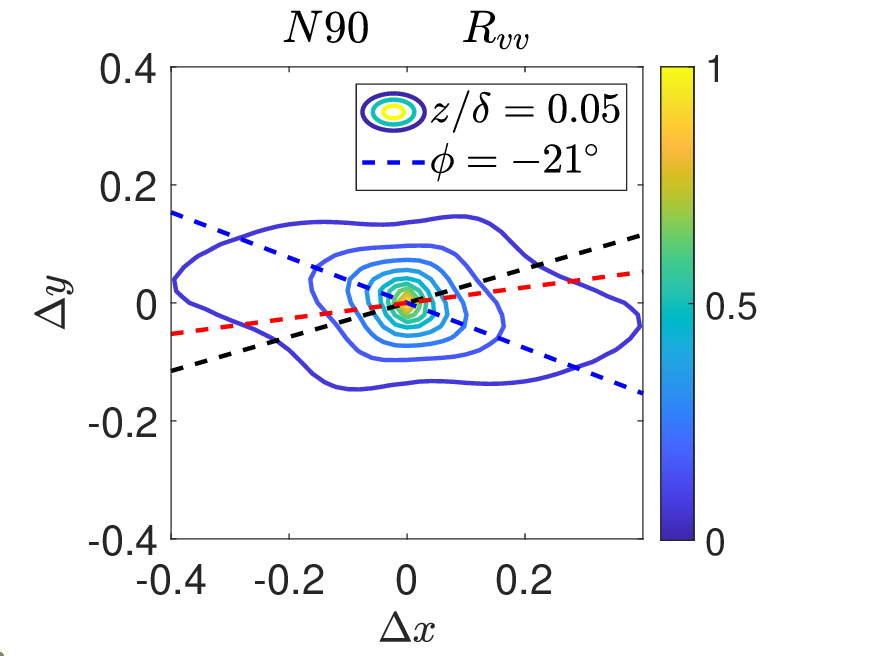}
            };
    \node[anchor=north west,
        xshift=-2mm,yshift=-2mm] at (image.north west) {{\rmfamily\fontsize{12}{13}\fontseries{l}\selectfont(j)}};
        \end{tikzpicture}}
    \subfloat[\label{selfc0.25vn90}]{
        \begin{tikzpicture}
        \node[anchor=north west, inner sep=0] (image) at (0,0) {
    \includegraphics[width=0.3\textwidth]{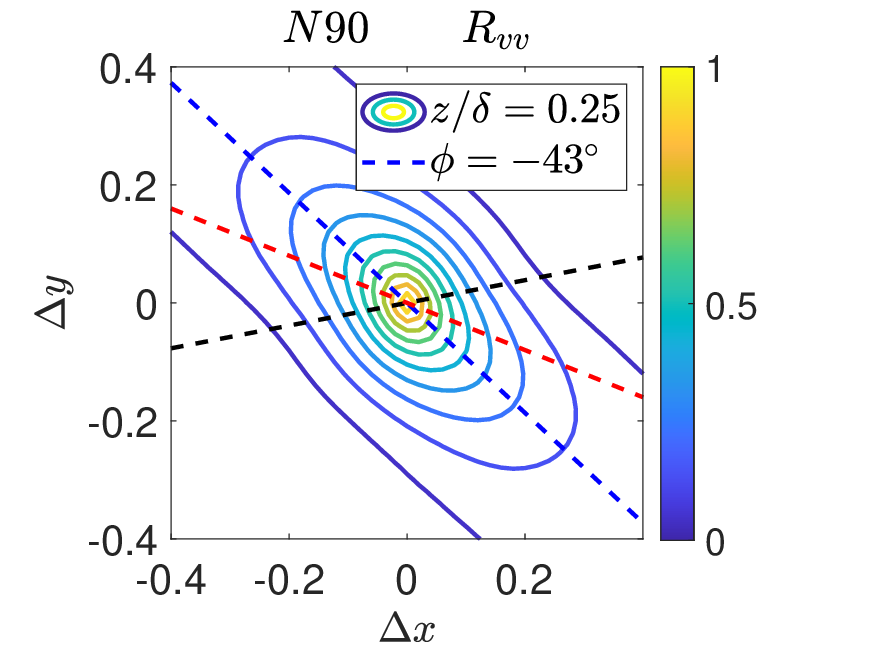}
            };
    \node[anchor=north west,
        xshift=-2mm,yshift=-2mm] at (image.north west) {{\rmfamily\fontsize{12}{13}\fontseries{l}\selectfont(k)}};
        \end{tikzpicture}}
     \subfloat[\label{selfc0.5vn90}]{
        \begin{tikzpicture}
        \node[anchor=north west, inner sep=0] (image) at (0,0) {
    \includegraphics[width=0.3\textwidth]{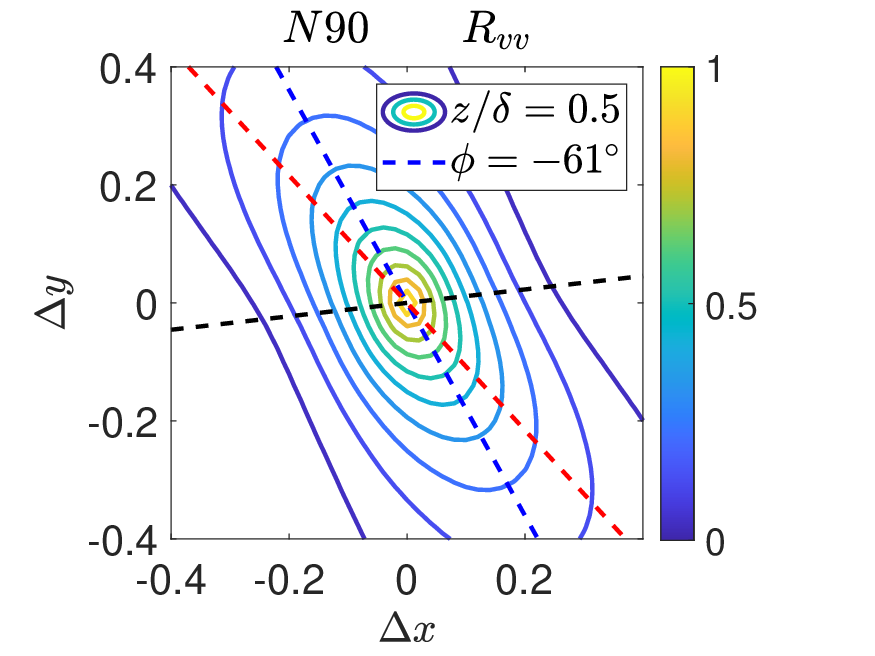}
            };
    \node[anchor=north west,
        xshift=-2mm,yshift=-2mm] at (image.north west) {{\rmfamily\fontsize{12}{13}\fontseries{l}\selectfont(l)}};
        \end{tikzpicture}}
  \vspace{-1.45cm}  
  
    \subfloat[\label{selfc0.05vs45}]{
        \begin{tikzpicture}
        \node[anchor=north west, inner sep=0] (image) at (0,0) {
    \includegraphics[width=0.3\textwidth]{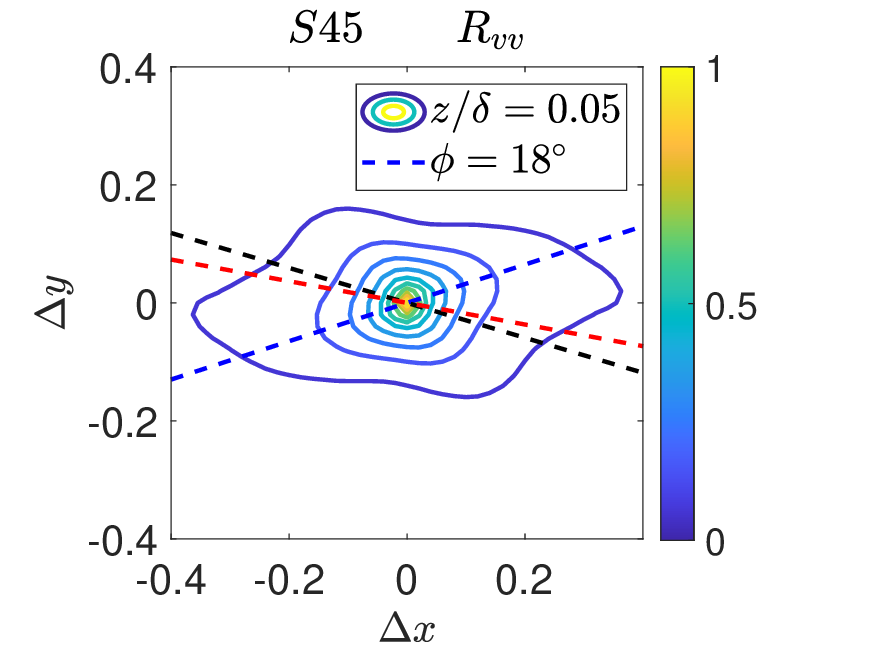}
            };
    \node[anchor=north west,
        xshift=-2mm,yshift=-2mm] at (image.north west) {{\rmfamily\fontsize{12}{13}\fontseries{l}\selectfont(m)}};
        \end{tikzpicture}}
    \subfloat[\label{selfc0.25vs45}]{
        \begin{tikzpicture}
        \node[anchor=north west, inner sep=0] (image) at (0,0) {
    \includegraphics[width=0.3\textwidth]{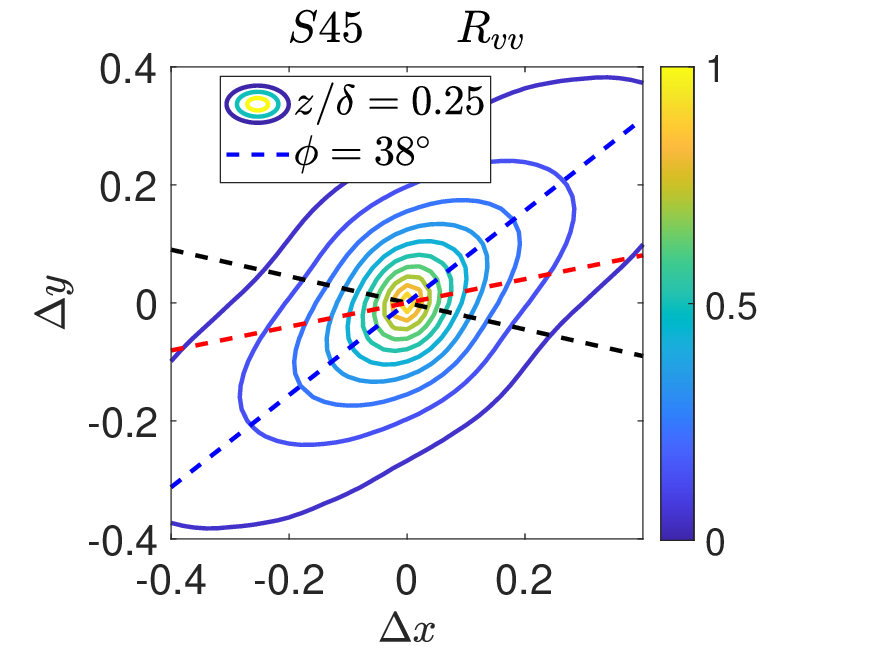}
            };
    \node[anchor=north west,
        xshift=-2mm,yshift=-2mm] at (image.north west) {{\rmfamily\fontsize{12}{13}\fontseries{l}\selectfont(n)}};
        \end{tikzpicture}}
    \subfloat[\label{selfc0.5vs45}]{
        \begin{tikzpicture}
        \node[anchor=north west, inner sep=0] (image) at (0,0) {
    \includegraphics[width=0.3\textwidth]{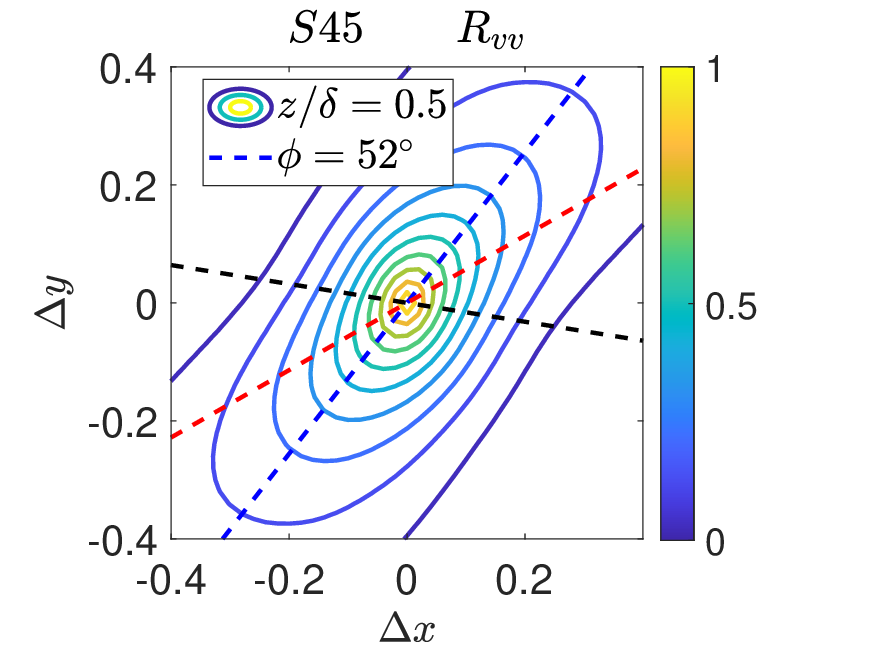}
            };
    \node[anchor=north west,
        xshift=-2mm,yshift=-2mm] at (image.north west) {{\rmfamily\fontsize{12}{13}\fontseries{l}\selectfont(o)}};
        \end{tikzpicture}}
    \caption{Two-point correlation contour maps of the spanwise velocity fluctuations in the $x-y$ plane at different heights and latitudes of the CNBLs and TNBL. The blue dashed line indicates the deflection direction. The black dashed line represents the mean wind direction, and the red dashed line is the mean shear direction {at the plane height}.}
    \label{fig:spanwise coherance}
\end{figure}

\begin{figure}
    \centering
    \subfloat[\label{selfc0.05w}]{
        \begin{tikzpicture}
        \node[anchor=north west, inner sep=0] (image) at (0,0) {
    \includegraphics[width=0.3\textwidth]{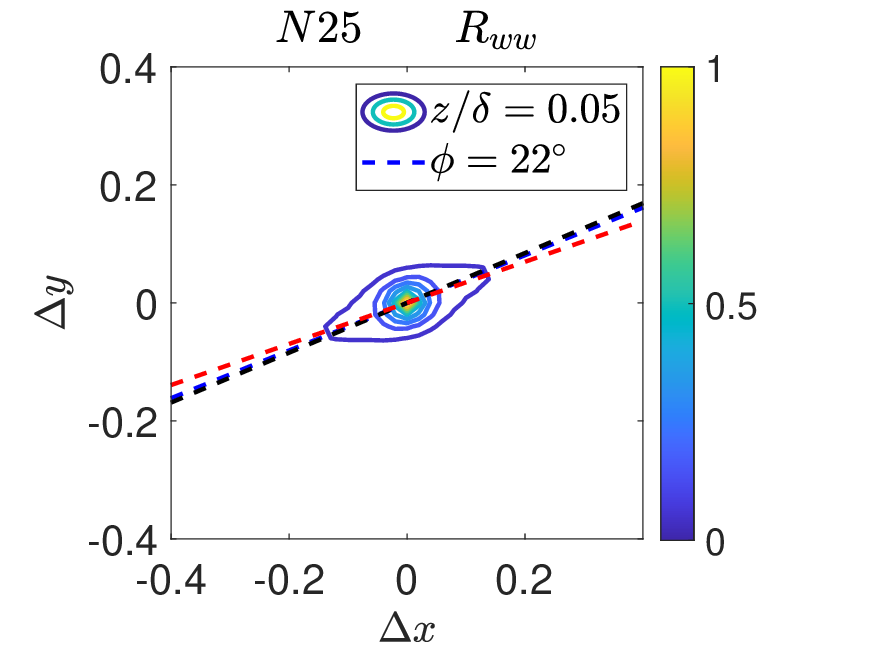}
            };
    \node[anchor=north west,
        xshift=-2mm,yshift=-2mm] at (image.north west) {{\rmfamily\fontsize{12}{13}\fontseries{l}\selectfont(a)}};
        \end{tikzpicture}}
    \subfloat[\label{selfc0.25w}]{
        \begin{tikzpicture}
        \node[anchor=north west, inner sep=0] (image) at (0,0) {
    \includegraphics[width=0.3\textwidth]{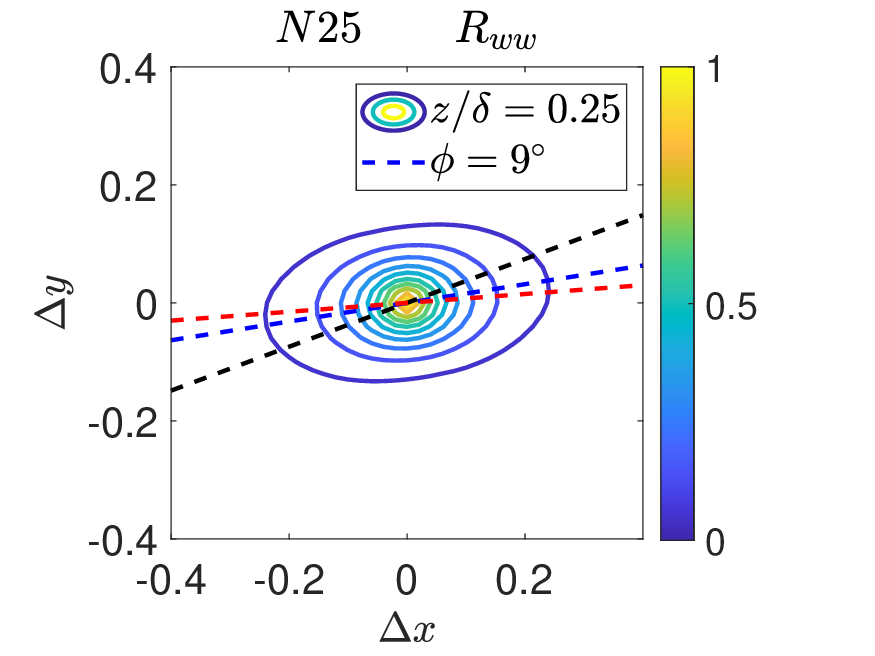}
            };
    \node[anchor=north west,
        xshift=-2mm,yshift=-2mm] at (image.north west) {{\rmfamily\fontsize{12}{13}\fontseries{l}\selectfont(b)}};
        \end{tikzpicture}}
    \subfloat[\label{selfc0.5w}]{
        \begin{tikzpicture}
        \node[anchor=north west, inner sep=0] (image) at (0,0) {
    \includegraphics[width=0.3\textwidth]{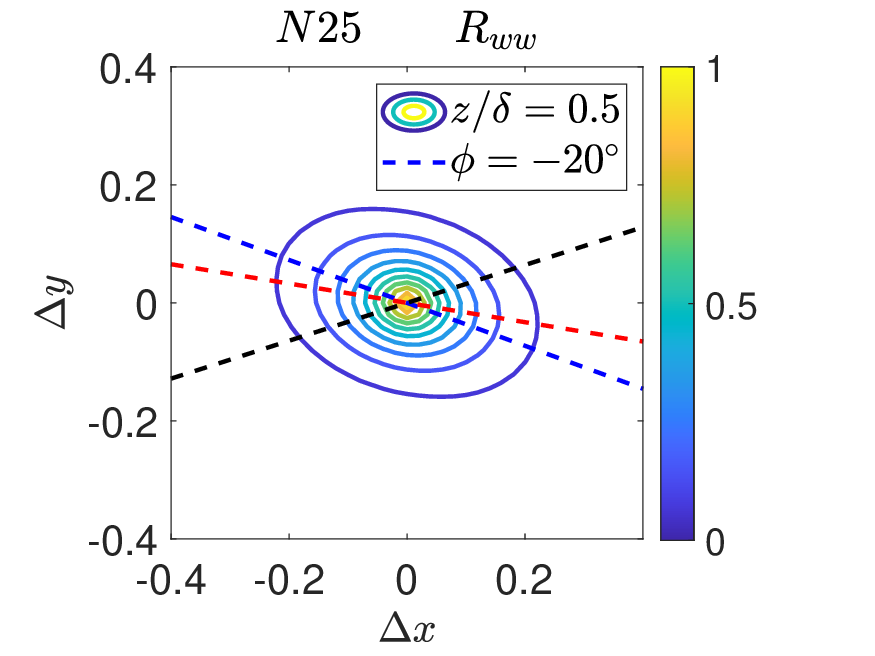}
            };
    \node[anchor=north west,
        xshift=-2mm,yshift=-2mm] at (image.north west) {{\rmfamily\fontsize{12}{13}\fontseries{l}\selectfont(c)}};
        \end{tikzpicture}}
\vspace{-1.45cm}  

 \centering
    \subfloat[\label{selfc0.05wn45}]{
        \begin{tikzpicture}
        \node[anchor=north west, inner sep=0] (image) at (0,0) {
    \includegraphics[width=0.3\textwidth]{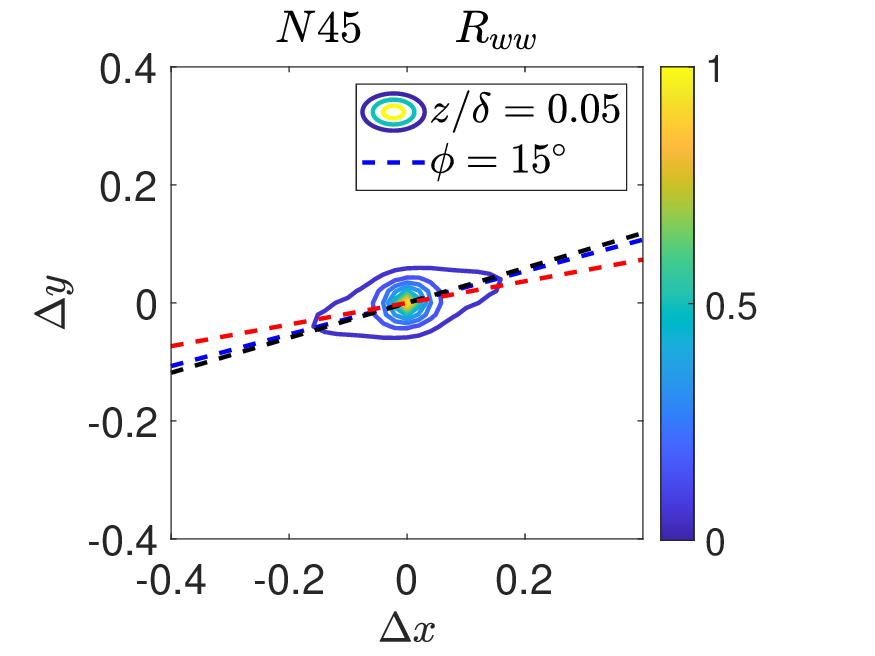}
            };
    \node[anchor=north west,
        xshift=-2mm,yshift=-2mm] at (image.north west) {{\rmfamily\fontsize{12}{13}\fontseries{l}\selectfont(d)}};
        \end{tikzpicture}}
   \subfloat[\label{selfc0.25wn45}]{
        \begin{tikzpicture}
        \node[anchor=north west, inner sep=0] (image) at (0,0) {
    \includegraphics[width=0.3\textwidth]{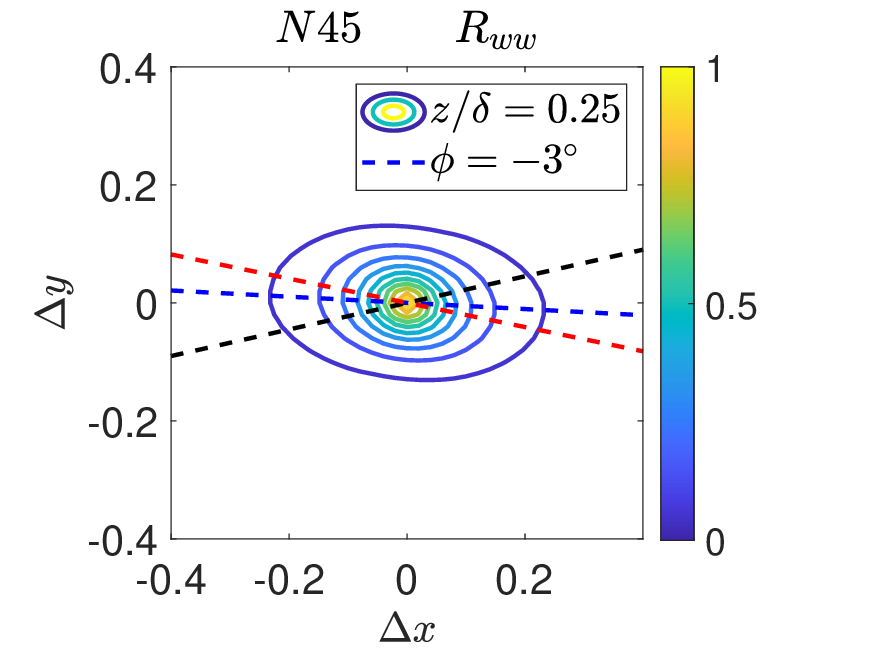}
            };
    \node[anchor=north west,
        xshift=-2mm,yshift=-2mm] at (image.north west) {{\rmfamily\fontsize{12}{13}\fontseries{l}\selectfont(e)}};
        \end{tikzpicture}}
    \subfloat[\label{selfc0.5wn45}]{
        \begin{tikzpicture}
        \node[anchor=north west, inner sep=0] (image) at (0,0) {
    \includegraphics[width=0.3\textwidth]{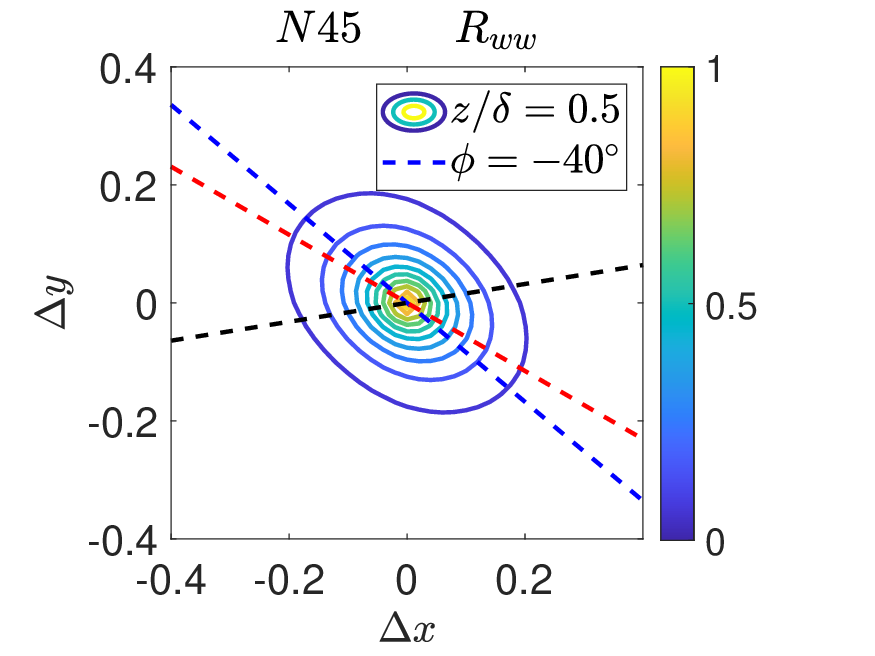}
            };
    \node[anchor=north west,
        xshift=-2mm,yshift=-2mm] at (image.north west) {{\rmfamily\fontsize{12}{13}\fontseries{l}\selectfont(f)}};
        \end{tikzpicture}}
  \vspace{-1.45cm}

  \centering
   \subfloat[\label{selfc0.05wn70}]{
        \begin{tikzpicture}
        \node[anchor=north west, inner sep=0] (image) at (0,0) {
    \includegraphics[width=0.3\textwidth]{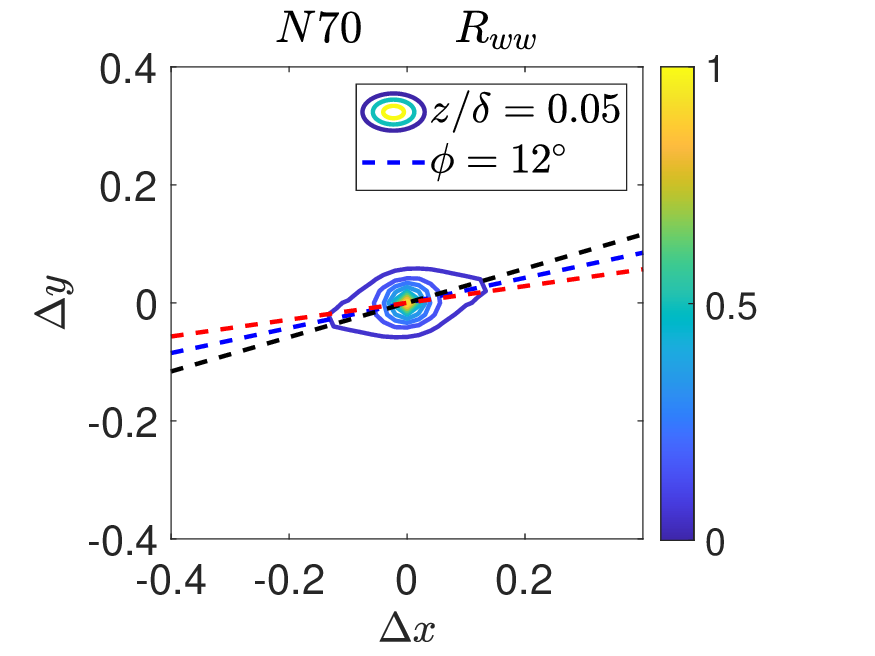}
            };
    \node[anchor=north west,
        xshift=-2mm,yshift=-2mm] at (image.north west) {{\rmfamily\fontsize{12}{13}\fontseries{l}\selectfont(g)}};
        \end{tikzpicture}}
     \subfloat[\label{selfc0.25wn70}]{
        \begin{tikzpicture}
        \node[anchor=north west, inner sep=0] (image) at (0,0) {
    \includegraphics[width=0.3\textwidth]{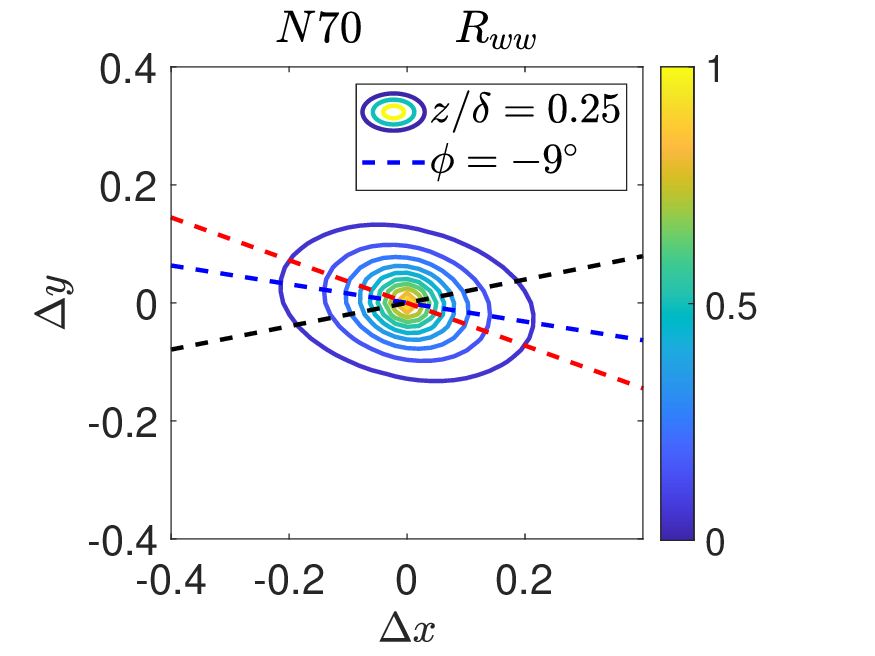}
            };
    \node[anchor=north west,
        xshift=-2mm,yshift=-2mm] at (image.north west) {{\rmfamily\fontsize{12}{13}\fontseries{l}\selectfont(h)}};
        \end{tikzpicture}}
     \subfloat[\label{selfc0.5wn70}]{
        \begin{tikzpicture}
        \node[anchor=north west, inner sep=0] (image) at (0,0) {
    \includegraphics[width=0.3\textwidth]{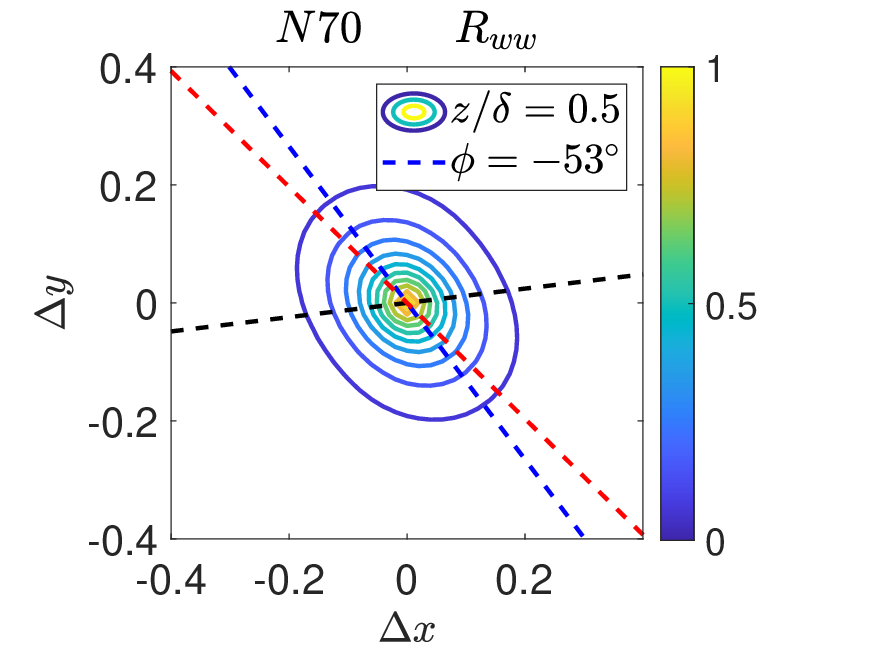}
            };
    \node[anchor=north west,
        xshift=-2mm,yshift=-2mm] at (image.north west) {{\rmfamily\fontsize{12}{13}\fontseries{l}\selectfont(i)}};
        \end{tikzpicture}}
 \vspace{-1.45cm}
   
         \centering
     \subfloat[\label{selfc0.05wn90}]{
        \begin{tikzpicture}
        \node[anchor=north west, inner sep=0] (image) at (0,0) {
    \includegraphics[width=0.3\textwidth]{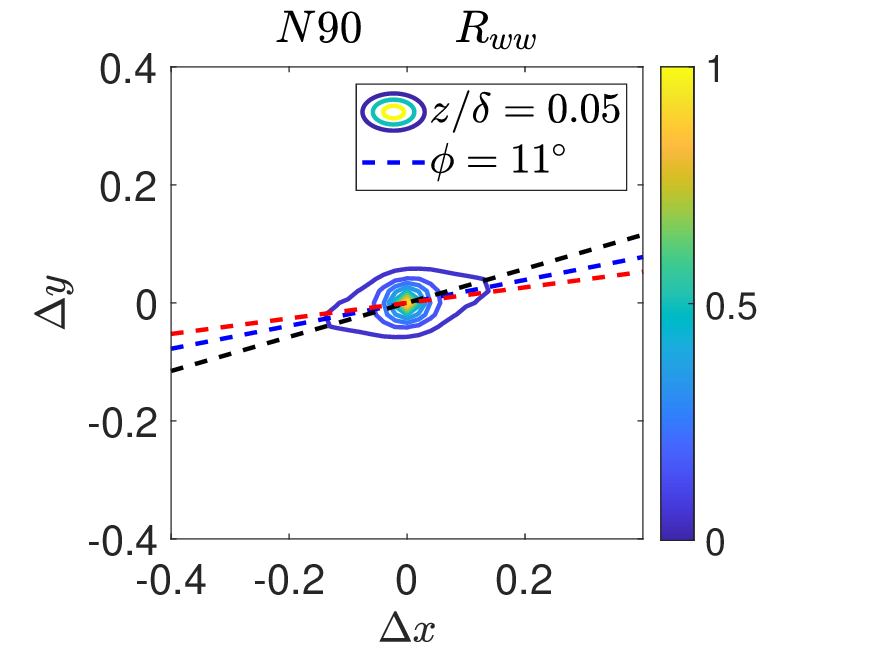}
            };
    \node[anchor=north west,
        xshift=-2mm,yshift=-2mm] at (image.north west) {{\rmfamily\fontsize{12}{13}\fontseries{l}\selectfont(j)}};
        \end{tikzpicture}}
    \subfloat[\label{selfc0.25wn90}]{
        \begin{tikzpicture}
        \node[anchor=north west, inner sep=0] (image) at (0,0) {
    \includegraphics[width=0.3\textwidth]{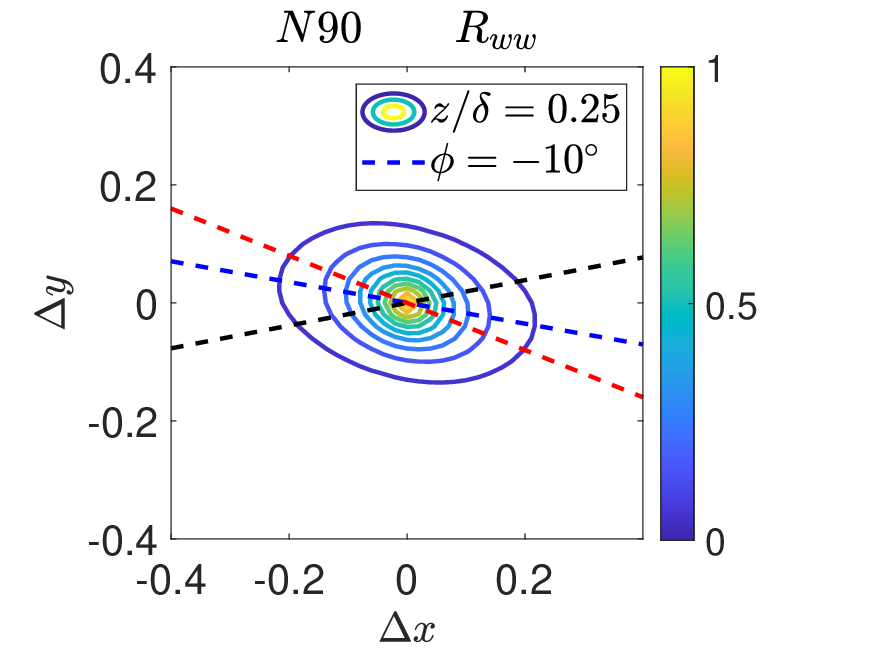}
            };
    \node[anchor=north west,
        xshift=-2mm,yshift=-2mm] at (image.north west) {{\rmfamily\fontsize{12}{13}\fontseries{l}\selectfont(k)}};
        \end{tikzpicture}}
     \subfloat[\label{selfc0.5wn90}]{
        \begin{tikzpicture}
        \node[anchor=north west, inner sep=0] (image) at (0,0) {
    \includegraphics[width=0.3\textwidth]{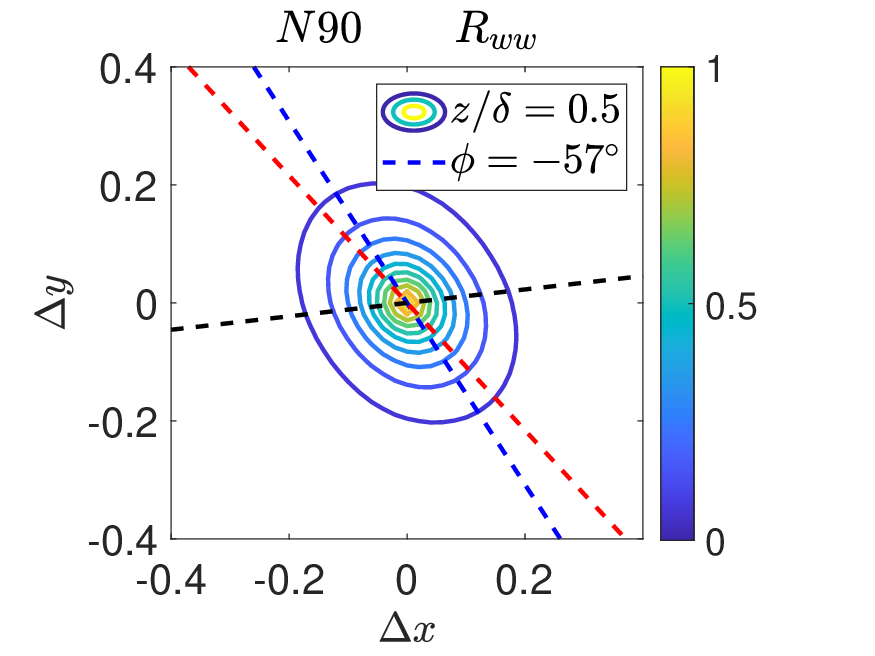}
            };
    \node[anchor=north west,
        xshift=-2mm,yshift=-2mm] at (image.north west) {{\rmfamily\fontsize{12}{13}\fontseries{l}\selectfont(l)}};
        \end{tikzpicture}}
  \vspace{-1.45cm}  
  
    \subfloat[\label{selfc0.05ws45}]{
        \begin{tikzpicture}
        \node[anchor=north west, inner sep=0] (image) at (0,0) {
    \includegraphics[width=0.3\textwidth]{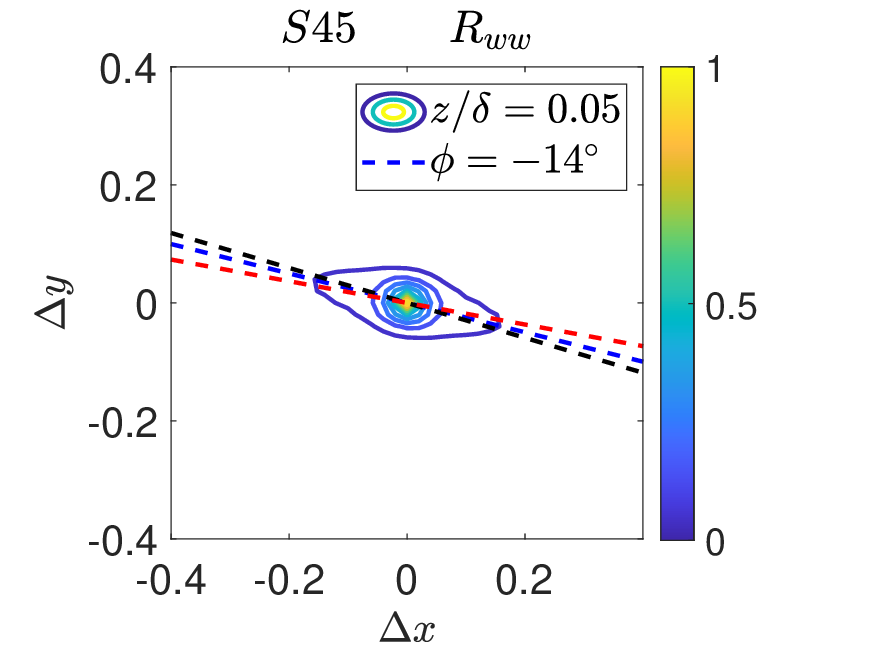}
            };
    \node[anchor=north west,
        xshift=-2mm,yshift=-2mm] at (image.north west) {{\rmfamily\fontsize{12}{13}\fontseries{l}\selectfont(m)}};
        \end{tikzpicture}}
    \subfloat[\label{selfc0.25ws45}]{
        \begin{tikzpicture}
        \node[anchor=north west, inner sep=0] (image) at (0,0) {
    \includegraphics[width=0.3\textwidth]{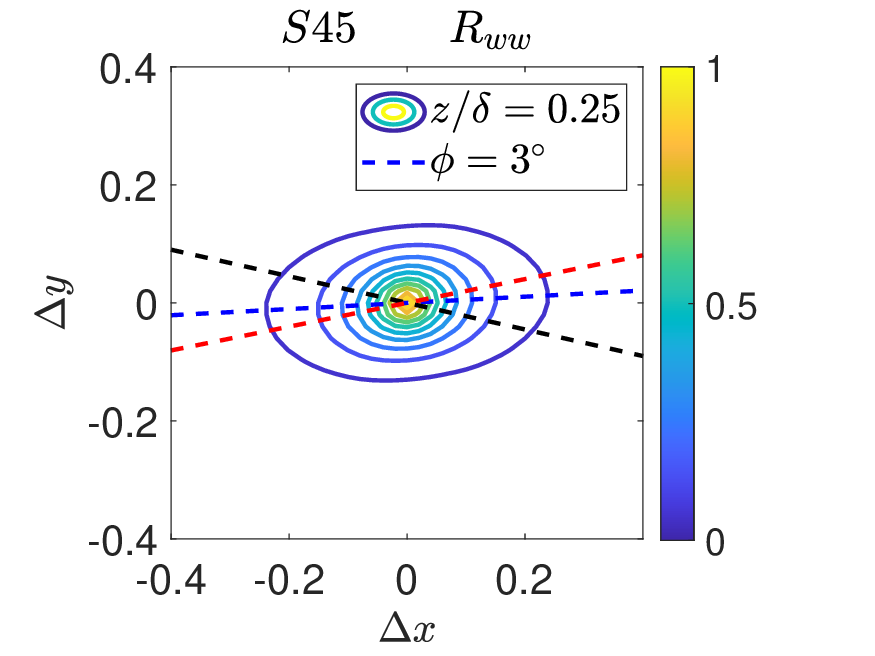}
            };
    \node[anchor=north west,
        xshift=-2mm,yshift=-2mm] at (image.north west) {{\rmfamily\fontsize{12}{13}\fontseries{l}\selectfont(n)}};
        \end{tikzpicture}}
    \subfloat[\label{selfc0.5ws45}]{
        \begin{tikzpicture}
        \node[anchor=north west, inner sep=0] (image) at (0,0) {
    \includegraphics[width=0.3\textwidth]{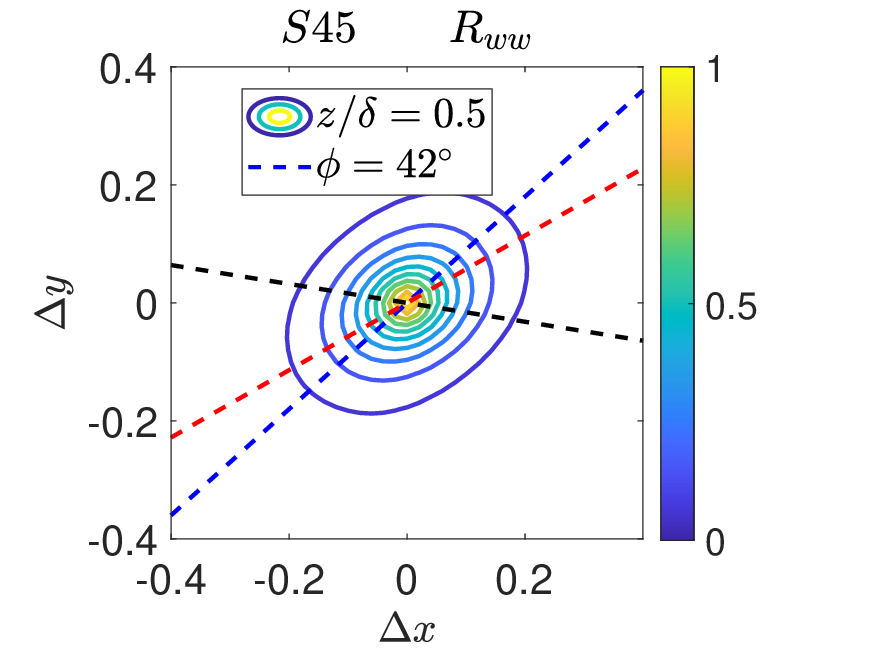}
            };
    \node[anchor=north west,
        xshift=-2mm,yshift=-2mm] at (image.north west) {{\rmfamily\fontsize{12}{13}\fontseries{l}\selectfont(o)}};
        \end{tikzpicture}}
    \caption{Two-point correlation contour maps of the vertical velocity fluctuations in the $x-y$ plane at different heights and latitudes of the CNBLs and TNBL. The blue dashed line indicates the deflection direction. The black dashed line represents the mean wind direction, and the red dashed line is the mean shear direction {at the plane height}.}
    \label{fig:vertical coherence}
\end{figure}

 As shown in Figs.~\ref{fig7} to \ref{fig9}, the Coriolis force causes the deflection of coherent velocity structures in the CNBLs.
 To quantify this effect, we use two-point correlations in the horizontal $x-y$ planes to characterize the deflection angle with respect to the longitudinal direction \citep{masonLargeEddySimulationsNeutralstaticstability1987,colemanNumericalStudyTurbulent1990a,zikanov2003large,shingaiStudyTurbulenceStructure2004a}.
Taking longitudinal velocity fluctuation as an example, the definition of the two-point correlation coefficient in the $x-y$ plane is
\begin{equation}\label{eq Ruu}
R_{uu}^{xy}(\Delta x, \Delta y, z) = \frac{
    \langle u'(x, y, z) \, u'(x + \Delta x, y + \Delta y, z) \rangle
}{
    \sigma_u(x, y, z)  \sigma_u(x + \Delta x, y + \Delta y, z),
}
\end{equation}
where $\sigma_u$ is the root-mean-square of the longitudinal velocity fluctuations, and the angle brackets represent the averaging in time and homogeneous directions.
{The direction of structure deflection is defined as the major orientation of the two-point velocity correlation distribution, which is determined as the straight zero-crossing line passing the largest correlation values at each $\Delta x$.}
 

Fig.~\ref{fig：streamwise coherence} displays the contour maps of the correlation coefficients of the streamwise velocity fluctuations at different heights in the CNBLs. 
{It is noted that the streamwise velocity fluctuation is the projection of the total velocity fluctuation along the mean wind velocity direction.}
One can see that the correlation contours have an elongated shape, and the major axis corresponds to the direction of the deflection, as indicated by the blue dashed line. 
{The figure also plots the directions of the mean wind (black dashed line) and mean wind shear (red dashed line). At $z/\delta=0.05$ of the CNBLs with different latitudes, the directions of the mean wind shear are always very close to the deflection directions of the streamwise velocity structures. The mean wind direction is also close to that of the streamwise velocity structures at $z/\delta=0.05$. At higher heights, they depart with each other evidently, and the deflection direction of the streamwise velocity structures is well between those of the mean wind velocity and mean wind shear.}

Fig.~\ref{fig:spanwise coherance} shows the correlation contour maps of the spanwise velocity fluctuations at different heights in the near-ground region of the CNBLs, where the variation of the structure deflection is the most significant. 
{Similar to the streamwise velocity, the correlation contours also have an elongated shape. 
While unlike the streamwise velocity structures, the deflection directions of the spanwise velocity structures are quite different from those of the mean wind velocity and mean wind shear in the CNBLs at different latitudes. }

The correlation contour maps of vertical velocity fluctuations at different heights in the CNBLs are shown in Fig.~\ref{fig:vertical coherence}. Compared with those of the streamwise and spanwise velocities, the contours of the vertical velocity correlations are much more compact but still have an elongated direction. 
{The variation of the deflection angle is similar to that of the streamwise velocity in that it is close to those of the mean wind velocity and mean wind shear at $z/\delta=0.05$ of the CNBLs with different latitudes, and deviates from each other if increasing the height. }

\begin{figure}
    \centering
    \subfloat[\label{selfc0.05g12}]{
        \begin{tikzpicture}
        \node[anchor=north west, inner sep=0] (image) at (0,0) {
    \includegraphics[width=0.3\textwidth]{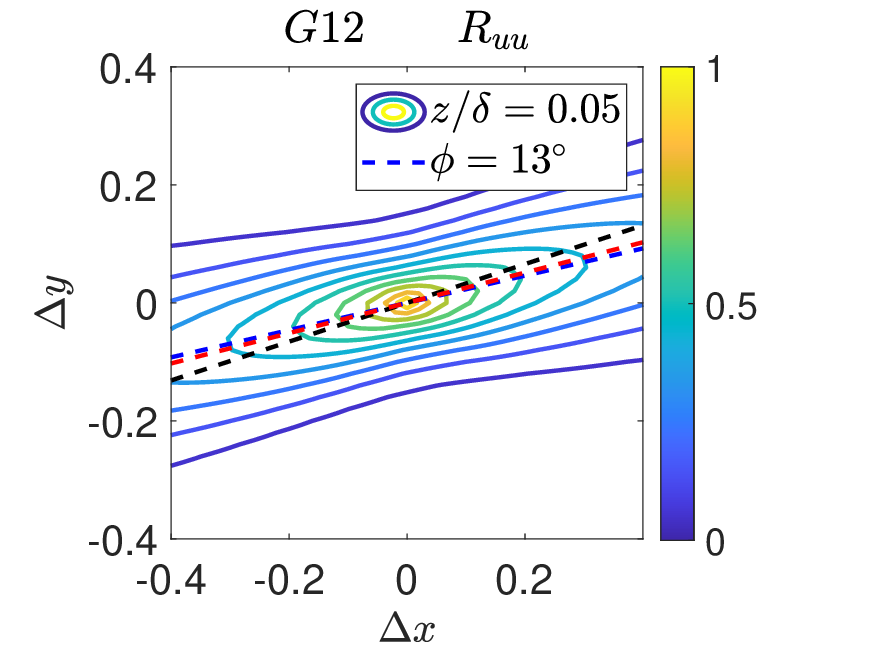}
            };
    \node[anchor=north west,
        xshift=-2mm,yshift=-2mm] at (image.north west) {{\rmfamily\fontsize{12}{13}\fontseries{l}\selectfont(a)}};
        \end{tikzpicture}}
    \subfloat[\label{selfc0.25g12}]{
        \begin{tikzpicture}
        \node[anchor=north west, inner sep=0] (image) at (0,0) {
    \includegraphics[width=0.3\textwidth]{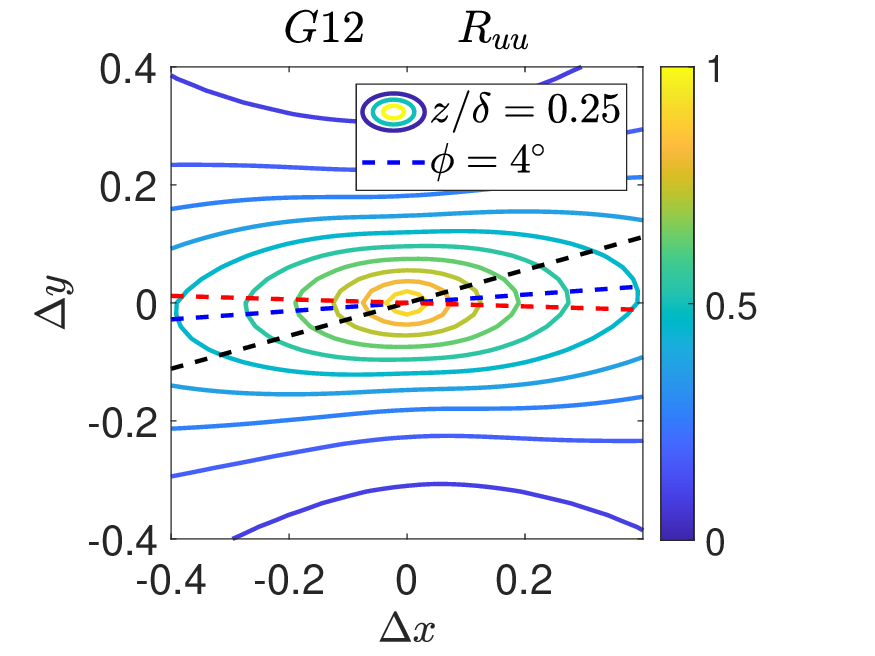}
            };
    \node[anchor=north west,
        xshift=-2mm,yshift=-2mm] at (image.north west) {{\rmfamily\fontsize{12}{13}\fontseries{l}\selectfont(b)}};
        \end{tikzpicture}}
    \subfloat[\label{selfc0.5g12}]{
        \begin{tikzpicture}
        \node[anchor=north west, inner sep=0] (image) at (0,0) {
    \includegraphics[width=0.3\textwidth]{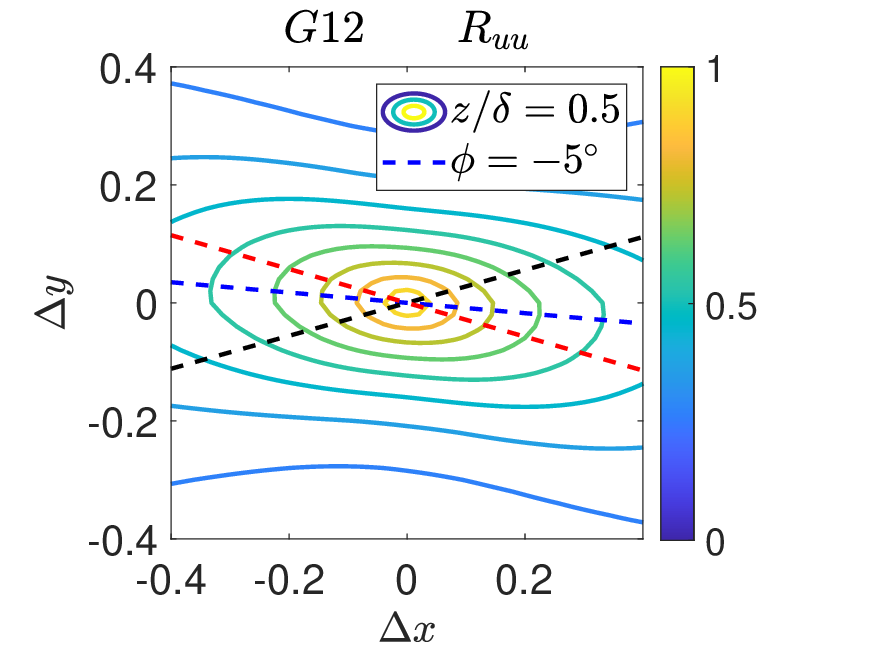}
            };
    \node[anchor=north west,
        xshift=-2mm,yshift=-2mm] at (image.north west) {{\rmfamily\fontsize{12}{13}\fontseries{l}\selectfont(c)}};
        \end{tikzpicture}}
\vspace{-1.45cm}  

 \centering
    \subfloat[\label{selfc0.05g16}]{
        \begin{tikzpicture}
        \node[anchor=north west, inner sep=0] (image) at (0,0) {
    \includegraphics[width=0.3\textwidth]{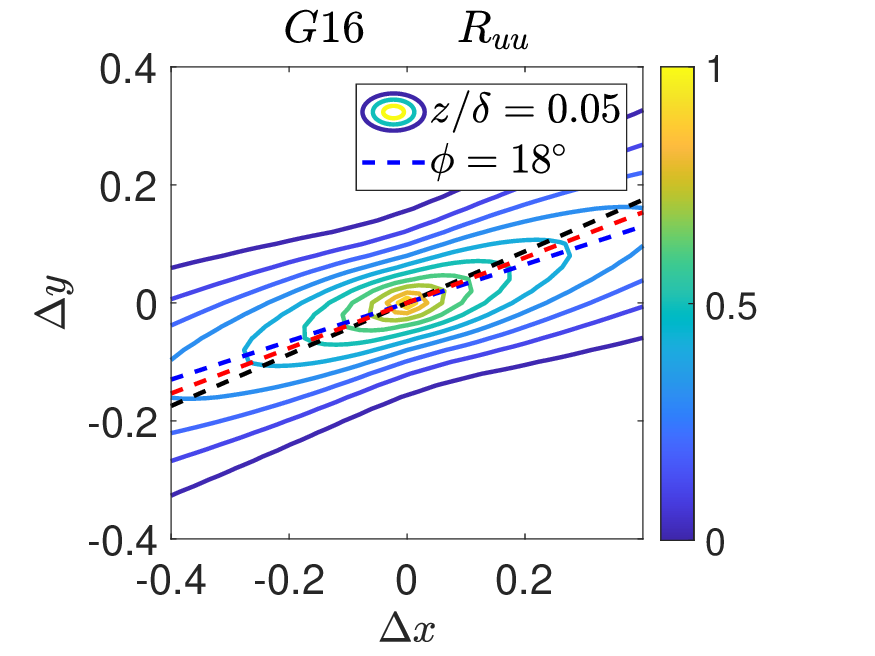}
            };
    \node[anchor=north west,
        xshift=-2mm,yshift=-2mm] at (image.north west) {{\rmfamily\fontsize{12}{13}\fontseries{l}\selectfont(d)}};
        \end{tikzpicture}}
   \subfloat[\label{selfc0.25g16}]{
        \begin{tikzpicture}
        \node[anchor=north west, inner sep=0] (image) at (0,0) {
    \includegraphics[width=0.3\textwidth]{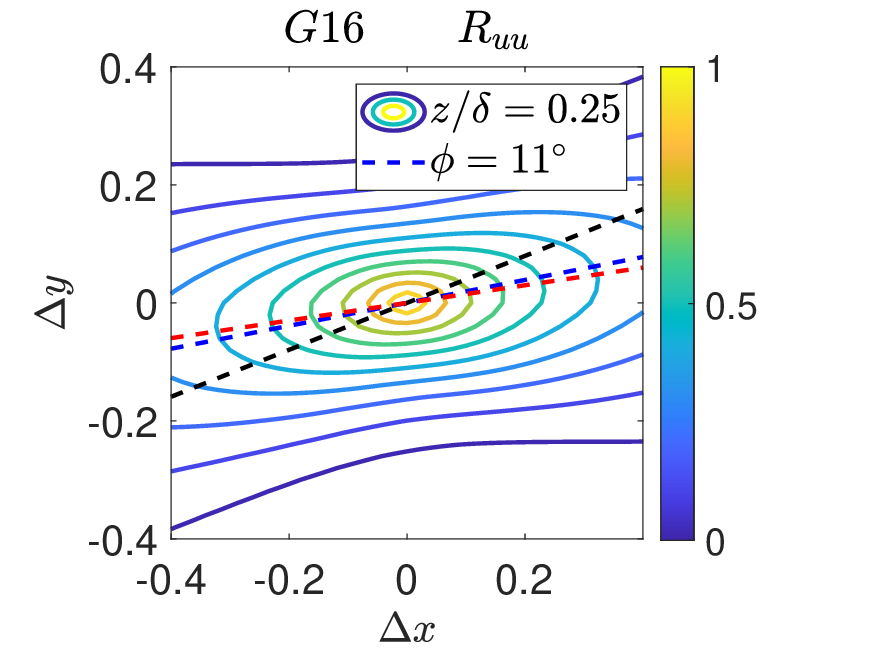}
            };
    \node[anchor=north west,
        xshift=-2mm,yshift=-2mm] at (image.north west) {{\rmfamily\fontsize{12}{13}\fontseries{l}\selectfont(e)}};
        \end{tikzpicture}}
    \subfloat[\label{selfc0.5g16}]{
        \begin{tikzpicture}
        \node[anchor=north west, inner sep=0] (image) at (0,0) {
    \includegraphics[width=0.3\textwidth]{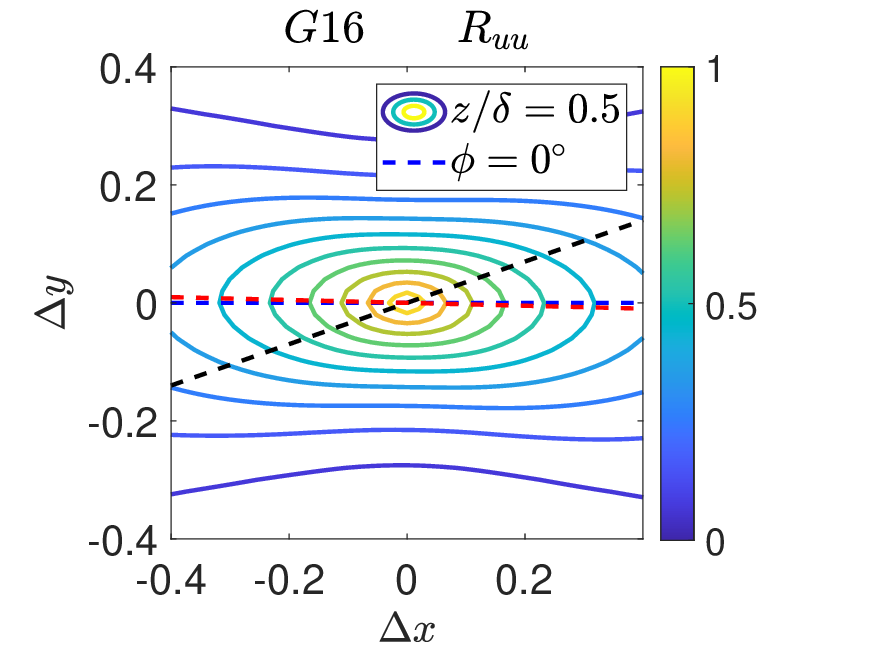}
            };
    \node[anchor=north west,
        xshift=-2mm,yshift=-2mm] at (image.north west) {{\rmfamily\fontsize{12}{13}\fontseries{l}\selectfont(f)}};
        \end{tikzpicture}}
    \caption{Two-point correlation contour maps of the streamwise velocity fluctuations in the $x-y$ plane at different heights with different geostrophic wind speeds. The blue dashed line indicates the deflection direction. The black dashed line represents the mean wind direction, and the red dashed line is the mean shear direction {at the plane height}.}
    \label{fig：G streamwise coherence}
\end{figure}

\begin{figure}
    \centering
    \subfloat[\label{selfc0.05vg12}]{
        \begin{tikzpicture}
        \node[anchor=north west, inner sep=0] (image) at (0,0) {
    \includegraphics[width=0.3\textwidth]{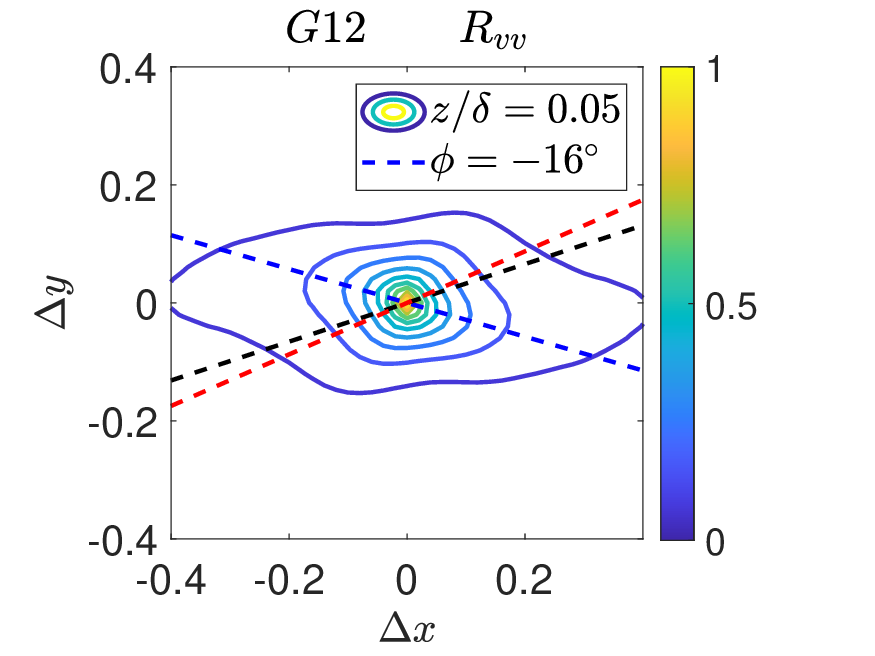}
            };
    \node[anchor=north west,
        xshift=-2mm,yshift=-2mm] at (image.north west) {{\rmfamily\fontsize{12}{13}\fontseries{l}\selectfont(a)}};
        \end{tikzpicture}}
    \subfloat[\label{selfc0.25vg12}]{
        \begin{tikzpicture}
        \node[anchor=north west, inner sep=0] (image) at (0,0) {
    \includegraphics[width=0.3\textwidth]{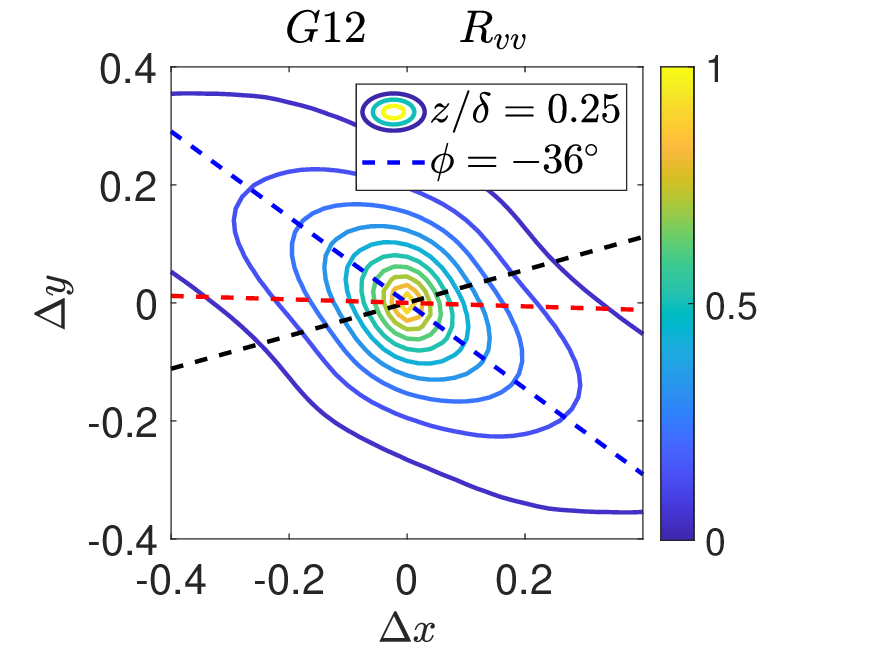}
            };
    \node[anchor=north west,
        xshift=-2mm,yshift=-2mm] at (image.north west) {{\rmfamily\fontsize{12}{13}\fontseries{l}\selectfont(b)}};
        \end{tikzpicture}}
    \subfloat[\label{selfc0.5vg12}]{
        \begin{tikzpicture}
        \node[anchor=north west, inner sep=0] (image) at (0,0) {
    \includegraphics[width=0.3\textwidth]{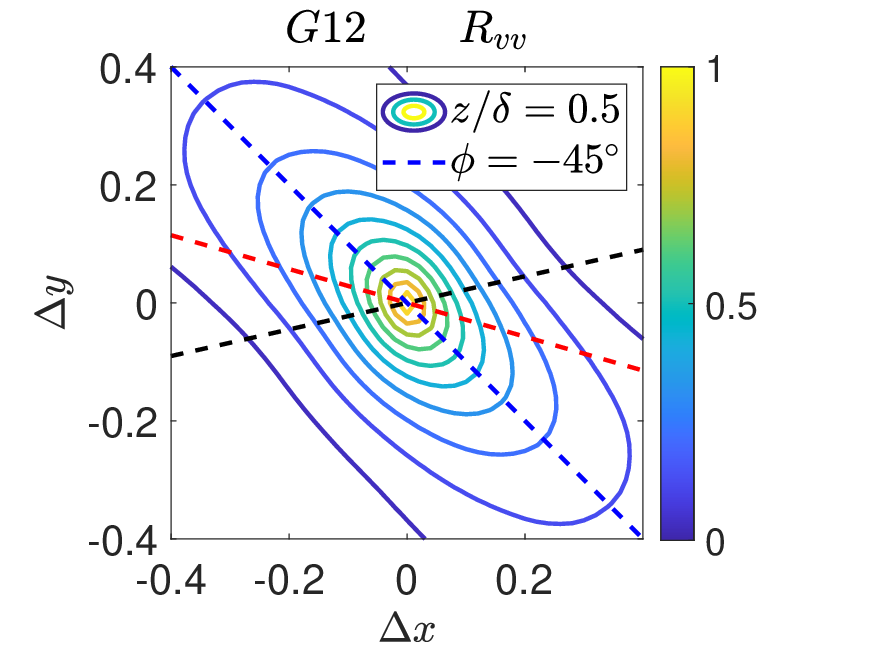}
            };
    \node[anchor=north west,
        xshift=-2mm,yshift=-2mm] at (image.north west) {{\rmfamily\fontsize{12}{13}\fontseries{l}\selectfont(c)}};
        \end{tikzpicture}}
\vspace{-1.45cm}  

 \centering
    \subfloat[\label{selfc0.05vg16}]{
        \begin{tikzpicture}
        \node[anchor=north west, inner sep=0] (image) at (0,0) {
    \includegraphics[width=0.3\textwidth]{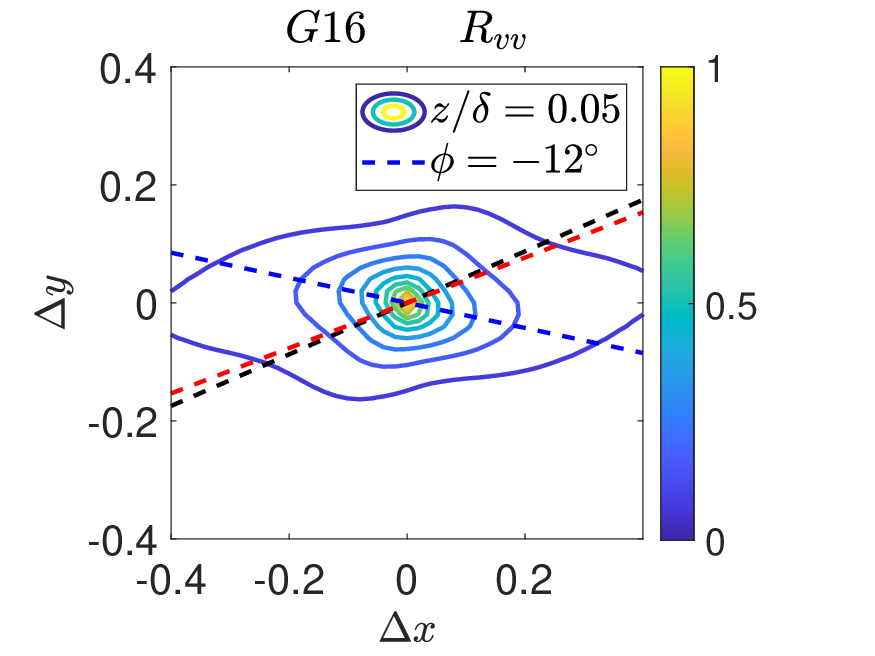}
            };
    \node[anchor=north west,
        xshift=-2mm,yshift=-2mm] at (image.north west) {{\rmfamily\fontsize{12}{13}\fontseries{l}\selectfont(d)}};
        \end{tikzpicture}}
   \subfloat[\label{selfc0.25vg16}]{
        \begin{tikzpicture}
        \node[anchor=north west, inner sep=0] (image) at (0,0) {
    \includegraphics[width=0.3\textwidth]{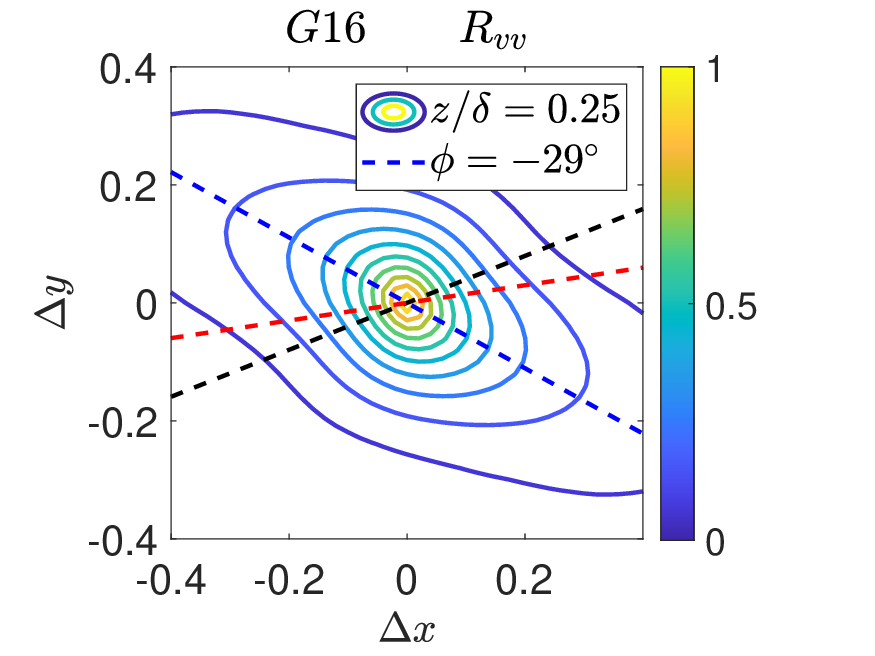}
            };
    \node[anchor=north west,
        xshift=-2mm,yshift=-2mm] at (image.north west) {{\rmfamily\fontsize{12}{13}\fontseries{l}\selectfont(e)}};
        \end{tikzpicture}}
    \subfloat[\label{selfc0.5vg16}]{
        \begin{tikzpicture}
        \node[anchor=north west, inner sep=0] (image) at (0,0) {
    \includegraphics[width=0.3\textwidth]{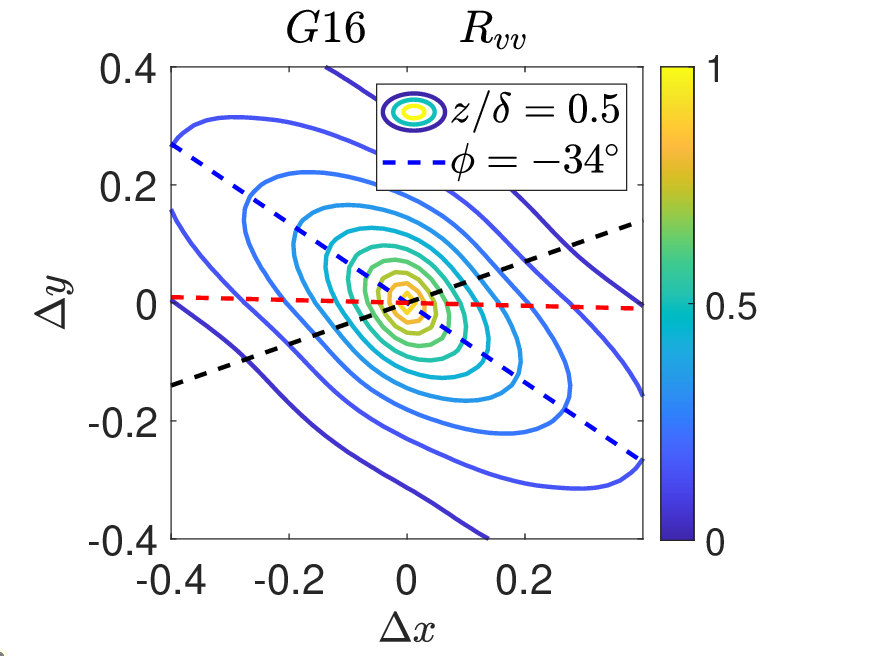}
            };
    \node[anchor=north west,
        xshift=-2mm,yshift=-2mm] at (image.north west) {{\rmfamily\fontsize{12}{13}\fontseries{l}\selectfont(f)}};
        \end{tikzpicture}}
    \caption{Two-point correlation contour maps of the spanwise velocity fluctuations in the $x-y$ plane at different heights with different geostrophic wind speeds. The blue dashed line indicates the deflection direction. The black dashed line represents the mean wind direction, and the red dashed line is the mean shear direction {at the plane height}.}
    \label{fig：G spanwise coherence}
\end{figure}

\begin{figure}
    \centering
    \subfloat[\label{selfc0.05wg12}]{
        \begin{tikzpicture}
        \node[anchor=north west, inner sep=0] (image) at (0,0) {
    \includegraphics[width=0.3\textwidth]{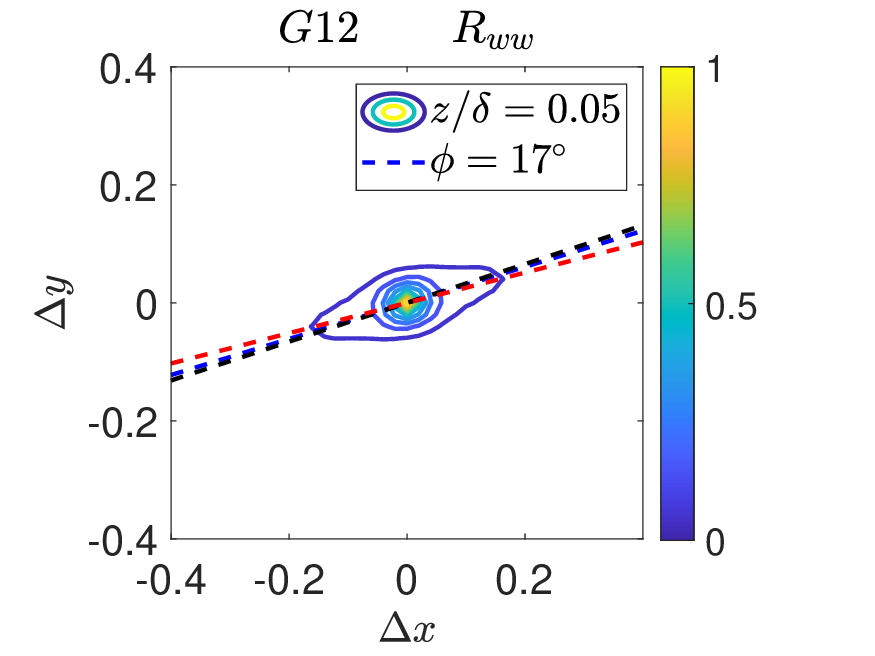}
            };
    \node[anchor=north west,
        xshift=-2mm,yshift=-2mm] at (image.north west) {{\rmfamily\fontsize{12}{13}\fontseries{l}\selectfont(a)}};
        \end{tikzpicture}}
    \subfloat[\label{selfc0.25wg12}]{
        \begin{tikzpicture}
        \node[anchor=north west, inner sep=0] (image) at (0,0) {
    \includegraphics[width=0.3\textwidth]{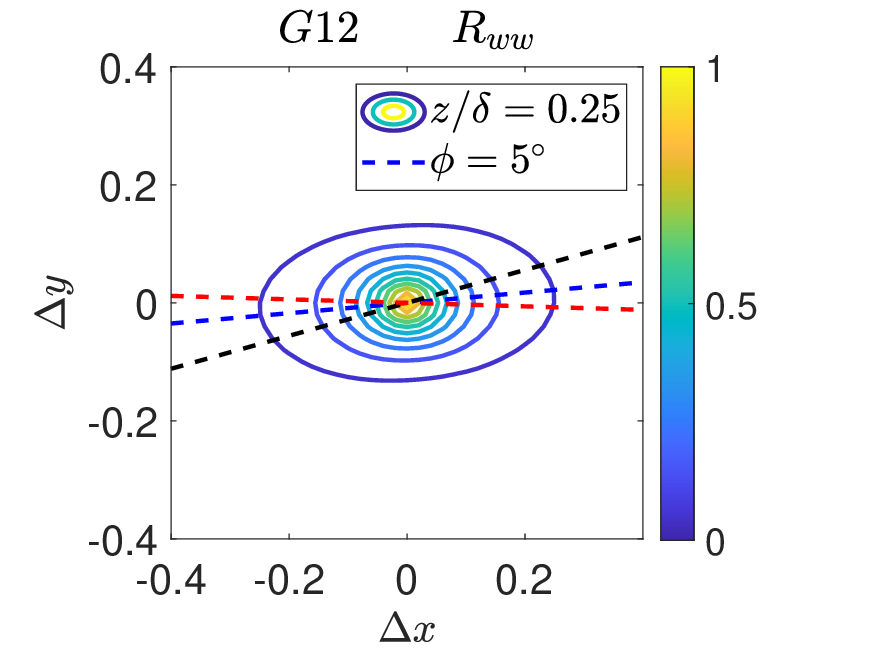}
            };
    \node[anchor=north west,
        xshift=-2mm,yshift=-2mm] at (image.north west) {{\rmfamily\fontsize{12}{13}\fontseries{l}\selectfont(b)}};
        \end{tikzpicture}}
    \subfloat[\label{selfc0.5wg12}]{
        \begin{tikzpicture}
        \node[anchor=north west, inner sep=0] (image) at (0,0) {
    \includegraphics[width=0.3\textwidth]{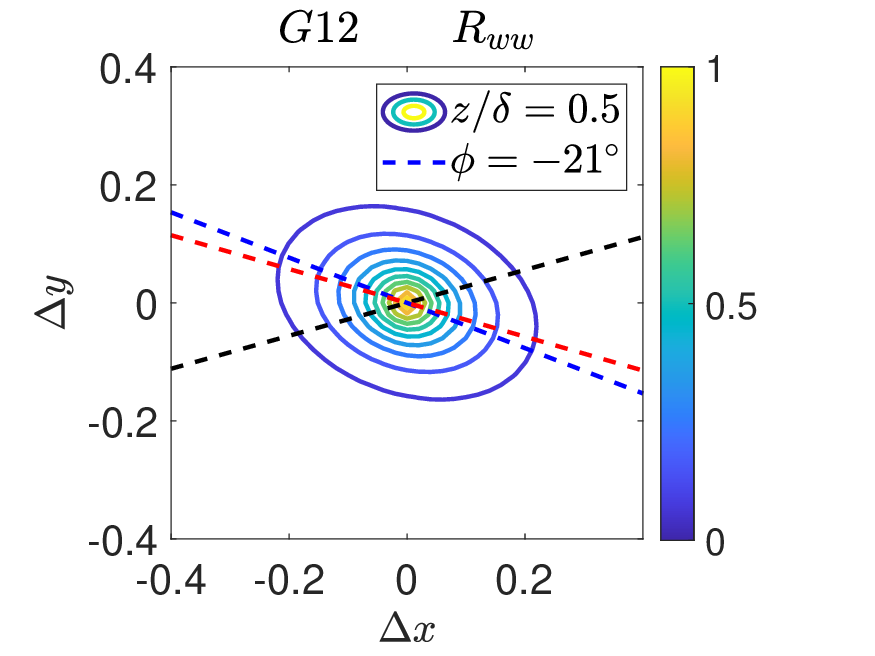}
            };
    \node[anchor=north west,
        xshift=-2mm,yshift=-2mm] at (image.north west) {{\rmfamily\fontsize{12}{13}\fontseries{l}\selectfont(c)}};
        \end{tikzpicture}}
\vspace{-1.45cm}  

 \centering
    \subfloat[\label{selfc0.05wg16}]{
        \begin{tikzpicture}
        \node[anchor=north west, inner sep=0] (image) at (0,0) {
    \includegraphics[width=0.3\textwidth]{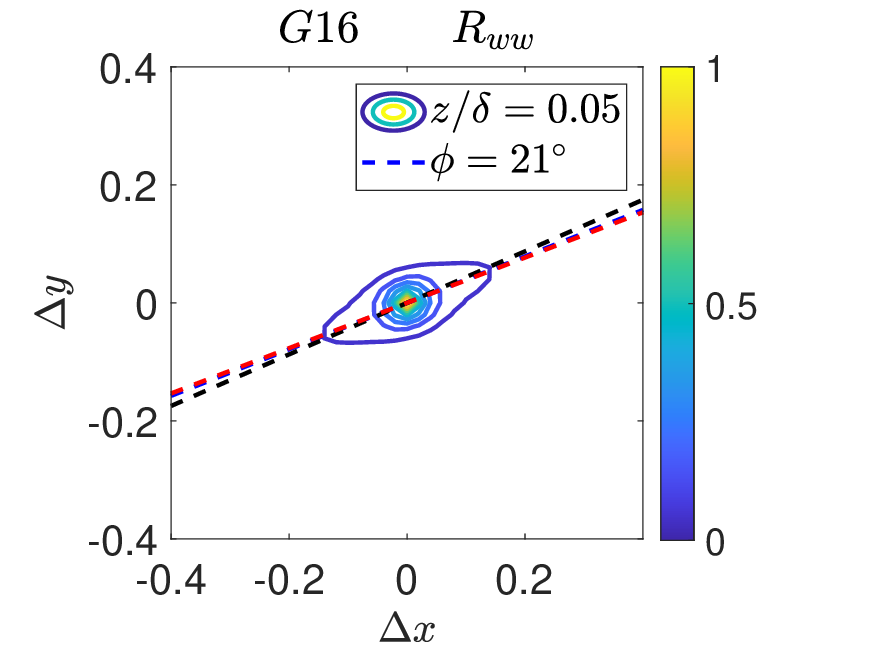}
            };
    \node[anchor=north west,
        xshift=-2mm,yshift=-2mm] at (image.north west) {{\rmfamily\fontsize{12}{13}\fontseries{l}\selectfont(d)}};
        \end{tikzpicture}}
   \subfloat[\label{selfc0.25wg16}]{
        \begin{tikzpicture}
        \node[anchor=north west, inner sep=0] (image) at (0,0) {
    \includegraphics[width=0.3\textwidth]{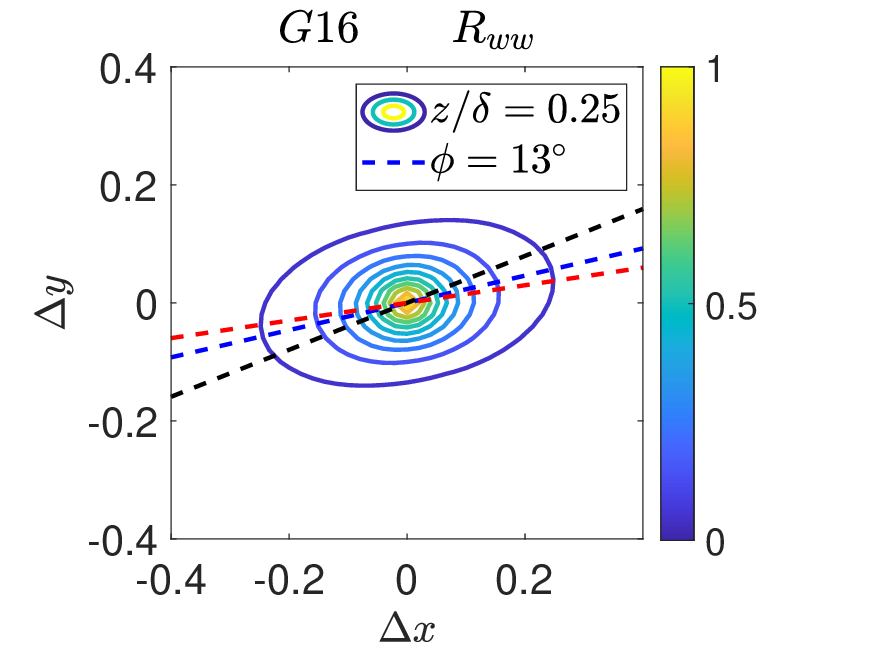}
            };
    \node[anchor=north west,
        xshift=-2mm,yshift=-2mm] at (image.north west) {{\rmfamily\fontsize{12}{13}\fontseries{l}\selectfont(e)}};
        \end{tikzpicture}}
    \subfloat[\label{selfc0.5wg16}]{
        \begin{tikzpicture}
        \node[anchor=north west, inner sep=0] (image) at (0,0) {
    \includegraphics[width=0.3\textwidth]{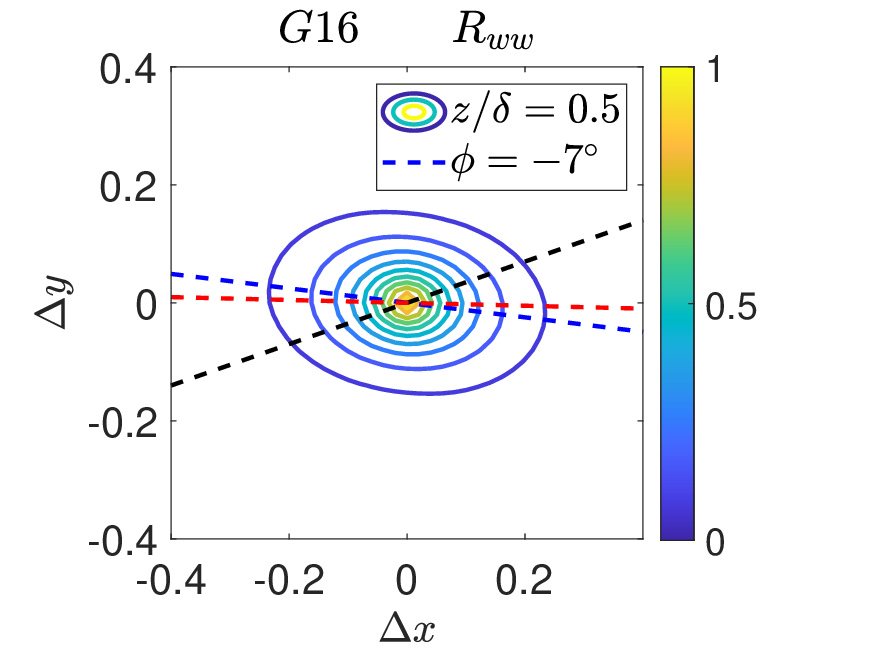}
            };
    \node[anchor=north west,
        xshift=-2mm,yshift=-2mm] at (image.north west) {{\rmfamily\fontsize{12}{13}\fontseries{l}\selectfont(f)}};
        \end{tikzpicture}}
    \caption{Two-point correlation contour maps of the vertical velocity fluctuations in the $x-y$ plane at different heights with different geostrophic wind speeds. The blue dashed line indicates the deflection direction. The black dashed line represents the mean wind direction, and the red dashed line is the mean shear direction {at the plane height}.}
    \label{fig：G vertical coherence}
\end{figure}

Fig.~\ref{fig：G streamwise coherence} shows the two-dimensional correlations of the streamwise velocity in the horizontal $x-y$ planes under different geostrophic wind speeds. 
In general, the shapes of the correlation contours are similar under the two geostrophic wind speeds, which are elongated in one direction as the major deflection axis. 
With a greater geostrophic wind speed, the deflection angle $\phi_u$ is more positive or less negative at the same height, implying a larger deflection due to the stronger Coriolis force. 
{The correlation contour maps of the spanwise velocity in the horizontal $x-y$ planes are shown in Fig.~\ref{fig：G spanwise coherence}.
It can be seen that, at the same height, the deflection angle $\phi_v$ is less negative with a greater geostrophic wind speed.}
Fig.~\ref{fig：G vertical coherence} shows the two-point correlation contour maps of the vertical flow velocity in the horizontal $x-y$ planes with different geostrophic wind speeds.
It shows that the general features of the correlation maps are similar, and the differences lie in the magnitudes of the deflection angle. 
With a greater geostrophic wind speed, the deflection angle is more positive or less negative at the same height.

\begin{figure}
    \centering
    \subfloat[\label{selfplanex}]{
        \begin{tikzpicture}
        \node[anchor=north west, inner sep=0] (image) at (0,0) {
    \includegraphics[width=0.3\textwidth]{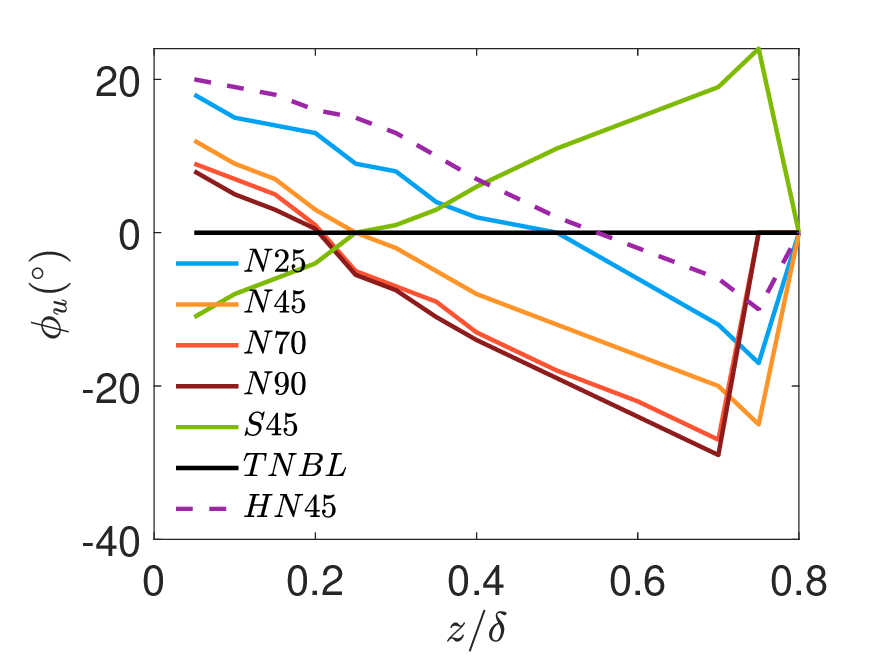}
            };
    \node[anchor=north west,
        xshift=-2mm,yshift=-2mm] at (image.north west) {{\rmfamily\fontsize{12}{13}\fontseries{l}\selectfont(a)}};
        \end{tikzpicture}}
    \subfloat[\label{selfplanex-theta}]{
        \begin{tikzpicture}
        \node[anchor=north west, inner sep=0] (image) at (0,0) {
    \includegraphics[width=0.3\textwidth]{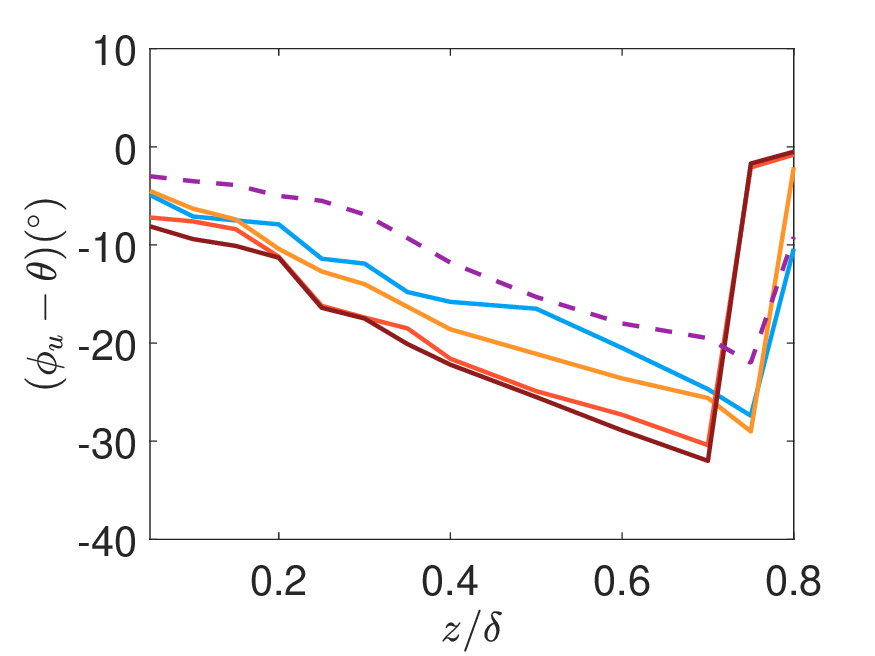}
            };
    \node[anchor=north west,
        xshift=-2mm,yshift=-2mm] at (image.north west) {{\rmfamily\fontsize{12}{13}\fontseries{l}\selectfont(b)}};
        \end{tikzpicture}}
    \subfloat[\label{selfplanex_theta}]{
        \begin{tikzpicture}
        \node[anchor=north west, inner sep=0] (image) at (0,0) {
    \includegraphics[width=0.3\textwidth]{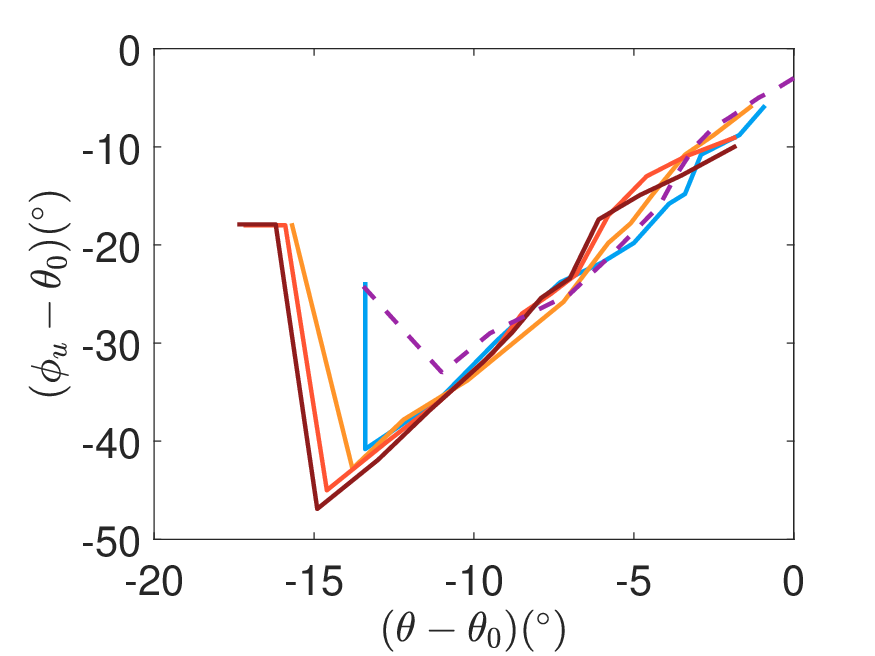}
            };
    \node[anchor=north west,
        xshift=-2mm,yshift=-2mm] at (image.north west) {{\rmfamily\fontsize{12}{13}\fontseries{l}\selectfont(c)}};
        \end{tikzpicture}}
        \vspace{-1.22cm} 
        
       \centering
    \subfloat[\label{selfplaney}]{
        \begin{tikzpicture}
        \node[anchor=north west, inner sep=0] (image) at (0,0) {
    \includegraphics[width=0.3\textwidth]{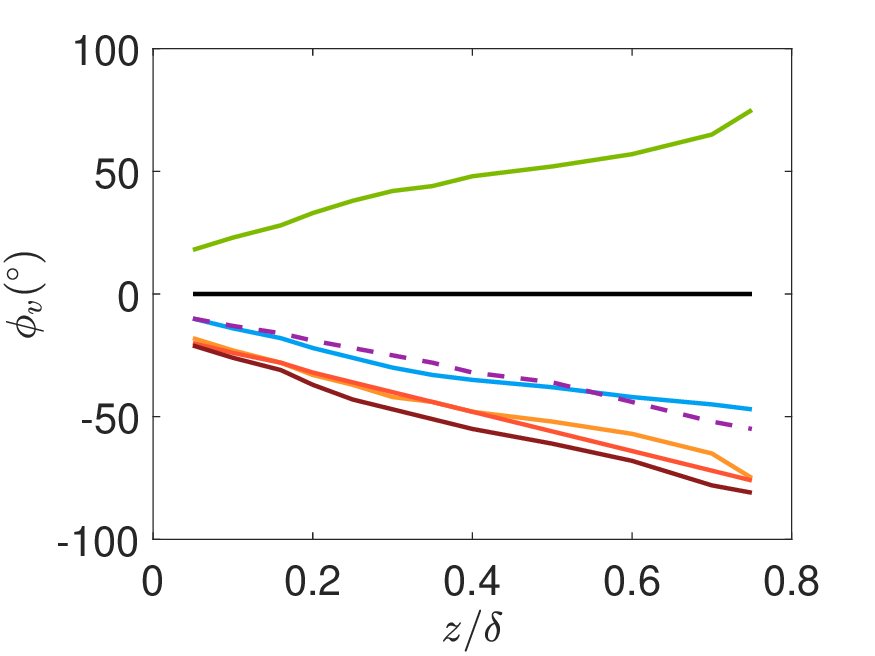}
            };
    \node[anchor=north west,
        xshift=-2mm,yshift=-2mm] at (image.north west) {{\rmfamily\fontsize{12}{13}\fontseries{l}\selectfont(d)}};
        \end{tikzpicture}}
    \subfloat[\label{selfplaney-theta}]{
        \begin{tikzpicture}
        \node[anchor=north west, inner sep=0] (image) at (0,0) {
    \includegraphics[width=0.3\textwidth]{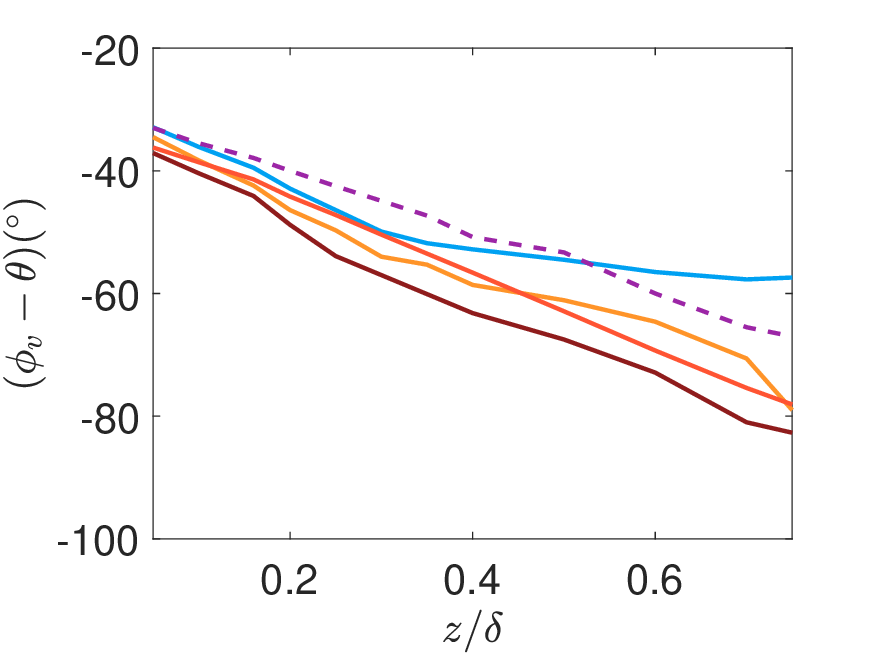}
            };
    \node[anchor=north west,
        xshift=-2mm,yshift=-2mm] at (image.north west) {{\rmfamily\fontsize{12}{13}\fontseries{l}\selectfont(e)}};
        \end{tikzpicture}}
            \subfloat[\label{selfplaney_theta}]{
        \begin{tikzpicture}
        \node[anchor=north west, inner sep=0] (image) at (0,0) {
    \includegraphics[width=0.3\textwidth]{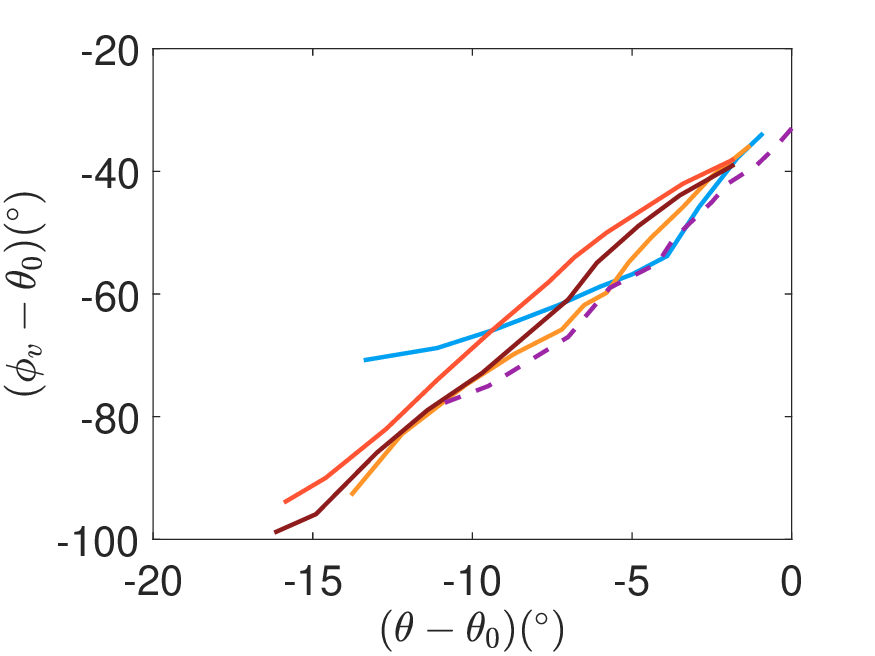}
            };
    \node[anchor=north west,
        xshift=-2mm,yshift=-2mm] at (image.north west) {{\rmfamily\fontsize{12}{13}\fontseries{l}\selectfont(f)}};
        \end{tikzpicture}}
        \vspace{-1.22cm} 

         \centering
    \subfloat[\label{selfplanez}]{
        \begin{tikzpicture}
        \node[anchor=north west, inner sep=0] (image) at (0,0) {
    \includegraphics[width=0.3\textwidth]{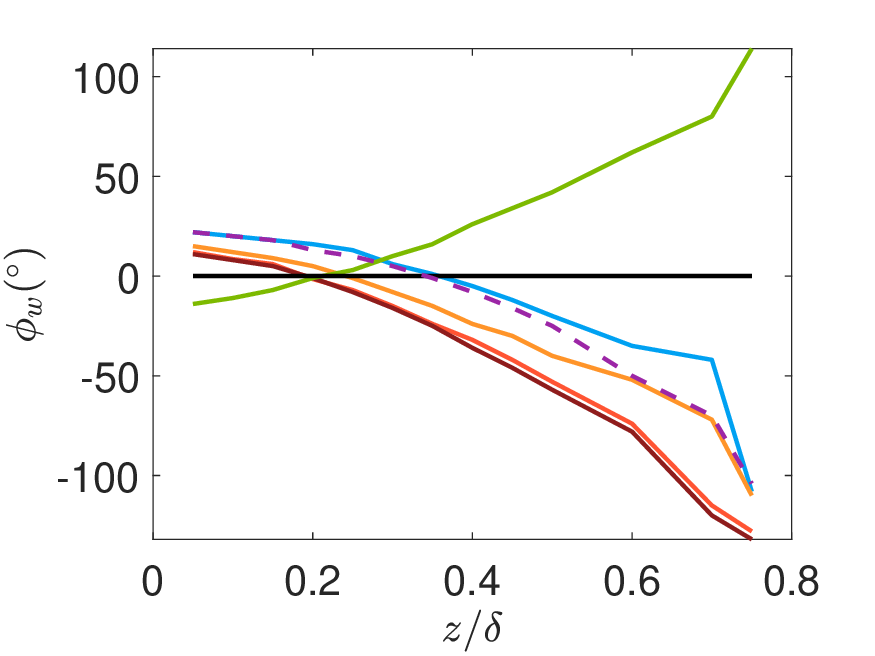}
            };
    \node[anchor=north west,
        xshift=-2mm,yshift=-2mm] at (image.north west) {{\rmfamily\fontsize{12}{13}\fontseries{l}\selectfont(g)}};
        \end{tikzpicture}}
    \subfloat[\label{selfplanez-theta}]{
        \begin{tikzpicture}
        \node[anchor=north west, inner sep=0] (image) at (0,0) {
    \includegraphics[width=0.3\textwidth]{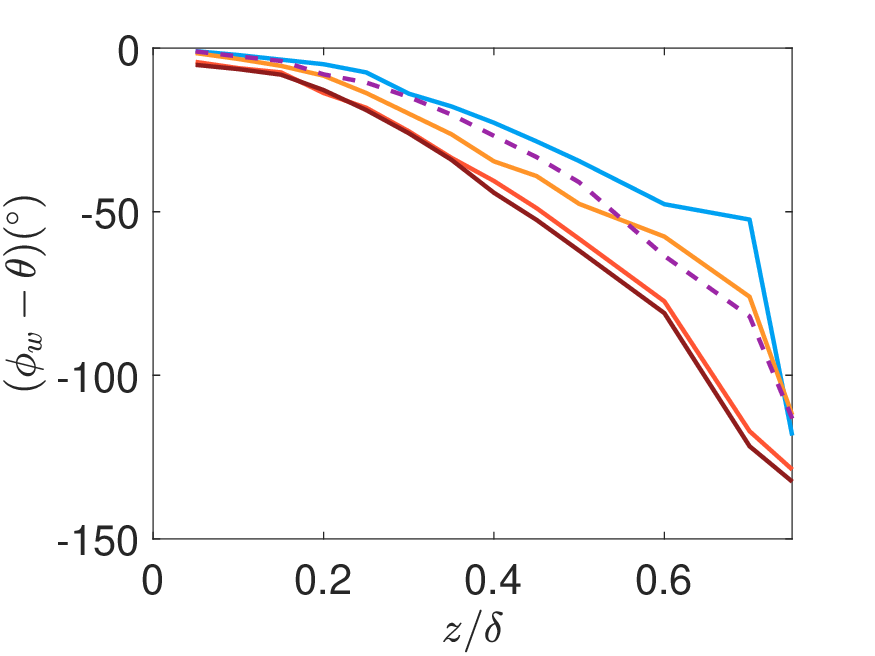}
            };
    \node[anchor=north west,
        xshift=-2mm,yshift=-2mm] at (image.north west) {{\rmfamily\fontsize{12}{13}\fontseries{l}\selectfont(h)}};
        \end{tikzpicture}}
            \subfloat[\label{selfplanez_theta}]{
        \begin{tikzpicture}
        \node[anchor=north west, inner sep=0] (image) at (0,0) {
    \includegraphics[width=0.3\textwidth]{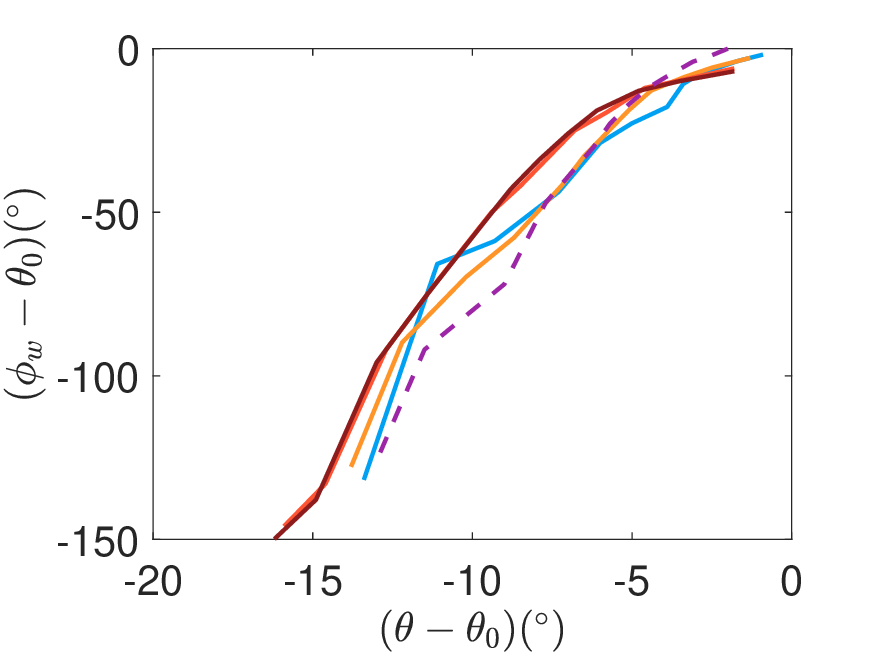}
            };
    \node[anchor=north west,
        xshift=-2mm,yshift=-2mm] at (image.north west) {{\rmfamily\fontsize{12}{13}\fontseries{l}\selectfont(i)}};
        \end{tikzpicture}}
        \vspace{-1cm} 
    \caption{Wall-normal variations of the deflection angles ($\phi_u$, $\phi_v$, $\phi_w$) and the correspondence to the wind veer angle ($\theta$) in the $x-y$ plane of the streamwise (a,b,c), spanwise (d,e,f) and vertical (g,h,i) velocity correlations in the CNBLs at different latitudes. {Here $\theta_0$ is the cross-isobaric angle.} }
    \label{fig:deflect_angle}
\end{figure}

{The wall-normal variations of the deflection angles of the velocity correlations in the CNBLs at different latitudes are shown in Fig.~\ref{fig:deflect_angle} (a,d,g). 
In the Northern Hemisphere, all the angles decrease with height, implying a clockwise variation of the deflection along the wall-normal direction. 
The deflection angle $\phi_u$ of the streamwise velocity varies from positive to negative with height.
The zero-crossing points at high latitudes (N70 and N90) are about $z/\delta \approx 0.2$, which is higher than lower latitudes and is $z/\delta \approx 0.5$ in the N25 CNBL. 
The magnitude of $\phi_u$ turns to decrease at $z/\delta>0.6\sim 0.7$ and approaches zero at $z/\delta \approx 0.8$.
The deflection angle $\phi_v$ of the spanwise velocity is observed to be always negative in the Northern Hemisphere, the magnitude of which increases with height. 
The variation of the deflection angle $\phi_w$ with height is similar to that of $\phi_u$ except for the returning to zero near the boundary-layer height.
The results of the Southern Hemisphere are the opposite to those of the Northern Hemisphere, and there is no deflection in the TNBL.}
{Fig.~\ref{fig:deflect_angle} (b,e,h) displays the wall-normal variations of the relative angles $\phi-\theta$, which characterize the departure of the coherent structure direction to the mean wind direction. For all three velocity components, the magnitudes of the angle differences increase with height, indicating a stronger and stronger departure from the mean wind direction. The angle difference is also larger at a higher latitude.}

{Fig.~\ref{fig:deflect_angle} (c,f,i) shows the relationships between the relative structure deflection angles ($\phi-\theta_0$) and the relative wind veer angle ($\theta-\theta_0$) by subtracting the cross-isobaric angle. One can find that the variations of $\phi-\theta_0$ with $\theta - \theta_0$ are well collapsed in the CNBLs, implying a probable universal relationship between structure deflection and mean wind veer.
The results of the structure deflection angles in the HN45 case are also presented in Fig.~\ref{fig:deflect_angle} to reveal the effect of boundary-layer height. 
It is found that the vertical variations of the relative structure deflection angles $\phi_u-\theta_0$, $\phi_v-\theta_0$ and $\phi_w-\theta_0$ of the HN45 case are quite similar to those of the other cases and the deflection angles still follow the universal relationship with the relative mean wind veer.}

\begin{figure}
    \centering
    \subfloat[\label{selfplanex2}]{
        \begin{tikzpicture}
        \node[anchor=north west, inner sep=0] (image) at (0,0) {
    \includegraphics[width=0.3\textwidth]{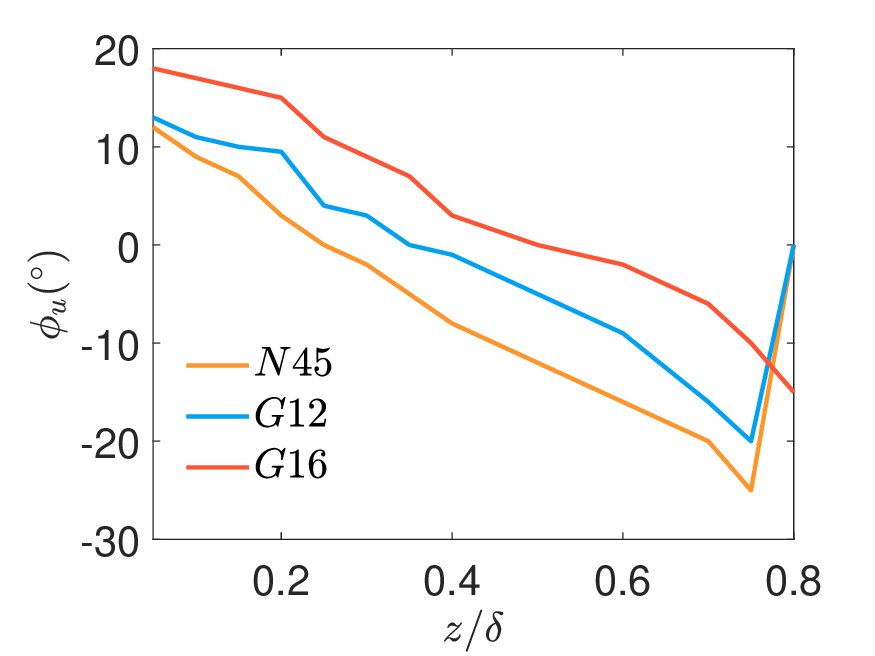}
            };
    \node[anchor=north west,
        xshift=-2mm,yshift=-2mm] at (image.north west) {{\rmfamily\fontsize{12}{13}\fontseries{l}\selectfont(a)}};
        \end{tikzpicture}}
    \subfloat[\label{selfplanex-theta2}]{
        \begin{tikzpicture}
        \node[anchor=north west, inner sep=0] (image) at (0,0) {
    \includegraphics[width=0.3\textwidth]{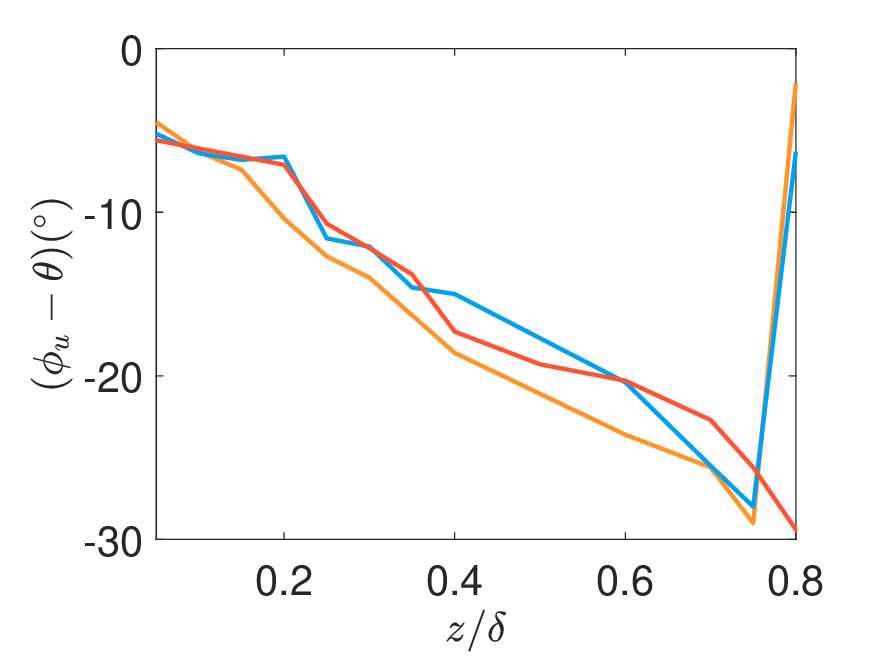}
            };
    \node[anchor=north west,
        xshift=-2mm,yshift=-2mm] at (image.north west) {{\rmfamily\fontsize{12}{13}\fontseries{l}\selectfont(b)}};
        \end{tikzpicture}}
    \subfloat[\label{selfplanex_theta2}]{
        \begin{tikzpicture}
        \node[anchor=north west, inner sep=0] (image) at (0,0) {
    \includegraphics[width=0.3\textwidth]{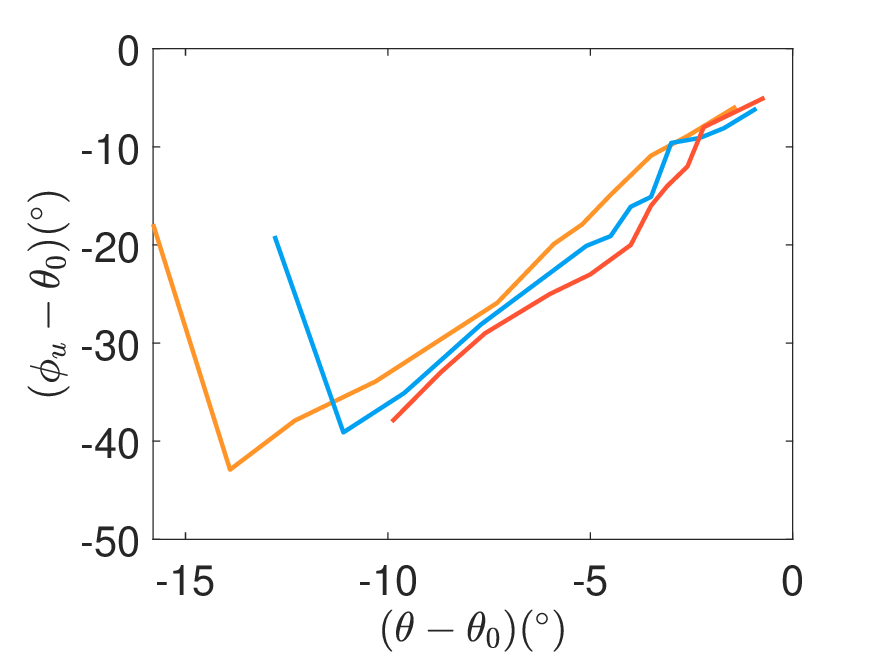}
            };
    \node[anchor=north west,
        xshift=-2mm,yshift=-2mm] at (image.north west) {{\rmfamily\fontsize{12}{13}\fontseries{l}\selectfont(c)}};
        \end{tikzpicture}}
        \vspace{-1.22cm} 
        
       \centering
    \subfloat[\label{selfplaney2}]{
        \begin{tikzpicture}
        \node[anchor=north west, inner sep=0] (image) at (0,0) {
    \includegraphics[width=0.3\textwidth]{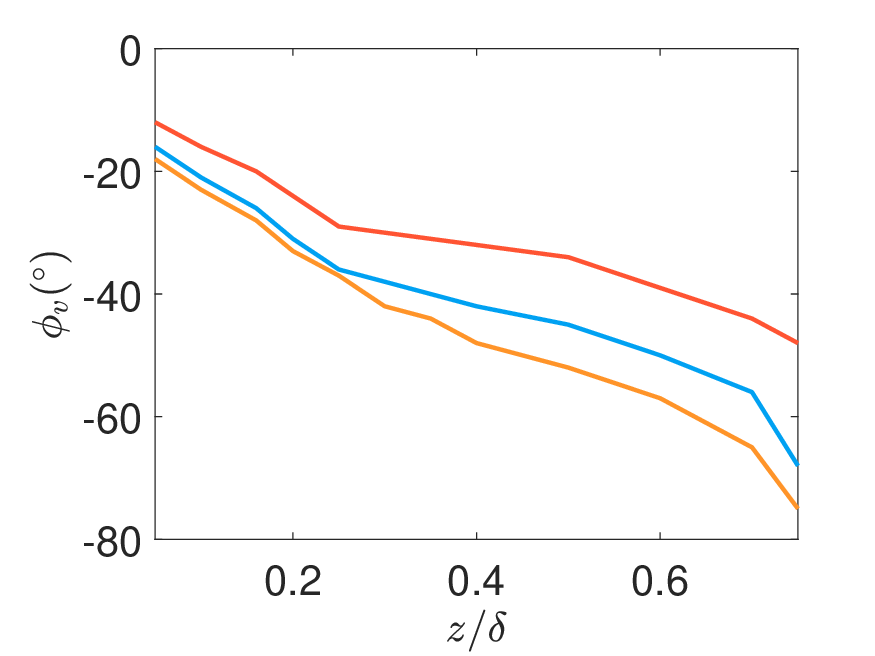}
            };
    \node[anchor=north west,
        xshift=-2mm,yshift=-2mm] at (image.north west) {{\rmfamily\fontsize{12}{13}\fontseries{l}\selectfont(d)}};
        \end{tikzpicture}}
    \subfloat[\label{selfplaney-theta2}]{
        \begin{tikzpicture}
        \node[anchor=north west, inner sep=0] (image) at (0,0) {
    \includegraphics[width=0.3\textwidth]{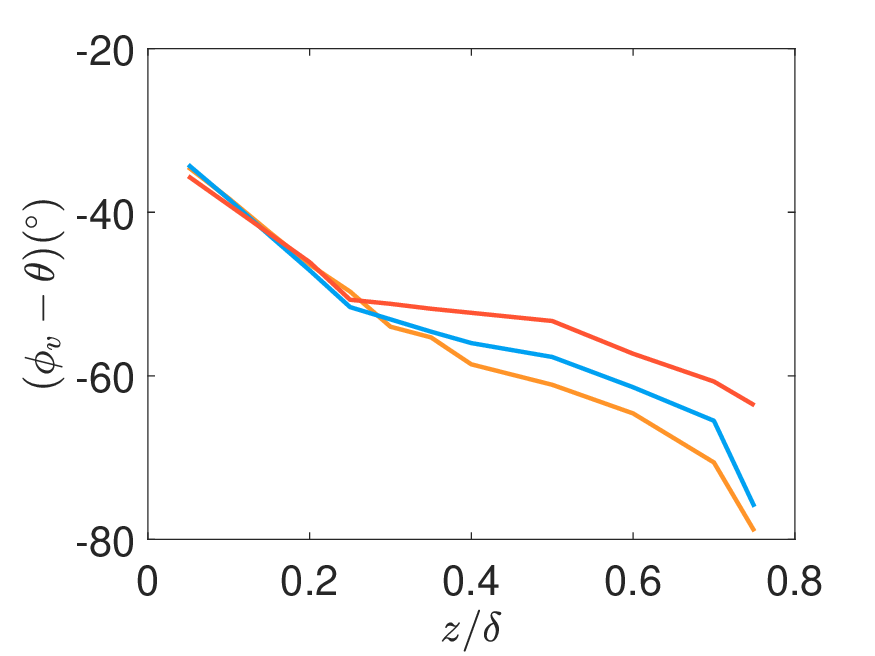}
            };
    \node[anchor=north west,
        xshift=-2mm,yshift=-2mm] at (image.north west) {{\rmfamily\fontsize{12}{13}\fontseries{l}\selectfont(e)}};
        \end{tikzpicture}}
    \subfloat[\label{selfplaney_theta2}]{
        \begin{tikzpicture}
        \node[anchor=north west, inner sep=0] (image) at (0,0) {
    \includegraphics[width=0.3\textwidth]{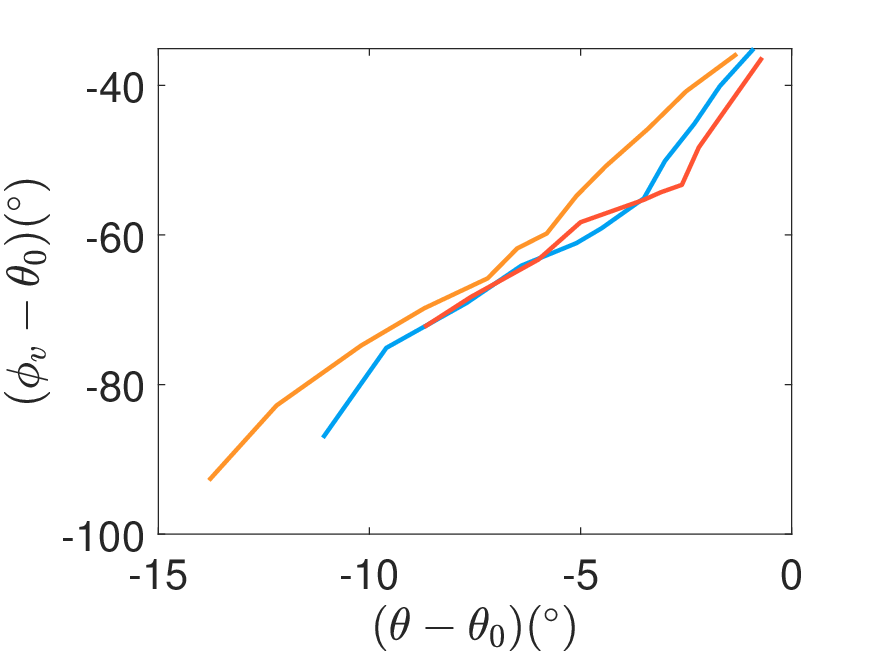}
            };
    \node[anchor=north west,
        xshift=-2mm,yshift=-2mm] at (image.north west) {{\rmfamily\fontsize{12}{13}\fontseries{l}\selectfont(f)}};
        \end{tikzpicture}}
        \vspace{-1.22cm} 

         \centering
    \subfloat[\label{selfplanez2}]{
        \begin{tikzpicture}
        \node[anchor=north west, inner sep=0] (image) at (0,0) {
    \includegraphics[width=0.3\textwidth]{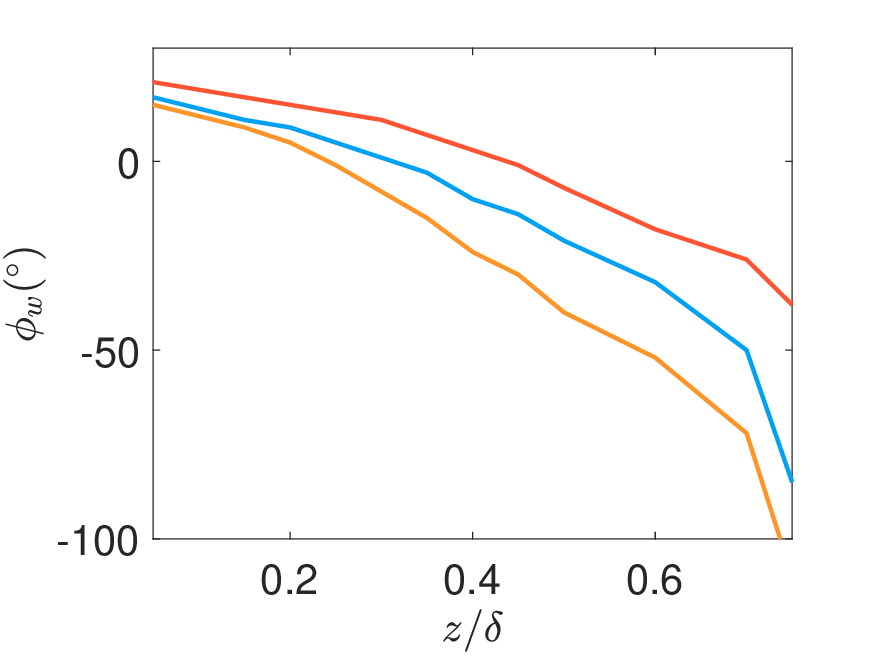}
            };
    \node[anchor=north west,
        xshift=-2mm,yshift=-2mm] at (image.north west) {{\rmfamily\fontsize{12}{13}\fontseries{l}\selectfont(g)}};
        \end{tikzpicture}}
    \subfloat[\label{selfplanez-theta2}]{
        \begin{tikzpicture}
        \node[anchor=north west, inner sep=0] (image) at (0,0) {
    \includegraphics[width=0.3\textwidth]{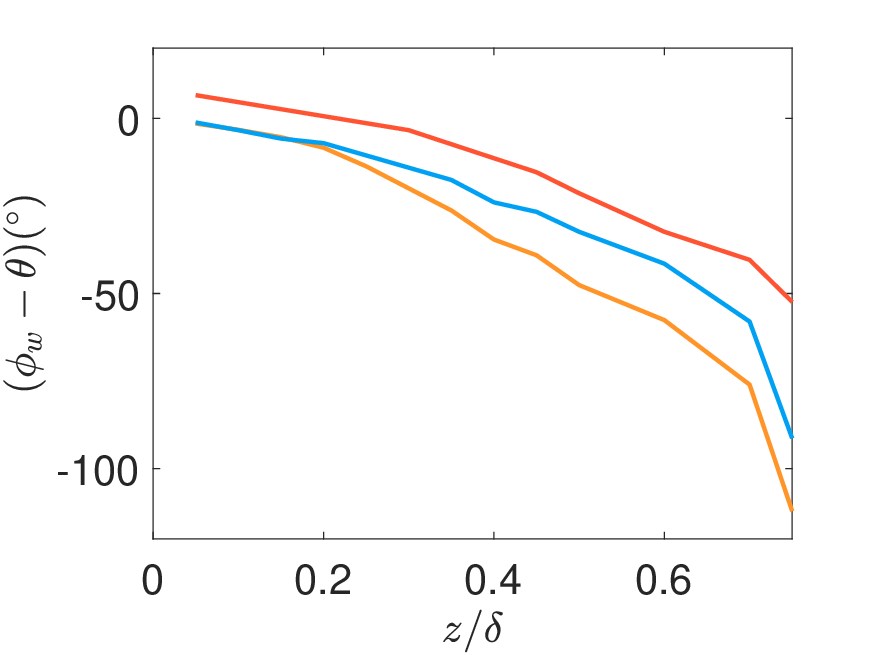}
            };
    \node[anchor=north west,
        xshift=-2mm,yshift=-2mm] at (image.north west) {{\rmfamily\fontsize{12}{13}\fontseries{l}\selectfont(h)}};
        \end{tikzpicture}}
    \subfloat[\label{selfplanez_theta2}]{
        \begin{tikzpicture}
        \node[anchor=north west, inner sep=0] (image) at (0,0) {
    \includegraphics[width=0.3\textwidth]{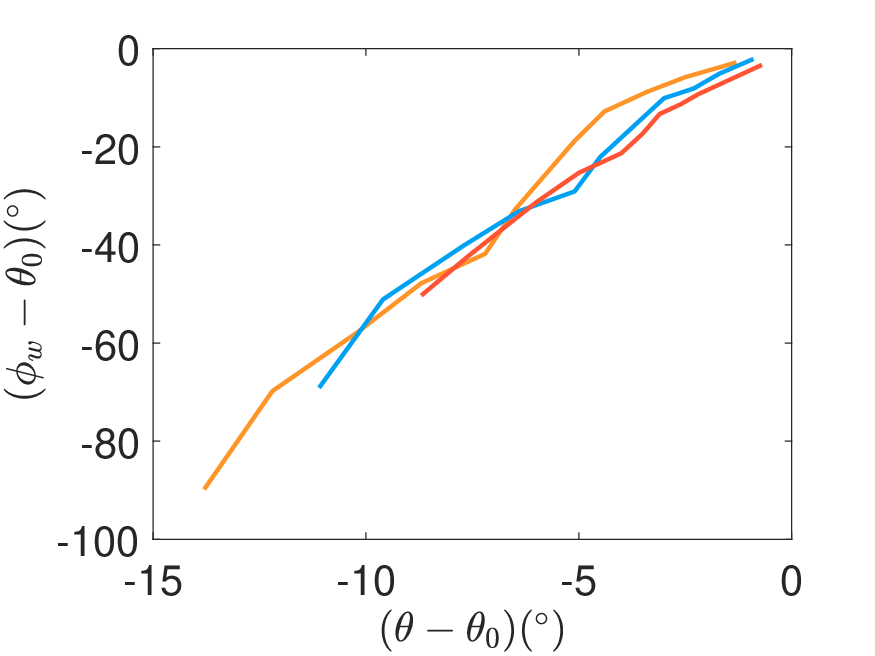}
            };
    \node[anchor=north west,
        xshift=-2mm,yshift=-2mm] at (image.north west) {{\rmfamily\fontsize{12}{13}\fontseries{l}\selectfont(i)}};
        \end{tikzpicture}}
        \vspace{-1cm} 
    \caption{{Wall-normal variations and correspondence to the wind veer angles of the deflection angles in the $x-y$ plane of the streamwise (a,b,c), spanwise (d,e,f) and vertical (g,h,i) velocity correlations in the CNBLs with different geostrophic wind speeds.}}
    \label{fig:deflect_angle2}
\end{figure}

Fig.~\ref{fig:deflect_angle2} shows the wall-normal variations of the deflection angles and their correspondence to the wind veer angles in the CNBLs with different geostrophic wind speeds.
For all velocity structures, a greater geostrophic wind speed will induce a more positive or less negative deflection angle due to the stronger Coriolis force.
The angle differences between $\phi$ and $\theta$ are also different with latitude.
{When the relative structure deflection angle $\phi-\theta_0$ is plotted against the relative wind veer angle $\theta - \theta_0$, the deflection angles under different geostrophic wind speeds are excellently collapsed, indicating a probable universal relationship with the mean wind veer regardless of latitude or geostrophic wind speed. }

\subsection{Structure inclinations}

A large number of experimental and numerical studies have revealed that the large-scale coherent structures in high-Reynolds-number wall-bounded turbulent flows are shallowly inclined to the horizontal \cite{brownLargeStructureTurbulent1977a,christensenStatisticalEvidenceHairpin2001,marusic2007reynolds,chauhan2013structure,liu2017variation,saleskyRevisitingInclinationLargescale2020,liScaledependentInclinationAngle2022,gibbsInclinationAnglesTurbulent2023,huangTheoreticalModelStructure2023}. 
To determine the inclination angle, the two-point correlation of longitudinal velocity in the $x-z$ plane is defined as
\begin{equation}\label{eq Ruz}
  R_{uu}^{xz}(\Delta x, y, \Delta z) = \frac{
    \langle u'(x, y, z_{ref}) \, u'(x + \Delta x, y, z_{ref} + \Delta z) \rangle
}{
    \sigma_{u}(x, y, z_{ref}) \sigma_{u}(x + \Delta x,y, z_{ref} + \Delta z)
},
\end{equation}
in which $z_{ref}/\delta=0.0078$ is the height of the reference point near the ground surface.
The mean inclination angle is calculated following \citet{liu2017variation} as 
\begin{equation}\label{eq gamma}
\gamma_u=\arctan(\langle \Delta z/\Delta x_p \rangle),
\end{equation}
in which $\Delta x_p$ is the streamwise delay of the peak $R_{uu}^{xz}$ to the reference point. Here, we just use the data in the surface layer at $z/\delta<0.2$ to calculate the inclination angle.

\begin{figure}
\centering
   \subfloat[\label{dip25}]{
        \begin{tikzpicture}
        \node[anchor=north west, inner sep=0] (image) at (0,0) {
    \includegraphics[width=0.45\textwidth]{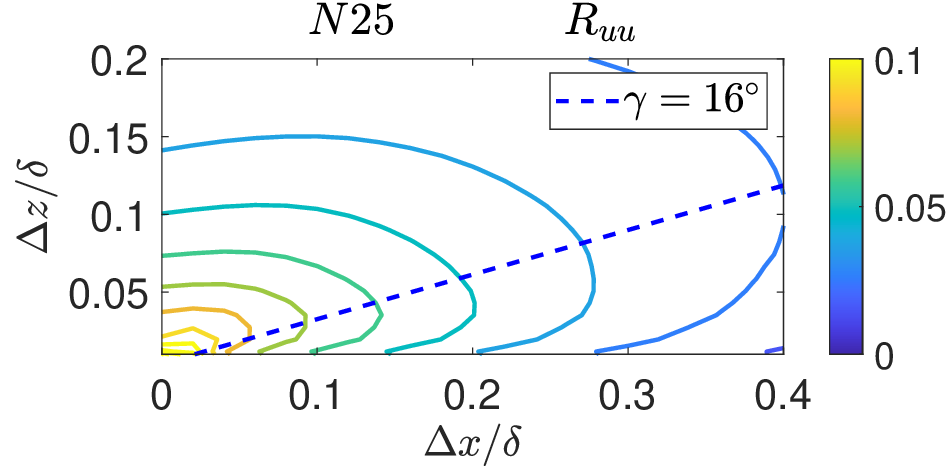}
            };
    \node[anchor=north west,
        xshift=-2mm,yshift=-2mm] at (image.north west) {{\rmfamily\fontsize{12}{13}\fontseries{l}\selectfont(a)}};
        \end{tikzpicture}}
    \subfloat[\label{dip45}]{
        \begin{tikzpicture}
        \node[anchor=north west, inner sep=0] (image) at (0,0) {
    \includegraphics[width=0.45\textwidth]{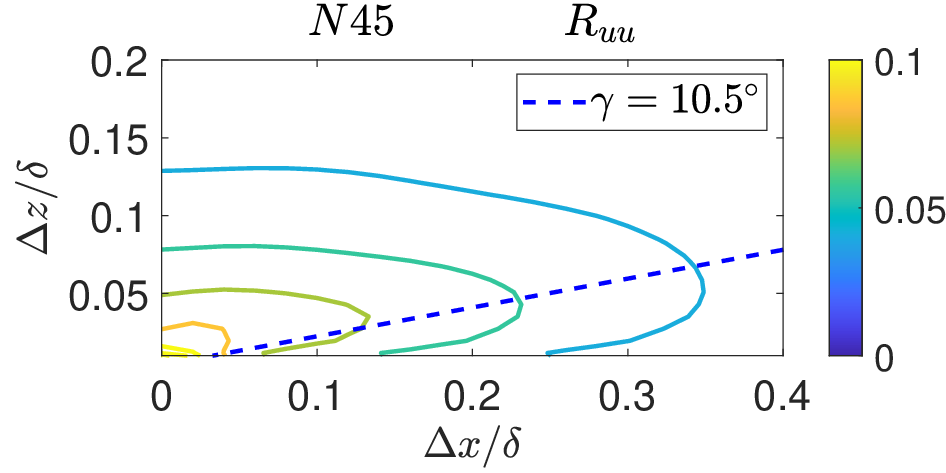}
            };
    \node[anchor=north west,
        xshift=-2mm,yshift=-2mm] at (image.north west) {{\rmfamily\fontsize{12}{13}\fontseries{l}\selectfont(b)}};
        \end{tikzpicture}}
        
 \vspace{-1.65cm}    
\centering
    \subfloat[\label{dip70}]{
        \begin{tikzpicture}
        \node[anchor=north west, inner sep=0] (image) at (0,0) {
    \includegraphics[width=0.45\textwidth]{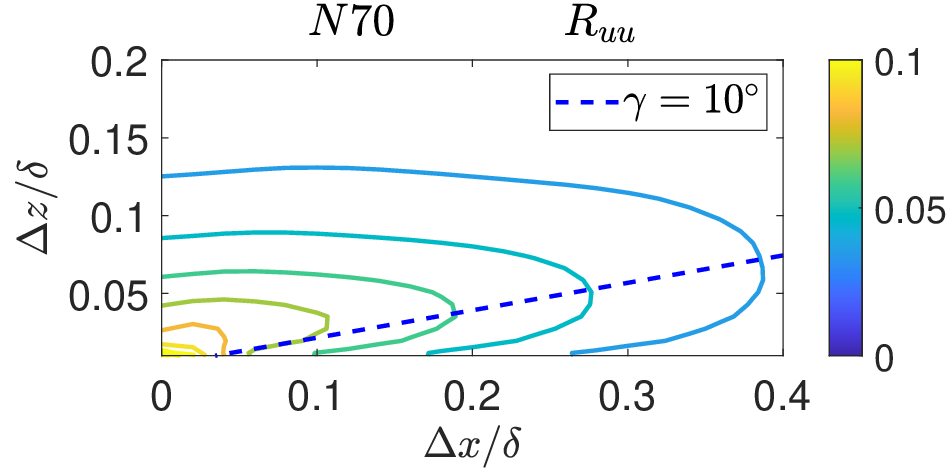}
            };
    \node[anchor=north west,
        xshift=-2mm,yshift=-2mm] at (image.north west) {{\rmfamily\fontsize{12}{13}\fontseries{l}\selectfont(c)}};
        \end{tikzpicture}}
   \subfloat[\label{dip90}]{
        \begin{tikzpicture}
        \node[anchor=north west, inner sep=0] (image) at (0,0) {
    \includegraphics[width=0.45\textwidth]{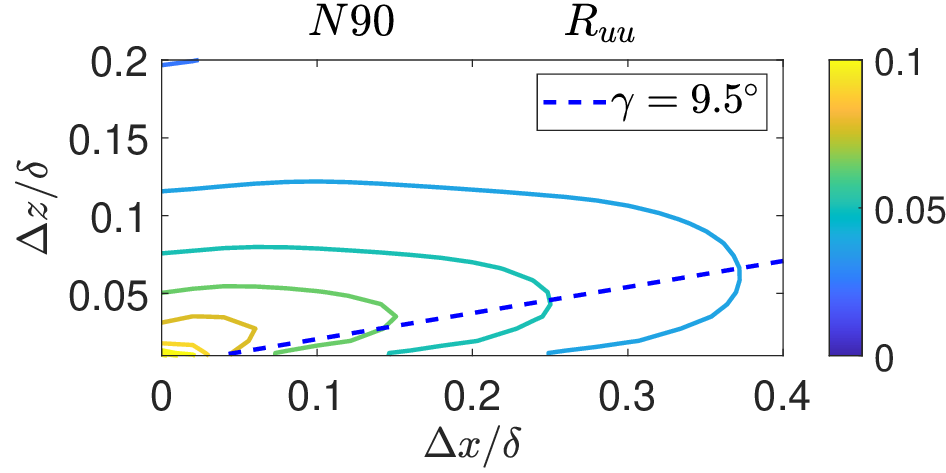}
            };
    \node[anchor=north west,
        xshift=-2mm,yshift=-2mm] at (image.north west) {{\rmfamily\fontsize{12}{13}\fontseries{l}\selectfont(d)}};
        \end{tikzpicture}}
        
\vspace{-1.65cm} 
 \centering
    \subfloat[\label{dipS45}]{
        \begin{tikzpicture}
        \node[anchor=north west, inner sep=0] (image) at (0,0) {
    \includegraphics[width=0.45\textwidth]{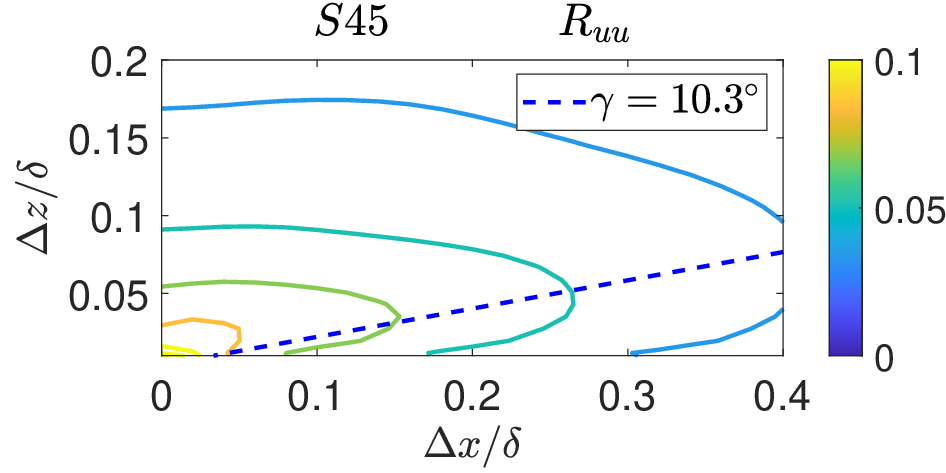}
            };
    \node[anchor=north west,
        xshift=-2mm,yshift=-2mm] at (image.north west) {{\rmfamily\fontsize{12}{13}\fontseries{l}\selectfont(e)}};
        \end{tikzpicture}}
   \subfloat[\label{dipTNBL}]{
        \begin{tikzpicture}
        \node[anchor=north west, inner sep=0] (image) at (0,0) {
    \includegraphics[width=0.45\textwidth]{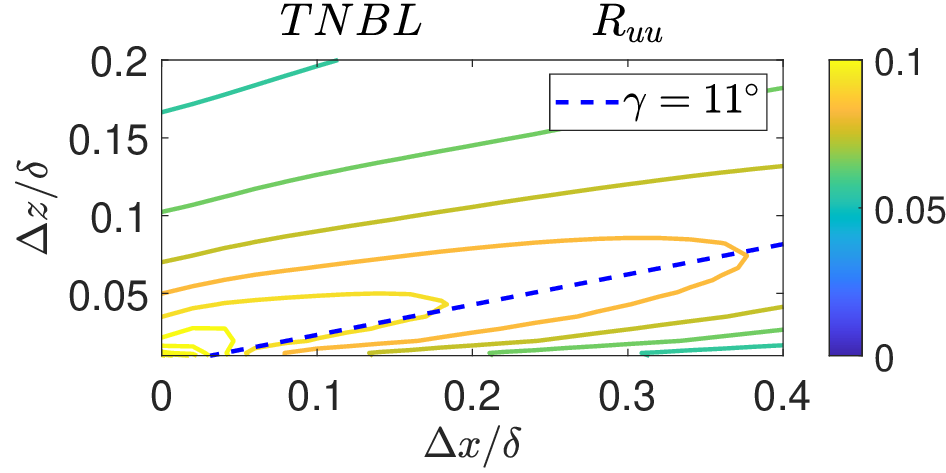}
            };
    \node[anchor=north west,
        xshift=-2mm,yshift=-2mm] at (image.north west) {{\rmfamily\fontsize{12}{13}\fontseries{l}\selectfont(f)}};
        \end{tikzpicture}}
    \caption{The $R_{uu}^{xz}$ correlation maps in the $x-z$ plane of the CNBLs at different latitudes and the TNBL.}
    \label{figdipu}
\end{figure}

\begin{figure}
 \centering
    \subfloat[\label{dipg12}]{
        \begin{tikzpicture}
        \node[anchor=north west, inner sep=0] (image) at (0,0) {
    \includegraphics[width=0.45\textwidth]{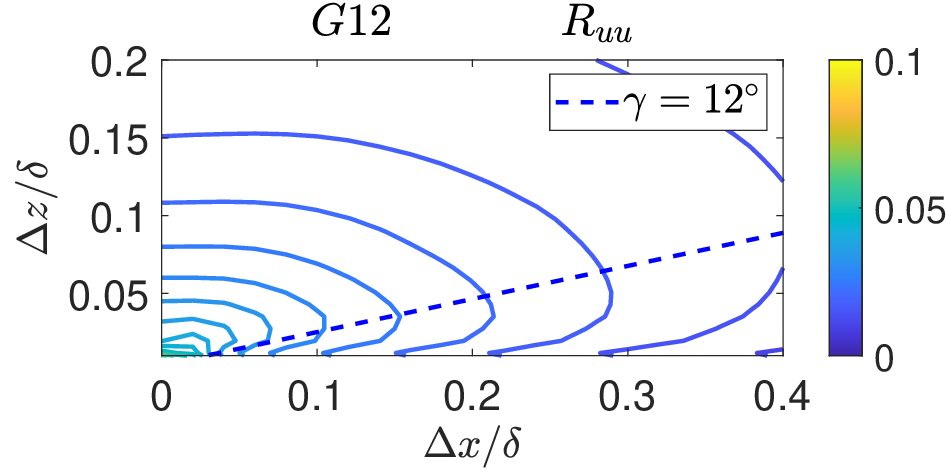}
            };
    \node[anchor=north west,
        xshift=-2mm,yshift=-2mm] at (image.north west) {{\rmfamily\fontsize{12}{13}\fontseries{l}\selectfont(a)}};
        \end{tikzpicture}}  
    \subfloat[\label{dipg16}]{
        \begin{tikzpicture}
        \node[anchor=north west, inner sep=0] (image) at (0,0) {
    \includegraphics[width=0.45\textwidth]{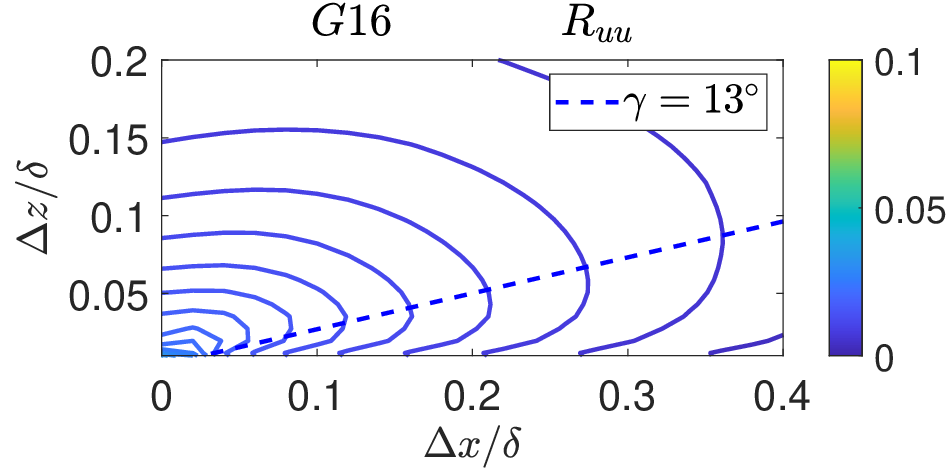}
            };
    \node[anchor=north west,
        xshift=-2mm,yshift=-2mm] at (image.north west) {{\rmfamily\fontsize{12}{13}\fontseries{l}\selectfont(c)}};
        \end{tikzpicture}} 
    \caption{Structure inclination angles with different geostrophic wind speeds.}
    \label{fig:G dip}
\end{figure}

Fig.~\ref{figdipu} displays the two-point correlation maps of the streamwise velocity fluctuation in the $x-z$ plane of the CNBLs at different latitudes and the TNBL. 
It is seen that the major axes of all the contours lean at a shallow angle with the streamwise direction due to the inclination of large-scale turbulent structures.
It is also not difficult to see that as latitude increases, the inclination of the structures decreases. 
The N25 CNBL has the largest $\gamma_u = 16^\circ$. This can be attributed to the largest deflection in the surface layer as shown in Fig.~\ref{fig:deflect_angle}.
The TNBL has shallow inclination angles of $\gamma_u=11^\circ$, which are close to those in the high-latitude CNBLs.
{Fig.~\ref{fig:G dip} displays the two-point correlation contour maps of the streamwise velocity in the $x-z$ plane of the CNBLs with different geostrophic wind speeds, from which the inclination angle of the coherent structures can be determined. 
One can clearly see that as the geostrophic wind speed increases, the inclination angle slightly increases. }

\begin{figure}
\centering
\subfloat{
        \begin{tikzpicture}
        \node[anchor=north west, inner sep=0] (image) at (0,0) {
    \includegraphics[width=0.45\textwidth]{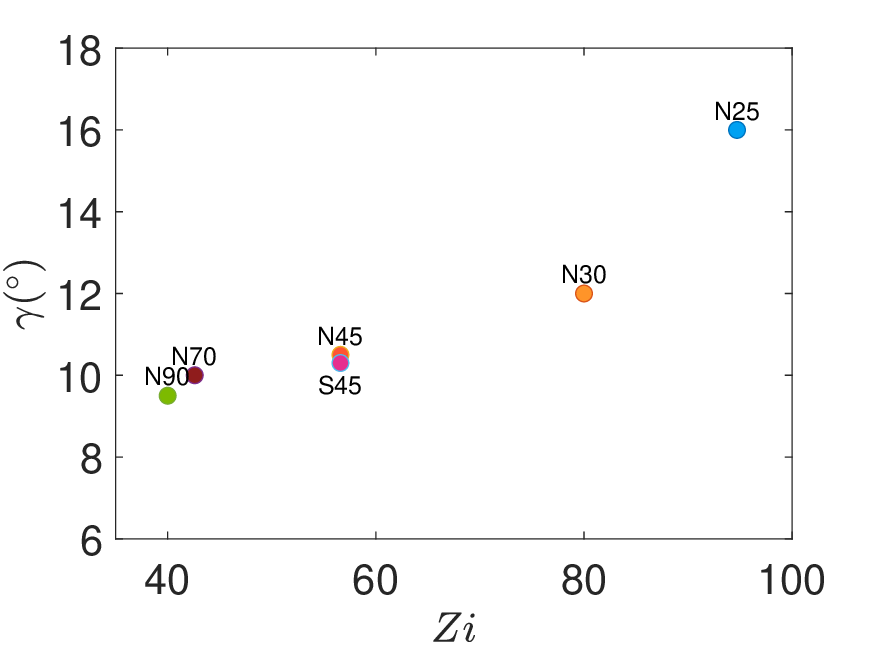}
            };
    \node[anchor=north west,
        xshift=-2mm,yshift=-2mm] at (image.north west) {{\rmfamily\fontsize{12}{13}\fontseries{l}\selectfont(a)}};
        \end{tikzpicture}}
\subfloat{
        \begin{tikzpicture}
        \node[anchor=north west, inner sep=0] (image) at (0,0) {
    \includegraphics[width=0.45\textwidth]{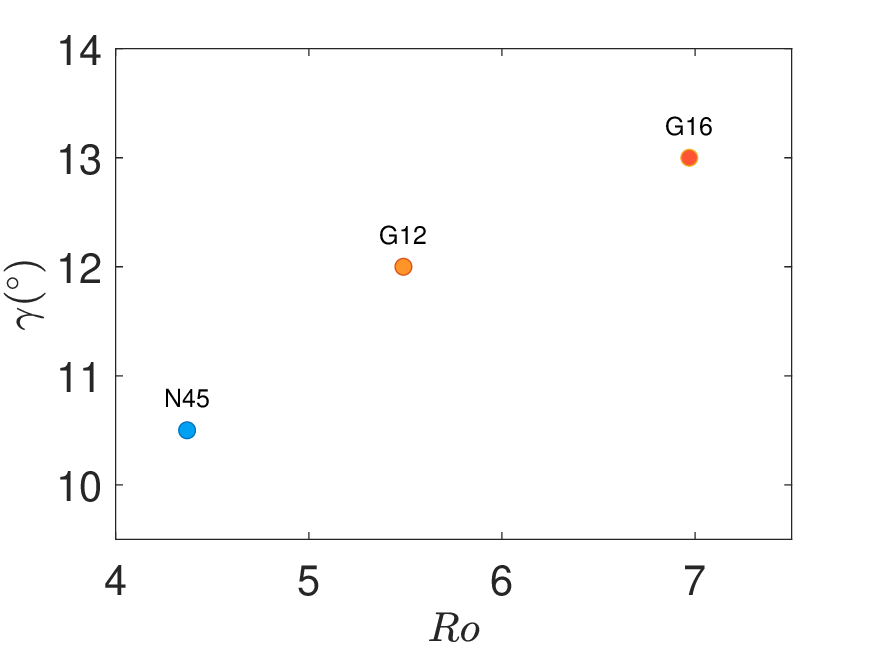}
            };
    \node[anchor=north west,
        xshift=-2mm,yshift=-2mm] at (image.north west) {{\rmfamily\fontsize{12}{13}\fontseries{l}\selectfont(b)}};
        \end{tikzpicture}}
\caption{Variation of the structure inclination angle with (a) $Zi$ and (b) $Ro$.}
\label{figsumdip}
\end{figure}


{We show the relationship of the structure inclination angles with the Zilitinkevich number and the Rossby number in Fig.~\ref{figsumdip}. 
It is seen that with the increase of $Zi$ (decreasing latitude), the structure inclination angle $\gamma_u$ also increases accordingly. At high latitudes, the variation in the inclination angle is small. At low latitudes, the inclination angle increases rapidly. 
}
As the $Ro$ increases (increasing geostrophic wind speed), the inclination angle of the streamwise velocity structure continues to increase from $10.5^\circ$ to $13.5^\circ$. 

 \section{Conclusions}
 \label{level:5}
 
In this work, we examined the effects of the Coriolis force on the characteristics of turbulent coherent structures in the CNBL utilizing LES under the traditional or ``$f$-plane'' approximation. The Coriolis force due to Earth's rotation varies with latitude or geostrophic wind speed. The deflection of turbulent coherent structures confirms similar observations in the literature. What is more, we quantified the structure deflections using two-point correlations of all three velocity components, as well as their dependence on latitude and geostrophic wind speed. 
In general, the deflection directions of coherent structures are different from the mean wind velocity and mean shear directions.
A striking finding is that a relative deflection angle $\phi-\theta_0$ under different conditions can be well collapsed when plotting against a relative wind veer angle $\theta-\theta_0$, suggesting a probable universal representation regardless of latitude or geostrophic wind speed. We found that the Coriolis force also affects the inclination angle of large-scale structures in the CNBL. At a lower latitude or with a stronger geostrophic wind, the inclination angle is larger. 
{This phenomenon may be attributed to the deflected coherent structures caused by the Coriolis force. 
The underlying physical mechanism for the structure deflection may be revealed through the analysis of the exact budget equations of two-point velocity correlations, which is left for future study.}

Finally, we should admit that the present findings are dependent on the traditional or ``$f$-plane'' approximation, in which the wall-normal component of the Coriolis force is neglected. Several authors have addressed the nontrivial role this component plays and suggested taking the full form of the Coriolis force \citep{leibovichInfluenceHorizontalComponent1985,esauCoriolisEffectCoherent2003,zikanov2003large,gerkemaGeophysicalAstrophysicalFluid2008,esauStructuringTurbulenceIts2013,howland2020influence,chewUnstableModeStratified2023}, which will be the next subject of our study.

\section*{Acknowledgement}
R.H. acknowledges the financial support from the National Natural Science Foundation of China (12388101 and 12472221), the Fundamental Research Funds for the Central Universities (lzujbky-2024-oy10), the Natural Science Foundation of Gansu Province (25JRRA636) and the
National Numerical Wind Tunnel project. L.L. acknowledges the financial support from the National Natural Science Foundation of China (12272371). 

\section*{Author Declarations}
The authors have no conflicts to disclose.

 \section*{Data Availability}
The data are available from the authors upon reasonable request.

\appendix

\section*{Appendix. Basic statistics}


\begin{figure}
    \centering
  \subfloat{
        \begin{tikzpicture}
            \node[anchor=north west, inner sep=0] (image) at (0,0) {
    \includegraphics[width=0.48\textwidth]{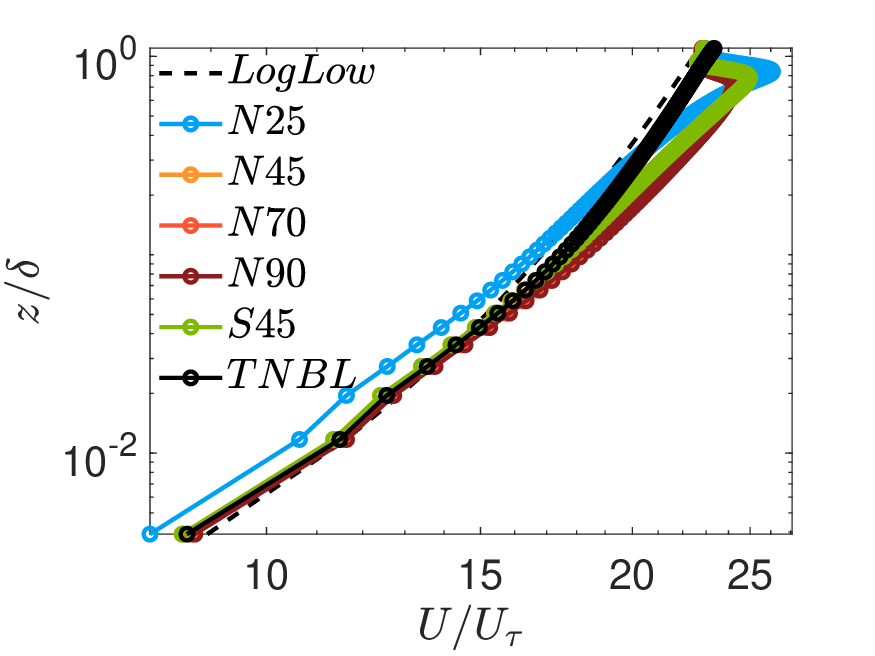}
            };
            \node[anchor=north west,
        xshift=-2mm,yshift=-2mm] at (image.north west) {{\rmfamily\fontsize{12}{13}\fontseries{l}\selectfont(a)}};
        \end{tikzpicture}}
    \hfill
     \subfloat{
        \begin{tikzpicture}
            \node[anchor=north west, inner sep=0] (image) at (0,0) {
    \includegraphics[width=0.48\textwidth]{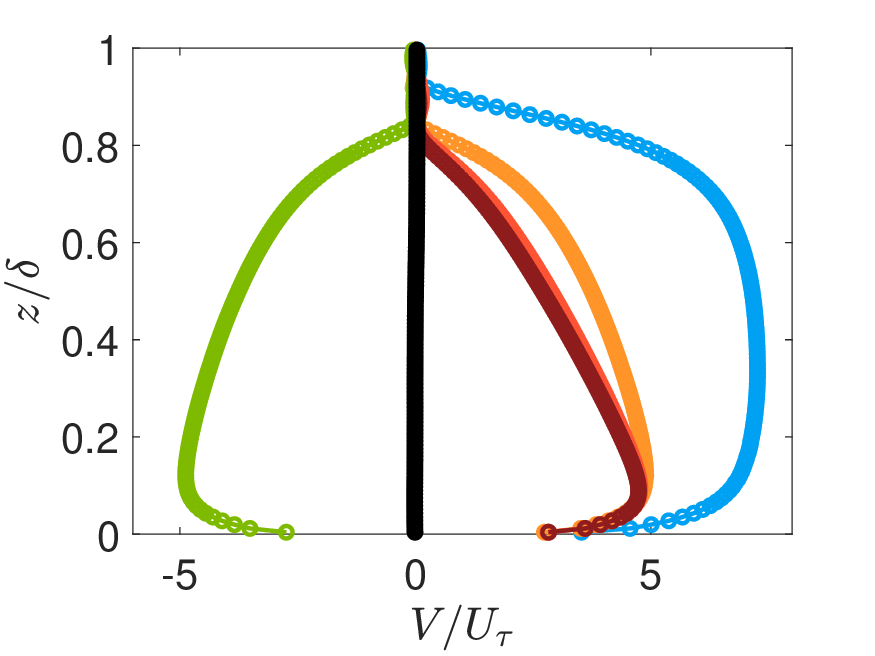}
            };
            \node[anchor=north west,
        xshift=-2mm,yshift=-2mm] at (image.north west) {{\rmfamily\fontsize{12}{13}\fontseries{l}\selectfont(b)}};
        \end{tikzpicture}}
    \vfill
     \subfloat{
        \begin{tikzpicture}
            \node[anchor=north west, inner sep=0] (image) at (0,0) {      \includegraphics[width=0.48\textwidth]{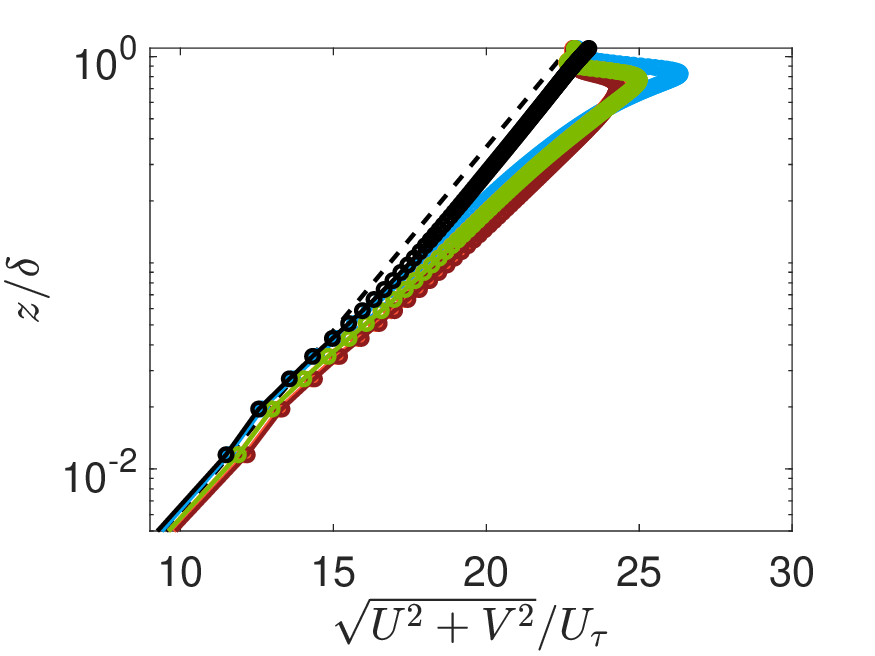}
            };
            \node[anchor=north west,
        xshift=-2mm,yshift=-2mm] at (image.north west) {{\rmfamily\fontsize{12}{13}\fontseries{l}\selectfont(c)}};
        \end{tikzpicture}}        
    \hfill
    \subfloat{
        \begin{tikzpicture}
            \node[anchor=north west, inner sep=0] (image) at (0,0) {
    \includegraphics[width=0.48\textwidth]{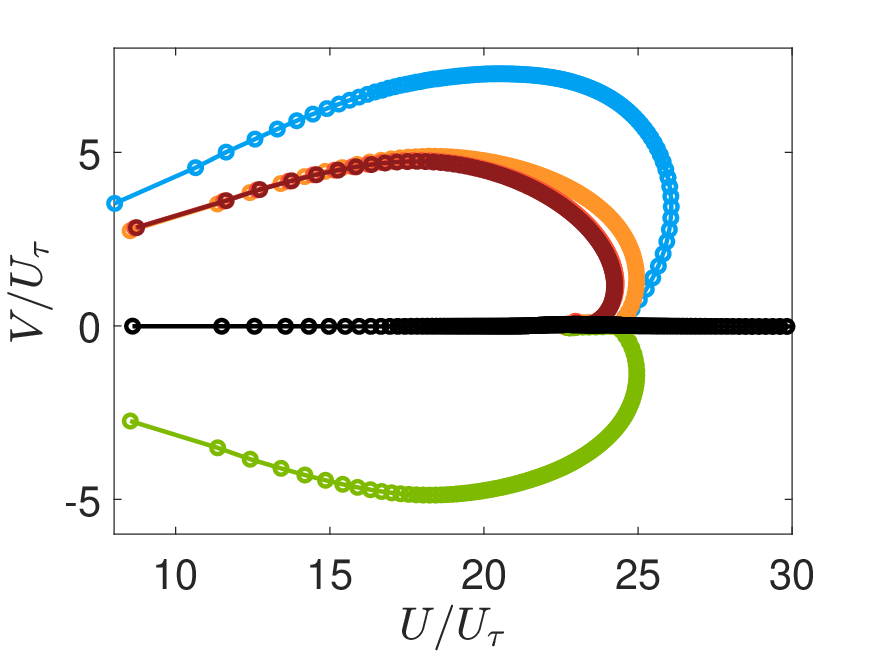}
            };
            \node[anchor=north west,
        xshift=-2mm,yshift=-2mm] at (image.north west) {{\rmfamily\fontsize{12}{13}\fontseries{l}\selectfont(d)}};
        \end{tikzpicture}}
        \vfill
    \subfloat{
        \begin{tikzpicture}
            \node[anchor=north west, inner sep=0] (image) at (0,0) {      \includegraphics[width=0.48\textwidth]{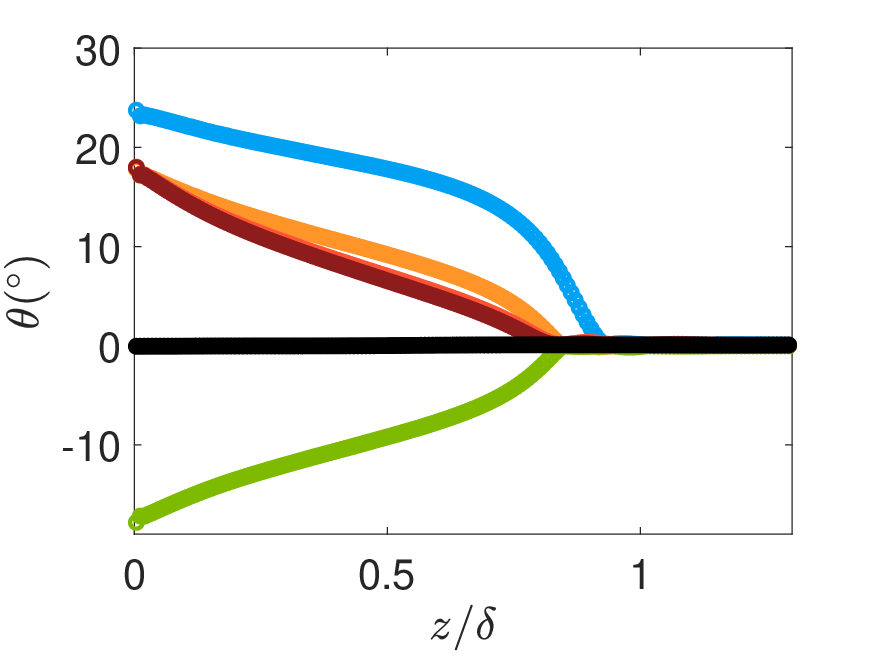}
            };
            \node[anchor=north west,
        xshift=-2mm,yshift=-2mm] at (image.north west) {{\rmfamily\fontsize{12}{13}\fontseries{l}\selectfont(e)}};
        \end{tikzpicture}}
    \caption{Comparison of (a) the mean streamwise wind speeds, (b) the mean spanwise wind speeds, (c) the total mean wind speeds, (d) the Ekman spirals, and (e) the wind veer angles to the streamwise direction (the geostrophic wind direction) in the CNBLs at different latitudes and the TNBL.}
    \label{fig5}
\end{figure}

{Fig.~\ref{fig5} (a) shows the mean longitudinal wind velocity profiles of the CNBLs and TNBL at different latitudes. It is seen that the Coriolis force increases the longitudinal wind speed of the CNBLs in almost the entire boundary layer depth except N25. The increase is stronger with a larger magnitude of the Coriolis force at a higher latitude. }
The geostrophic wind eventually restricts the mean longitudinal wind speeds, causing them to collapse at the top of the boundary layer. 
The above results can be understood through the geostrophic balance relationship (\ref{eq:geostr_bal}), which indicates that a larger $U$ is needed when $f_c$ is smaller above the boundary layer, resulting in a sharper velocity gradient near the bottom surface where the Coriolis force has a much weaker effect.
From Fig.~\ref{fig5} (b), one can see that the mean lateral wind speed is larger at a lower latitude. This phenomenon can also be understood from (\ref{eq:geostr_bal}) as a smaller $f_c$ requires a larger $V$ to satisfy the geostrophic balance.
The mean lateral wind speed of the TNBL case is always zero due to the absence of the Coriolis force.
Fig.~\ref{fig5} (c) displays the profiles of the total mean wind speed, which are quite similar among the CNBLs. The mean wind velocity in the TNBL is the smallest and very close to the classical logarithmic law. However, evident departure from the log law is observed for the CNBLs at $z/\delta>0.1$.
Fig.~\ref{fig5} (d) depicts the Ekman spiral hodogram \citep{ekman1905influence}.  
The spiral is wider at a lower latitude since the relaxation of the geostrophic balance is slower.
In the TNBL, there is no spiral but a straight line, as the mean lateral wind speed is zero.
Fig.~\ref{fig5} (e) shows the profiles of the wind veer angle in the CNBLs and TNBL, which is defined as the angle of the total wind velocity vector to the longitudinal direction, \emph{i.e.} $\theta = \arctan(V/U)$. 
This variation is qualitatively similar to the mean lateral wind velocity, namely, at higher latitudes, the veer angle is smaller, and its development across the boundary layer is faster.
Moreover, one can see that the magnitudes of the wind speed are the same for cases N45 and S45, with opposite signs of the mean lateral wind velocity and the veer angle due to the reversed direction of the tangential Coriolis force and the omission of the normal Coriolis force (the traditional or the ``$f$-plane'' approximation). 

\begin{figure}
    \centering
    \subfloat[\label{fig6a}]{
        \begin{tikzpicture}
        \node[anchor=north west, inner sep=0] (image) at (0,0) {
    \includegraphics[width=0.48\textwidth]{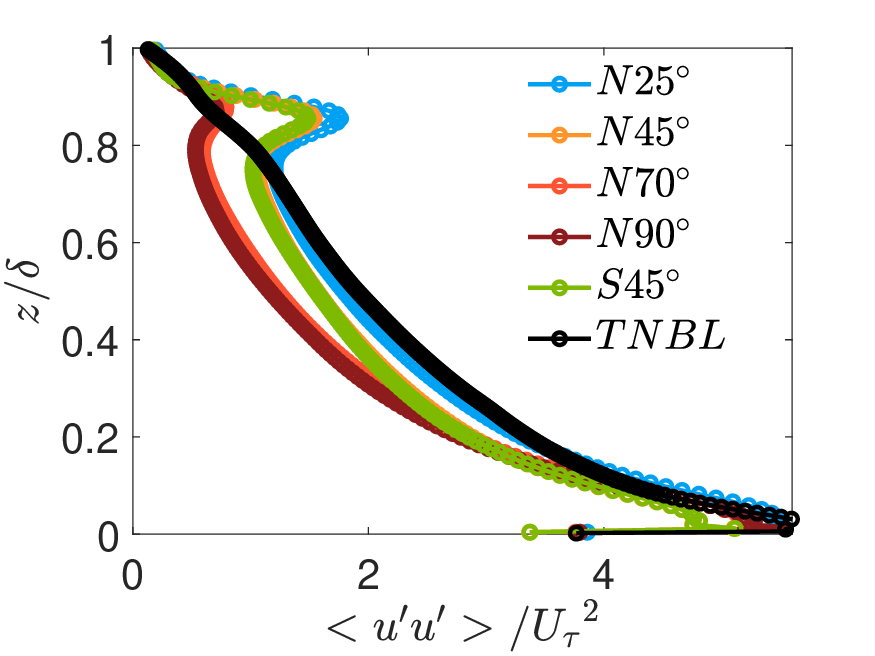}
            };
    \node[anchor=north west,
        xshift=-2mm,yshift=-2mm] at (image.north west) {{\rmfamily\fontsize{12}{13}\fontseries{l}\selectfont(a)}};
        \end{tikzpicture}} 
        \hfill
     \subfloat[\label{fig6b}]{
        \begin{tikzpicture}
            \node[anchor=north west, inner sep=0] (image) at (0,0) {      \includegraphics[width=0.48\textwidth]{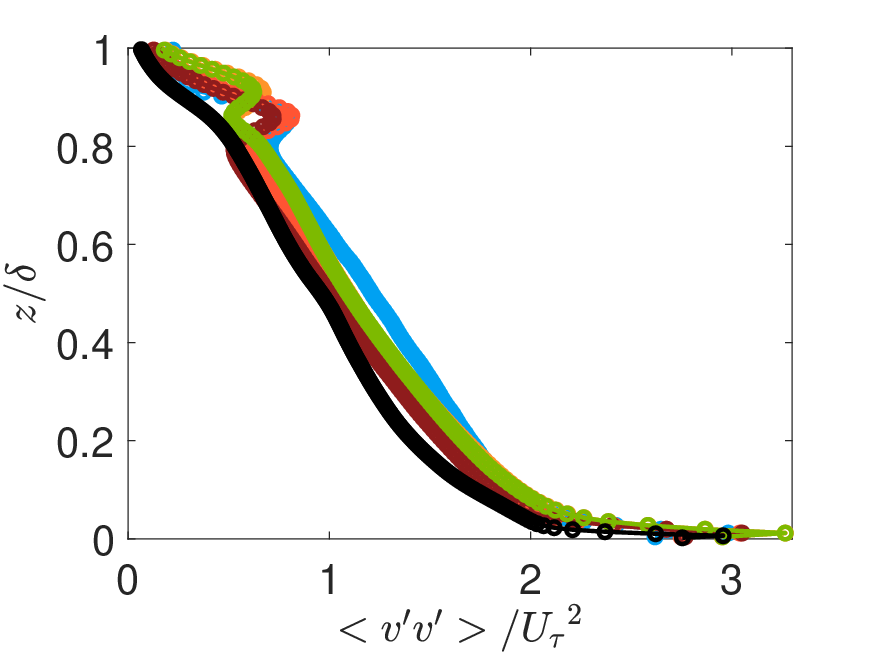}
            };
            \node[anchor=north west,
        xshift=-2mm,yshift=-2mm] at (image.north west) {{\rmfamily\fontsize{12}{13}\fontseries{l}\selectfont(b)}};
        \end{tikzpicture}}        
     \vfill    
     \subfloat[\label{fig6c}]{
        \begin{tikzpicture}
            \node[anchor=north west, inner sep=0] (image) at (0,0) {      \includegraphics[width=0.48\textwidth]{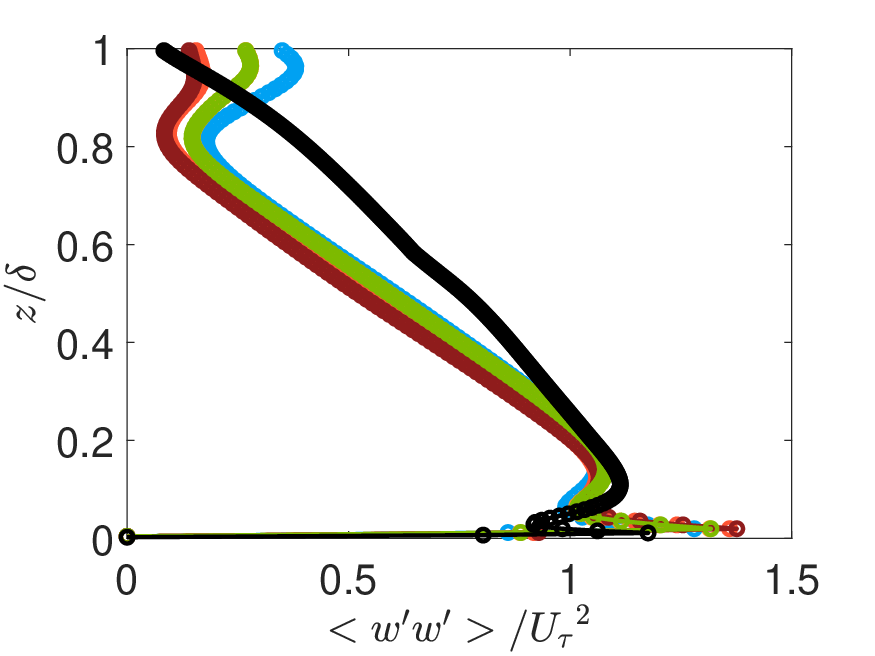}
            };
            \node[anchor=north west,
        xshift=-2mm,yshift=-2mm] at (image.north west) {{\rmfamily\fontsize{12}{13}\fontseries{l}\selectfont(c)}};
        \end{tikzpicture}}
    \subfloat[\label{fig6d}]{
        \begin{tikzpicture}
            \node[anchor=north west, inner sep=0] (image) at (0,0) {      \includegraphics[width=0.48\textwidth]{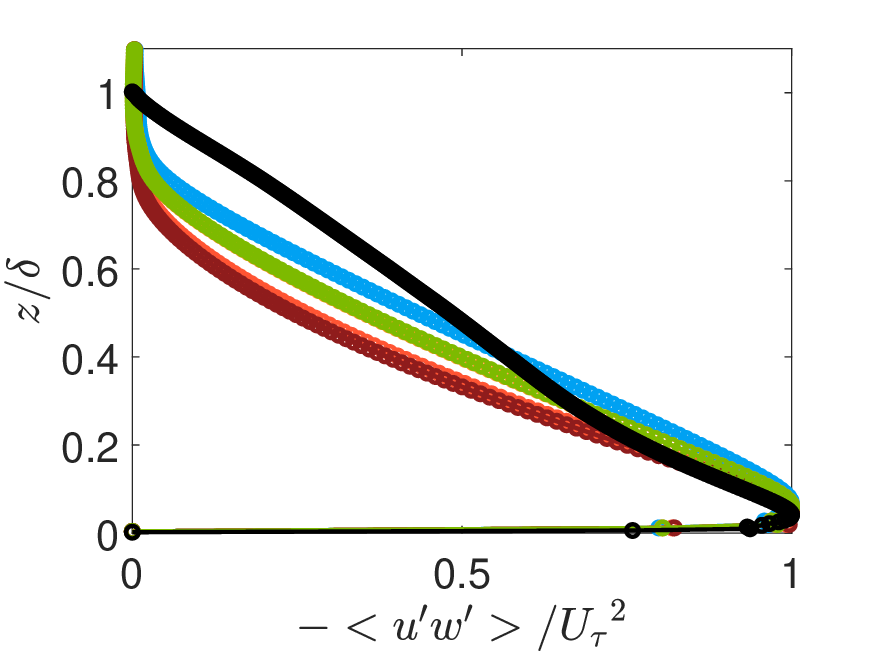}
            };
            \node[anchor=north west,
        xshift=-2mm,yshift=-2mm] at (image.north west) {{\rmfamily\fontsize{12}{13}\fontseries{l}\selectfont(d)}};
        \end{tikzpicture}}
         \vfill    
    \caption{Comparison of (a) the longitudinal velocity variance, (b) the lateral velocity variance, and (c) the vertical velocity variance, {and (d) the longitudinal Reynolds shear stress} in the CNBLs at different latitudes and the TNBL.}
    \label{fig6}
\end{figure}

\begin{figure}
    \centering
  \subfloat[\label{u2}]{
        \begin{tikzpicture}
            \node[anchor=north west, inner sep=0] (image) at (0,0) {
    \includegraphics[width=0.45\textwidth]{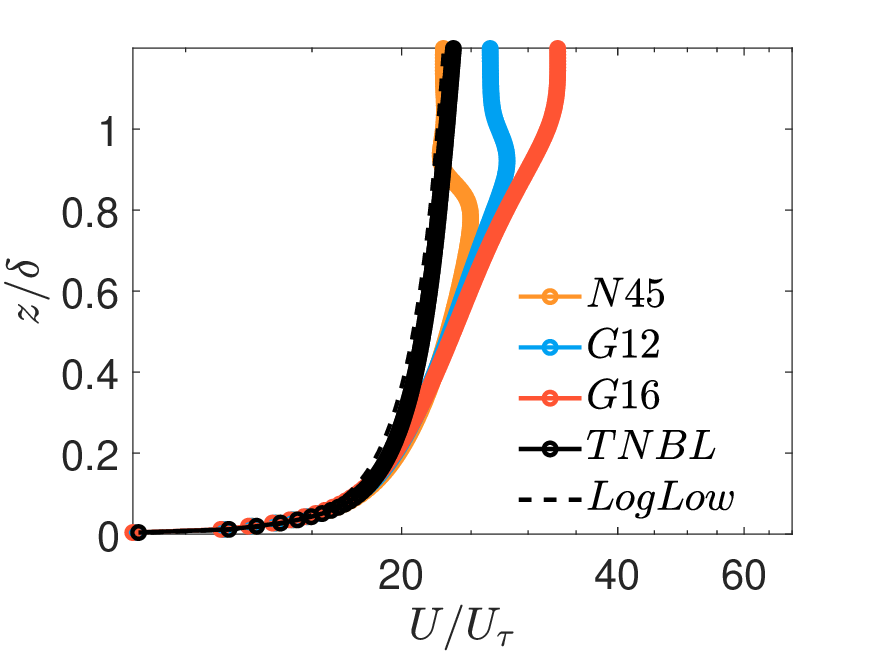}
            };
            \node[anchor=north west,
        xshift=-2mm,yshift=-2mm] at (image.north west) {{\rmfamily\fontsize{12}{13}\fontseries{l}\selectfont(a)}};
        \end{tikzpicture}}
    \hfill
     \subfloat[\label{v2}]{
        \begin{tikzpicture}
            \node[anchor=north west, inner sep=0] (image) at (0,0) {
    \includegraphics[width=0.45\textwidth]{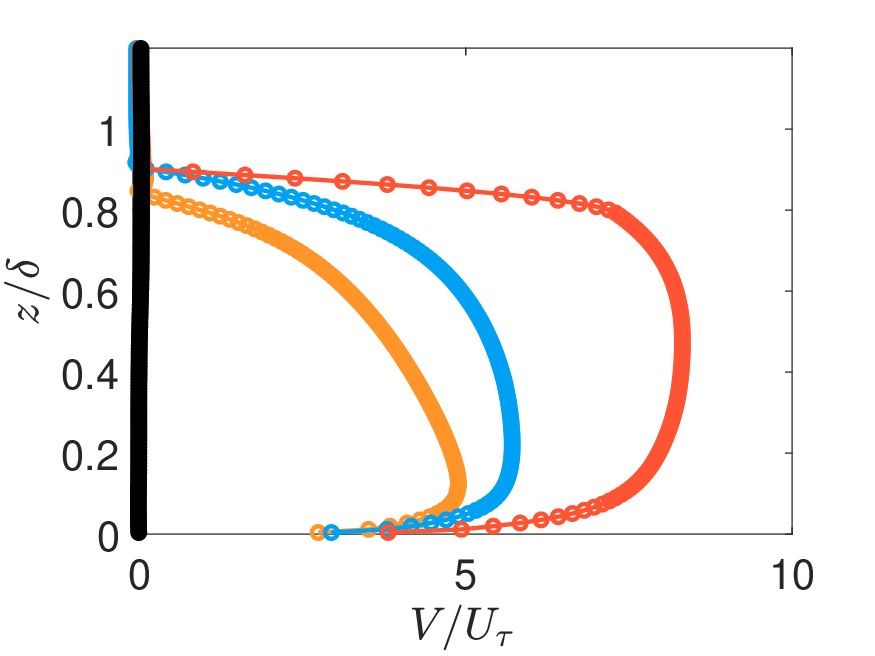}
            };
            \node[anchor=north west,
        xshift=-2mm,yshift=-2mm] at (image.north west) {{\rmfamily\fontsize{12}{13}\fontseries{l}\selectfont(b)}};
        \end{tikzpicture}}
    \vfill
     \vspace{-0.8cm}  
     \subfloat[\label{resultant2}]{
        \begin{tikzpicture}
            \node[anchor=north west, inner sep=0] (image) at (0,0) {      \includegraphics[width=0.45\textwidth]{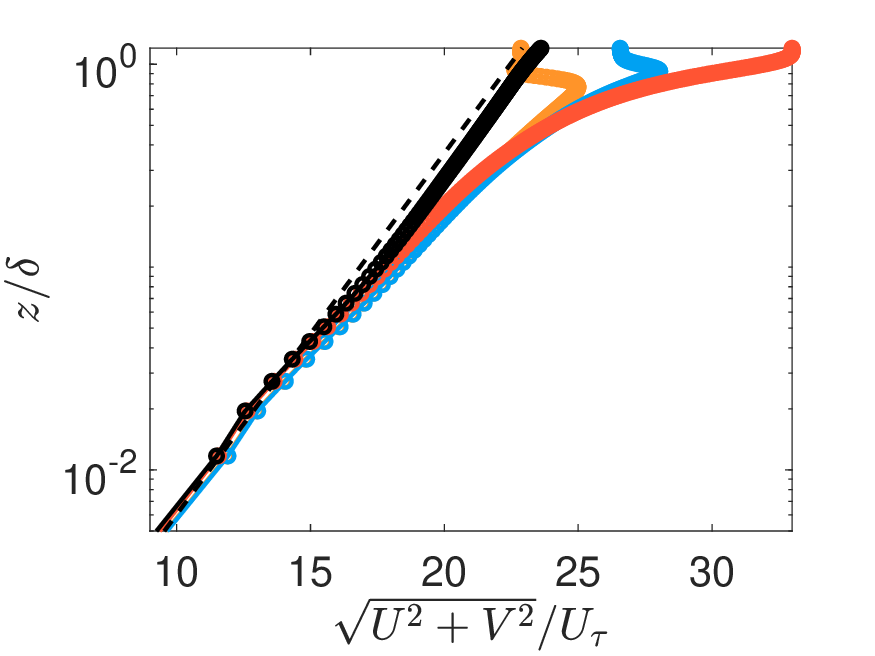}
            };
            \node[anchor=north west,
        xshift=-2mm,yshift=-2mm] at (image.north west) {{\rmfamily\fontsize{12}{13}\fontseries{l}\selectfont(c)}};
        \end{tikzpicture}}        
    \hfill
    \subfloat[\label{spiral2}]{
        \begin{tikzpicture}
            \node[anchor=north west, inner sep=0] (image) at (0,0) {
    \includegraphics[width=0.45\textwidth]{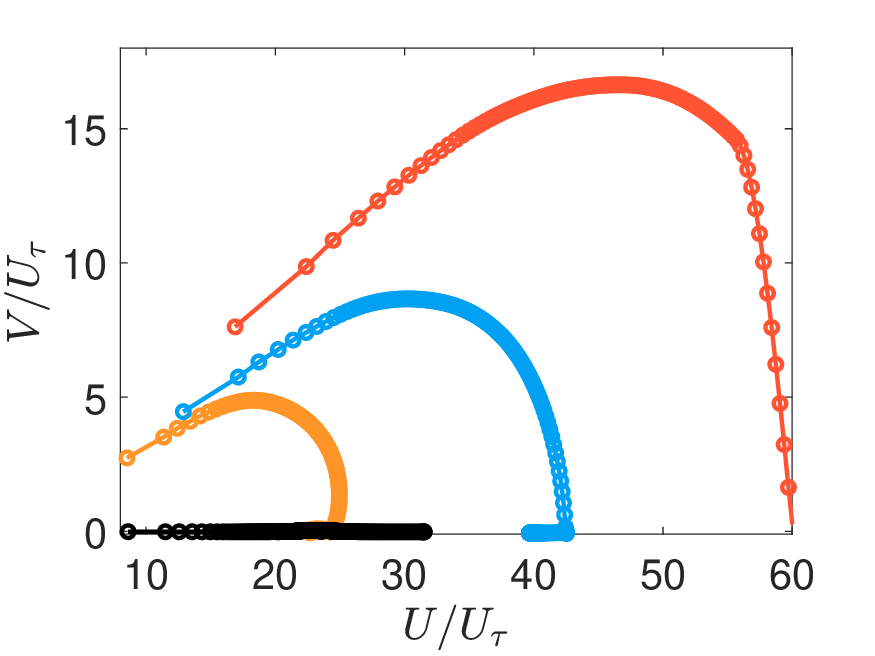}
            };
            \node[anchor=north west,
        xshift=-2mm,yshift=-2mm] at (image.north west) {{\rmfamily\fontsize{12}{13}\fontseries{l}\selectfont(d)}};
        \end{tikzpicture}}
        \vfill
        \vspace{-0.8cm} 
    \subfloat[\label{deflection2}]{
        \begin{tikzpicture}
            \node[anchor=north west, inner sep=0] (image) at (0,0) {      \includegraphics[width=0.45\textwidth]{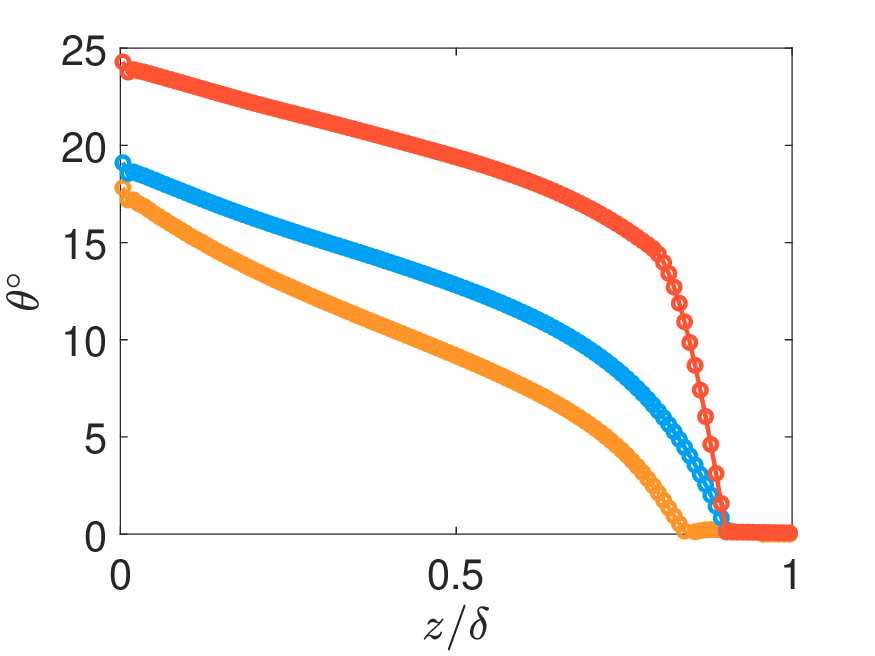}
            };
            \node[anchor=north west,
        xshift=-2mm,yshift=-2mm] at (image.north west) {{\rmfamily\fontsize{12}{13}\fontseries{l}\selectfont(e)}};
        \end{tikzpicture}}
    \caption{Comparisons of (a) the mean longitudinal wind speeds, (b) the mean lateral wind speeds, (c) the mean total wind speeds, (d) the Ekman spirals, and (e) the wind veer angles in the CNBLs with different geostrophic wind speeds and the TNBL.}
    \label{fig:mean_geo}
\end{figure}

\begin{figure}
\centering
    \subfloat[\label{uu2}]{
        \begin{tikzpicture}
        \node[anchor=north west, inner sep=0] (image) at (0,0) {
    \includegraphics[width=0.45\textwidth]{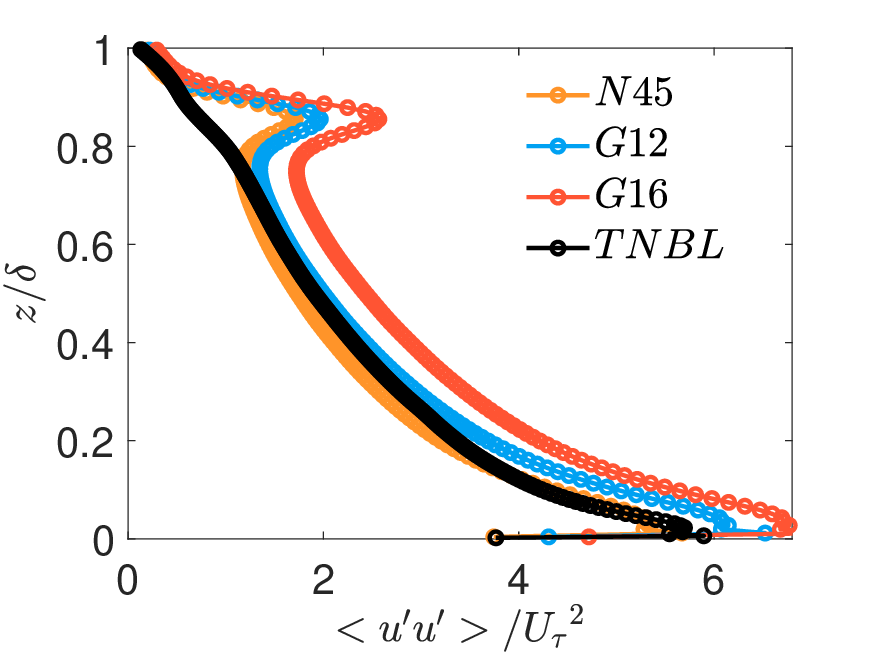}
            };
    \node[anchor=north west,
        xshift=-2mm,yshift=-2mm] at (image.north west) {{\rmfamily\fontsize{12}{13}\fontseries{l}\selectfont(a)}};
        \end{tikzpicture}}
   \subfloat[\label{vv2}]{
        \begin{tikzpicture}
        \node[anchor=north west, inner sep=0] (image) at (0,0) {
    \includegraphics[width=0.45\textwidth]{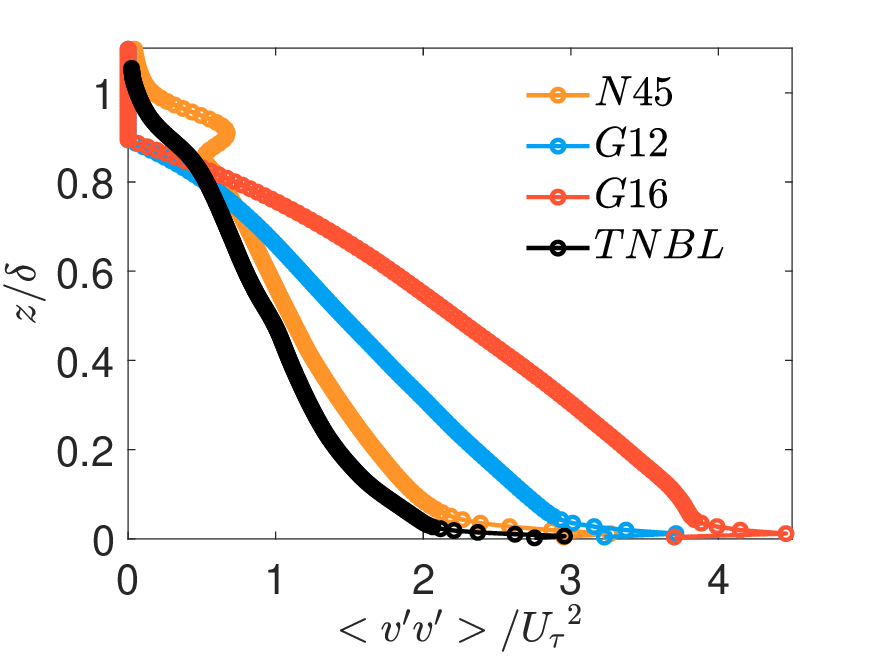}
            };
    \node[anchor=north west,
        xshift=-2mm,yshift=-2mm] at (image.north west) {{\rmfamily\fontsize{12}{13}\fontseries{l}\selectfont(b)}};
        \end{tikzpicture}}
        
        \vspace{-0.8cm} 
        \centering
   \subfloat[\label{ww2}]{
        \begin{tikzpicture}
        \node[anchor=north west, inner sep=0] (image) at (0,0) {
    \includegraphics[width=0.45\textwidth]{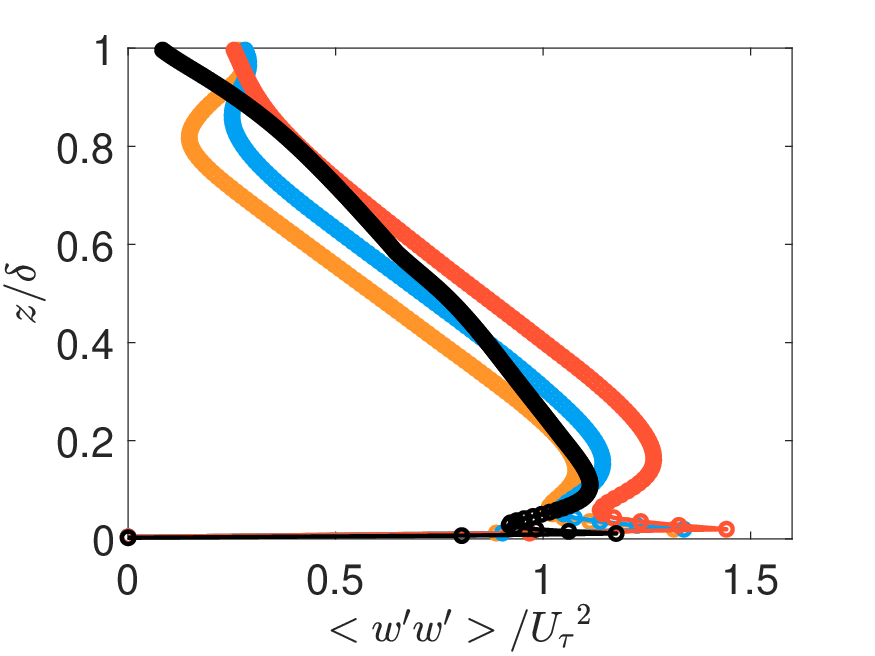}
            };
    \node[anchor=north west,
        xshift=-2mm,yshift=-2mm] at (image.north west) {{\rmfamily\fontsize{12}{13}\fontseries{l}\selectfont(c)}};
        \end{tikzpicture}}
     \subfloat[\label{taux2}]{
         \begin{tikzpicture}
         \node[anchor=north west, inner sep=0] (image) at (0,0) {
     \includegraphics[width=0.45\textwidth]{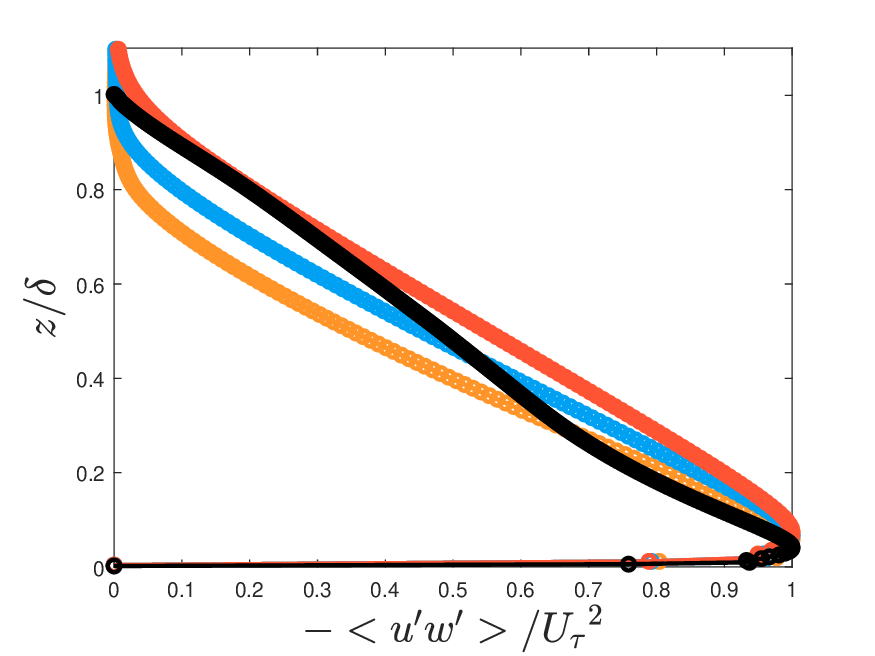}
             };
     \node[anchor=north west,
         xshift=-2mm,yshift=-2mm] at (image.north west) {{\rmfamily\fontsize{12}{13}\fontseries{l}\selectfont(d)}};
         \end{tikzpicture}}        
    \caption{Comparisons of (a) the longitudinal velocity variance, (b) the lateral velocity variance, and (c) the vertical velocity variance, {and (d) the longitudinal Reynolds shear stress} in the CNBLs with different geostrophic wind speeds and the TNBL.}
    \label{fig:flow_var_geo}
\end{figure}

Fig.~\ref{fig6} shows the profiles of wind velocity variances and Reynolds shear stress at different latitudes and the TNBL. 
Fig.~\ref{fig6} (a) displays the profiles of the longitudinal velocity variance, and one can see that there is an abrupt peak in the capping inversion layer of the CNBLs. 
Below the inversion layer, the longitudinal velocity variances in the low-to-mid-latitude CNBLs (N25, N45, and S45) and TNBL are much larger than those in the high-latitude CNBLs (N70 and N90). 
From Fig.~\ref{fig6} (b), it is seen that the lateral velocity variance in the lowest-latitude CNBL (N25) is slightly larger than those at higher latitudes, while that in the TNBL is the smallest. The lateral velocity variances in the mid- and high-latitude CNBLs are quite close to each other.
In contrast, the vertical velocity variance is the strongest in the TNBL, and those in the CNBLs are smaller and close to each other, as shown in Fig.~\ref{fig6} (c).
In summary, we found that the Coriolis force can suppress the magnitudes of the longitudinal and vertical velocity fluctuations while augmenting the lateral velocity variance in the CNBLs compared to the TNBL. 
{The possible mechanisms are discussed as follows. 
First, Fig.~\ref{fig6} (d) depicts that the Coriolis force causes a reduction in the longitudinal Reynolds shear stress, which can further induce a reduction in the longitudinal turbulence production as well as the longitudinal velocity variances.
Second, in canonical wall-bounded flows, lateral turbulence fluctuations are generated only from the longitudinal velocity component due to pressure-strain interactions. However, in CNBLs, the non-zero mean lateral velocity caused by the Coriolis force can produce additional turbulent fluctuations in the lateral direction, which may be a mechanism for the enhanced lateral velocity fluctuations observed in CNBLs.
The precise mechanisms need further quantitative study via the exact turbulent kinetic energy budget equations.}

Fig.~\ref{fig:mean_geo} displays the basic statistics of the CNBLs with different geostrophic wind speeds.
As shown in Fig.~\ref{fig:mean_geo} (a) and (c), the dimensionless mean longitudinal and total wind speeds normalized by the friction velocity basically meet the logarithmic law at $z/\delta<0.1$ in the CNBLs and the TNBL.
In the upper ABL ($z/\delta>0.6$), the mean longitudinal and total wind speeds increase with the geostrophic wind speed due to the top boundary conditions. 
The mean longitudinal wind velocity of the TNBL is the smallest.
With a larger geostrophic wind speed, the magnitude of the mean lateral wind velocity also increases, as demonstrated in Fig.~\ref{fig:mean_geo} (b). The mean lateral wind velocity of the TNBL is always zero throughout the boundary layer.
The augmented mean lateral mean wind speed in the CNBLs can be attributed to the stronger Coriolis force in the momentum equation (\ref{eq 3}). 
Consequently, the Ekman spiral is wider, and the wind veer angle is larger with a greater geostrophic wind speed, as in Fig.~\ref{fig:mean_geo} (d) and (e).

The flow velocity variances and longitudinal Reynolds shear stress in the three directions are shown in Fig.~\ref{fig:flow_var_geo}. 
It can be seen that all the flow velocity variances increase with the geostrophic wind speed. The mechanism may be the stronger turbulence production due to larger mean wind speed, Reynolds shear stress and energy transfer among the velocity components, which can be revealed through analysis of turbulence kinetic energy budgets of the CNBLs in the future. The longitudinal and lateral flow velocity variances in the TNBL are the smallest. The circumstance in the vertical velocity variance is more complex. At $z/\delta<0.1$, the vertical velocity variance in the TNBL is smaller than that in the CNBLs, while its decreasing rate is much lower at $z/\delta>0.1$, resulting in a larger magnitude in some wall-normal ranges. This is consistent with the observation in Fig.~\ref{fig6}.

\bibliography{manuscript}

\end{document}